\documentclass[pre,aps,nofootinbib,twocolumn]{revtex4-1}
\usepackage{amsmath}
\usepackage{amssymb}
\usepackage{graphicx}
\usepackage[]{natbib}
\input{tcilatex}

\begin{document}

\title{Bosonic Spectral Function and The Electron-Phonon Interaction in HTSC
Cuprates}
\author{$^{1}$E. G. Maksimov, $^{2,3}$M. L. Kuli\'{c}, $^{4}$O. V. Dolgov}

\affiliation{$^{1}$ Lebedev Physical Institute, 119991 Moscow, Russia \\
$^{2}$Institute for Theoretical Physics, Goethe-University D-60438
Frankfurt
am Main, Germany \\
$^{3}$Max-Born-Institut f\"{u}r Nichtlineare Optik und
Kurzzeitspektroskopie, 12489 Berlin,Germany\\
$^{4}$Max-Planck-Institut f\"{u}r Festk\"{o}rperphysik,70569
Stuttgart, Germany}

\begin{abstract}
In \textit{Part I} we discuss accumulating experimental evidence related to
the structure and origin of the bosonic spectral function $\alpha
^{2}F(\omega )$ in high-temperature superconducting (HTSC) cuprates \textit{%
at and near optimal doping}. Some global properties of $\alpha ^{2}F(\omega
) $, such as number and positions of peaks, are extracted by combining
optics, neutron scattering, ARPES and tunnelling measurements. These methods
give convincing evidence for strong electron-phonon interaction ($EPI$) with
$1<\lambda _{ep}\lesssim 3$ in cuprates \textit{near optimal doping}. Here
we clarify how these results are in favor of the Eliashberg-like theory for
HTSC cuprates near optimal doping.

In \textit{Part II} we discuss some theoretical ingredients - such as strong
EPI, strong correlations - which are necessary to explain the experimental
results related to the mechanism of d-wave pairing in optimally doped
cuprates. These comprise the Migdal-Eliashberg theory for EPI in strongly
correlated systems which give rise to the forward scattering peak. The
latter is further supported by the weakly screened Madelung interaction in
the ionic-metallic structure of layered cuprates. In this approach EPI is
responsible for the strength of pairing while the residual Coulomb
interaction (by including spin fluctuations) triggers the d-wave pairing.
\end{abstract}

\date{\today }
\maketitle

\part{Experimental Evidence for Strong EPI}

\section{Introduction}

In spite of an unprecedented intensive experimental and theoretical study
after the discovery of high-temperature superconductivity (HTSC) in cuprates
there is, even twenty-three years after, no consensus on the pairing
mechanism in these materials. At present there are two important
experimental facts which are not under dispute: (1) the critical temperature
$T_{c}$ in cuprates is high, with the maximum $T_{c}^{\max }\sim 160$ $K$ in
the Hg-1223 compounds; (2) the pairing in cuprates is d-wave like, i.e. $%
\Delta (\mathbf{k},\omega )\approx \Delta _{s}(k,\omega )+\Delta _{d}(\omega
)(\cos k_{x}-\cos k_{y})$ with $\Delta _{s}<0.1\Delta _{d}$. On the contrary
there is a dispute concerning the scattering mechanism which governs normal
state properties and pairing in cuprates. To this end, we stress that in the
HTSC cuprates, a number of properties can be satisfactorily explained by
assuming that the quasi-particle dynamics is governed by some electron-boson
scattering and in the superconducting state bosonic quasi-particles are
responsible for Cooper pairing. Which bosonic quasi-particles are dominating
in the cuprates is the subject which will be discussed in this work. It is
known that the electron-boson (phonon) scattering is well described by the
Migdal-Eliashberg theory if the adiabatic parameter $A\equiv $ $\alpha \cdot
\lambda (\omega _{B}/W_{b})$ fulfills the condition $A\ll 1$, where $\lambda
$ is the electron-boson coupling constant, $\omega _{B}$ is the
characteristic bosonic energy and $W_{b}$ is the electronic band width and $%
\alpha $ depends on numerical approximations \citep{Migdal}. The
important characteristic of the electron-boson scattering is the
Eliashberg spectral function $\alpha
^{2}F(\mathbf{k},\mathbf{k}^{\prime },\omega )$ (or its average
$\alpha ^{2}F(\omega )$) which characterizes scattering of
quasi-particle from $\mathbf{k}$ to $\mathbf{k}^{\prime }$ by
exchanging bosonic energy $\omega $. Therefore, in systems with
electron-boson scattering the knowledge of this function is of
crucial importance. There are at least two approaches differing in
assumed pairing bosons in the HTSC cuprates. The \textit{first
one} is based on the electron-phonon interaction
(EPI), with the main proponents in \citep{MaksimovReview}, \citep{KulicReview}%
, \citep{Falter}, \citep{Alexandrov},
\citep{GunnarssonReview2008}, where mediating bosons are
\textit{phonons }and where the average spectral
function $\alpha ^{2}F(\omega )$ is similar to the phonon density of states $%
F_{ph}(\omega )$. Note, $\alpha ^{2}F(\omega )$ is not the product of two
functions although sometimes one defines the function $\alpha ^{2}(\omega
)=\alpha ^{2}F(\omega )/F(\omega )$ which should approximate the energy
dependence of the strength of the EPI coupling. There are numerous
experimental evidence in cuprates for the importance of the EPI scattering
mechanism with a rather large coupling constant in the normal scattering
channel $1<\lambda _{ep}\lesssim 3$, which will be discussed in detail
below. In the EPI approach $\alpha ^{2}F_{ph}(\omega )$ is extracted from
tunnelling measurements in conjunction with IR optical measurements. We
stress again that the Migdal-Eliashberg theory is well justified framework
for EPI since in most superconductors the condition $A<1$ is fulfilled. The
HTSC cuprates are on the borderline and it is a natural question - under
which condition can high T$_{c}$ be realized in the non-adiabatic limit $%
A\sim 1$? The \textit{second approach} \citep{Pines} assumes that
EPI is too weak to be responsible for high $T_{c}$ in cuprates and
it is based on a phenomenological model for spin-fluctuation
interaction ($SFI$) as the dominating scattering mechanism, i.e.
it is a non-phononic mechanism. In this (phenomenological)
approach the spectral function is proportional to
the imaginary part of the spin susceptibility $Im\chi (\mathbf{k}-\mathbf{k}%
^{\prime },\omega )$, i.e. $\alpha ^{2}F(\mathbf{k},\mathbf{k}^{\prime
},\omega )\sim Im\chi (\mathbf{k}-\mathbf{k}^{\prime },\omega )$. NMR
spectroscopy and magnetic neutron scattering give evidence that in HTSC
cuprates $\chi (\mathbf{q},\omega )$ is peaked at the antiferromagnetic wave
vector $Q=(\pi /a,\pi /a)$ and this property is very favorable for d-wave
pairing. The $SFI$ theory roots basically on the strong electronic repulsion
on Cu atoms, which is usually studied by the Hubbard model or its (more
popular) derivative the t-J model. Regarding the possibility to explain high
T$_{c}$ \textit{solely by strong correlations}, as it is reviewed in \citep%
{PatrickLee}, we stress two facts. First, at present there is no
viable theory as well as experimental facts which can justify
these (non-phononic) mechanisms of pairing with some exotic
pairing mechanism such as RVB pairing \citep{PatrickLee},
fractional statistics and anyon superconductivity, etc. Therefore
we shall not discuss these, in theoretical sense, very interesting
approaches. Second, the central question in these non-phononic
approaches is - do models based solely on the Hubbard Hamiltonian
show up superconductivity at sufficiently high critical
temperatures ($T_{c}\sim 100$ $K$) ? Although the answer on this
important question is not definitely settled there are a number of
\textit{numerical studies} of these models which offer rather
convincing negative answers. For instance, the sign-free
variational Monte Carlo algorithm in the 2D repulsive ($U>0$)
Hubbard model gives \textit{no evidence for superconductivity with
high T}$_{c}$, neither the BCS- nor
Berezinskii-Kosterlitz-Thouless (BKT)-like \citep{ImadaMC}. At the
same time, similar calculations show that there is a strong
tendency to superconductivity in the attractive ($U<0$) Hubbard
model for the same strength of $U$, i.e. at finite temperature in
the 2D model with $U<0$ the BKT superconducting transition is
favored. Concerning the possibility of HTSC in the $t-J$ model,
various numerical calculations such as Monte Carlo calculations of
the Drude spectral weight \citep{ScalapinoDrudeWeight} and high
temperature expansion for the pairing susceptibility
\citep{Pryadko} have shown that there is no superconductivity at
temperatures characteristic for cuprates and if it exists $T_{c}$
must be rather low - few Kelvins. These numerical results tell us
that the lack of high $T_{c}$\ (even in $2D$
BKT phase) in the repulsive ($U>0$)\ single-band Hubbard model and in the $%
t-J$ model is not only due to thermodynamical $2D$-fluctuations (which at
finite T suppress and destroy superconducting phase coherence in large
systems) but it is mostly due to an \textit{inherent ineffectiveness of
strong correlations to produce solely high }$T_{c}$ \textit{in cuprates}.
These numerical results signal that the simple single-band Hubbard and its
derivative the t-J model are insufficient to explain solely the pairing
mechanism in cuprates and some additional ingredients must be included.

Since $EPI$ is rather strong in cuprates, then it must be
accounted for. As it will be argued in the following, the
experimental support for the importance of EPI in cuprates comes
from optics, tunnelling, and recent ARPES measurements
\citep{ShenReview}. It is worth mentioning that recent ARPES
activity was a strong impetus for renewed experimental and
theoretical studies of EPI in cuprates. However, in spite of
accumulating experimental evidence for importance of EPI with
$\lambda _{ep}>1$, there are occasionally reports which doubt its
importance in cuprates. This is the case with recent
interpretation of some optical measurements in terms of SFI only
\citep{Carbotte}, \citep{HwangTimusk1}, \citep{HwangTimusk2} and
with LDA-DFT (local density approximation-density functional
theory) band structure calculations \citep{BohnenCohen},
\citep{Giuistino}, where both claim that EPI is negligibly small,
i.e. $\lambda _{ep}<0.3$. The inappropriateness of these
calculations will be discussed in the following Sections.

The paper is organized as follows. In \textit{Part I} we will mainly discuss
experimental results in \textit{cuprates at and near optimal doping} by
giving also minimal theoretical explanations which are related to the\textit{%
\ bosonic spectral function} $\alpha ^{2}F(\omega )$ as well to the
transport spectral function $\alpha _{tr}^{2}F(\omega )$ and their relations
to EPI. The reason that we study only cuprates at and near optimal doping
is, that in these systems there are rather well defined quasi-particles -
although strongly interacting, while in highly underdoped systems the
superconductivity is perplexed and possibly masked by other phenomena, such
as pseudogap effects, formation of small polarons, interaction with spin and
(possibly charge) order parameters, pronounced inhomogeneities of the
scattering centers, etc. As ARPES experiments confirm there are no polaronic
effects in systems at and near optimal doping, while there are pronounced
polaronic effects due to EPI in undoped and very underdoped HTSC \citep%
{Alexandrov}, \citep{GunnarssonReview2008}. In this work we
consider mainly those direct one-particle and two-particles probes
of low energy quasi-particle excitations and scattering rates
which give information on
the structure of the spectral functions $\alpha ^{2}F(\mathbf{k},\mathbf{k}%
^{\prime },\omega )$ and $\alpha _{tr}^{2}F(\omega )$ in systems near
optimal doping. These are angle-resolved photoemission ($ARPES$), various
arts of tunnelling spectroscopy such as superconductor/insulator/ normal
metal ($SIN$) junctions and break junctions, scanning-tunnelling microscope
spectroscopy ($STM$), infrared ($IR$) and Raman optics, inelastic neutron
and x-ray scattering, etc. We shall argue that these direct probes give
evidence for a rather strong EPI in cuprates. Some other experiments on EPI
are also discussed in order to complete the arguments for the importance of
EPI in cuprates. The detailed contents of Part I is the following. In
\textit{Section II} we discuss some prejudices related to the strength of $%
EPI$ as well as on the Fermi-liquid behavior of HTSC cuprates. We argue that
any non-phononic mechanism of pairing should have very large bare critical
temperature $T_{c0}\gg T_{c}$ in the presence of the large EPI coupling
constant, $\lambda _{ep}\geq 1$, if the EPI spectral function is weakly
momentum dependent, i.e. if $\alpha ^{2}F(\mathbf{k},\mathbf{k}^{\prime
},\omega )\approx \alpha ^{2}F(\omega )$ like in low temperature
superconductors. The fact that EPI is large in the normal state of cuprates
and the condition that it must conform with d-wave pairing implies
inevitably that EPI in cuprates must be \textit{strongly momentum dependent}%
. In \textit{Section III} we discuss \textit{direct and indirect
experimental evidence for the importance of EPI} in cuprates and for the
weakness of SFI in cuprates. These are:

(\textbf{A}) \textit{Magnetic neutron scattering} \textit{measurements - }%
These measurements provide dynamic spin susceptibility $\chi (\mathbf{q}%
,\omega )$ which is in the $SFI$ \textit{phenomenological approach} \citep%
{Pines} related to the Eliashberg spectral function, i.e. $\alpha ^{2}F_{sf}(%
\mathbf{k},\mathbf{k}^{\prime },\omega )\sim g_{sf}^{2}Im\chi (\mathbf{q}=%
\mathbf{k}-\mathbf{k}^{\prime },\omega )$. We stress that such an approach
can be theoretically justified only in the weak coupling limit, $g_{sf}\ll
W_{b}$, where $W_{b}$ is the band width and $g_{sf}$ is the phenomenological
SFI coupling constant. Here we discuss experimental results on YBCO which
give evidence for strong rearrangement (with respect to $\omega $) of $%
Im\chi (\mathbf{q},\omega )$ (with $\mathbf{q}$ at and near $\mathbf{Q}=(\pi
,\pi )$) by doping toward the optimal doped HTSC \citep{Bourges}, \citep%
{ReznikNewIMNS}. It turns out that in the optimally doped cuprates with $%
T_{c}=92.5$ $K$ $Im\chi (\mathbf{Q},\omega )$ is \textit{drastically
suppressed} compared to that in slightly underdoped ones with $T_{c}=91$ $K$%
. This fact implies that the SFI coupling constant $g_{sf}$ must be small.

\textbf{(\textit{B})} \textit{Optical conductivity} \textit{measurements} -
From these measurements one can extract the transport relaxation rate $%
\gamma _{tr}(\omega )$ and indirectly an approximative shape of the
transport spectral function $\alpha _{tr}^{2}F(\omega )$. In the case of
systems near optimal doping we discuss the following questions: (i) the
physical and quantitative difference between the optical relaxation rate $%
\gamma _{tr}(\omega )$ and the quasi-particle relaxation rate $\gamma
(\omega )$. It was shown in the past that by equating these two (unequal)
quantities is dangerous and brings incorrect results concerning the
quasi-particle dynamics in most metals by including HTSC cuprates too \citep%
{MaksimovReview}, \citep{KulicReview}, \citep{Allen},
\citep{DolgovShulga}, \citep{Shulga}, \citep{KulicAIP}; (ii)
methods of extraction of the transport spectral function $\alpha
_{tr}^{2}F(\omega )$. Although these methods give at finite
temperature $T$ a blurred $\alpha _{tr}^{2}F(\omega )$ which is
(due to the ill-defined methods) temperature dependent, it turns
out that the width and the shape of the extracted $\alpha
_{tr}^{2}F(\omega )$ are in favor of $EPI$; (iii) the restricted
sum-rule for the optical weight as a function of $T$ which can be
explained by strong $EPI$ \citep{MaksKarakoz1},
\citep{MaksKarakoz2}; (iv) good agreement with experiments of the $T$%
-dependence of the resistivity $\rho (T)$ in optimally doped YBCO, where $%
\rho (T)$ is calculated by using the spectral function from tunnelling
experiments. Recent femtosecond time-resolved optical spectroscopy in $%
La_{2-x}Sr_{x}CuO_{4}$ which gives additional evidence for
importance of EPI \citep{Kusar2008} will be shortly discussed.

\textbf{(\textit{C})} \textit{ARPES} \textit{measurements and EPI} - From
these measurements the self-energy $\Sigma (\mathbf{k},\omega )$ is
extracted as well as some properties of $\alpha ^{2}F(\mathbf{k},\mathbf{k}%
^{\prime },\omega )$. Here we discuss the following items: (i) existence of
the nodal and anti-nodal kinks in optimally and slightly underdoped
cuprates, as well as the structure of the ARPES self-energy ($\Sigma (%
\mathbf{k},\omega )$) and its isotope dependence, which are all due to EPI;
(ii) appearance of different slopes of $\Sigma (\mathbf{k},\omega )$ at low (%
$\omega \ll \omega _{ph}$) and high energies ($\omega \gg \omega _{ph}$)
which can be explained by strong EPI; (iii) formation of small polarons in
the undoped HTSC was interpreted to be due to strong EPI - this gives rise
to phonon side bands which are clearly seen in ARPES of undoped HTSC \citep%
{GunnarssonReview2008}.

(\textbf{D}) \textit{Tunnelling spectroscopy - }It is well known that this
method is of an immense importance in obtaining the spectral function $%
\alpha ^{2}F(\omega )$ from tunnelling conductance. In this part
we discuss the following items: (i) the extracted Eliashberg
spectral function $\alpha ^{2}F(\omega )$ with the coupling
constant $\lambda ^{(tun)}=2-3.5$ from the tunnelling conductance
of break-junctions in optimally doped YBCO and Bi-2212
\citep{TunnelingVedeneev}-\citep{PonomarevTunnel} which gives that
the maxima of $\alpha ^{2}F(\omega )$ coincide with the maxima in
the phonon density of states $F_{ph}(\omega )$; (ii) existence of
\textit{eleven peaks} in $-d^{2}I/dV^{2}$ in superconducting
$La_{1.84}Sr_{0.16}CuO_{4}$ films \citep{Chaudhari}, where these
peaks match precisely with the peaks in the intensity of the
existing phonon Raman scattering data \citep{SugaiRaman}; (iii)
the presence of the dip in dI/dV in STM which shows the pronounced
oxygen isotope effect and important role of these phonons.

\textbf{(\textit{E})} \textit{Inelastic} \textit{neutron and x-ray scattering%
} \textit{measurements} - From these experiments one can extract the phonon
density of state $F_{ph}(\omega )$ and in some cases strengths of the
quasi-particle coupling with various phonon modes. These experiments give
sufficient evidence for quantitative inadequacy of LDA-DFT calculations in
HTSC cuprates. Here we argue, that the \textit{large softening and broadening%
} of the half-breathing $Cu-O$ bond-stretching phonon, of apical oxygen
phonons and of oxygen $B_{1g}$ buckling phonons (in LSCO, BISCO,YBCO) can
not be explained by LDA-DFT. It is curious that the magnitude of the
softening can be partially obtained by LDA-DFT but the calculated widths of
some important modes are an order of magnitude smaller than the neutron
scattering data show. This remarkable fact confirms additionally \textit{the}
\textit{inadequacy of LDA-DFT in strongly correlated systems} and a more
sophisticated many body theory for EPI is needed. The problem of EPI will be
discussed in more details in \textit{Part II }.

In \textit{Section IV} brief summary of the \textit{Part I} is given. Since
\textit{we are dealing with electron-boson scattering in cuprate near
optimal doping}, then in \textit{Appendix} \textit{A} (and in Part II) we
introduce the reader briefly into the Migdal-Eliashberg theory for
superconductors (and normal metals) where the quasi-particle spectral
function $\alpha ^{2}F(\mathbf{k},\mathbf{k}^{\prime },\omega )$ and the
transport spectral function $\alpha _{tr}^{2}F(\omega )$ are defined.

Finally, one can pose a question - do the experimental results of the above
enumerated spectroscopic methods allow a building of a satisfactory and
physically reasonable microscopic theory for basic scattering and pairing
mechanism in cuprates? The posed question is very modest compared with a
much stringent request for the \textit{theory of everything} - which would
be able to explain all properties of HTSC materials. Such an ambitious
project is not realized even in those low-temperature conventional
superconductors where it is definitely proved that in most materials the
pairing is due to EPI and many properties are well accounted for by the
Migdal-Eliashberg theory. Let us mention only two examples: - First, the
experimental value for the coherence peak in the microwave response $\sigma
_{s}(T<T_{c},\omega =const)$ at $\omega =17$ $GHz$ in the superconducting $%
Nb $ is much higher than the theoretical value obtained by the
strong coupling Eliashberg theory \citep{Marsiglio1994}. So to
say, the theory explains the coherence peak at $17$ $GHz$ in $Nb$
qualitatively but not quantitatively. However, the measurements at
higher frequency $\omega \sim 60 $ $GHz$ are in agreement with the
Eliashberg theory \citep{Klein1994}. Then one can say that instead
of the theory of everything we deal with a satisfactory theory,
which allows us qualitative and in many aspects quantitative
explanation of phenomena in superconducting state. - Second
example is the experimental boron (B) isotope effect in $MgB_{2}$ ($%
T_{c}\approx 40$ $K$) which is much smaller than the theoretical
value, i.e. $\alpha _{B}^{\exp }\approx 0.3<\alpha _{B}^{th}=0.5$,
although the pairing is due solely by EPI for boron vibrations
\citep{MgB2Isotop}. Since the theory of everything is impossible
in the complex materials such as HTSC cuprates in \textit{Part I}
we shall not discuss those phenomena which need much more
microscopic details and/or more sophisticated many-body theory.
These are selected by chance: (\textit{i}) large ratio $2\Delta
/T_{c}$ which is on optimally doped YBCO and BISCO $\approx 5$ and
$7$, respectively, while in underdoped BISCO one has even
$(2\Delta /T_{c})\approx 20$; (\textit{ii}) peculiarities of the
coherence peak in the microwave response $\sigma (T)$ in HTSC
cuprates, which is peaked at $T$ much smaller than $T_{c}$,
contrary to the case of LTSC where it occurs near $T_{c}$;
(\textit{iii}) the dependence of $T_{c}$ on the number of
$CuO_{2}$ in the unit cell; (\textit{iv}) temperature dependence
of the Hall coefficient; (\textit{v}) distribution of states in
the vortex core, etc.

The microscopic theory of the mechanism for superconducting pairing in HTSC
cuprates will be discussed in \textit{Part II. }In Section V we introduce an
\textit{ab initio many-body theory} of superconductivity which is based on
the fundamental (microscopic) Hamiltonian and the many-body technique. This
theory can in principle calculate measurable properties of materials such as
the critical temperature $T_{c}$, critical fields, dynamic and transport
properties, etc. However, although this method is in principle exact, which
needs only some fundamental constants $e,\hbar ,m_{e},M_{ion},k_{B\text{ }}$
and the chemical composition of superconducting materials, it was
practically never realized in practice due to the complexity of many-body
interactions - electron-electron and electron-lattice, as well as of
structural properties. Fortunately, the problem can be simplified by using
the fact that superconductivity is a low-energy phenomenon characterized by
\textit{very small energy parameters} $(T_{c}/E_{F},$ $\Delta /E_{F},$ $%
\omega _{ph}/E_{F})\ll 1$. It turns out, that one can integrate high-energy
electronic processes (which are not changed by the appearance of
superconductivity) and then solve the\textit{\ low-energy problem} by the
(so called) strong-coupling Migdal-Eliashberg theory. It turns out that in
such an approach the physics is separated into: (1) solving an \textit{ideal
band-structure Hamiltonian with the nonlocal exact crystal potential}
(sometimes called excitation potential) $V_{IBS}(\mathbf{r},\mathbf{r}%
^{\prime })$ ($IBS$ - \textit{ideal band structure}) which includes the
static self-energy ($\Sigma _{c0}^{(h)}(\mathbf{r},\mathbf{r}^{\prime
},\omega =0)$) due to high-energy electronic processes, i.e. $V_{IBS}(%
\mathbf{r},\mathbf{r}^{\prime })=[V_{e-i}(\mathbf{r})+V_{H}(\mathbf{r}%
)]\delta (\mathbf{r}-\mathbf{r}^{\prime })+\Sigma _{c0}^{(h)}(\mathbf{r},%
\mathbf{r}^{\prime },\omega =0)$, with $V_{e-i}$, $V_{H}$ the electron-ion
and Hartree potential, respectively; (2) solving the low-energy Eliashberg
equations. However, the calculation of the (excited) potential $V_{IBS}(%
\mathbf{r},\mathbf{r}^{\prime })$ and the real EPI coupling $g_{ep}(\mathbf{r%
},\mathbf{r}^{\prime })=\delta V_{IBS}(\mathbf{r},\mathbf{r}^{\prime
})/\delta \mathbf{R}_{n}$, which include high-energy many-body electronic
processes - for instance large Hubbard U effects, is extremely difficult at
present, especially in strongly correlated systems such as HTSC cuprates.
Due to this difficulty the calculations of the EPI coupling in the past was
usually based on the LDA-DFT method which will be discussed in Section VI in
the contest of HTSC cuprates, where the nonlocal potential is replaced by
the \textit{local potential} $V_{LDA}(\mathbf{r})$ - the ground state
potential, and the real EPI coupling by the "local" LDA one $g_{ep}(\mathbf{r%
})=\delta V_{LDA}(\mathbf{r})/\delta \mathbf{R}_{n}$. Since the
exchange-correlation effects enter $V_{LDA}(\mathbf{r})=V_{e-i}(\mathbf{r}%
)+V_{H}(\mathbf{r})+V_{XC}(\mathbf{r})$ via the local exchange-correlation
potential $V_{XC}(\mathbf{r})$ it is clear that the LDA-DFT method describes
strong correlations scarcely and it is inadequate in HTSC cuprates (and
other strongly correlated systems such as heavy fermions) wher one needs an
approach beyond the LDA-DFT method. In Section VII we discuss a \textit{%
minimal theoretical model} for HTSC cuprates which takes into account a
minimal number of electronic orbitals and strong correlations in a
controllable manner \citep{KulicReview}. This theory treats the \textit{%
interplay of EPI and strong correlations }in systems with finite doping in a
systematic and controllable way. The minimal model can be further reduced in
some parameter range to the single-band $t-J$ model, which allows the
approximative calculation of the excited potential $V_{IBS}(\mathbf{r},%
\mathbf{r}^{\prime })$ and the non-local EPI coupling $g_{ep}(\mathbf{r},%
\mathbf{r}^{\prime })$. As a result one obtains the momentum dependent EPI
coupling $g_{ep}(\mathbf{k}_{F},\mathbf{q})$ which is for small hole-doping (%
$\delta <0.3$) strongly peaked at small transfer momenta (more precisely at $%
q=0$) - the \textit{forward scattering peak}. In the framework of this
minimal model it is possible to explain some important properties and
resolve some puzzling experimental results, for instance: \textbf{(}\textit{%
a)} Why is d-wave pairing realized in the presence of strong EPI? \textbf{(}%
\textit{b}\textbf{)} Why is the transport coupling constant ($\lambda _{tr}$%
) rather smaller than the pairing one $\lambda $, i.e. $\lambda
_{tr}\lesssim \lambda /3$? (\textit{c}) Why is the mean-field (one-body)
LDA-DFT approach unable to give reliable values for the EPI coupling
constant in cuprates and how many-body effects help; (\textit{d}) Why is
d-wave pairing robust in the presence of non-magnetic impurities and
defects? (e) Why are the ARPES nodal and antinodal kinks differently
renormalized in the superconducting states, etc? In spite of the encouraging
successes of this minimal model, at least in a qualitative explanation of
numerous important properties of HTSC cuprates, we are still at present
stage far from a fully microscopic theory of HTSC cuprates which is able to
explain high $T_{c}$. In that respect at the and of Section VII we discuss
possible improvements of the present minimal model in order to obtain at
least a semi-quantitative theory for HTSC cuprates.

Finally, we would like to point out that in real HTSC materials
there are numerous experimental evidence for nanoscale
inhomogeneities. For instance recent STM experiments show rather
large gap dispersion at least on the surface of BISCO crystals
\citep{Davis} giving rise for a pronounced
inhomogeneity of the superconducting order parameter, i.e. $\Delta (\mathbf{k%
},\mathbf{R})$ where $\mathbf{k}$ is the relative momentum of the Cooper
pair and $\mathbf{R}$ is the center of mass of Cooper pairs. One possible
reason for the inhomogeneity of $\Delta (\mathbf{k},\mathbf{R})$ and
disorder on the atomic scale can be due to extremely high doping level of $%
\sim (10-20)$ $\%$ in HTSC cuprates which is many orders of
magnitude larger than in standard semiconductors ($10^{21}$ vs
$10^{15}$ carrier concentration). There are some claims that high
$T_{c}$ is exclusively due to these inhomogeneities (of an
extrinsic or intrinsic origin) which may effectively increase
pairing potential \citep{Phillips}, while some others try to
explain high $T_{c}$ solely within the inhomogeneous Hubbard or
$t-J$ model. Here we shall not discuss this interesting problem
but mention only that the concept of T$_{c}$ increase by
inhomogeneity is ill-defined, since
the increase of $T_{c}$ is defined with respect to the average value $\bar{T}%
_{c}$. However, $\bar{T}_{c}$ is experimentally not well defined quantity
and the hypothesis of an increase of $T_{c}$ by material inhomogeneities
cannot be tested at all. In studying and analyzing \textit{HTSC cuprates
near optimal doping} we assume that basic effects are realized in nearly
homogeneous systems and inhomogeneities are of secondary role, which deserve
to be studied and discussed separately.

\section{ EPI vs non-phononic mechanisms}

Concerning the high $T_{c}$ in cuprates, two dilemmas have been dominating
after its discovery: (\textit{i}) which interaction is responsible for
strong quasi-particle scattering in the normal state - this question is
related also to the dilemma Fermi vs non-Fermi liquid; (\textit{ii}) what is
the mediating (gluing) boson responsible for the superconducting pairing,
i.e. there is a dilemma \textit{phonons or non-phonons}? In the last
twenty-three years, the scientific community was overwhelmed by numerous
proposed pairing mechanisms, most of which are hardly verifiable in HTSC\
cuprates.

\textit{1. Fermi vs non-Fermi liquid in cuprates}

After discovery of HTSC in cuprates there was a large amount of evidence on
strong scattering of quasi-particles which contradicts the canonical
(popular but narrow) definition of the Fermi liquid, thus giving rise to
numerous proposals of the so called non-Fermi liquids, such as Luttinger
liquid, RVB theory, marginal Fermi liquid, etc. In our opinion there is no
need for these radical approaches in explaining basic physics in cuprates at
least \textit{in optimally, slightly underdoped and overdoped} metallic and
superconducting HTSC cuprates. Here we give some clarifications related to
the dilemma of Fermi vs non-Fermi liquid. The definition of the \textit{%
canonical Fermi liquid} (based on the Landau work) in interacting Fermi
systems comprises the following properties: (1) there are quasi-particles
with charge $q=\pm e$, spin $s=1/2$ and low-laying energy excitations $\xi _{%
\mathbf{k}}(=\epsilon _{\mathbf{k}}-\mu )$ which are much larger than their
inverse life-times, i.e. $\xi _{\mathbf{k}}\gg 1/\tau _{\mathbf{k}}\sim \xi
_{\mathbf{k}}^{2}/W_{b}$. Since the level width $\Gamma =2/\tau _{\mathbf{k}%
} $ of the quasi-particle is negligibly small, this means that the excited
states of the Fermi liquid are placed in one-to-one correspondence with the
excited states of the free Fermi gas; (2) at $T=0$ $K$ there is an energy
level with the Fermi surface at which $\xi _{\mathbf{k}_{F}}=0$ and the
Fermi quasi-particle distribution function $n_{F}(\xi _{\mathbf{k}})$ has
finite jump at $k_{F}$; (3) the number of quasi-particles under the Fermi
surface is equal to the total number of conduction particles (we omit here
other valence and core electrons) - the Luttinger theorem; (4) the
interaction between quasi-particles are characterized with a few (Landau)
parameters which describe low-temperature thermodynamics and transport
properties. Having this definition in mind one can say that if fermionic
quasi-particles interact with some bosonic excitation, for instance with
phonons, and if the coupling is sufficiently strong, then the former are not
described by the canonical Fermi liquid since at energies and temperatures
of the order of the characteristic (Debye) temperature $k_{B}\Theta
_{D}(\equiv \hbar \omega _{D})$ (for the Debye spectrum $\sim \Theta _{D}/5$%
), i.e. for $\xi _{\mathbf{k}}\sim \Theta _{D}$ one has $\tau _{\mathbf{k}%
}^{-1}\gtrsim \xi _{\mathbf{k}}$ and the quasi-particle picture (in the
sense of the Landau definition) is broken down. In that respect an
electron-boson system can be classified as a \textit{non-canonical Fermi
liquid }for sufficiently strong electron-boson coupling. It is nowadays well
known that for instance Al, Zn are weak coupling systems since for $\xi _{%
\mathbf{k}}\sim \Theta _{D}$ one has $\tau _{\mathbf{k}}^{-1}\ll \xi _{%
\mathbf{k}}$ and they are well described by the Landau theory. However, in
(the non-canonical) cases, where for higher energies $\xi _{\mathbf{k}}\sim
\Theta _{D}$ one has $\tau _{\mathbf{k}}^{-1}\gtrsim \xi _{\mathbf{k}}$, the
electron-phonon system is satisfactory described by the \textit{%
Migdal-Eliashberg theory and the Boltzmann theory}, where thermodynamic and
transport properties depend on the spectral function $\alpha ^{2}F_{sf}(%
\mathbf{k},\mathbf{k}^{\prime },\omega )$ and its higher momenta. Since in
HTSC cuprates the electron-boson (phonon) coupling is strong and $T_{c}$ is
large, i.e. of the order of characteristic boson energies ($\omega _{B}$), $%
T_{c}\sim \omega _{B}/5$, then it is natural that in the normal state (at $%
T> $ $T_{c}$) we deal with a strong interacting non-canonical Fermi liquid
which is for modest non-adiabaticity parameter $A<1$ described by the
Migdal-Eliashberg theory, at least qualitatively and semi-quantitatively. In
order to justify this statement we shall in the following elucidate some
properties in more details by studying optical, ARPES, tunnelling and other
experiments in HTSC oxides.

\textit{2}. \textit{Is there limitation of the strength of EPI?}

In spite of reach experimental evidence in favor of strong EPI in
HTSC oxides there was a disproportion in the research activity
(especially theoretical) in the past, since the investigation of
the SFI mechanism of pairing prevailed in the literature. This
trend was partly due to an incorrect statement in \citep{Cohen} on
the possible upper limit of T$_{c}$ in the phonon mechanism of
pairing. Since in the past we have discussed this
problem thoroughly in numerous papers - for the recent one see \citep%
{MaksimovDolgov2007}, we shall outline here the main issue and results only.

It is well known that in an electron-ion crystal, besides the attractive
EPI, there is also repulsive Coulomb interaction. In case of an isotropic
and homogeneous system with weak quasi-particle interaction, the effective
potential $V_{eff}(\mathbf{k},\omega )$ in the leading approximation looks
like as for two external charges ($e$) embedded in the medium with the
\textit{total longitudinal dielectric function} $\varepsilon _{tot}(\mathbf{k%
},\omega )$ ($\mathbf{k}$ is the momentum and $\omega $ is the
frequency) \citep{Kirzhnitz}, \citep{Ginzburg}, i.e.
\begin{equation}
V_{eff}(\mathbf{k},\omega )=\frac{V_{ext}(\mathbf{k})}{\varepsilon _{tot}(%
\mathbf{k},\omega )}=\frac{4\pi e^{2}}{k^{2}\varepsilon _{tot}(\mathbf{k}%
,\omega )}.  \label{Veff}
\end{equation}%
In case of strong interaction between quasi-particles, the state of embedded
quasi-particles changes significantly due to interaction with other
quasi-particles, giving rise to $V_{eff}(\mathbf{k},\omega )\neq 4\pi
e^{2}/k^{2}\varepsilon _{tot}(\mathbf{k},\omega )$. In that case $V_{eff}$
depends on other (than $\varepsilon _{tot}(\mathbf{k},\omega )$) response
functions. However, in the case when Eq.(\ref{Veff}) holds, i. e. when the
weak-coupling limit is realized, $T_{c}$ is given by $T_{c}\approx \bar{%
\omega}\exp (-1/(\lambda _{ep}-\mu ^{\ast })$ \citep{Kirzhnitz}, \citep%
{Ginzburg}, \citep{AllenMitrovic}. Here, $\lambda _{ep}$ is the
EPI coupling constant, $\bar{\omega}$ is an average phonon
frequency and $\mu ^{\ast }$
is the Coulomb pseudo-potential, $\mu ^{\ast }=\mu /(1+\mu \ln E_{F}/\bar{%
\omega})$ ($E_{F}$ is the Fermi energy). The couplings $\lambda _{ep}$ and $%
\mu $ are expressed by $\varepsilon _{tot}(\mathbf{k},\omega =0)$
\begin{equation*}
\mu -\lambda _{ep}=\langle N(0)V_{eff}(\mathbf{k},\omega =0)\rangle
\end{equation*}%
\begin{equation}
=N(0)\int_{0}^{2k_{F}}\frac{kdk}{2k_{F}^{2}}\frac{4\pi e^{2}}{%
k^{2}\varepsilon _{tot}(\mathbf{k},\omega =0)},  \label{NVeff}
\end{equation}%
where $N(0)$ is the density of states at the Fermi surface and
$k_{F}$ is the Fermi momentum - see more in
\citep{MaksimovReview}. In \citep{Cohen} it was claimed that
lattice stability of the system with respect to the charge
density wave formation implies the condition $\varepsilon _{tot}(\mathbf{k}%
,\omega =0)>1$ for all $\mathbf{k}$. If this were correct then from Eq.(\ref%
{NVeff}) it follows that $\mu >\lambda _{ep}$, which limits the
maximal value of T$_{c}$ to the value $T_{c}^{\max }\approx
E_{F}\exp (-4-3/\lambda _{ep})$. In typical metals $E_{F}<(1-10)$
$eV$ and if one accepts the statement in \citep{Cohen} that
$\lambda _{ep}\leq \mu (\leq 0.5)$, one obtains $T_{c}\sim (1-10)$
$K$. \ The latter result, if it would be correct, means that EPI
is ineffective in producing not only high-T$_{c}$
superconductivity but also low-temperature superconductivity (LTS with $%
T_{c}\lesssim 20$ $K$). However, this result is in conflict first of all
with experimental results in LTSC, where in numerous systems one has $\mu
\leq \lambda _{ep}$ and $\lambda _{ep}>1$. For instance, $\lambda
_{ep}\approx 2.6$ is realized in $PbBi$ alloy which is definitely much
higher than $\mu (<1)$, etc.

\begin{center}
\begin{figure}[!tbp]
\resizebox{.4\textwidth}{!} {\includegraphics*[ width=6cm]{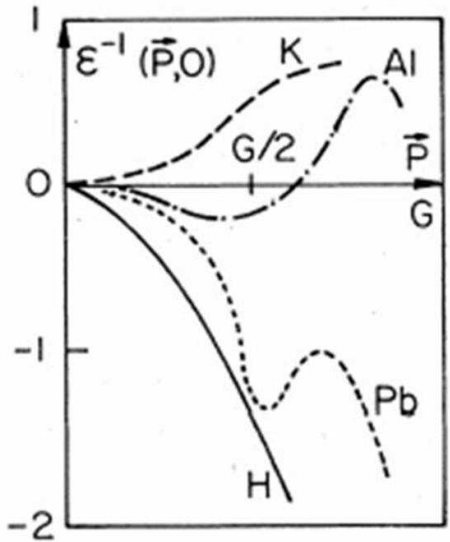}}
\caption{Inverse total static dielectric function $\protect\varepsilon^{-1}(%
\mathbf{p})$ for normal metals (K, Al, Pb) and metallic H in $\mathbf{p}%
=(1,0,0)$ direction. $\mathbf{G}$ is the reciprocal lattice vector.}
\label{Epsilon-k}
\end{figure}
\end{center}

Moreover, the basic theory tells us that $\varepsilon
_{tot}(\mathbf{k}\neq 0,\omega )$ is not the response function
\citep{Kirzhnitz}, \citep{Ginzburg} (contrary to the assumption in
\citep{Cohen}). Namely, if a small external potential $\delta
V_{ext}(\mathbf{k},\omega )$ is applied to the system (of
electrons and ions in solids) it induces screening by charges of
the medium and the total potential is given by $\delta
V_{tot}(\mathbf{k},\omega )=\delta V_{ext}(\mathbf{k},\omega
)/\varepsilon _{tot}(\mathbf{k},\omega )$ which means that
$1/\varepsilon _{tot}(\mathbf{k},\omega )$ is the response
function. The latter obeys the Kramers-Kronig dispersion relation
which implies the following stability condition \citep{Kirzhnitz},
\citep{Ginzburg}
\begin{equation}
1/\varepsilon _{tot}(\mathbf{k},\omega =0)<1\text{, }\mathbf{k}\neq 0,
\label{inv-eps}
\end{equation}%
i.e. either
\begin{equation}
\varepsilon _{tot}(\mathbf{k}\neq 0,\omega =0)>1  \label{eps-pos}
\end{equation}%
or
\begin{equation}
\varepsilon _{tot}(\mathbf{k}\neq 0,\omega =0)<0.  \label{eps-neg}
\end{equation}%
This important theorem invalidates the restriction on the maximal value of T$%
_{c}$ in the EPI mechanism given in \citep{Cohen}. We stress that
the condition $\varepsilon _{tot}(\mathbf{k}\neq 0,\omega =0)<0$
is not in conflict with the lattice stability at all. For
instance, in inhomogeneous systems such as crystal, the total
longitudinal dielectric function is
matrix in the space of reciprocal lattice vectors ($\mathbf{Q}$), i.e. $\hat{%
\varepsilon}_{tot}(\mathbf{k+Q},\mathbf{k+Q}^{\prime },\omega )$, and $%
\varepsilon _{tot}(\mathbf{k},\omega )$ is defined by $\varepsilon
_{tot}^{-1}(\mathbf{k},\omega )=\hat{\varepsilon}_{tot}^{-1}(\mathbf{k+0},%
\mathbf{k+0},\omega )$. In dense metallic systems with one ion per cell
(such as metallic hydrogen) and with the electronic dielectric function $%
\varepsilon _{el}(\mathbf{k},0)$ and the macroscopic total
dielectric function $\varepsilon _{tot}(\mathbf{k},0)$ is given by
\citep{DKM}
\begin{equation}
\varepsilon _{tot}(\mathbf{k},0)=\frac{\varepsilon _{el}(\mathbf{k},0)}{%
1-1/\varepsilon _{el}(\mathbf{k},0)G_{ep}(\mathbf{k})}.  \label{eps-tot}
\end{equation}%
At the same time the energy of the longitudinal phonon $\omega _{l}(\mathbf{k%
})$ is given by
\begin{equation}
\omega _{l}^{2}(\mathbf{k})=\frac{\Omega _{p}^{2}}{\varepsilon _{el}(\mathbf{%
k},0)}[1-\varepsilon _{el}(\mathbf{k},0)G_{ep}(\mathbf{k})],  \label{ph-fr}
\end{equation}%
where $\Omega _{p}^{2}$ is the ionic plasma frequency, $G_{ep}$ is
the local (electric) field correction - see Ref. \citep{DKM}. The
right condition for lattice stability requires that $\omega
_{l}^{2}(\mathbf{k})>0$, which implies that for $\varepsilon
_{el}(\mathbf{k},0)>0$ one has $\varepsilon
_{el}(\mathbf{k},0)G_{ep}(\mathbf{k})<1$. The latter condition
gives automatically $\varepsilon _{tot}(\mathbf{k},0)<0$.
Furthermore, the calculations \citep{DKM} show that in the
\textit{metallic hydrogen} (H)\ crystal, $\varepsilon
_{tot}(\mathbf{k},0)<0$ for all $\mathbf{k\neq 0}$.
Note, that in metallic H the EPI coupling constant is very large, i.e $%
\lambda _{ep}\approx 7$ and T$_{c}$ may reach very large value
$T_{c}\approx 600$ $K$ \citep{SavrasMaksH}. Moreover, the analyzes
of crystals with more
ions per unit cell \citep{DKM} gives that $\varepsilon _{tot}(\mathbf{k\neq 0}%
,0)<0$ is \textit{more a rule than an exception }- see Fig.~\ref{Epsilon-k}.
The physical reason for $\varepsilon _{tot}(\mathbf{k\neq 0},0)<0$ are local
field effects described by $G_{ep}(\mathbf{k})$. Whenever the local electric
field $\mathbf{E}_{loc}$ acting on electrons (and ions) is different from
the average electric field $\mathbf{E}$, i.e. $\mathbf{E}_{loc}\neq \mathbf{E%
}$, there are corrections to $\varepsilon _{tot}(\mathbf{k},0)$ which may
lead to $\varepsilon _{tot}(\mathbf{k},0)<0$.

The above analysis tells us that in real crystals $\varepsilon _{tot}(%
\mathbf{k},0)$ \textit{can be negative} in the large portion of the
Brillouin zone giving rise to $\lambda _{ep}-\mu >0$ in Eq.(\ref{NVeff}).
This means that analytic properties of the dielectric function $\varepsilon
_{tot}$ \textit{do not limit }$T_{c}$\textit{\ in the phonon mechanism of
pairing}. This result does not mean that there is no limit on T$_{c}$ at
all. We mention in advance that the local field effects play important role
in HTSC oxides, due to their layered structure with very unusual\textit{\
ionic-metallic binding}, thus opening a possibility for large $EPI$.

In conclusion, we point out that there are no serious theoretical and
experimental arguments for ignoring EPI in HTSC cuprates. To this end it is
necessary to answer several important questions which are related to
experimental findings in HTSC cuprates (oxides): (1) if EPI is important for
pairing in HTSC cuprates and if superconductivity is of $d-wave$ type, how
are these two facts compatible? (2) Why is the transport EPI coupling
constant $\lambda _{tr}$ (entering resistivity) rather smaller than the
pairing EPI coupling constant $\lambda _{ep}(>1)$ (entering T$_{c}$), i.e.
why one has $\lambda _{tr}(\approx 0.6-1.4)\ll \lambda _{ep}(\sim 2-3.5)$?
(3) If EPI is ineffective for pairing in HTSC oxides, in spite of $\lambda
_{ep}>1$, why it is so?

\textit{3. Is a non-phononic pairing realized in HTSC?}

Regarding EPI one can pose a question - whether it contributes significantly
to d-wave pairing in cuprates? Surprisingly, despite numerous experiments in
favor of EPI, there is a believe that EPI is irrelevant for pairing \citep%
{Pines}. This belief is mainly based first, on the above discussed incorrect
lattice stability criterion related to the sign of $\varepsilon _{tot}(%
\mathbf{k},0)$, which implies small EPI and second, on the well established
experimental fact that d-wave pairing is realized in cuprates \citep%
{TsuiKirtley}, which is believed to be incompatible with EPI. Having in mind
that EPI in HTSC at and near optimal dopimng is strong with $2<\lambda
_{ep}<3.5$ (see below), we assume for the moment that the leading pairing
mechanism in cuprates, which gives d-wave pairing, is due to some
non-phononic mechanism. For instance, let us assume an \textit{exitonic}
mechanism due to the high energy pairing boson ($\Omega _{nph}\gg \omega
_{ph}$) and with the bare critical temperature $T_{c0}$ and look for the
effect of EPI on $T_{c}$. If EPI is approximately \textit{isotropic}, like
in most LTSC materials, then it would be very detrimental for d-wave
pairing. In the case of dominating \textit{isotropic} EPI in the normal
state and the exitonic-like pairing, then near $T_{c}$ the linearized
Eliashberg equations have an approximative form for a weak non--phonon
interaction (with the large characteristic frequency $\Omega _{nph}$)
\begin{equation*}
Z(\omega _{n})\Delta _{n}(\mathbf{k})\approx \pi T_{c}\sum_{m}^{\Omega
_{_{nph}}}\sum_{\mathbf{q}}V_{nph}(\mathbf{k},\mathbf{q},n,m)\frac{\Delta
_{m}(\mathbf{q})}{\left\vert \omega _{n^{\prime }}\right\vert }
\end{equation*}%
\begin{equation}
Z(\omega _{n})\approx 1+\Gamma _{ep}/\omega _{n}.  \label{LinElia}
\end{equation}%
For pure d-wave pairing with the pairing potential $V_{nph}=V_{nph}(\theta _{%
\mathbf{k}},\theta _{\mathbf{q}})\cdot \Theta (\Omega _{nph}-\left\vert
\omega _{n}\right\vert )\Theta (\Omega _{nph}-\left\vert \omega _{n^{\prime
}}\right\vert )$ with $V_{nph}(\mathbf{k},\mathbf{q})=V_{0}\cdot
Y_{d}(\theta _{\mathbf{k}})Y_{d}(\theta _{\mathbf{q}})$ and $Y_{d}(\theta _{%
\mathbf{k}})=\pi ^{-1/2}\cos 2\theta _{\mathbf{k}}$ one obtains $\Delta _{n}(%
\mathbf{k})=\Delta _{d}\cdot \Theta (\Omega _{nph}-\left\vert
\omega _{n}\right\vert )Y_{d}(\theta _{\mathbf{k}})$ and the
equation for T$_{c}$ - see \citep{MaksimovReview}
\begin{equation}
\ln \frac{T_{c}}{T_{c0}}\approx \Psi (\frac{1}{2})-\Psi (\frac{1}{2}+\frac{%
\Gamma _{ep}}{2\pi T_{c}}).  \label{gama}
\end{equation}%
Here $\Psi $ is the di-gamma function. At temperatures near $T_{c}$ one has $%
\Gamma _{ep}\approx 2\pi \lambda _{ep}T_{c}$ $\ $and the solution of Eq. (%
\ref{gama}) is approximately $T_{c}\approx T_{c0}\exp \{-\lambda _{ep}\}$
with $T_{c0}\approx \Omega _{nph}\exp \{-\lambda _{nph}\}$, $\lambda
_{nph}=N(0)V_{0}$. This means that for $T_{c}^{\max }\sim 160$ $K$ and $%
\lambda _{ep}>1$ the bare $T_{c0}$ due to the non-phononic interaction must
be very large, i.e. $T_{c0}>500$ $K$.

Concerning other non-phononic mechanisms, such as the SFI one, the effect of
EPI in the framework of Eliashberg equations was studied numerically in \citep%
{Licht}. The latter is based on Eqs.(\ref{Z-Eli}-\ref{Fi-Eli}) in Appendix
A. with the kernels in the normal and superconducting channels $\lambda _{%
\mathbf{kp}}^{Z}(i\nu _{n})$ and $\lambda _{\mathbf{kp}}^{\Delta }$,
respectively. Usually, the spin-fluctuation kernel $\lambda _{sf,\mathbf{kp}%
}(i\nu _{n})$ is taken in the FLEX approximation
\citep{ScalapinoReview}. The calculations \citep{Licht} confirm
the very detrimental effect of the isotropic \
($\mathbf{k}$-independent) EPI on d-wave pairing due to SFI. For
the bare SFI critical temperature $T_{c0}\sim 100$ $K$ and for
$\lambda _{ep}>1$ the calculations give very small (renormalized)
critical temperature $T_{c}\ll 100$ $K$. These results tell us
that a more realistic pairing interaction must be operative in
cuprates and that EPI must be \textit{strongly momentum dependent
}\citep{Kulic1}. Only in that case is strong EPI conform with
d-wave pairing, either as its main cause or as a supporter of a
non-phononic mechanism. In \textit{Part II} we shall argue that
the strongly momentum dependent EPI is important scattering
mechanism in cuprates providing the strength of the pairing
mechanism, while the residual Coulomb interaction (by including
weaker SFI) triggers it to d-wave pairing.

\section{Experimental evidence for strong EPI}

In the following we discuss some important experiments which give
evidence for strong electron-phonon interaction (EPI) in cuprates.
However, before doing it, we shall discuss some indicative
\textit{inelastic magnetic neutron scattering measurements} (IMNS)
in cuprates whose results in fact seriously doubt in the
effectiveness of the phenomenological SFI mechanism of pairing
which is advocated in \citep{Pines}, \citep{DahmScalapinoO66}.
First, the experimental results related to the pronounced
imaginary part of the susceptibility $Im\chi
(\mathbf{k},k_{z},\omega )$ in the normal state at and near the AF
wave vector $\mathbf{k}=\mathbf{Q}=(\pi ,\pi )$ were interpreted
in a number of papers as a support for the SFI mechanism for
pairing \citep{Pines}, \citep{DahmScalapinoO66}. Second, the
existence of the so called magnetic resonance peak of $Im\chi
(\mathbf{k},k_{z},\omega )$ (at some energies $\omega <2\Delta $)
in the superconducting state was also interpreted in a number of
papers either as the origin of superconductivity or as a mechanism
strongly affecting superconducting gap at the ant-nodal point.

\subsection{Magnetic neutron scattering and the spin fluctuation spectral
function}

\textit{a. Huge rearrangement of the SFI spectral function and little change
of }$T_{c}$\textit{\ }

\begin{figure}[!tbp]
\resizebox{.5\textwidth}{!} {\includegraphics*[
width=6cm]{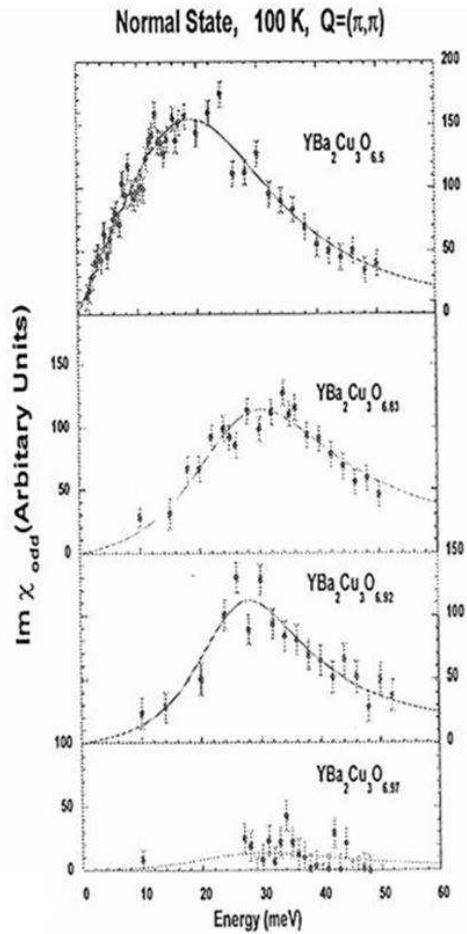}}
\caption{Magnetic spectral function $Im\protect\chi ^{(-)}(\mathbf{k},%
\protect\omega )$ in the normal state of $YBa_{2}Cu_{3}O_{6+x}$ at $T=100$ $%
K $ and at $Q=(\protect\pi ,\protect\pi )$. $100$ counts in the vertical
scale corresponds to $\protect\chi _{max}^{(-)}\approx 350\protect\mu %
_{B}^{2}/eV$. From Ref. \protect\citep{Bourges}.} \label{SuscFigN}
\end{figure}

Before discussing experimental results in cuprates on the imaginary part of
the spin susceptibility $Im\chi (\mathbf{k},\omega )$ we point out that in
(phenomenological) theories based on spin fluctuation interaction (SFI) the
quasi-particle self-energy $\hat{\Sigma}_{sf}(\mathbf{k},\omega _{n})$ ($%
\omega _{n}$ is the Matsubara frequency and $\hat{\tau}_{0}$ is the Nambu
matrix) in the normal and superconducting state and the effective
(repulsive) pairing potential $V_{sf}(\mathbf{k},\omega )$ (where $i\omega
_{n}\rightarrow \omega +i\eta $) are assumed in the form \citep{Pines}%
\begin{equation*}
\hat{\Sigma}_{sf}(\mathbf{k},\omega _{n})=\frac{T}{N}\sum_{\mathbf{k}%
^{\prime },m}V_{sf}(\mathbf{k-k}^{\prime },\omega _{n}-\omega _{m})\hat{\tau}%
_{0}\hat{G}(\mathbf{k}^{\prime },\omega _{m})\hat{\tau}_{0},
\end{equation*}%
\begin{equation}
V_{sf}(\mathbf{k},\omega _{n})=g_{sf}^{2}\int_{-\infty }^{\infty }\frac{d\nu
}{\pi }\frac{Im\chi (\mathbf{q},\nu +i0^{+})}{\nu -i\omega _{n}}.
\label{Vsf}
\end{equation}%
Although the form of $V_{sf}$ can not be justified theoretically, except in
the weak coupling limit ($g_{sf}\ll W_{b}$) only, it is used in the analysis
of the quasi-particle properties in the normal and superconducting state of
cuprates where the spin susceptibility (spectral function) Im$\chi (\mathbf{q%
},\omega )$ is strongly peaked at and near the AF wave vector $\mathbf{Q}%
=(\pi /a,\pi /a)$. Can the pairing mechanism in HTSC cuprates be explained
by such a phenomenology and what is the prise for it? The best answer is to
look at the experimental results related to inelastic magnetic neutron
scattering (IMNS) experiments which give $Im\chi (\mathbf{q},\omega )$. In
that respect very indicative and impressive IMNS measurements on $%
YBa_{2}Cu_{3}O_{6+x}$, which are done by Bourges group
\citep{Bourges},
demonstrate clearly that the normal state susceptibility $Im\chi ^{(odd)}(%
\mathbf{q},\omega )$ (the odd part of the spin susceptibility in the bilayer
system) at $\mathbf{q}=\mathbf{Q}=(\pi ,\pi )$ is strongly dependent on the
hole doping as it is shown in Fig.~\ref{SuscFigN}.

The most pronounced result for our discussion is that by varying doping
\textit{there is a huge rearrangement} of $Im\chi ^{(odd)}(\mathbf{Q},\omega
)$ in the normal state, especially in the energy (frequency) region which
might be important for superconducting pairing, let say $0$ $meV<\omega <60$
$meV$. This is clearly seen in the last two curves in Fig. \ref{SuscFigN}
where this \textit{rearrangement is very pronounced}, while at the same time
there is only \textit{small variation of the critical temperature} $T_{c}$.
It is seen in Fig.~\ref{SuscFigN} that in the underdoped $%
YBa_{2}Cu_{3}O_{6.92}$\ crystal Im$\chi ^{(odd)}(\mathbf{Q},\omega )$ and $S(%
\mathbf{Q})=N(\mu )g_{sf}^{2}\int_{0}^{60}d\omega Im\chi ^{(odd)}(\mathbf{Q}%
,\omega )$ are much larger than that in the near optimally doped $%
YBa_{2}Cu_{3}O_{6.97}$, i.e. one has $S_{6.92}(\mathbf{Q})\gg S_{6.97}(%
\mathbf{Q})$, although the difference in the corresponding critical
temperatures $T_{c}$ is very small, i.e. $T_{c}^{(6.92)}=91$ $K$ (in $%
YBa_{2}Cu_{3}O_{6.92}$) and $T_{c}^{(6.97)}=92.5$ $K$ (in $%
YBa_{2}Cu_{3}O_{6.97}$). This \textit{pronounced rearrangement and
suppression }of Im$\chi ^{(odd)}(\mathbf{Q},\omega )$ (in the normal state
of YBCO) by doping toward the optimal doping but with negligible change in $%
T_{c}$ is \textit{strong evidence} that the $SFI$ pairing
mechanism is not the dominating one in HTSC cuprates. This
insensitivity of $T_{c}$, if interpreted in term of the $SFI$
coupling constant $\lambda _{sf}(\sim g_{sf}^{2})$, means that the
latter is small, i.e. $\lambda _{sf}^{(\exp )}\ll 1$. We stress
that the explanation of high T$_{c}$ in cuprates by the SFI
phenomenological theory \citep{Pines} assumes very large SFI
coupling energy with $g_{sf}^{(th)}\approx 0.7$ $eV$ while the
frequency(energy) dependence of $Im\chi (\mathbf{Q},\omega )$ is
extracted from the fit of the NMR relaxation rate
$T_{1}^{-1}$which gives $T_{c}^{(NMR)}\approx 100$ $K$
\citep{Pines}. To this point, the NMR measurements (of
$T_{1}^{-1}$) give that there is an \textit{anti-correlation}
between the decrease of the NMR
spectral function $I_{\mathbf{Q}}=\lim_{\omega \rightarrow 0}Im\chi ^{(NMR)}(%
\mathbf{Q},\omega )/\omega $ and the increase of $T_{c}$ by
increasing doping toward the optimal one - see \citep{KulicReview}
and References therein. The latter result additionally disfavors
the SFI model of pairing \citep{Pines} since the strength of
pairing interaction is little affected by SFI. Note, that if
instead of taking $Im\chi (\mathbf{Q},\omega )$ from NMR
measurements one takes it from IMNS measurements, as it was done in \citep%
{Levin}, than for the same value $g_{sf}^{(th)}$one obtains much smaller $%
T_{c}$. For instance, by taking the experimental values for $Im\chi
^{(IMNS)}(\mathbf{Q},\omega )$ in underdoped $YBa_{2}Cu_{3}O_{6.6}$ with $%
T_{c}\approx 60$ $K$ one obtains $T_{c}^{(IMNS)}<T_{c}^{(NMR)}/3$ \citep%
{Levin}, while $T_{c}^{(IMNS)}\rightarrow 50$ $K$ for $g_{sf}^{(th)}\gg 1$.
The situation is even worse if one trays to \textit{fit the resistivity}
with $Im\chi ^{(IMNS)}(\mathbf{Q},\omega )$ in $YBa_{2}Cu_{3}O_{6.6}$ since
this fit gives $T_{c}^{(IMNS)}<7$ $K$. These results point to a deficiency
of the SFI phenomenology (at least that based on Eq.(\ref{Vsf})) to describe
pairing in HTSC cuprates.

Having in mind the results in \citep{Levin}, the recent
theoretical
interpretation in \citep{DahmScalapinoO66} of IMNS experiments \citep%
{HinkovYBCO66} and ARPES measurements \citep{BorisenkoARPES} on
the underdoped $YBa_{2}Cu_{3}O_{6.6}$ in terms of the SFI
phenomenology deserve to be commented. The IMNS experiments
\citep{HinkovYBCO66} give evidence for the "hourglass" spin
excitation spectrum (in the superconducting state) for the momenta
$\mathbf{q}$ at, near and far from $\mathbf{Q}$, which is richer
than the common spectrum with magnetic resonance peaks measured at $\mathbf{Q%
}$. In \citep{DahmScalapinoO66} the self-energy of electrons due
to their interaction with spin excitations is calculated by using
Eq. (\ref{Vsf}) with $g_{sf}^{2}=(3/2)\tilde{U}^{2}$ and $Im\chi
(\mathbf{q},\omega )$ taken from \citep{HinkovYBCO66}. However, in
order to fit the ARPES self-energy and
low-energy kinks (see discussion in Section C) the authors of \citep%
{DahmScalapinoO66} use \textit{very large value} $\tilde{U}=1.59$ $eV$, i.e.
much larger than the one used in \citep{Levin}. Such a large value of $\tilde{%
U}$ has been used earlier within the Monte Carlo simulation and
the fit of the Hubbard model \citep{Maier}. In our opinion this
value for $\tilde{U}$ is unrealistically large in the case of
strongly correlated systems where spin-fluctuations are governed
by the effective electron exchange interaction $J_{Cu-Cu}\lesssim
0.15$ $eV$ \citep{Anderson2007}. This implies
that $\tilde{U}\ll 1$ $eV$ and $T_{c}\ll 10$ $K$. Note, that this value for $%
J_{Cu-Cu}$\ comes out also from the theory of strongly correlated electrons
in the three-band Emery model with $J_{Cu-Cu}\approx (4t_{pd}^{4}/(\Delta
_{dp}+U_{pd})^{2})[(1/U_{d})+2/(U_{p}+2\Delta )]$ - for parameters see Part
II Section VII. \

Concerning the problem related to the rearrangement of the SFI spectral
function $Im\chi (\mathbf{Q},\omega )$ in $YB_{2}Cu_{3}O_{6+x}$ \citep%
{Bourges} we would like to stress, that in spite that the latter
results were obtained ten years ago they are not disputed by the
new IMNS measurements \citep{ReznikNewIMNS} on high quality
samples of the same compound (where much longer counting times
were used in order to reduce statistical errors). In fact the
results in \citep{Bourges} \textit{are}
\textit{confirmed} in \citep{ReznikNewIMNS} where the magnetic intensity $I(%
\mathbf{q},\omega )(\sim Im\chi (\mathbf{q},\omega ))$ (for $\mathbf{q}$ at
and in the broad range of $\mathbf{Q}$) for the optimally doped $%
YBa_{2}Cu_{3}O_{6.95}$ (with $T_{c}=93$ $K$) is \textit{at least three times
smaller} than in the underoped $YBa_{2}Cu_{3}O_{6.6}$ with $T_{c}=60$ $K$.
This result is again very indicative sign of the weakness of SFI since such
a huge reconstruction would decrease $T_{c}$ in the optimally doped $%
YBa_{2}Cu_{3}O_{6.95}$ if analyzed in the framework of the
phenomenological SFI theory based on Eq. (\ref{Vsf}). It also
implies that due to the suppression of $Im\chi (\mathbf{q},\omega
)$ by increasing doping toward the optimal one a straightforward
extrapolation of the theoretical approach in
\citep{DahmScalapinoO66} to the explanation of $T_{c}$ in the
optimally doped $YBa_{2}Cu_{3}O_{6.95}$ would require an increase
of $\tilde{U}$ to the value even larger than $4$ $eV$, what is\
highly improbable.

\textit{b. Ineffectiveness of the magnetic resonance peak }

\begin{figure}[tbp]
\resizebox{.5\textwidth}{!} {\includegraphics*[
width=6cm]{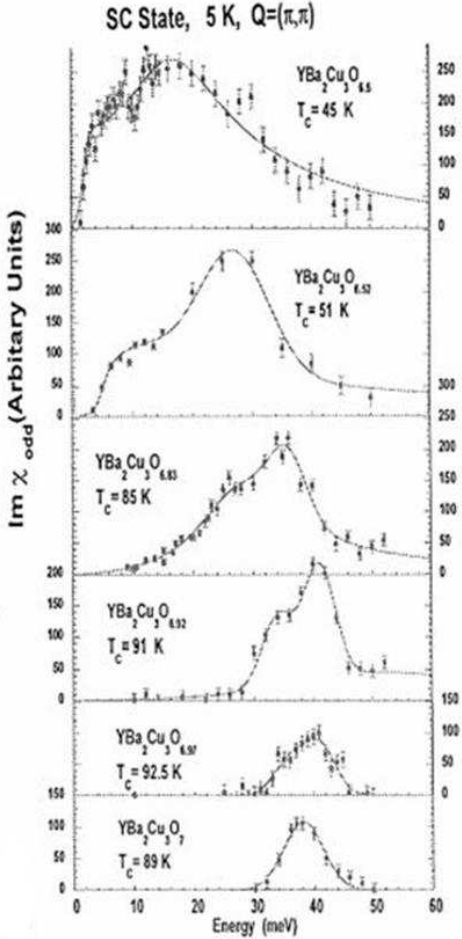}}
\caption{Magnetic spectral function $Im\protect\chi ^{(-)}(\mathbf{k},%
\protect\omega )$ in the superconducting state of $YBa_{2}Cu_{3}O_{6+x}$ at $%
T=5$ $K$ and at $Q=(\protect\pi ,\protect\pi )$. $100$ counts in the
vertical scale corresponds to $\protect\chi _{max}^{(-)}\approx 350\protect%
\mu _{B}^{2}/eV$. From Ref. \protect\citep{Bourges}.}
\label{SuscFigS}
\end{figure}

A less direct argument for \textit{smallness of the SFI coupling constant,
i.e.} $g_{sf}^{\exp }\leq 0.2$ $eV$ and $g_{sf}^{\exp }\ll g_{sf}$ comes
from other experiments related to the magnetic resonance peak in the
superconducting state, and this will be discussed next. In the
superconducting state of optimally doped YBCO and BISCO, $Im\chi (\mathbf{Q}%
,\omega )$ is significantly suppressed at low frequencies except near the
resonance energy $\omega _{res}\approx 41$ $meV$ where a pronounced narrow
peak appears - the \textit{magnetic} \textit{resonance peak}. We stress that
there is no magnetic resonance peak in some families of HTSC cuprates, for
instance in LSCO, and consequently one can question the importance of the
resonance peak in the scattering processes. The experiments tell us that the
relative intensity of this peak (compared to the total one) is small, i.e. $%
I_{0}\sim (1-5)\%$ - see Fig~\ref{SuscFigS}. In underdoped cuprates this
peak is present also in the normal state as it is seen in Fig~\ref{SuscFigN}%
.

After the discovery of the resonance peak there were attempts to
relate it first, to the origin of the superconducting condensation
energy and second, to the kink in the energy dispersion or the
peak-dimp structure in the ARPES spectral function. In order that
the condensation energy is due to the magnetic resonance it is
necessary that the peak intensity $I_{0}$ is small
\citep{Kivelson}. $I_{0}$ is obtained approximately by equating
the condensation energy $E_{con}\approx N(0)\Delta ^{2}/2$ with
the change of the magnetic energy $E_{mag}$ in the superconducting
state, i.e. $\delta E_{mag}\approx 4I_{0}\cdot E_{mag}$
\begin{equation}
E_{mag}=J\iint \frac{d\omega d^{2}k}{(2\pi )^{3}}(1-\cos k_{x}-\cos k_{y})S(%
\mathbf{k},\omega ).  \label{Emag}
\end{equation}%
By taking $\Delta \approx 2T_{c}$ and the realistic value $N(0)\sim
1/(10J)\sim 1$ $states/eV\cdot spin$, one obtains $I_{0}\sim
10^{-1}(T_{c}/J)^{2}\sim 10^{-3}$. However, such a small intensity can not
be responsible for the anomalies in ARPES and optical spectra since it gives
rise to small coupling constant $\lambda _{sf,res}$ for the interaction of
holes with the resonance peak, i.e. $\lambda _{sf,res}\approx
(2I_{0}N(0)g_{sf}^{2}/\omega _{res})\ll 1$. Such a small coupling does not
affect superconductivity at all. Moreover, by studying the width of the
resonance peak one can extract order of magnitude of the SFI coupling
constant $g_{sf}$. Since the magnetic resonance disappears in the normal
state of the optimally doped YBCO it can be qualitatively understood by
assuming that its broadening scales with the resonance energy $\omega _{res}$%
, i.e. $\gamma _{res}<\omega _{res}$, where the line-width is given by $%
\gamma _{res}=4\pi (N(0)g_{sf})^{2}\omega _{res}$
\citep{Kivelson}. This condition limits the SFI coupling to
$g_{sf}<0.2$ $eV$. We stress that the obtained $g_{sf}$ is
\textit{much smaller} (at least by factor three) than that assumed
in the phenomenological spin-fluctuation theory \citep{Pines}
and \citep{DahmScalapinoO66} where $g_{sf}\sim 0.6-0.7$ $eV$ and $\tilde{U}%
\approx 1.6$ $eV$, but much larger than estimated in \citep{Kivelson} (where $%
g_{sf}<0.02$ $eV$). The smallness of $g_{sf}$ comes out also from
the analysis of the antiferromagnetic state in underdoped metals
of LSCO and YBCO \citep{KulicKulic}, where the small (ordered)
magnetic moment $\mu (<0.1 $ $\mu _{B})$ points to an itinerant
antiferromagnetism with small coupling constant $g_{sf}<0.2$ $eV$.
The conclusion from this analysis is that in the optimally doped
YBCO the sharp magnetic resonance is a consequence of the onset of
superconductivity\textit{\ }and not its cause. There is also one
principal reason against the pairing due to the magnetic resonance
peak at least in optimally doped cuprates. Since the intensity of
the magnetic resonance near $T_{c}$ is vanishingly small, though
not affecting pairing at the second order phase transition at
$T_{c}$, then if it would be solely the origin for
superconductivity the phase transition at T$_{c}$ would be
\textit{first order}, contrary to experiments. Recent ARPES
experiments give evidence that the magnetic resonance cannot be
related to the kinks in ARPES spectra \citep{Lanzara},
\citep{Valla} - see the discussion below.

Finally, we would like to point out that the recent magnetic neutron
scattering measurements on optimally-doped large-volume crystals $%
Bi_{2}Sr_{2}CaCu_{2}O_{8+\delta }$ \citep{Tranquada2009}, where
the absolute value of $Im\chi (\mathbf{q},\omega )$ is measured,
are questioning the interpretation of the electronic magnetism in
cuprates in terms of itinerant magnetism. This experiment shows a
lack of temperature dependence of the
local spin susceptibility $Im\chi (\omega )=\sum_{q}Im\chi (\mathbf{q}%
,\omega )$ across the superconducting transition $T_{c}=91$ $K$, i.e there
is only a minimal change in $Im\chi (\omega )$ between $10$ $K$ and $100$ $K$%
. Note, if the magnetic excitations were due to itinerant quasi-particles we
should have seen dramatic changes of $Im\chi (\omega )$ as a function of $T$
over the whole energy range. This $T$-independence of $Im\chi (\omega )$
strongly opposes many theoretical results in \citep{Carbotte}, \citep%
{HwangTimusk1}, \citep{HwangTimusk2} which assume that the bosonic
spectral function is proportional to $Im\chi (\omega )$ that can
be extracted from optic measurements. This procedure gives that
$Im\chi (\omega )$ is strongly $T$-dependent contrary to the
experimental results in \citep{Tranquada2009} - see more in
Subsection B on optical conductivity.

\subsection{Optical conductivity and EPI}

Optical spectroscopy gives information on \textit{optical conductivity} $%
\sigma (\omega )$ and on two-particle excitations, from which one can
indirectly extract the transport spectral function $\alpha _{tr}^{2}F(\omega
)$. Since this method probes bulk sample (on the skin depth), contrary to
ARPES\ and tunnelling methods which probe tiny regions ($10-15$ \AA ) near
the sample surface, this method is indispensable. However, one should be
careful not over-interpreting the experimental results since $\sigma (\omega
) $ \textit{is not a directly} \textit{measured quantity} but it is derived
from the reflectivity $R(\omega )=\left\vert (\sqrt{\varepsilon _{ii}(\omega
)}-1)/(\sqrt{\varepsilon _{ii}(\omega )}+1)\right\vert ^{2}$ with the
transversal dielectric tensor $\varepsilon _{ii}(\omega )=\varepsilon
_{ii,\infty }+\varepsilon _{ii,latt}+4\pi i\sigma _{ii}(\omega )/\omega $.
Here, $\varepsilon _{ii,\infty }$ is the high frequency dielectric function,
$\varepsilon _{ii,latt}$ describes the contribution of the lattice
vibrations and $\sigma _{ii}(\omega )$ describes the optical (dynamical)
conductivity of conduction carriers. Since $R(\omega )$ is usually measured
in the limited frequency interval $\omega _{\min }<\omega <\omega _{\max }$
some physical modelling for $R(\omega )$ is needed in order to guess it
outside this range - see more in reviews \citep{MaksimovReview}, \citep%
{KulicReview}. This was the reason for numerous misinterpretations of optic
measurements in cuprates, that will be uncover below. An illustrative
example for \ this claim is large dispersion in the reported value of $%
\omega _{pl}$ - from $0.06$ to $25$ $eV$, i.e. almost three orders
of magnitude. However, it turns out that $IR$ measurements of
$R(\omega )$ in conjunction with elipsometric measurements of
$\varepsilon _{ii}(\omega )$ at high frequencies allows more
reliable determination of $\sigma (\omega )$
\citep{BozovicPlasma}.

1. \textit{Transport and quasi-particle relaxation rates }

The widespread misconception in studying the quasi-particle
scattering in cuprates was an ad hoc assumption that the
\textit{transport relaxation rate} $\gamma _{tr}(\omega )$ is
equal to the \textit{quasi-particle relaxation rate} $\gamma
(\omega )$, in spite of the well known fact that the inequality
$\gamma _{tr}(\omega )\neq \gamma (\omega )$ holds in a broad
frequency (energy) region \citep{Allen}. This (incorrect)
assumption was one of the main arguments against EPI as relevant
scattering mechanism in cuprates. Although we have discussed this
problem several times before, we do it again due to the importance
of this subject.

\begin{figure}[!tbp]
\resizebox{.5\textwidth}{!} {\includegraphics*[
width=6cm]{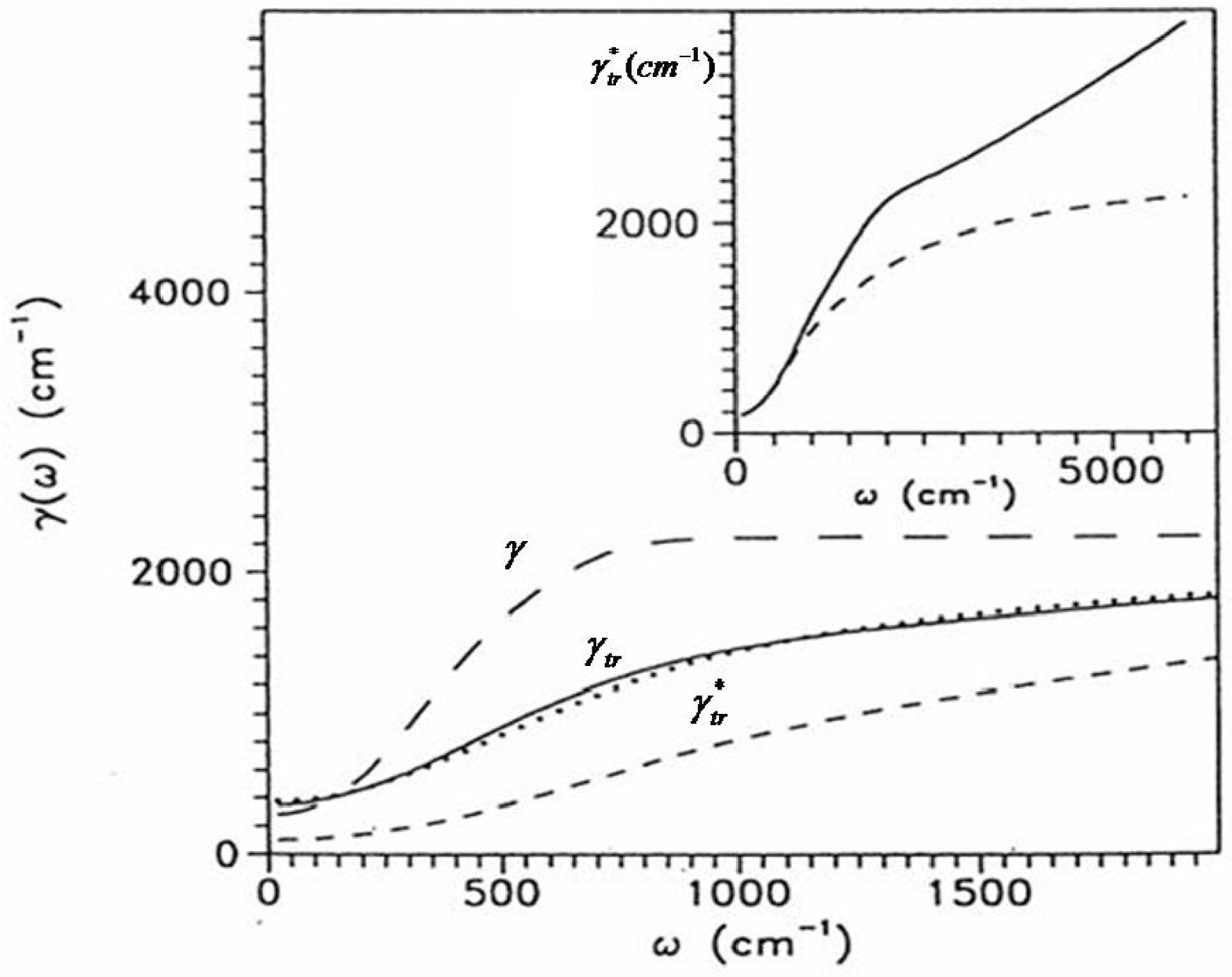}} {\includegraphics*[
width=8cm]{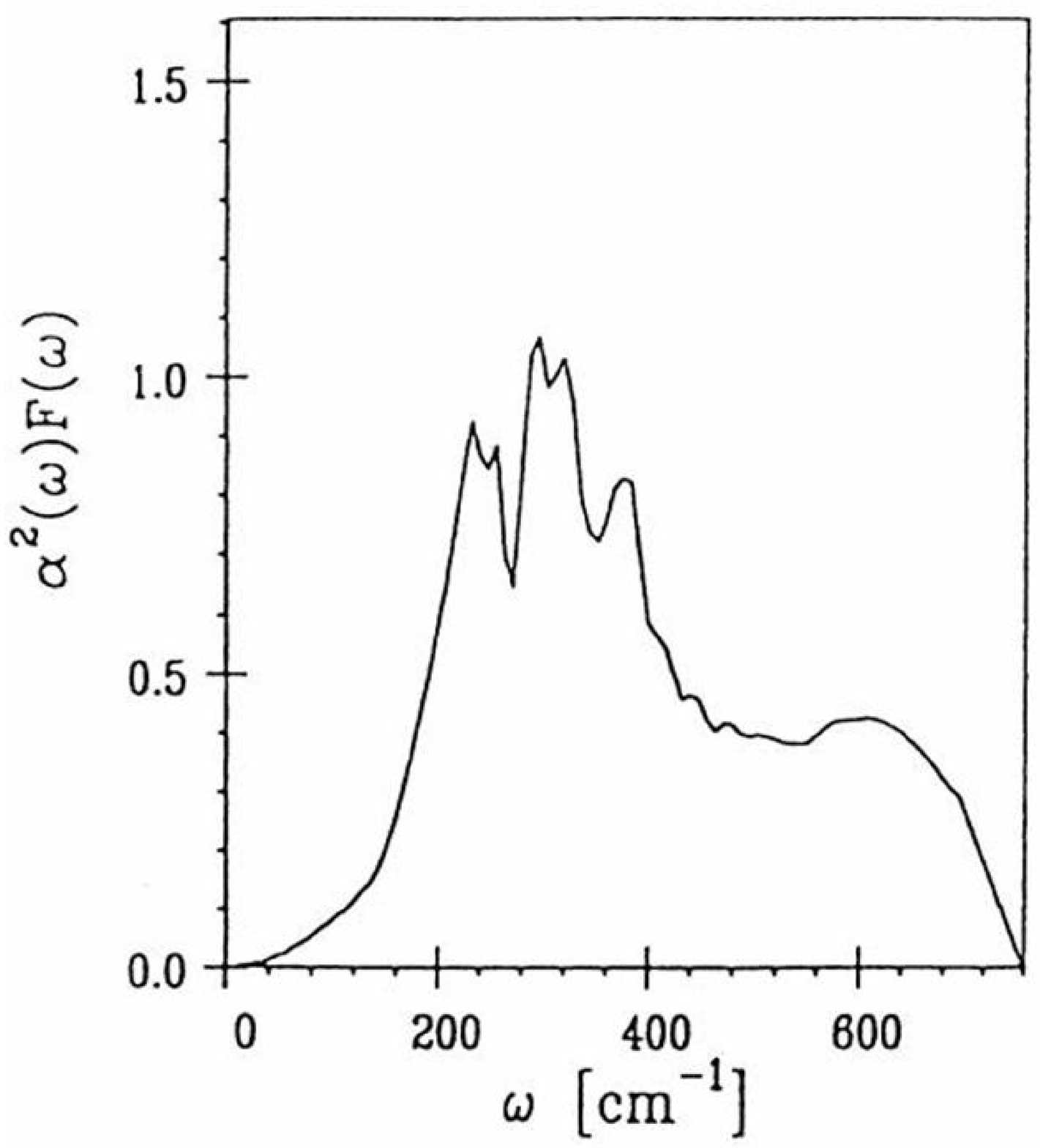}}
\caption{(top) Scattering rates $\protect\gamma (\protect\omega ,T) $, $%
\protect\gamma _{tr}(\protect\omega ,T)$ and $\protect\gamma _{tr}^{\ast }$
- from top to bottom, for the Eliashberg function in ({bottom}). From
\protect\citep{DolgovShulga}. (bottom) Eliashberg spectral function $\protect%
\alpha _{ep}^{2}(\protect\omega )F_{ph}(\protect\omega )$ obtained
from tunnelling experiments on break junctions
\protect\citep{TunnelingVedeneev}. Inset shows $\protect\gamma
_{tr}^{\ast }$ with (full line) and without (dashed line)
inter-band transitions \protect\citep{MaksimovReview}.}
\label{Rates}
\end{figure}

The dynamical conductivity $\sigma (\omega )$ consists of two parts, i.e. $%
\sigma (\omega )=\sigma ^{inter}(\omega )+\sigma ^{intra}(\omega )$ where $%
\sigma ^{inter}(\omega )$ describes \textit{inter-band}
\textit{transitions} which contribute at higher than intra-band
energies, while $\sigma ^{intra}(\omega )$ is due to
\textit{intra-band} transitions which are relevant at low energies
$\omega <(1-2)$ $eV$. (Note, that in the $IR$ measurements the
frequency is usually given in $cm^{-1}$, where the following
conversion holds $1cm^{-1}=29.98$ $GHz=0.123985$ $meV=1.44$ $K$.)
The experimental data for $\sigma (\omega )=\sigma _{1}+i\sigma
_{2}$ in cuprates are usually processed by the generalized
(extended) Drude formula \citep{Allen}, \citep{DolgovShulga},
\citep{Shulga}, \citep{Schlesinger},
\begin{equation}
\sigma (\omega )=\frac{\omega _{p}^{2}}{4\pi }\frac{1}{\gamma _{tr}(\omega
)-i\omega m_{tr}(\omega )/m_{\infty }}\equiv \frac{1}{\tilde{\omega}%
_{tr}(\omega )},  \label{Drude}
\end{equation}%
which is an useful representation for systems with single band
electron-boson scattering which is justified in HTSC cuprates.
However, this procedure is inadequate for interpreting optical
data in multi-band systems such as new high-temperature
superconductors Fe-based pnictides since even in absence of the
inelastic intra- and inter-band scattering the effective optic
relaxation rate may be strongly frequency dependent
\citep{DolgKulSDW}.
(The usefulness of introducing the optic relaxation $\tilde{\omega}%
_{tr}(\omega )$ will be discussed below and in Appendix B.) Here, $i=a,b$
enumerates the plane axis, $\omega _{p}$, $\gamma _{tr}(\omega ,T)$ and $%
m_{op}(\omega )$ are the electronic plasma frequency, the
transport (optical) scattering rate and the optical mass,
respectively. Very frequently it is analyzed the quantity $\gamma
_{tr}^{\ast }(\omega ,T)$ given by \citep{Schlesinger}
\begin{equation}
\gamma _{tr}^{\ast }(\omega ,T)=\frac{m_{\infty }}{m_{tr}(\omega )}\gamma
_{tr}(\omega ,T)=\frac{Im\sigma (\omega )}{\omega Re\sigma (\omega )},
\label{gamma-star}
\end{equation}%
which is determined from the half-width of the Drude-like expression for $%
\sigma (\omega )$ and is independent of $\omega _{p}^{2}$. In the weak
coupling limit $\lambda _{ep}<1$, the formula for conductivity given in
Appendix B Eqs. (\ref{Sigma-tr}-\ref{K2}) can be written in the form of Eq.(%
\ref{Drude}) where $\gamma _{tr}$ reads
\citep{DolgovShulga}-\citep{Shulga}
\begin{equation*}
\gamma _{tr}(\omega ,T)=\pi \sum_{l}\int_{0}^{\infty }d\nu \alpha
_{tr,l}^{2}F_{l}(\nu )[2(1+2n_{B}(\nu ))
\end{equation*}%
\begin{equation}
-2\frac{\nu }{\omega }-\frac{\omega +\nu }{\omega }n_{B}(\omega +\nu )+\frac{%
\omega -\nu }{\omega }n_{B}(\omega -\nu )].  \label{Gamma-tr}
\end{equation}%
Here $n_{B}(\omega )$ is the Bose distribution function. For completeness we
give also the explicit form of the transport mass $m_{tr}(\omega )$ see \citep%
{MaksimovReview}, \citep{KulicReview}, \citep{Allen},
\citep{DolgovShulga}, \citep{Shulga}, .
\begin{equation*}
\frac{m_{tr}(\omega )}{m_{\infty }}=1+\frac{2}{\omega }\sum_{l}\int_{0}^{%
\infty }d\Omega \alpha _{tr,l}^{2}F_{l}(\Omega )ReK(\frac{\omega }{2\pi T},%
\frac{\Omega }{2\pi T}).
\end{equation*}%
with the Kernel $K(x,y)=(i/y)+\{((y-x)/x)[\psi (1-ix+iy)-\psi
(1+iy)]\}-\{y\rightarrow -y\}$ where $\psi $ is the di-gamma function. In
the presence of impurity scattering one should add $\gamma _{imp,tr}$ to $%
\gamma _{tr}$. It turns out that Eq.(\ref{Gamma-tr}) holds within a few
percents also for large $\lambda _{ep}(>1)$. Note, that $\alpha
_{tr,l}^{2}F_{l}(\nu )\neq \alpha _{l}^{2}F_{l}(\nu )$ and the index $l$
\textit{enumerates all scattering bosons} - phonons, spin fluctuations, etc.
For comparison, the quasi-particle scattering rate $\gamma (\omega ,T)$ is
given by%
\begin{equation*}
\gamma (\omega ,T)=2\pi \int\limits_{0}^{\infty }d\nu \alpha ^{2}F(\nu
)\{2n_{B}(\nu )
\end{equation*}%
\begin{equation}
+n_{F}(\nu +\omega )+n_{F}(\nu -\omega )\}+\gamma ^{imp},  \label{gamma-n}
\end{equation}%
where $n_{F}$ is the Fermi distribution function. For completeness we give
also the expression for the quasi-particle effective mass $m^{\ast }(\omega
) $
\begin{equation*}
\frac{m^{\ast }(\omega )}{m}=1+\frac{1}{\omega }\sum_{l}\int\limits_{0}^{%
\infty }d\Omega \alpha _{l}^{2}F_{l}(\Omega )
\end{equation*}%
\begin{equation}
\times Re\{\psi (\frac{1}{2}+i\frac{\omega +\Omega }{2\pi T})-\psi (\frac{1}{%
2}-i\frac{\omega -\Omega }{2\pi T})\}.  \label{qp-mass}
\end{equation}%
The term $\gamma ^{imp}$ is due to the impurity scattering. By comparing Eq.(%
\ref{Gamma-tr}) and Eq.(\ref{gamma-n}), it is seen that $\gamma _{tr}$ and $%
\gamma $ are different quantities, i.e. $\gamma _{tr}\neq \gamma $, where
the former describes the \textit{relaxation of Bose particles (excited
electron-hole pairs)} while the latter one describes \textit{the relaxation
of Fermi particles}. This difference persists also at $T=0$ $K$ \ where one
has (due to simplicity we omit in the following summation over $l$ ) \citep%
{Allen}
\begin{equation}
\gamma _{tr}(\omega )=\frac{2\pi }{\omega }\int_{0}^{\omega }d\nu (\omega
-\nu )\alpha _{tr}^{2}(\nu )F(\nu )  \label{Gamma-tr-0}
\end{equation}%
and
\begin{equation}
\gamma (\omega )=2\pi \int_{0}^{\omega }d\nu \alpha ^{2}(\nu )F(\nu ).
\label{Gamma-qp-0}
\end{equation}%
In the case of EPI with the constant electronic density of states, the above
equations give that $\gamma _{ep}(\omega )=const$ for $\omega >\omega
_{ph}^{\max }$ while $\gamma _{ep,tr}(\omega )$ (as well as $\gamma
_{ep,tr}^{\ast }$) is monotonic growing for $\omega >\omega _{ph}^{\max }$,
where $\omega _{ph}^{\max }$ is the maximal phonon frequency. So, the
growing of $\gamma _{ep,tr}(\omega )$ (and $\gamma _{ep,tr}^{\ast }$) for $%
\omega >\omega _{ph}^{\max }$ is ubiquitous and natural for EPI
scattering and has nothing to do with some exotic scattering
mechanism. This behavior is clearly seen by comparing $\gamma
(\omega ,T)$, $\gamma _{tr}(\omega ,T)$ and $\gamma _{tr}^{\ast }$
which are calculated for the EPI spectral function $\alpha
_{ep}^{2}(\omega )F_{ph}(\omega )$ extracted from tunnelling
experiments in YBCO (with $\omega _{ph}^{\max }\sim 80$ $meV$)
\citep{TunnelingVedeneev} - see Fig.~\ref{Rates}.

The results shown in Fig.~\ref{Rates} clearly demonstrate the physical
difference between two scattering rates $\gamma _{ep}$ and $\gamma _{ep,tr}$
(and $\gamma _{tr}^{\ast }$). It is also seen that $\gamma _{tr}^{\ast
}(\omega ,T)$ is even more linear function of $\omega $ than $\gamma
_{tr}(\omega ,T)$. From these calculations one concludes that the
quasi-linearity of $\gamma _{tr}(\omega ,T)$ (and $\gamma _{tr}^{\ast }$) is
not in contradiction with the EPI scattering mechanism but it is in fact a
natural consequence of EPI. We stress that such behavior of $\gamma _{ep}$
and $\gamma _{ep,tr}$ (and $\gamma _{ep,tr}^{\ast }$), shown in Fig.~\ref%
{Rates}, is in fact not exceptional for HTSC cuprates but it is \textit{%
generic for many metallic systems}, for instance 3D metallic oxides, low
temperature superconductors such as $Al$, $Pb$, etc. - see more in \citep%
{MaksimovReview}, \citep{KulicReview} and References therein.

\begin{figure}[!tbp]
\resizebox{.4\textwidth}{!} {\includegraphics*[
width=6cm]{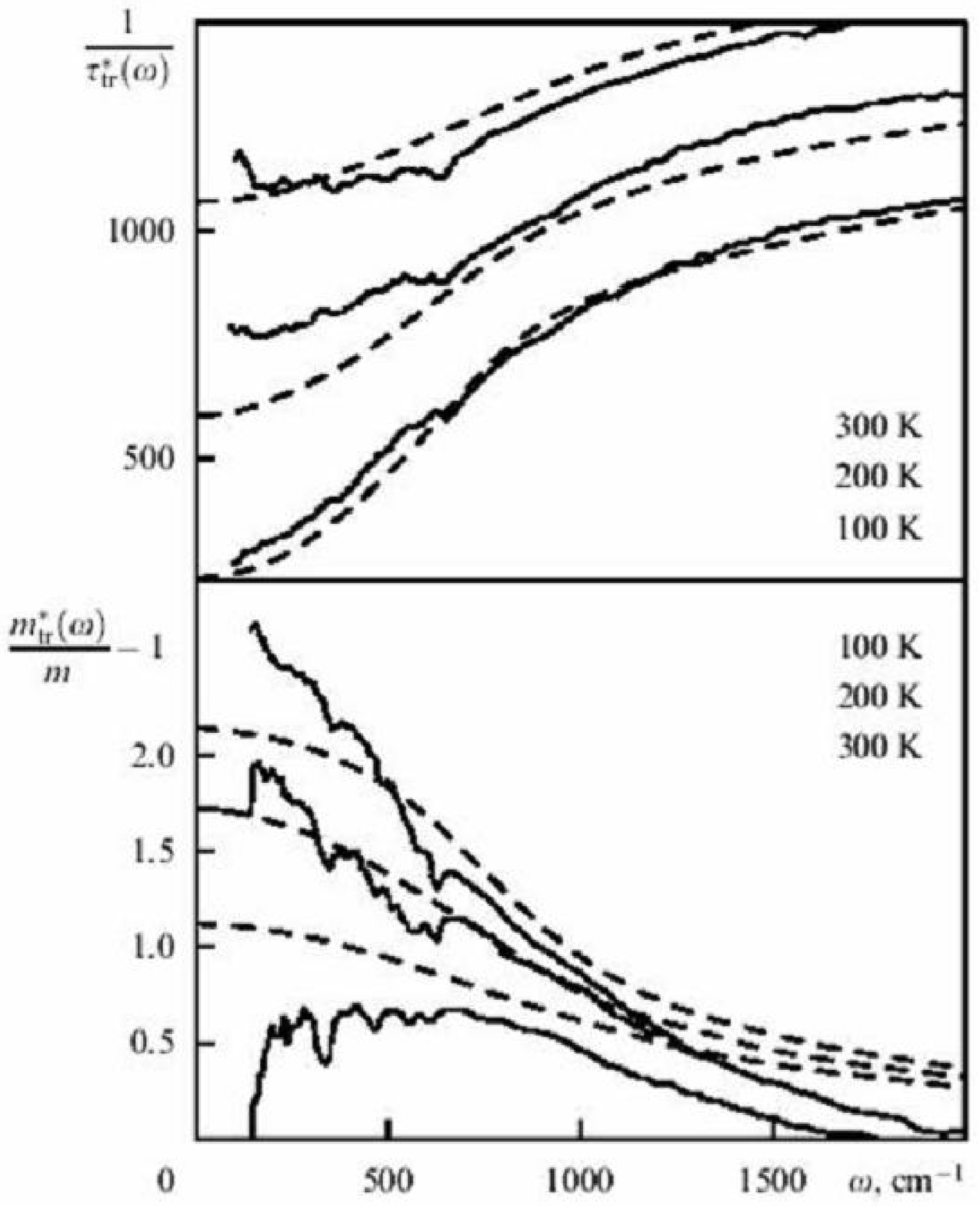}} {\includegraphics*
[width=8cm]{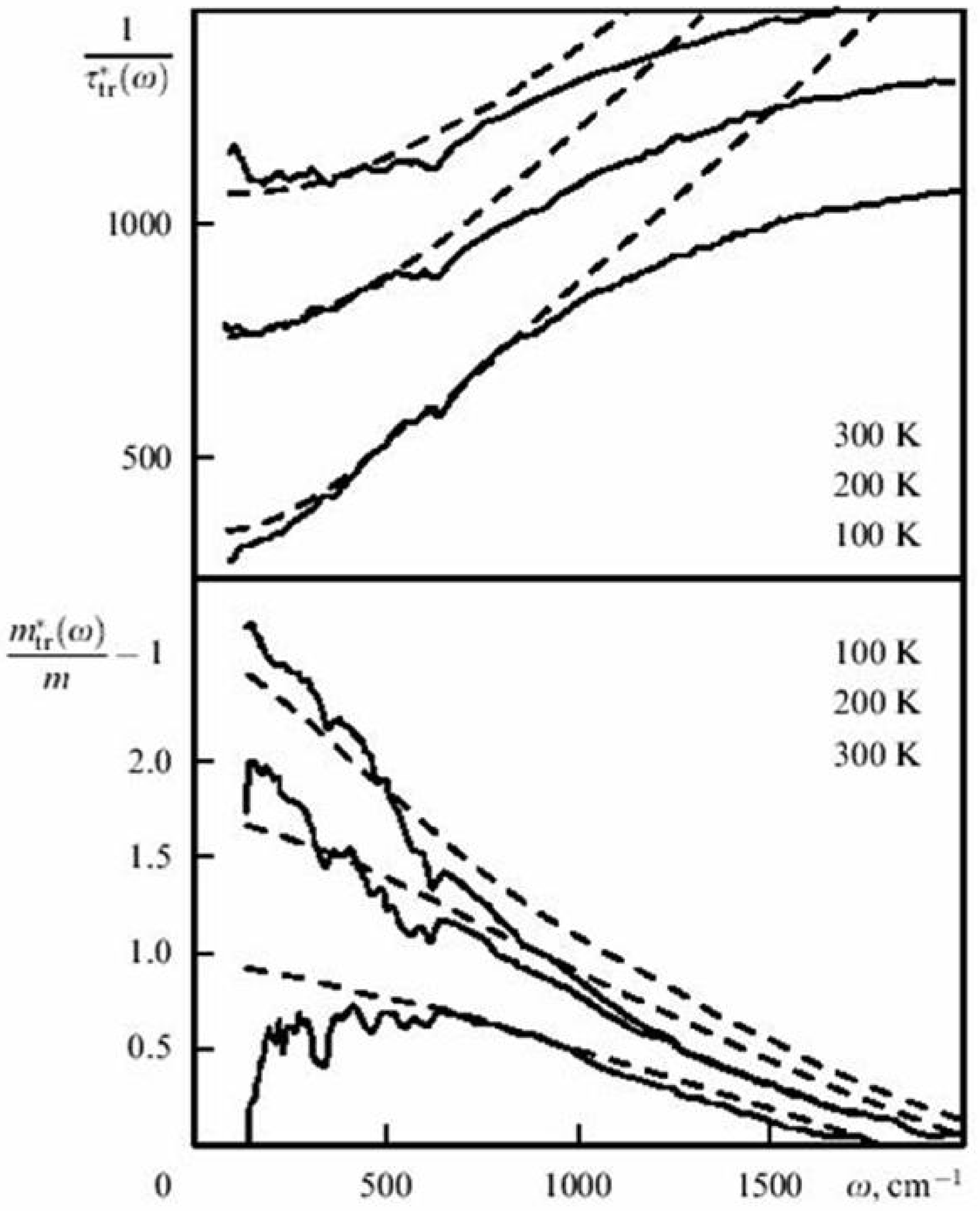}}
\caption{(top) Experimental transport scattering rate $\protect\gamma %
_{tr}^{\ast }$ (solid lines) for BISCO and the theoretical curve by using
Eq. (\protect\ref{Sigma-tr}) and transport mass $m _{tr}^{\ast }$ with $%
\protect\alpha^{2}F(\protect\omega)$ due to EPI which is described
in text (dashed lines). (bottom) Comparison with the marginal
Fermi liquid theory - dashed lines. From
\protect\citep{MaksimovReview}.} \label{MaksRev15-16}
\end{figure}

Let us discuss briefly the experimental results for $R(\omega )$
and $\gamma _{tr}^{\ast }(\omega ,T)$ and compare these with
theoretical predictions obtained by using a single band model with
$\alpha _{ep}^{2}(\omega )F_{ph}(\omega )$ extracted from the
tunnelling data with the EPI coupling constant $\lambda
_{ep}\gtrsim 2$ \citep{TunnelingVedeneev}. In the case of YBCO the
agreement between measured and calculated $R(\omega )$ is very
good up to energies $\omega <6000$ $cm^{-1}$ which confirms the
importance of EPI in scattering processes. For higher energies,
where a mead-infrared peak
appears, it is necessary to account for inter-band transitions \citep%
{MaksimovReview}. In optimally doped $Bi_{2}Sr_{2}CaCu_{2}O_{6}$
($Bi-2212$) \citep{Romero92} the experimental results for $\gamma
_{tr}^{\ast }(\omega ,T) $ are explained theoretically by assuming
that the EPI spectral function
$\alpha _{ep}^{2}(\omega )F(\omega )\sim F_{ph}(\omega )$, where $%
F_{ph}(\omega )$ is the phononic density of states in BISCO while $\alpha
_{ep}^{2}(\omega )\sim \omega ^{1.6}$, $\lambda _{ep}=1.9$ and $\gamma
_{imp}\approx 320$ $cm^{-1}$ - see Fig. ~\ref{MaksRev15-16}(top). The
agreement is rather good. At the same time the fit of $\gamma _{tr}^{\ast
}(\omega ,T)$ by the marginal Fermi liquid phenomenology fails as it is
evident in Fig.~\ref{MaksRev15-16}(bottom).

\begin{figure}[!tbp]
\resizebox{.5\textwidth}{!} {\includegraphics*[
width=6cm]{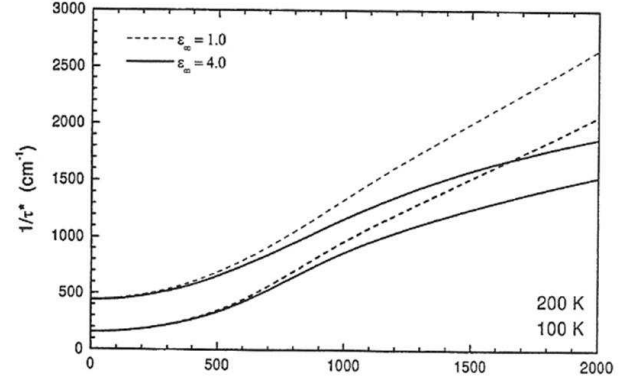}}
\caption{Dependence of $\protect\gamma _{tr}^{\ast }(\protect\omega ,T)$ on $%
\protect\varepsilon _{\infty }$ in $Bi_{2}Sr_{2}CaCu_{2}O_{8}$ for different
temperatures: $\protect\varepsilon _{\infty }=4$ (solid lines) and $\protect%
\varepsilon _{\infty }=1$ (dashed lines). On the horizontal is $\protect%
\omega$ in units $cm^{-1}$. From \protect\citep{Kaufmann}.}
\label{GammaEpsilon}
\end{figure}

Now we will comment the so called pronounced linear behavior of $\gamma
_{tr}(\omega ,T)$ (and $\gamma _{tr}^{\ast }(\omega ,T)$) which was one of
the main arguments for numerous inadequate conclusions regarding the
scattering and pairing bosons and EPI. We stress again that the measured
quantity is reflectivity $R(\omega )$ and derived ones are $\sigma (\omega )$%
, $\gamma _{tr}(\omega ,T)$ and $m_{tr}(\omega )$, which are very sensitive
to the value of the dielectric constant $\varepsilon _{\infty }$. This
sensitivity is clearly demonstrated in Fig.~\ref{GammaEpsilon} for Bi-2212
where it is seen that $\gamma _{tr}(\omega ,T)$ (and $\gamma _{tr}^{\ast
}(\omega ,T)$) for $\varepsilon _{\infty }=1$ is linear up to much higher $%
\omega $ than in the case $\varepsilon _{\infty }>1$.

However, in some experiments \citep{Puchkov}, \citep{TimuskOld} the extracted $%
\gamma _{tr}(\omega ,T)$ (and $\gamma _{tr}^{\ast }(\omega ,T)$) is linear
up to very high $\omega \approx 1500$ $cm^{-1}$. This means that the ion
background and inter-band transitions (contained in $\varepsilon _{\infty }$%
) are not properly taken into account since too small $\varepsilon _{\infty
}(\gtrsim 1)$ is assumed. The recent elipsometric measurements on YBCO \citep%
{BorisMPI} give the value $\varepsilon _{\infty }\approx 4-6$, which gives
much less spectacular linearity in the relaxation rates $\gamma _{tr}(\omega
,T)$ (and $\gamma _{tr}^{\ast }(\omega ,T)$) than it was the case
immediately after the discovery of HTSC cuprates, where much smaller $%
\varepsilon _{\infty }$ was assumed.

Furthermore, we would like to comment on two points concerning $\sigma $, $%
\gamma _{tr}$, $\gamma $ and their interrelations. First, the
parametrization of $\sigma (\omega )$ with the generalized Drude formula in
Eq.(\ref{Drude}) and its relation to the transport scattering rate $\gamma
_{tr}(\omega ,T)$ and the transport mass $m_{tr}(\omega ,T)$ is useful if we
deal with electron-boson scattering in a single band problem. In \citep%
{Shulga}, \citep{DolgKulSDW} it is shown that $\sigma (\omega )$
of a two-band model with only elastic impurity scattering can be
represented by the generalized (extended) Drude formula with
$\omega $ and $T$ dependence
of effective parameters $\gamma _{tr}^{eff}(\omega ,T)$, $%
m_{tr}^{eff}(\omega ,T)$ despite the fact that the inelastic electron-boson
scattering is absent. To this end we stress that the single-band approach is
justified for a number of HTSC cuprates such as LSCO, BISCO etc. Second, at
the beginning we said that $\gamma _{tr}(\omega ,T)$ and $\gamma (\omega ,T)$
are physically different quantities and it holds $\gamma _{tr}(\omega
,T)\neq \gamma (\omega ,T)$. In order to give the physical picture and
qualitative explanation for this difference we assume that $\alpha
_{tr}^{2}F(\nu )\approx \alpha ^{2}F(\nu )$. In that case the renormalized
quasi-particle frequency $\tilde{\omega}(\omega )=Z(\omega )\omega =\omega
-\Sigma (\omega )$ and the transport one $\tilde{\omega}_{tr}(\omega )$ -
defined in Eq.(\ref{Drude}), are related and at $T=0$ they are given by \citep%
{Allen}, \citep{Shulga}
\begin{equation}
\tilde{\omega}_{tr}(\omega )=\frac{1}{\omega }\int_{0}^{\omega }d\omega
^{\prime }2\tilde{\omega}(\omega ^{\prime }).  \label{tr-qp}
\end{equation}%
(For the definition of $Z(\omega )$ see \textit{Appendix} \textit{A}.) It
gives the relation between $\gamma _{tr}(\omega )$ and $\gamma (\omega )$, $%
m_{tr}(\omega )$ and $m^{\ast }(\omega )$, respectively
\begin{equation}
\gamma _{tr}(\omega )=\frac{1}{\omega }\int_{0}^{\omega }d\omega ^{\prime
}\gamma (\omega ^{\prime })  \label{gammatr-gamma}
\end{equation}%
\begin{equation}
\omega m_{tr}(\omega )=\frac{1}{\omega }\int_{0}^{\omega }d\omega ^{\prime
}2\omega ^{\prime }m^{\ast }(\omega ^{\prime }).  \label{mtr-m}
\end{equation}%
The physical meaning of Eq.(\ref{tr-qp}) is the following: in optical
measurements one photon with the energy $\omega $ is absorbed and two
excited particles (electron and hole) are created above and below the Fermi
surface. If the electron has energy $\omega ^{\prime }$ and the hole $\omega
-\omega ^{\prime }$, then they relax as quasi-particles with the
renormalized frequency $\tilde{\omega}$. Since $\omega ^{\prime }$ takes
values $0<\omega ^{\prime }<\omega $ then the optical relaxation $\tilde{%
\omega}_{tr}(\omega )$ is the energy-averaged
$\tilde{\omega}(\omega )$ according to Eq.(\ref{tr-qp}). The
factor $2$ is due to the two quasi-particles - electron and hole.
At finite $T$, the generalization reads \citep{Allen},
\citep{Shulga}
\begin{equation}
\tilde{\omega}_{tr}(\omega )=\frac{1}{\omega }\int_{0}^{\infty }d\omega
^{\prime }[1-n_{F}(\omega ^{\prime })-n_{F}(\omega -\omega ^{\prime })]2%
\tilde{\omega}(\omega ^{\prime }).  \label{omtr-T}
\end{equation}

2. \textit{Inversion of the optical data and }$\alpha _{tr}^{2}(\omega
)F(\omega )$\textit{\ }

\begin{figure}[!tbp]
\resizebox{.4\textwidth}{!} {\includegraphics*[
width=6cm]{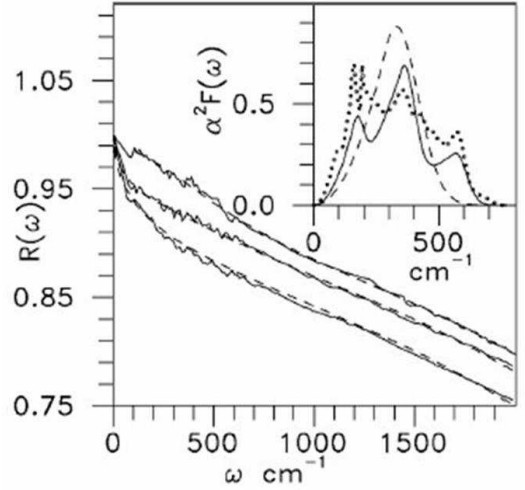}}
\caption{Experimental (solid lines) and calculated (dashed lines) of $R(%
\protect\omega)$ in optimally doped YBCO
\protect\citep{Schutzmann} at T=100, 200, 300 K (from top to
bottom). Inset: the two (solid and dashed lines) reconstructed
$\protect\alpha _{tr}^{2}(\protect\omega )F(\protect\omega )$ at
T=100 K. The phonon density of states $F(\protect\omega )$ -
dotted line in the inset. From \protect\citep{DolgovShulga}.}
\label{ShulgaFig1}
\end{figure}

\begin{figure}[!tbp]
\resizebox{.4\textwidth}{!} {\includegraphics*[
width=6cm]{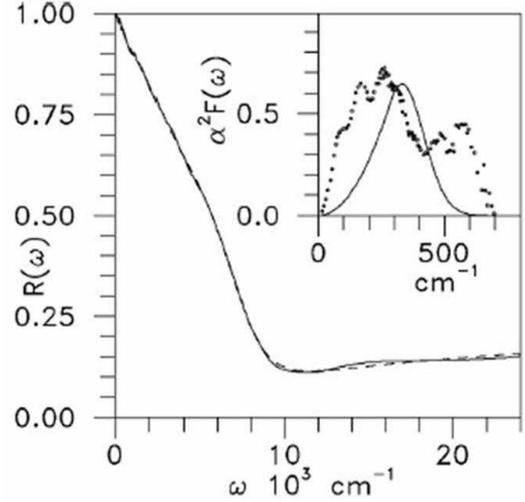}}
\caption{Experimental (solid line) and calculated (dashed line) of $R(%
\protect\omega )$ in optimally doped BISCO \protect\citep{Kamaras}
at T=100
K. Inset: the reconstructed $\protect\alpha _{tr}^{2}(\protect\omega )F(%
\protect\omega )$ - solid line. The phonon density of states $F(\protect%
\omega )$ - dotted line. From \protect\citep{DolgovShulga}.}
\label{ShulgaFig2}
\end{figure}

In principle, the transport spectral function $\alpha _{tr}^{2}(\omega
)F(\omega )$ can be extracted from $\sigma (\omega )$ (or $\gamma
_{tr}(\omega )$) only at $T=0$ $K$, which follows from Eq.( \ref{Gamma-tr-0}%
)
\begin{equation}
\alpha _{tr}^{2}(\omega )F(\omega )=\frac{\omega _{p}^{2}}{8\pi ^{2}}\frac{%
\partial ^{2}}{\partial \omega ^{2}}[\omega Re\frac{1}{\sigma (\omega )}],
\label{spect-func}
\end{equation}%
or equivalently $\alpha _{tr}^{2}(\omega )F(\omega )=(1/2\pi )\partial
^{2}(\omega \gamma _{tr}(\omega ))/\partial \omega ^{2}$. However, real
measurements are performed at finite $T$ (at $T>T_{c}$ which is rather high
in HTSC cuprates) and \textit{the inversion procedure is an ill-posed problem%
} since $\alpha _{tr}^{2}(\omega )F(\omega )$ is the de-convolution of the
inhomogeneous Fredholm integral equation of the first kind with the
temperature dependent Kernel $K_{2}(\omega ,\nu ,T)$ - see Eq.(\ref{Gamma-tr}%
). It is known that an ill-posed mathematical problem is very sensitive to
input since experimental data contain less information than one needs. This
procedure can cause \textit{first}, that the fine structure of $\alpha
_{tr}^{2}(\omega )F(\omega )$ \textit{gets blurred} (most peaks are washed
out) in the extraction procedures and \textit{second}, the \textit{extracted}
$\alpha _{tr}^{2}(\omega )F(\omega )$ \textit{is temperature dependent} even
when the true $\alpha _{tr}^{2}(\omega )F(\omega )$ is $T$ independent. This
artificial $T$-dependence is especially pronounced in HTSC cuprates because $%
T_{c}(\sim 100$ $K)$ is very high. In the context of HTSC
cuprates, this problem was first studied in \citep{DolgovShulga},
\citep{Shulga} where this picture is confirmed by the following
results: (\textit{1}) the extracted shape of $\alpha
_{tr}^{2}(\omega )F(\omega )$ in $YBa_{2}Cu_{3}O_{7-x}$\ (and
other cuprates) is not unique and it is temperature dependent,
i.e. at higher $T>T_{c}$ the peak structure is smeared and only a
single peak (slightly shifted to higher $\omega $) is present. For
instance, the experimental data of $R(\omega )$ in YBCO were
reproduced by two different spectral functions $\alpha
_{tr}^{2}(\omega )F(\omega )$, one with single
peak and the other one with three peaks structure as it is shown in Fig.~\ref%
{ShulgaFig1}, where all spectral functions give almost identical $R(\omega )$%
. The similar situation is \textit{realized in optimally doped} BISCO as it
is seen in Fig.~\ref{ShulgaFig2} where again different functions $\alpha
^{2}(\omega )F(\omega )$ reproduce very well curves for $R(\omega )$ and $%
\sigma (\omega )$. However, it is important to stress that the obtained
width of the extracted $\alpha _{tr}^{2}(\omega )F(\omega )$ in both
compounds coincide with the width of the phonon density of states $%
F_{ph}(\omega )$ \citep{DolgovShulga}, \citep{Shulga}, \citep{Kaufmann}. (%
\textit{2}) The upper energy bound for $\alpha _{tr}^{2}(\omega
)F(\omega )$ is extracted in \citep{DolgovShulga}, \citep{Shulga}
and it coincides approximately with the maximal phonon frequency
in cuprates $\omega
_{ph}^{\max }\lesssim 80$ $meV$ as it is seen in Figs. \ref{ShulgaFig1}-\ref%
{ShulgaFig2}.

These results demonstrate the importance of EPI in cuprates \citep%
{DolgovShulga}, \citep{Shulga}. We point out that the width of
$\alpha _{tr}^{2}(\omega )F(\omega )$ which is extracted from the
optical measurements \citep{DolgovShulga}, \citep{Shulga}
coincides with the width of the quasi-particle spectral function
$\alpha ^{2}(\omega )F(\omega )$ obtained in \textit{tunnelling
and ARPES spectra} (which we shall discuss below), i.e. both
functions are spread over the energy interval $0<\omega <\omega
_{ph}^{\max }(\lesssim 80$ $meV)$. Since in cuprates this interval
coincides with the width in the phononic density of states
$F(\omega )$ and since the maxima of $\alpha ^{2}(\omega )F(\omega
)$ and $F(\omega )$ almost coincide, this is further evidence for
the importance of EPI.

To this end, we would like to comment two aspects which appear
from time to time in the literature. \textit{First}, in some
reports \citep{Carbotte}, \citep{HwangTimusk1},
\citep{HwangTimusk2} it is assumed that $\alpha _{tr}^{2}(\omega
)F(\omega )$ of cuprates can be extracted also in the
superconducting state by using Eq. (\ref{spect-func}). However, Eq. (\ref%
{spect-func}) holds exclusively in the normal state (at $T=0$)
since $\sigma (\omega )$ can be described by the generalized
(extended) Drude formula in Eq. (\ref{Drude}) only in the normal
state. Such an approach does not hold in the superconducting state
since the dynamical conductivity depends not only on the
electron-boson scattering but also on coherence factors and on the
momentum and energy dependent order parameter $\Delta
(\mathbf{k},\omega )$. In such a case it is unjustified to extract
$\alpha _{tr}^{2}(\omega )F(\omega )$ from Eq. (\ref{spect-func}).
\textit{Second}, if $R(\omega )$ (and $\sigma (\omega )$) in
cuprates are due to some other bosonic scattering which is
pronounced up to much higher energies $\omega _{c}\gg \omega
_{ph}^{\max }$, this should be seen in the extracted spectral
function $\alpha _{tr}^{2}(\omega )F(\omega )$. Such an assumption
is made, for instance, in the phenomenological spin-fluctuation
approach \citep{Pines}
where it is assumed that $\alpha ^{2}(\omega )F(\omega )\sim g_{sf}^{2}$Im$%
\chi (\omega )$ where Im$\chi (\omega )$ is extended up to the
large energy cutoff $\omega _{c}\approx 400$ $meV$. This
assumption is in conflict with the above theoretical and
experimental analysis which shows that solely EPI can describe
$R(\omega )$ very well and that the contribution from higher
energies $\omega \gg \omega _{ph}^{\max }$ must be small and
therefore irrelevant for pairing \citep{DolgovShulga},
\citep{Shulga}, \citep{Kaufmann}. The finding of the importance of
EPI is also confirmed by tunnelling measurements - see discussion
in Subsection D. Despite the experimentally established facts,
that the energy width of the extracted $\alpha _{tr}^{2}(\omega
)F(\omega )$ coincides with the phononic range, which favor
EPI, some reports appeared recently claiming that SFI dominates and that $%
\alpha _{tr}^{2}(\omega )F(\omega )\sim g_{sf}^{2}Im\chi (\omega )$ where $%
Im\chi (\omega )=\int d^{2}k\chi (\mathbf{k},\omega )$
\citep{HwangTimusk1}, \citep{HwangTimusk2}. This claim is based on
reanalyzing of some IR measurements \citep{HwangTimusk1},
\citep{HwangTimusk2} and the transport
spectral function $\alpha _{tr}^{2}(\omega )F(\omega )$ is extracted in \citep%
{HwangTimusk1} by using the maximum entropy method in solving the Fredholm
equation. However, in order to exclude negative values in the extracted $%
\alpha _{tr}^{2}(\omega )F(\omega )$, which is an artefact and due
to the chosen method, in \citep{HwangTimusk1} it is assumed that
$\alpha _{tr}^{2}(\omega )F(\omega )$ has a rather large tail at
large energies - up to 400 meV. It turns out that even such an
assumption in extracting $\alpha
_{tr}^{2}(\omega )F(\omega )$ does not reproduce the experimental curve $%
Im\chi (\omega )$ \citep{Vignolle} in some important respects:
(\textit{i})
the relative heights of the two peaks in the extracted spectral function $%
\alpha _{tr}^{2}(\omega )F(\omega )$ at lower temperatures are
opposite to the experimental curve $Im\chi (\omega )$
\citep{Vignolle} - see Fig. 1 in \citep{HwangTimusk1}.
(\textit{ii}) the strong temperature dependence of the
extracted $\alpha _{tr}^{2}(\omega )F(\omega )$ found in \citep{HwangTimusk1}%
, \citep{HwangTimusk2} is in fact not an intrinsic property of the
spectral function but it is an artefact due to the high
sensitivity of the extraction procedure on temperature. As it is
already explained before, this is due to the ill-posed problem of
solving the Fredholm integral equation of the first kind with
strong $T$-dependent kernel. \textit{Third}, in fact the extracted
spectral weight $I_{B}(\omega )=\alpha _{tr}^{2}(\omega )F(\omega )$ in \citep%
{HwangTimusk1} has much smaller values at larger frequencies ($\omega >100$ $%
meV$) than it is the case for the measured Im$\chi (\omega )$, i.e. $%
(I_{B}(\omega >100$ $meV)/I_{B}(\omega _{\max }))\ll Im\chi (\omega >100$ $%
meV)/Im\chi (\omega _{\max })$ - see Fig. 1 in \citep{HwangTimusk1}. \textit{%
Fourth}, the recent magnetic neutron scattering measurements on
optimally-doped \textit{large-volume} crystals $Bi_{2}Sr_{2}CaCu_{2}O_{8+%
\delta }$ \citep{Tranquada2009} (where the absolute value of $Im\chi (\mathbf{%
q},\omega )$ is measured) are not only questioning the theoretical
interpretation of magnetism in HTSC cuprates in terms of itinerant
magnetism but they also oppose the finding in
\citep{HwangTimusk1}-\citep{HwangTimusk2}. This experiment shows
that the local spin susceptibility $Im\chi (\omega
)=\sum_{q}Im\chi (\mathbf{q},\omega )$ is temperature independent
across the superconducting transition $T_{c}=91$ $K$, i.e there is
only a minimal
change in $Im\chi (\omega )$ between $10$ $K$ and $100$ $K$. This $T$%
-independence of $Im\chi (\omega )$ strongly opposes the (above
discussed) results in \citep{Carbotte}, \citep{HwangTimusk1},
\citep{HwangTimusk2}, where the fit of optic measurements give
strong $T$-dependence of $Im\chi (\omega ) $.

\textit{Fifth}, the transport coupling constant $\lambda _{tr}$
extracted\ in \citep{HwangTimusk1} is too large, i.e. $\lambda
_{tr}>3$ contrary to the
previous findings that $\lambda _{tr}<1.5-2$ \citep{DolgovShulga}, \citep%
{Shulga}, \citep{Kaufmann}. Since in HTSC one has $\lambda
>\lambda _{tr}$ this would probably give $\lambda \approx 6$ what
is not confirmed by other experiments. \textit{Sixth}, the
interpretation of $\alpha _{tr}^{2}(\omega )F(\omega )$ in LSCO
and BISCO solely in terms of Im$\chi (\omega )$ is in
contradiction with the magnetic neutron scattering in the
optimally doped and slightly underdoped YBCO \citep{Bourges}. The
latter was discussed in\ Subsection A, where it is shown that
Im$\chi (\mathbf{Q},\omega )$ is small in the normal state and its
magnitude is even below the experimental noise. This means that if
the assumption that $\alpha _{tr}^{2}(\omega )F(\omega )\approx
g_{sf}^{2}$Im$\chi (\omega )$ were correct then the contribution
to Im$\chi (\omega )$ from the momenta $0<k\ll Q$ would be
dominant which is very detrimental for d-wave superconductivity.

Finally, we point out that very similar (to cuprates) properties, of $\sigma
(\omega )$, $R(\omega )$ (and $\rho (T)$ and electronic Raman spectra) were
observed in 3D isotropic metallic oxides $La_{0.5}Sr_{0.5}CoO_{3}$ and $%
Ca_{0.5}Sr_{0.5}RuO_{3}$ which are non-superconducting
\citep{Bozovic} and in
$Ba_{1-x}K_{x}BiO_{3}$ which superconducts below $T_{c}\simeq 30$ $K$ at $%
x=0.4$. This means that in all of these materials the scattering mechanism
might be of similar origin. Since in these compounds there are no signs of
antiferromagnetic fluctuations (which are present in cuprates), then EPI
plays important role also in other oxides.

3. \textit{Restricted optical sum-rule }

The \textit{restricted} \textit{optical sum-rule }was studied
intensively in HTSC cuprates. It shows some peculiarities not
present in low-temperature superconductors. It turns out that the
restricted spectral weight $W(\Omega _{c},T)$ is strongly
temperature dependent in the normal and superconducting state,
that was interpreted either to be due to EPI \citep{MaksKarakoz1},
\citep{MaksKarakoz2} or to some non-phononic mechanisms
\citep{Hirsch}. In the following we demonstrate that the
temperature dependence of $W(\Omega _{c},T)=W(0)-\beta T^{2}$ in
the normal state can be explained in a natural way by the
$T$-dependence of the EPI transport relaxation rate $\gamma
_{tr}^{ep}(\omega ,T)$ \citep{MaksKarakoz1}, \citep{MaksKarakoz2}.
Since the problem of the restricted sum-rule has attracted much
interest it will be considered here in some details. In fact,
there are two kinds of sum rules related to $\sigma (\omega )$.
The first one is the \textit{total sum rule} which in the normal
state reads

\begin{equation}
\int_{0}^{\infty }\sigma _{1}^{N}(\omega )d\omega =\frac{\omega _{pl}^{2}}{8}%
=\frac{\pi ne^{2}}{2m},  \label{TSR}
\end{equation}
while in the \textit{superconducting state} it is given by the
Tinkham-Ferrell-Glover (TFG) sum-rule

\begin{equation}
\int_{0}^{\infty }\sigma _{1}^{S}(\omega )d\omega =\frac{c^{2}}{8\lambda
_{L}^{2}}+\int_{+0}^{\infty }\sigma _{1}^{S}(\omega )d\omega =\frac{\omega
_{pl}^{2}}{8}.  \label{TSR-SC}
\end{equation}%
Here,$\ n$ is the total electron density, $e$ is the electron charge, $m$ is
the bare electron mass and $\lambda _{L}$ is the London penetration depth.
The first (singular) term $c^{2}/8\lambda _{L}^{2}$ in Eq.(\ref{TSR-SC}) is
due to the superconducting condensate which contributes $\sigma
_{1,cond}^{S}(\omega )=(c^{2}/4\lambda _{L}^{2})\delta (\omega )$. The total
sum rule represents the fundamental property of matter - the conservation of
the electron number. In order to calculate it one should use the total
Hamiltonian $\hat{H}_{tot}=\hat{T}_{e}+\hat{H}_{int}$ where all electrons,
electronic bands and their interactions $\hat{H}_{int}$ (Coulomb, EPI, with
impurities, etc.) are accounted for. Here, $T_{e}$ is the \textit{kinetic
energy of bare electrons}
\begin{equation}
\hat{T}_{e}=\sum_{\sigma }\int d^{3}x\hat{\psi}_{\sigma }^{\dagger }(x)\frac{%
\mathbf{\hat{p}}^{2}}{2m}\hat{\psi}_{\sigma }(x)=\sum_{\mathbf{p},\sigma }%
\frac{\mathbf{p}^{2}}{2m_{e}}\hat{c}_{\mathbf{p}\sigma }^{\dagger }\hat{c}_{%
\mathbf{p}\sigma }.  \label{Kin-energy}
\end{equation}

The \textit{partial sum rule} is related to the energetics solely in the
\textit{conduction (valence) band\ }which is described by the Hamiltonian of
the conduction (valence) band electrons
\begin{equation}
\hat{H}_{v}=\sum_{\mathbf{p},\sigma }\xi _{\mathbf{p}}\hat{c}_{v,\mathbf{p}%
\sigma }^{\dagger }\hat{c}_{v,\mathbf{p}\sigma }+\hat{V}_{v,c}.  \label{Hval}
\end{equation}%
$\hat{H}_{v}$ contains the band-energy with the dispersion $\epsilon _{%
\mathbf{p}}$ ($\xi _{\mathbf{p}}=\epsilon _{\mathbf{p}}-\mu $) and the
effective Coulomb interaction of the valence electrons $\hat{V}_{v,c}$. In
this case the \textit{partial sum-rule} in the normal state reads \citep%
{Maldague} (for general form of $\epsilon _{\mathbf{p}}$)
\begin{equation}
\int_{0}^{\infty }\sigma _{1,v}^{N}(\omega )d\omega =\frac{\pi e^{2}}{2V}%
\sum_{\mathbf{p}}\frac{\langle \hat{n}_{v,\mathbf{p}}\rangle _{H_{v}}}{m_{%
\mathbf{p}}}  \label{Total}
\end{equation}%
where the number operator $\hat{n}_{v,\mathbf{p}}=\sum_{\sigma }\hat{c}_{%
\mathbf{p}\sigma }^{\dagger }\hat{c}_{\mathbf{p}\sigma }$; $1/m_{\mathbf{p}%
}=\partial ^{2}\epsilon _{\mathbf{p}}/\partial p_{x}^{2}$ is the momentum
dependent reciprocal mass and $V$ is volume. In practice, measurements are
performed up to finite frequency and the integration over $\omega $ goes up
to some cutoff frequency $\Omega _{c}$ (of the order of the band plasma
frequency). In this case the restricted sum-rule has the form

\begin{equation*}
W(\Omega _{c},T)=\int_{0}^{\Omega _{c}}\sigma _{1,v}^{N}(\omega )d\omega
\end{equation*}%
\begin{equation}
=\frac{\pi }{2}\left[ K^{d}+\Pi (0)\right] -\int_{0}^{\Omega _{c}}\frac{%
Im\Pi (\omega )}{\omega }d\omega .  \label{rest-sum}
\end{equation}%
where $K^{d}$ is the diamagnetic Kernel given by Eq.(\ref{app-rest-sum})
below and $\Pi (\omega )$ is the paramagnetic (current-current) response
function. In the perturbation theory without vertex correction $\Pi (i\omega
_{n})$ (at the Matsubara frequency $\omega _{n}=\pi T(2n+1)$)\ is given by
\citep{MaksKarakoz1}, \citep{MaksKarakoz2}%
\begin{equation*}
\Pi (i\omega )=2\sum_{\mathbf{p}}\left( \frac{\partial \epsilon _{\mathbf{p}}%
}{\partial \mathbf{p}}\right) ^{2}\sum_{\omega _{m}}G(\mathbf{p},i\omega
_{n}+i\omega _{m})G(\mathbf{p},i\omega _{m}),
\end{equation*}%
where $G(\mathbf{p},i\omega _{n})=(i\omega _{n}-\xi _{\mathbf{p}}-\Sigma (%
\mathbf{p},i\omega _{n}))^{-1}$ is the electron Green's function. In the
case when the inter-band gap $E_{g}$ is the largest scale in the problem,
i.e. when $W_{b}<\Omega _{c}<E_{g}$, in this region one has approximately Im$%
\Pi (\omega )\approx 0$ and the limit $\Omega _{c}\rightarrow \infty $ in
Eq.(\ref{rest-sum}) is justified. In that case one has $\Pi (0)\approx
\int_{0}^{\Omega _{c}}($Im$\Pi (\omega )/\omega )d\omega $ which gives the
\textit{approximate formula} for $W(\Omega _{c},T)$
\begin{equation*}
W(\Omega _{c},T)=\int_{0}^{\Omega _{c}}\sigma _{1,v}^{N}(\omega )d\omega
\approx \frac{\pi }{2}K^{d}
\end{equation*}%
\begin{equation}
=e^{2}\pi \sum_{\mathbf{p}}\frac{\partial ^{2}\epsilon _{\mathbf{p}}}{%
\partial \mathbf{p}^{2}}n_{\mathbf{p}},  \label{app-rest-sum}
\end{equation}%
where $n_{\mathbf{p}}=\left\langle \hat{n}_{v,\mathbf{p}}\right\rangle $ is
the quasi-particle distribution function in the interacting system. Note,
that $W(\Omega _{c},T)$ is cutoff dependent while $K^{d}$ in Eq.(\ref%
{app-rest-sum}) does not depend on the cutoff energy $\Omega _{c}$. So, one
should be careful not to over-interpret the experimental results in cuprates
by this formula. In that respect the best way is to calculate $W(\Omega
_{c},T)$ by using the exact result in Eq.(\ref{rest-sum}) which apparently
depends on $\Omega _{c}$. However, Eq.(\ref{app-rest-sum}) is useful for
appropriately chosen $\Omega _{c}$, since it allows us to obtain
semi-quantitative results. In most papers\ related to the restricted
sum-rule in HTSC cuprates it was assumed, due to simplicity, the \textit{%
tight-binding model with nearest neighbors} (n.n.) with the energy $\epsilon
_{\mathbf{p}}=-2t(\cos p_{x}a+\cos p_{y}a)$ which gives $1/m_{\mathbf{p}%
}=-2ta^{2}\cos p_{x}a$. It is straightforward to show that in this case Eq.(%
\ref{app-rest-sum}) is reduced to a simpler one
\begin{equation*}
W(\Omega _{c},T)=\int_{0}^{\Omega _{c}}\sigma _{1,v}^{N}(\omega )d\omega
\end{equation*}%
\begin{equation}
\approx \frac{\pi e^{2}a^{2}}{2V}\langle -T_{v}\rangle ,  \label{Partial}
\end{equation}%
where $\langle T_{v}\rangle _{H_{v}}=\sum_{\mathbf{p}}\epsilon _{\mathbf{p}%
}\langle n_{v}\rangle _{H_{v}}$ is the average kinetic energy of the band
electrons, $a$ is the $Cu-Cu$ lattice distance and $V$ is the volume of the
system. In this approximation $W(\Omega _{c},T)$ is a direct measure of the
average band (kinetic) energy. In\textit{\ }the\textit{\ superconducting
state} the partial band sum-rule reads
\begin{equation*}
W_{s}(\Omega _{c},T)=\frac{c^{2}}{8\lambda _{L}^{2}}+\int_{+0}^{\Omega
_{c}}\sigma _{1,v}^{S}(\omega )d\omega
\end{equation*}%
\begin{equation}
=\frac{\pi e^{2}a^{2}}{2V}\langle -T_{v}\rangle _{s}.  \label{Partial-sc}
\end{equation}%
In order to introduce the reader to (the complexity of) the problem of the $%
T $-dependence of $W(\Omega _{c},T)$, let us consider the electronic system
in the normal state and in absence of quasi-particle interaction. In that
case one has $n_{\mathbf{p}}=f_{\mathbf{p}}$ ($f_{\mathbf{p}}$ is the Fermi
distribution function) and $W_{n}(\Omega _{c},T)$ increases with the
decrease temperature, i.e. $W_{n}(\Omega _{c},T)=W_{n}(0)-\beta _{b}T^{2}$
where $\beta _{b}\sim 1/W_{b}$. To this end, let us mention in advance that
the experimental value $\beta _{\exp }$ is much larger than $\beta _{b}$,
i.e. $\beta _{\exp }\gg \beta _{b}$ thus telling us that the simple
Sommerfeld-like smearing of $f_{\mathbf{p}}$ by the temperature effects
cannot explain quantitatively the T-dependence of $W(\Omega _{c},T)$. We
stress that the smearing of $f_{\mathbf{p}}$ by temperature lowers the
spectral weight compared to that at $T=0$ $K$, i.e. $W_{n}(\Omega
_{c},T)<W_{n}(\Omega _{c},0)$. In that respect it is not surprising that
there is a lowering of $W_{s}(\Omega _{c},T)$ in the BCS superconducting
state, $W_{s}^{BCS}(\Omega _{c},T\ll T_{c})<W_{n}(\Omega _{c},T\ll T_{c})$
since at low temperatures $f_{\mathbf{p}}$ is smeared mainly due to the
superconducting gap, i.e. $f_{\mathbf{p}}=[1-(\xi _{\mathbf{p}}/E_{\mathbf{p}%
})th(E_{\mathbf{p}}/2T)]/2$, $E_{\mathbf{p}}=\sqrt{\xi _{\mathbf{p}%
}^{2}+\Delta ^{2}}$, $\xi _{\mathbf{p}}=\epsilon _{\mathbf{p}}-\mu $. The
maximal decrease of $W_{s}(\Omega _{c},T)$ is at $T=0$.

\begin{figure}[!tbp]
\resizebox{.5\textwidth}{!} {\includegraphics*[
width=6cm]{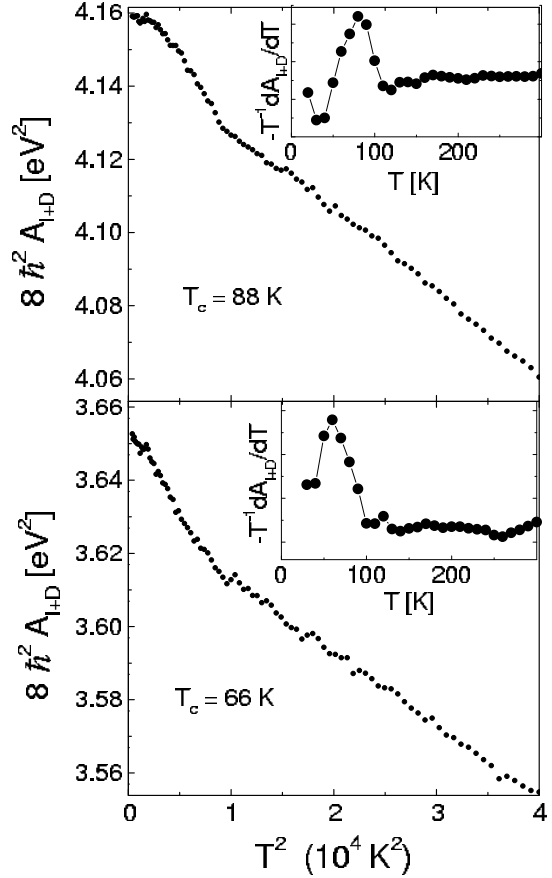}} \caption{Measured spectral weight
$W_{s}(\Omega _{c},T)$($\sim A_{l+D}$ in figures) for
$\protect\omega _{c}\approx 1.25eV$ in two underdoped $Bi2212$
(with $T_{c}=88$ $K$ and $T_{c}=66$ $K$). From
\protect\citep{Molegraaf}.} \label{Weight-NonBCS}
\end{figure}

Let us enumerate and discuss the \textit{main experimental results for }$%
W(\Omega _{c},T)$ in HTSC cuprates: \textbf{1}. in the \textit{normal state}
($T>T_{c}$) of most cuprates, one has $W_{n}(\Omega _{c},T)=W_{n}(0)-\beta
_{ex}T^{2}$ with $\beta _{\exp }\gg \beta _{b}$, i.e. $W_{n}(\Omega _{c},T)$
is increasing by decreasing $T$, even at $T$ below $T^{\ast }$ - the
temperature for the opening of the pseudogap. The increase of $W_{n}(\Omega
_{c},T)$ from room temperature down to $T_{c}$ is no more than $5$ $\%$.\
\textbf{2}. In the \textit{superconducting state} ($T<T_{c}$) of some
underdoped and optimally doped Bi-2212 compounds \citep{Carbone}, \citep%
{Molegraaf} (and underdoped Bi-2212 films \citep{Santander2003})
there is \textit{an effective increase} of $W_{s}(\Omega _{c},T)$
with respect to that in the normal state, i.e. $W_{s}(\Omega
_{c},T)>W_{n}(\Omega _{c},T)$
for $T<T_{c}$. This is a \textit{non-BCS behavior} which is shown in Fig. ~%
\ref{Weight-NonBCS}. Note, that in the tight binding model the effective
band (kinetic) energy $\left\langle T_{v}\right\rangle $ is negative ($%
\left\langle T_{v}\right\rangle <0$) and in the standard BCS case Eq.(\ref%
{Partial-sc}) gives that $W_{s}(T<Tc)$ \textit{decreases} due to the
increase of $\left\langle T_{v}\right\rangle $. Therefore the experimental
increase of $W_{s}(T<Tc)$ by decreasing $T$ is called the non-BCS behavior.
The latter means a lowering of the kinetic energy $\left\langle
T_{v}\right\rangle $ which is frequently interpreted to be due either to
strong correlations or to a Bose-Einstein condensation (BEC) of the
preformed tightly bound Cooper pair-bosons, for instance bipolarons \citep%
{Vidmar}. It is known that in the latter case the kinetic energy of bosons
is decreased below the BEC critical temperature $T_{c}$. In \citep%
{DeutscherSumRule} it is speculated that the latter case might be realized
in in underdoped cuprates.

\begin{figure}[!tbp]
\resizebox{.5\textwidth}{!} {\includegraphics*[
width=6cm]{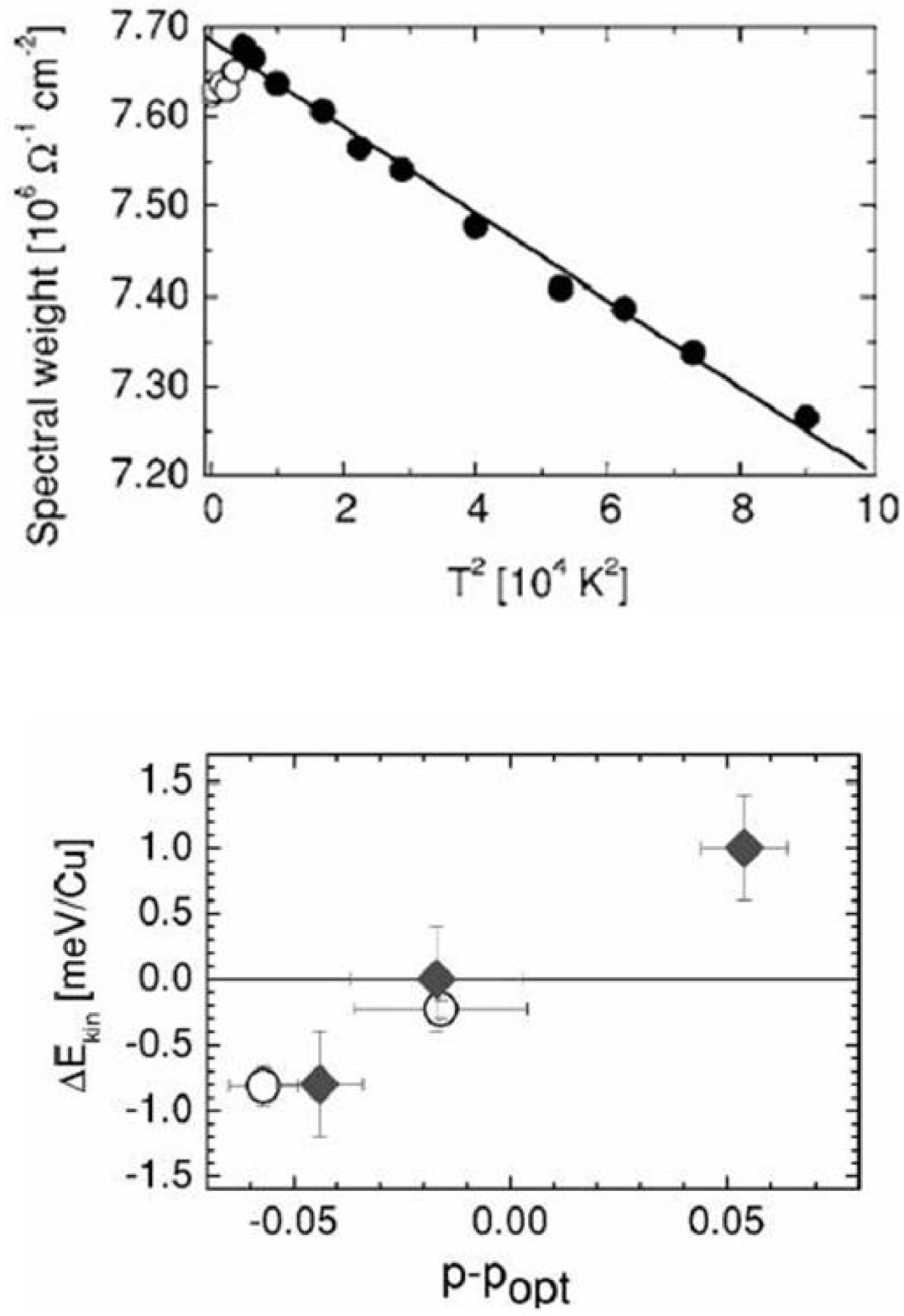}}
\caption{(top) Spectral weight $W_{n}(\Omega _{c},T)$ of the overdoped $%
Bi2212 $ for $\Omega _{c}=1 eV$. Closed symbols - normal state. Open symbols
- superconducting state. (bottom) Change of the kinetic energy $\Delta
E_{kin}=E_{kin,S}-E_{kin,N}$ in $meV$ per Cu site vs the charge $p$ per Cu
with respect to the optimal value $p_{opt}$. From \protect\citep%
{DeutscherOptics}.}
\label{Weight-BCS}
\end{figure}

However, in some optimally doped and in most overdoped cuprates, there is a
decrease of $W_{s}(\Omega _{c},T)$ at $T<T_{c}$ ($W_{s}(\Omega
_{c},T)<W_{n}(\Omega _{c},T)$) which is the BCS-like behavior \citep%
{DeutscherOptics} as it is seen in Fig.~\ref{Weight-BCS}

We stress that the non-BCS behavior of $W_{s}(\Omega _{c},T)$ for
underdoped (and in some optimally doped) systems was obtained by
assuming that $\Omega _{c}\approx (1-1.2)$ $eV$. However, in Ref.
\citep{BorisMPI} these results
were questioned and the conventional BCS-like behavior was observed ($%
W_{s}(\Omega _{c},T)<W_{n}(\Omega _{c},T)$) in the optimally doped YBCO and
slightly underdoped Bi-2212 by using larger cutoff energy $\Omega _{c}=1.5$ $%
eV$. This discussion demonstrates how risky is to make definite
conclusions on some fundamental physics based on the parameter
(such as the cutoff energy $\Omega _{c}$) dependent analysis.
Although the results obtained in \citep{BorisMPI} looks very
trustfully, it is fair to say that the issue of the reduced
spectral weight in the superconducting state of cuprates is still
unsettled and under dispute. In overdoped Bi-2212 films, the
BCS-like behavior $W_{s}(\Omega _{c},T)<W_{n}(\Omega _{c},T)$ was
observed, while in
LSCO it was found that $W_{s}(\Omega _{c},T)\approx const$, i.e. $%
W_{s}(\Omega _{c},T<T_{c})\approx W_{n}(\Omega _{c},T_{c})$.

\begin{figure}[!tbp]
\resizebox{.4 \textwidth}{!} {\includegraphics*[
width=6cm]{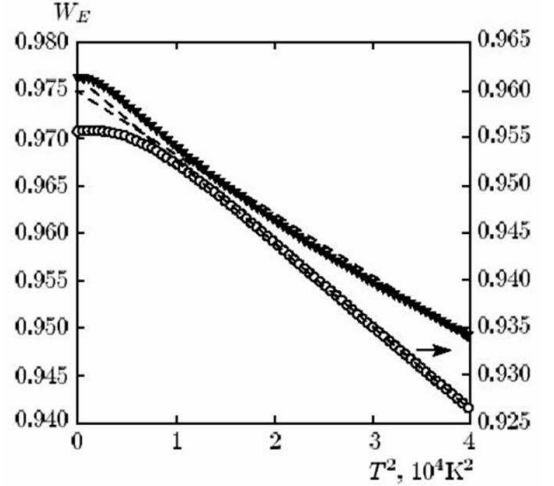}} \caption{Spectral weight $W(\Omega
_{c},T)$ in the normal state for Einstein phonons with $\Omega
_{E}=200$ $K$ (full triangles) and $\Omega _{E}=400$ $K$ (open
circles, left axis). Dashed lines is $T^{2}$ asymptotic. From
\protect\citep{MaksKarakoz2}.} \label{MaksKarakoz4}
\end{figure}

The first question is - how to explain the strong temperature dependence of $%
W(\Omega _{c},T)$ in the normal state? In \citep{MaksKarakoz1}, \citep%
{MaksKarakoz2} $W(T)$ is explained solely in the framework of the
EPI physics where the EPI relaxation $\gamma _{ep}(T)$ plays the
main role in the $T$-dependence of $W(\Omega _{c},T)$. The main
theoretical results of \citep{MaksKarakoz1}, \citep{MaksKarakoz2}
are the following: the calculations of $W(T)$\ based on exact
Eq.(\ref{app-rest-sum}) give that for $\Omega _{c}\gg \Omega _{D}$
(the Debye energy) the difference in spectral weights of the
normal and superconducting state is small, i.e. $W_{n}(\Omega
_{c},T)\approx W_{s}(\Omega _{c},T)$ since $W_{n}(\Omega
_{c},T)-W_{s}(\Omega _{c},T)\sim \Delta ^{2}/\Omega _{c}^{2}$. (2)
In the
case of large $\Omega _{c}$ the calculations based on Eq.(\ref{app-rest-sum}%
) gives
\begin{equation}
W(\Omega _{c},T)\approx \frac{\omega _{pl}^{2}}{8}\left[ 1-\frac{\gamma (T)}{%
W_{b}}-\frac{\pi ^{2}}{2}\frac{T^{2}}{W_{b}^{2}}\right] .  \label{Wapp}
\end{equation}%
In the case of EPI, one has $\gamma =\gamma _{ep}(T)+\gamma _{imp}$ where $%
\gamma _{ep}(T)=\int_{0}^{\infty }dz\alpha ^{2}(z)F(z)\coth (z/2T)$. It
turns out that for $\alpha ^{2}(\omega )F(\omega )$ shown in Fig.~\ref{Rates}%
, one obtains: (i) $\gamma _{ep}(T)\sim T^{2}$ in the temperature interval $%
100$ $K<T<200$ $K$ as it is seen in Fig.~\ref{MaksKarakoz4} for
the T-dependence of $W(\Omega _{c},T)$ \citep{MaksKarakoz1},
\citep{MaksKarakoz2}; (ii) the second term in Eq.(\ref{Wapp}) is
much larger than the last one (the Sommerfeld-like term). For the
EPI coupling constant $\lambda
_{ep,tr}=1.5$ one obtains rather good agreement between the theory in \citep%
{MaksKarakoz1}, \citep{MaksKarakoz2} and experiments \citep{BorisMPI}, \citep%
{Carbone}, \citep{Molegraaf}. At lower temperatures, $\gamma
_{ep}(T)$ deviates from the $T^{2}$ behavior and this deviation
depends on the structure of the spectrum in $\alpha ^{2}(\omega
)F(\omega )$. It is seen in Fig.~\ref{MaksKarakoz4} that for a
softer Einstein spectrum (with $\Omega _{E}=200$ $K$), $W(\Omega
_{c},T)$ lies above the curve with the $T^{2}$
asymptotic behavior, while the curve with a harder phononic spectrum (with $%
\Omega _{E}=400$ $K$) lies below it.
This result means that different behavior of $W(\Omega _{c},T)$ in
the superconducting state of cuprates for different doping might
be simply related to different contributions of low and high
frequency phonons. We stress that such a behavior of $W(\Omega
_{c},T)$ was observed in experiments \citep{BorisMPI},
\citep{Carbone}, \citep{Molegraaf}. To summarize, the above
analysis demonstrates that the theory based on EPI is able to
explain in a satisfactory way the strange temperature behavior of
$W(\Omega _{c},T)$ above and below $T_{c}$ in systems at and near
optimal doping and that \textit{there is no need to invoke exotic
scattering mechanisms}. In that respect we would like to stress
that at present we still do not have fully microscopic theory
which comprises strong correlations and EPI and
which is able to predict the complex behavior of $W(T)$ as a function of $T$%
, doping etc.

4. $\alpha ^{2}(\omega )F(\omega )$ \textit{and the} \textit{in-plane
resistivity }$\rho _{ab}(T)$

\begin{figure}[!tbp]
\resizebox{.4\textwidth}{!} {\includegraphics*[
width=6cm]{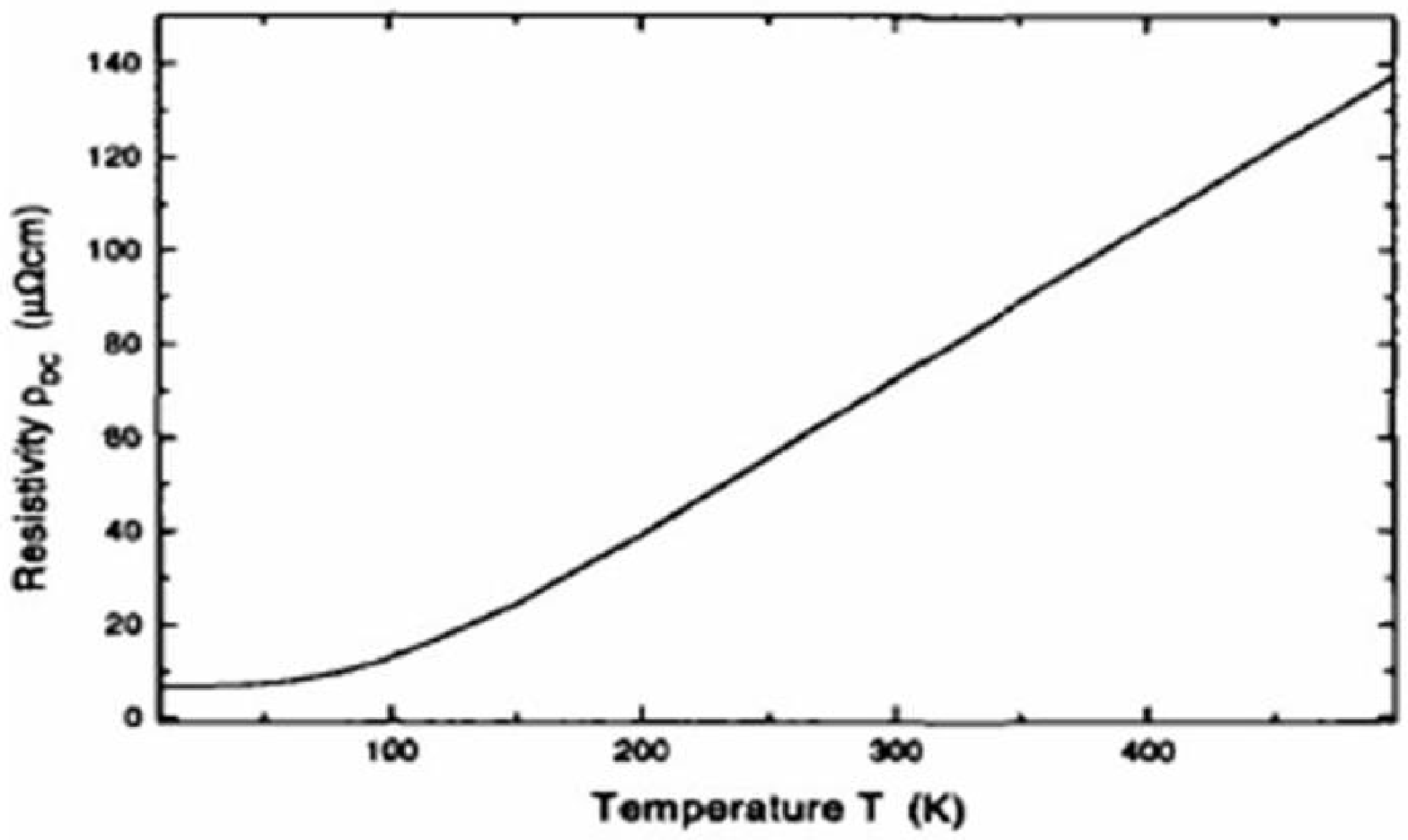}} {\includegraphics*
[width=8cm]{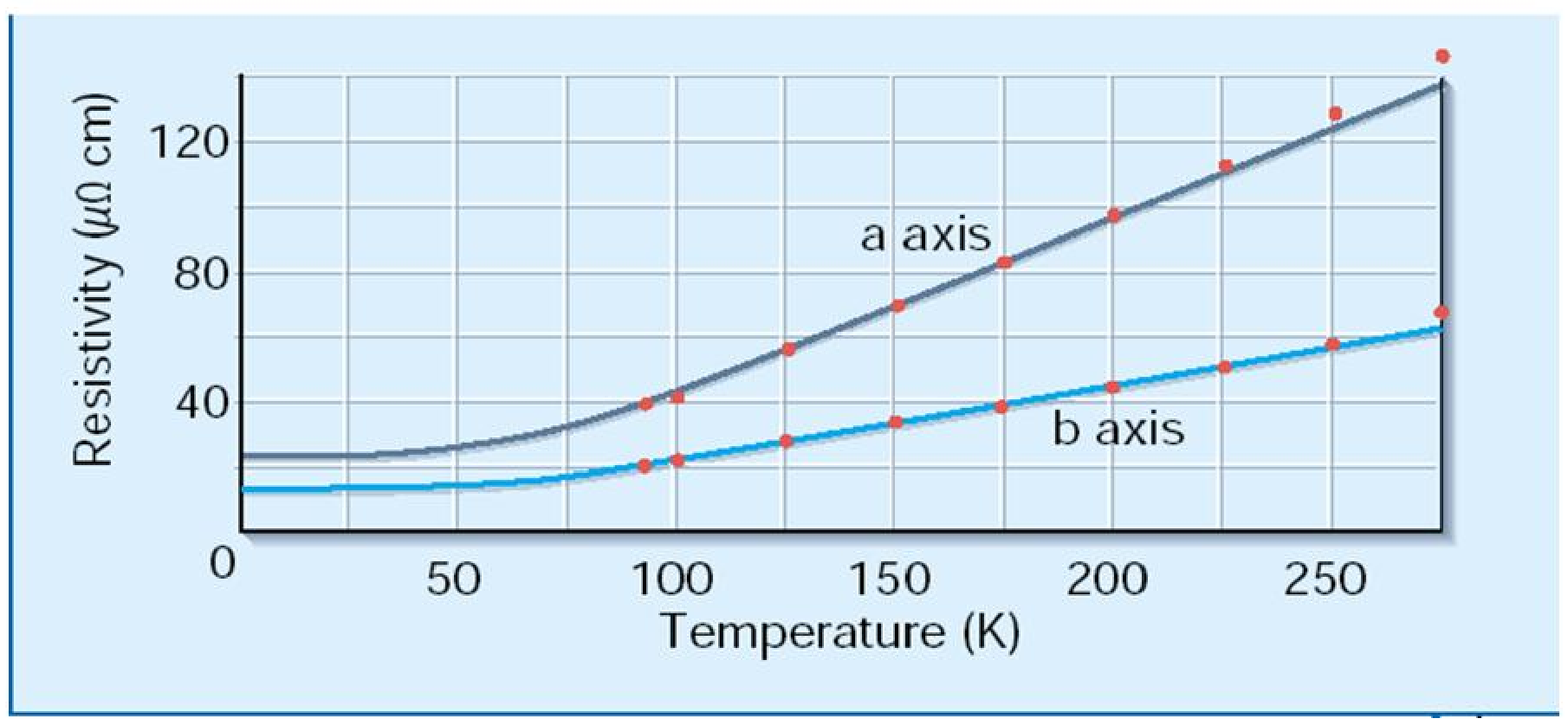}}
\caption{(top) Calculated resistivity $\protect\rho (T)$ for the EPI
spectral function $\protect\alpha _{tr}^{2}(\protect\omega )F(\protect\omega %
)$ in \protect\citep{KMS}. (bottom) Measured resistivity in a(x)-
and b(y)-crystal direction of YBCO \protect\citep{Friedman} and
calculated Bloch-Gr\"{u}neisen curve (points) for $\protect\lambda
^{ep}=1$, \protect\citep{AllenKinky}.} \label{Res-T}
\end{figure}

The temperature dependence of the in-plane resistivity $\rho _{ab}(T)$ in
cuprates is a direct consequence of the quasi-$2D$ motion of quasi-particles
and of the inelastic scattering which they experience. At present, there is
no consensus on the origin of the linear temperature dependence of the
in-plane resistivity $\rho _{ab}(T)$ in the normal state. Our intention is
not to discuss this problem, but only to demonstrate that the EPI\ spectral
function $\alpha ^{2}(\omega )F(\omega )$, which is obtained from tunnelling
experiments in cuprates (see Subsection D), is able to explain the
temperature dependence of $\rho _{ab}(T)$ in optimally doped $YBCO$. In the
Boltzmann theory$\ \rho _{ab}(T)$ is given by
\begin{equation}
\rho _{ab}(T)=\rho _{imp}+\frac{4\pi }{\omega _{p}^{2}}\gamma _{tr}(T)
\label{inv-sig}
\end{equation}%
\begin{equation}
\gamma _{tr}(T)=\frac{\pi }{T}\int_{0}^{\infty }d\omega \frac{\omega }{\sinh
^{2}(\omega /2T)}\alpha _{tr}^{2}(\omega )F(\omega ),  \label{g-tr-T}
\end{equation}%
where $\rho _{imp}$ is the impurity scattering. Since $\rho (T)=1/\sigma
(\omega =0,T)$ and having in mind that the dynamical conductivity $\sigma
(\omega ,T)$ in $YBCO$ (at and near the optimal doping) is satisfactory
explained by the EPI scattering, then it is to expect that $\rho _{ab}(T)$
is also dominated by EPI in some temperature region $T>T_{c}$. This is
indeed \textit{confirmed in the optimally doped} $YBCO$, where $\rho _{imp}$
is chosen appropriately and the spectral function $\alpha _{tr}^{2}(\omega
)F(\omega )$ is taken from the tunnelling experiments in \citep%
{TunnelingVedeneev}. The very good agreement with the experimental
results \citep{KMS} is shown in Fig. \ref{Res-T}. We stress that
in the case of EPI
there is always a temperature region where $\gamma _{tr}(T)\sim T$ for $%
T>\alpha \Theta _{D}$, $\alpha <1$ depending on the shape of $\alpha
_{tr}^{2}(\omega )F(\omega )$ (for the simple Debye spectrum $\alpha \approx
0.2$). In the linear regime one has $\rho (T)\simeq \rho _{imp}+8\pi
^{2}\lambda _{ep,tr}(k_{B}T/\hbar \omega _{p}^{2})=\rho _{imp}+\rho ^{\prime
}T$.

\begin{figure}[!tbp]
\resizebox{.5\textwidth}{!} {\includegraphics*[
width=6cm]{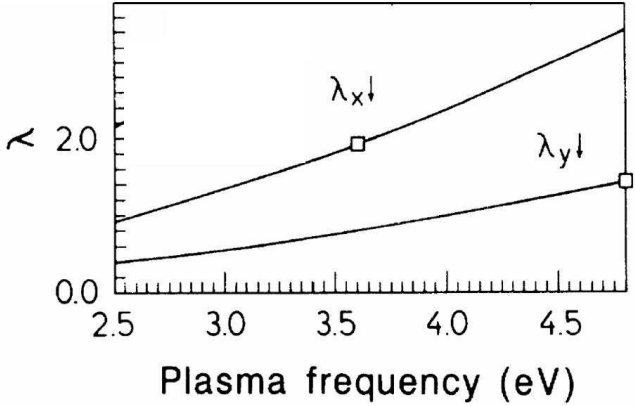}}
\caption{Transport EPI spectral function coupling constant in YBCO as a
function of plasma frequency $\protect\omega _{p}$ as derived from the
experimental slope of resistivity $\protect\rho ^{\prime }(T)$. $\protect%
\lambda _{x}$ for $\protect\rho _{x}^{\prime }(T)=0.6\protect\mu
\Omega \mathrm{cm}/K$ and $\protect\lambda _{y}$ for $\protect\rho
_{y}^{\prime }(T)=0.25\protect\mu \Omega \mathrm{cm}/K$
\protect\citep{Friedman}. Squares are LDA values
\protect\citep{MazinDolgov}.} \label{lam-plas}
\end{figure}

There is an experimental constraint on $\lambda _{tr}$, i.e. $\lambda
_{tr}\approx 0.25\omega _{pl}^{2}(eV)\rho ^{\prime }(\mu \Omega $ $cm/K)$,
which imposes a limit on it. For instance, for $\omega _{pl}\approx (2-3)$ $%
eV$ \citep{Bozovic} and $\rho ^{\prime }\approx 0.6$ in the
oriented YBCO films and $\rho ^{\prime }\approx 0.3-0.4$ in single
crystals of BSCO, one obtains $\lambda _{tr}\approx 0.6-1.4$. In
case of YBCO single crystals, there is a pronounced anisotropy in
$\rho _{ab}(T)$ \citep{Friedman} which gives $\rho _{x}^{\prime
}(T)=0.6$ $\mu \Omega \mathrm{cm}/K$ and $\rho _{y}^{\prime
}(T)=0.25$ $\mu \Omega \mathrm{cm}/K$. The function $\lambda
_{tr}(\omega _{pl})$ is shown in Fig.\ref{lam-plas} where the
plasma frequency $\omega _{pl}$ can be calculated by LDA-DFT and
also extracted from the width ($\sim $ $\omega _{pl}^{\ast })$ of
the Drude peak at small frequencies, where $\omega
_{pl}=\sqrt{\varepsilon _{\infty }}\omega _{pl}^{\ast }$. We
stress that the rather good agreement of theoretical and
experimental results for $\rho _{ab}(T)$ \textit{should be not
over-interpreted} in the sense that the above rather simple
electron-phonon approach can explain the resistivity in HTSC for
various doping. In fact this in highly underdoped systems where
$\rho _{ab}(T)$ is very different from the behavior in Fig.
\ref{Res-T}. In this case one should certainly
take into account polaronic effects \citep{Alexandrov}, \citep%
{GunnarssonReview2008}, strong correlations, etc. and in these cases the
simplified Migdal-Eliashberg theory is not sufficient. The above analysis on
the resistivity in optimally doped system demonstrates only, that if in Eq.(%
\ref{g-tr-T}) one uses the EPI spectral function $\alpha ^{2}(\omega
)F(\omega )$ obtained from the tunnelling experiments (and optics) one
obtains the correct $T$-dependence of $\rho _{a,b}(T)$ for some optimally
doped cuprates, which is an additional evidence for the importance of EPI.

Concerning the linear (in $T$) resistivity we would like to point out that
there is some evidence that linear resistivity is observed in HTSC cuprates
sometimes at temperatures $T<0.2$ $\Theta _{D}$ \citep{KondoResist}, \citep%
{MengResist}. In that respect, it was shown by
\citep{PickettResist} that in two-dimensional systems with a broad
interval of phonon spectra the quasi-linear behavior of $\rho
_{ab}(T)$ is realized even at $T<0.2$ $\Theta _{D}$. The
quasi-linear behavior of the resistivity at $T\ll 0.2$ $\Theta
_{D}$ has been observed in
Bi$_{2}$(Sr$_{0.97}$Pr$_{0.003}$)$_{2}$CuO$_{6}$
\citep{King} inLSCO and $1$-layer Bi-2201 \citep{KondoResist}, \citep%
{MengResist}, \citep{MartinResist}, \citep{VedeneevResistPhysica}, \citep%
{VedeneevResistPRB}, all systems with rather small $T_{c}\approx
10$ $K$. We would like to emphasize here that some of these
observations are contradictory. For example, the results obtained
in the group of Vedeneev \citep{VedeneevResistPhysica},
\citep{VedeneevResistPRB} show that some samples demonstrate the
quasi-linear behavior of the resistivity up to $T=10$ $K$ but some
others with approximately the same $T_{c}$ have the usual
Bloch-Gr\"{u}neisen type behavior characteristic for EPI. In that
respect it is very unlikely that the linear resistivity up to
$T=10$ $K$ can be simply explained in the standard way by
interactions of electrons with some known bosons \textit{either by
phonons or spin-fluctuations} (magnons). The
question, why in some cuprates the linear resistivity is observed up to $%
T=10 $ $K$ is still a mystery and its explanation is a challenge for all
kinds of the electron-boson scattering, not only for EPI. In that respect it
is interesting to mention that the existence of the forward scattering peak
in EPI (with the width $q_{c}\ll k_{F}$), which is due to strong
correlations, may give rise to the linear behavior of $\rho (T)$ down to
very low temperatures $T\sim \Theta _{D}/30$ \citep{KulicReview}, \citep%
{VarelogResist}, \citep{KulicDolgovResistAIP} - see more in Part
II, Section VII.D item 6..
We shall argue in Subsection D that if one interprets the tunnelling
experiments in systems near optimal doping \citep{TunnelingVedeneev}-\citep%
{Tsuda} in the framework of the Eliashberg theory one obtains large EPI
coupling constant $\lambda _{ep}\approx 2-3.5$ which implies that $\lambda
_{tr}\sim (\lambda /3)$. This means that EPI is reduced much more in
transport properties than in the self-energy, which is due to some reasons
that shall be discussed in \textit{Part II}. Such a large reduction of $%
\lambda _{tr}$ cannot be obtained within the LDA-DFT band
structure calculations which means that $\lambda _{ep}$ and
$\lambda _{tr}$ contain renormalization which do not enter in the
$LDA-DFT$ theory. In \textit{Part II} we shall argue that the
strong suppression of $\lambda _{tr}$ may have its origin in
strong electronic correlations \citep{Kulic1}, \citep{Kulic2}
and in the long-range Madelung energy \citep{MaksimovReview}, \citep%
{KulicReview}.

4. \textit{Femtosecond time-resolved optical spectroscopy}

The femtosecond time-resolved optical spectroscopy\textit{\
}(FTROS) has been developed in the last couple of years and
applied to HTSC cuprates. In this method a femtosecond ($1$
$fs=10^{-15}\sec $) laser pump excites in materials electron-hole
pairs via interband transitions. These hot carriers release their
energy via electron-electron (with the relaxation time $\tau
_{ee}$) and electron-phonon scattering reaching states near the
Fermi energy within $10-100$ $fs$ - see \citep{MihailovicKabanov}.
The typical energy density of the laser pump pulses with the
wavelength $\lambda \approx 810$
nm ($\hbar \omega =1.5$ $eV$) was around $F\sim 1\mu J/cm^{2}$ (the \textit{%
excitation fluence} $F$) which produces approximately $3\times 10^{10}$
carriers per pulse (by assuming that each photon produces $\hbar \omega
/\Delta $ carriers, $\Delta $ is the superconducting gap). By measuring
photoinduced changes of the reflectivity in time, i.e. $\Delta R(t)/R_{0}$,
one can extract information on the relaxation dynamics of the low-laying
electronic excitations. Since $\Delta R(t)$ relax to equilibrium the fit
with exponential functions is used
\begin{equation}
\frac{\Delta R(t)}{R_{0}}=f(t)\left[ Ae^{-\frac{t}{\tau _{A}}}+Be^{-\frac{t}{%
\tau _{B}}}+...\right] ,  \label{ftr}
\end{equation}%
where $f(t)=H(t)[1-\exp \{-t/\tau _{ee}\}]$ ($H(t)$ is the Heavyside
function) describes the finite rise-time. The parameters $A$, $B$ depends on
the fluence $F$. This method was used in studying the superconucting phase
of $La_{2-x}Sr_{x}CuO_{4}$, with $x=0.1$, $0.15$ and $T_{c}=30$ $K$ and $38$
$K$ respectively \citep{Kusar2008}. In that case one has $A\neq 0$ for $T$$<$$%
T_{c}$ and $A=0$ for $T>T_{c}$, while the signal $B$ was present also at $%
T>T_{c}$. It turns out that the signal $A$ is related to the quasi-particle
recombination across the superconducting gap $\Delta (T)$ and has a
relaxation time of the order $\tau _{A}>10$ $ps$ at $T=4.5$ $K$. At the so
called threshold fluence ($F_{T}=4.2\pm 1.7$ $\mu J/cm^{2}$ for $x=0.1$ and $%
F_{T}=5.8\pm 2.3$ $\mu J/cm^{2}$ for $x=0.15$) the vaporization (destroying)
of the superconducting phase occurs, where the parameter $A$ saturates. This
vaporization process takes place at the time scala $\tau _{r}\approx 0.8$ $%
ps $. The external fluence is distributed in the sample over the \textit{%
excitation volume} which is proportional to the optical penetration depth $%
\lambda _{op}$($\approx 150$ $nm$ at $\lambda \approx 810$ $nm$) of the
pump. The energy densities stored in the excitation volume at the
vaporization threshold for $x=0.1$ and $x=0.15$ are $U_{p}=F_{T}/\lambda
_{op}=2.0\pm 0.8$ $K/Cu$ and $2.6\pm 1.0$ $K/Cu$, respectively. The
important fact is that $U_{p}$ is much larger than the superconducting
condensation energy which is $U_{cond}\approx 0.12$ $K/Cu$ for $x=0.1$ and $%
U_{cond}\approx 0.3$ $K/Cu$ for $x=0.15$, i.e. $U_{p}\gg U_{cond}$. This
means that the energy difference $U_{p}-U_{cond}$ must be stored elsewhere
on the time scale $\tau _{r}$. The only present reservoir which can absorb
the difference in energy are the bosonic baths of phonons and spin
fluctuations. The energy required to heat the spin reservoir from $T=4.5$ $K$
to $T_{c}$ is $U_{sf}=\int_{T}^{T_{c}}C_{sf}(T)dT$. The measured specific
heat $C_{sf}(T)$ in $La_{2}CuO_{4}$ \citep{Kusar2008} gives very small value $%
U_{sf}\approx 0.01$ $K$. In the case of the phonon reservoir on obtains $%
U_{ph}=\int_{T}^{T_{c}}C_{ph}(T)dT=9$ $K/Cu$ for $x=0.1$ and $28$ $K/Cu$ for
$x=0.15$, where $C_{ph}$ is the phonon specific heat. Since $U_{sf}\ll
U_{p}-U_{cond}$ the spin reservoir cannot absorb the rest energy $%
U_{p}-U_{cond}$. The situation is opposite with phonons since
$U_{ph}\gg U_{p}-U_{cond}$ \ and phonon can absorb the rest energy
in the excitation volume. The complete vaporization dynamics can
be described in the framework of the Rothwarf-Taylor model which
describes approaching of electrons and phonons to
quasi-equilibrium on the time scale of 1 ps \citep{KabanovPRL}. We
shall not go into details but only summarize by quoting the
conclusion in \citep{KabanovPRL}, that only phonon-mediated
vaporization is consistent with the experiments, thus ruling out
spin-mediated quasi-particle recombination and pairing in HTSC
cuprates. The FTROS method tell us that at least for
non-equilibrium processes EPI\ is more important than SFI. It
gives also some opportunities for obtaining the strength of EPI
but at present there is no reliable analysis.

\textit{In conclusion}, optics and resistivity measurements in the normal
state of cuprates give evidence that EPI is important while the spin
fluctuation scattering is weaker than it is believed. However, some
important questions related to the transport properties remain to be
answered: (\textbf{i}) what are the values of $\lambda _{tr}$ and $\omega
_{pl}$; (\textbf{ii}) what is the reason that $\lambda _{tr}\ll \lambda $ is
realized in cuprates; (\textbf{iii}) what is the role of Coulomb scattering
in $\sigma (\omega )$ and $\rho (T)$. Later on we shall argue that ARPES
measurements in cuprates give evidence for an appreciable contribution of
Coulomb scattering at higher frequencies, where $\gamma (\omega )\approx
\gamma _{0}+\lambda _{c}\omega $ for $\omega >\omega _{\max }^{ph}$ with $%
\lambda _{c}\sim 1$. One should stress, that despite the fact that EPI is
suppressed in transport properties it is sufficiently strong in the
quasi-particle self-energy, as it comes out from tunnelling measurements
discussed below.

\subsection{ARPES and the EPI self-energy}

The angle-resolved photoemission spectroscopy (ARPES) is nowadays
one of leading spectroscopy methods in the solid state physics
\citep{ShenReview}. In some favorable conditions it provides
direct information on the one-electron removal spectrum in a
complex many body system. The method involves shining light
(photons) with energies between $E_{i}=5-1000$ $eV$
on samples and by detecting momentum ($\mathbf{k}$)- and energy($\omega $%
)-distribution of the outgoing electrons. The resolution of ARPES has been
significantly increased in the last decade with the energy resolution of $%
\Delta E\approx 1-2$ $meV$ (for photon energies $\sim 20$ $eV$) and angular
resolution of $\Delta \theta \lesssim 0.2{{}^{\circ }}$. On the other side
the ARPES method is surface sensitive technique, since the average escape
depth ($l_{esc}$)\ of the outgoing electrons is of the order of $l_{esc}\sim
10$ \AA , depending on the energy of incoming photons. Therefore, very good
surfaces are needed in order that the results to be representative for bulk
samples. The most reliable studies were done on the bilayer $%
Bi_{2}Sr_{2}CaCu_{2}O_{8}$ ($Bi-2212$) and its single layer counterpart $%
Bi_{2}Sr_{2}CuO_{6}$ ($Bi2201$), since these materials contain weakly
coupled $BiO$ planes with the longest inter-plane separation in the
cuprates. This results in a \textit{natural cleavage} plane making these
materials superior to others in ARPES experiments. After a drastic
improvement of sample quality in other families of HTSC materials, the ARPES
technique has became an important method in theoretical considerations. The
ARPES can indirectly give information on the momentum and energy dependence
of the pairing potential. Furthermore, the electronic spectrum of the (above
mentioned) cuprates is highly \textit{quasi-2D} which allows rather
unambiguous determination of the initial state momentum from the measured
final state momentum, since the component parallel to the surface is
conserved in photoemission. In this case, the ARPES probes (under some
favorable conditions) directly the single particle spectral function $A(%
\mathbf{k},\omega )$. In the following we discuss mainly those
ARPES experiments which give evidence for the importance of the
EPI in cuprates - see more in \citep{ShenReview}.

ARPES measures a nonlinear response function of the electron system and it
is usually analyzed in the so-called \textit{three-step model}, where the
total photoemission intensity $I_{tot}(\mathbf{k},\omega )\approx I_{1}\cdot
I_{2}\cdot I_{3}$ is the product of three independent terms: (\textbf{1}) $%
I_{1}$ - describes optical excitation of the electron in the bulk; (\textbf{2%
}) $I_{2}$ - describes the scattering probability of the travelling
electrons; (\textbf{2}) $I_{3}$ - the transmission probability through the
surface potential barrier. The central quantity in the three-step model is $%
I_{1}(\mathbf{k},\omega )$ and it turns out that for $\mathbf{k=k}%
_{\parallel }$ it can be written in the form $I_{1}(\mathbf{k},\omega
)\simeq I_{0}(\mathbf{k},\upsilon )f(\omega )A(\mathbf{k},\omega )$ \citep%
{ShenReview} with $I_{0}(\mathbf{k},\upsilon )\sim \mid \langle \psi
_{f}\mid \mathbf{pA\mid }\psi _{i}\rangle \mid ^{2}$ and the quasi-particle
spectral function $A(\mathbf{k},\omega )=-$\textrm{Im}$G(\mathbf{k},\omega
)/\pi $%
\begin{equation}
A(\mathbf{k},\omega )=-\frac{1}{\pi }\frac{Im\Sigma (\mathbf{k},\omega )}{%
[\omega -\xi (\mathbf{k})-Re\Sigma (\mathbf{k},\omega )]^{2}+Im\Sigma ^{2}(%
\mathbf{k},\omega )}.  \label{A}
\end{equation}%
Here, $\langle \psi _{f}\mid \mathbf{p\cdot A\mid }\psi _{i}\rangle $ is the
dipole matrix element which depends on $\mathbf{k}$, polarization and energy
$E_{i}$ of the incoming photons. The knowledge of the matrix element is of a
great importance and its calculation from first principles was done in \citep%
{Bansil}. $f(\omega )$ is the Fermi function, $G$ and $\Sigma
=Re\Sigma +iIm\Sigma $ are the quasi-particle Green's function and
the self-energy, respectively. We summarize and comment here some
important ARPES results which were obtained in the last several
years and which confirm the existence of the Fermi surface and
importance of EPI in the quasi-particle scattering
\citep{ShenReview}.

\textit{ARPES in the normal state }

\begin{figure}[!tbp]
\resizebox{.45\textwidth}{!} {
\includegraphics*[width=6cm]{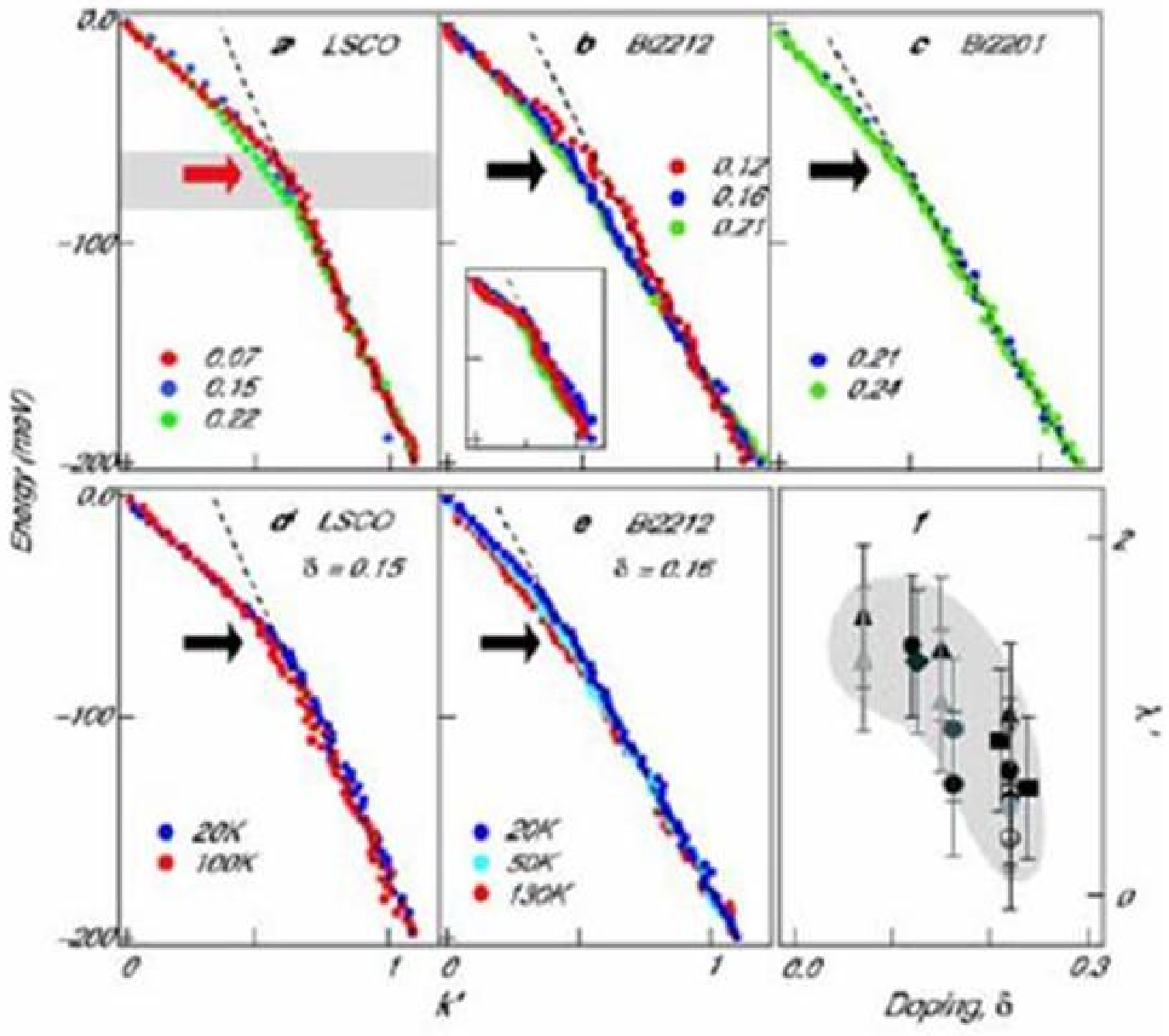}} {\includegraphics*[
width=8cm]{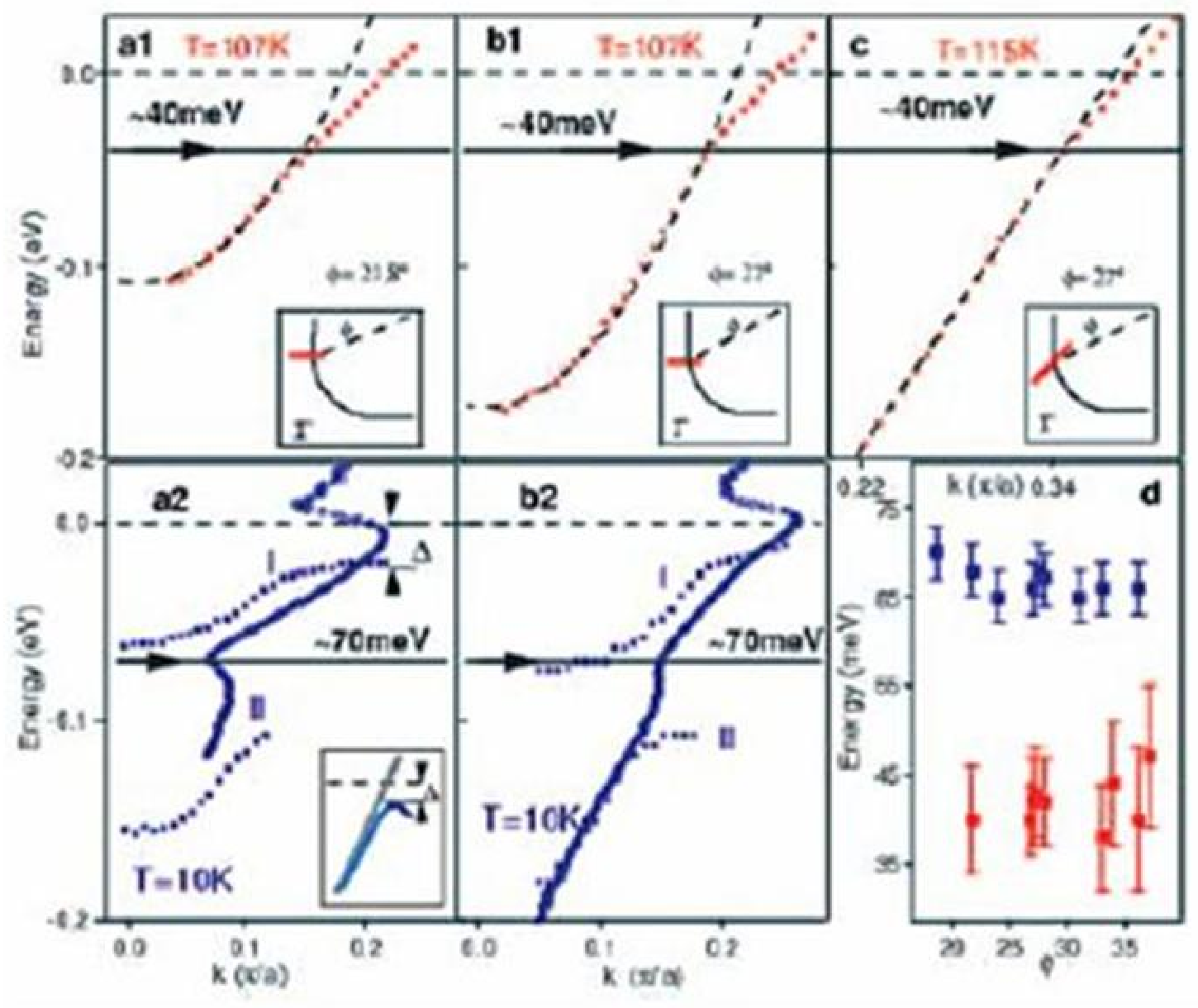}}
\caption{(top) Quasi-particle dispersion of $Bi2212$, $Bi2201$ and $LSCO$
along the \textit{nodal} direction, plotted vs the momentum $k$ for $(a)-(c)$
different doping, and $(d)-(e)$ different $T$; black arrows indicate the
kink energy; the red arrow indicates the energy of the $q=(\protect\pi ,0)$
oxygen stretching phonon mode; inset of $(e)$- T-dependent $\Sigma ^{\prime
} $ for optimally doped $Bi2212$; $(f)$ - doping dependence of the effective
coupling constant $\protect\lambda ^{\prime }$ along $(0,0)-(\protect\pi ,%
\protect\pi )$ for the different HTSC oxides. From Ref. \protect\citep%
{Lanzara}. (bottom) Quasi-particle dispersion $E(k)$ in the normal state
(a1, b1, c), at 107 K and 115 K, along various directions $\protect\phi $
around the \textit{anti-nodal} point. The kink at $E=40meV$ is shown by the
horizontal arrow. (a2 and b2) is $E(k)$ in the superconducting state at 10 K
with the shifted kink to $70meV$. (d) kink positions as a function of $%
\protect\phi $ in the anti-nodal region. From Ref.
\protect\citep{Cuk}.} \label{ARPESLanzFig}
\end{figure}

($\mathbf{N1}$) There is well defined Fermi surface in the metallic state of
\textit{optimally and near optimally} doped cuprates with the topology
predicted by the LDA-DFT. However, the bands are narrower than LDA-DFT
predicts which points to a strong quasi-particle renormalization. ($\mathbf{%
N2}$) The spectral lines are broad with $\mid $Im$\Sigma (\mathbf{k},\omega
)\mid \sim \omega $ (or $\sim T$ for $T>\omega $) which tells us that the
quasi-particle liquid is a non-canonical Fermi liquid for larger values of $%
T,\omega $. ($\mathbf{N3}$) There is a bilayer band splitting in $Bi-2212$
(at least in the over-doped state) what is also predicted by LDA-DFT. In the
case when the coherent hopping $t^{\perp }$ between two layers in the
bilayer dominates, then the anti-bonding and bonding bands $\xi _{\mathbf{k}%
}^{a,b}=\xi _{\mathbf{k}}\pm t_{\mathbf{k}}^{\perp }$ with $t_{\mathbf{k}%
}^{\perp }=[t^{\perp }(\cos ^{2}k_{x}-\cos ^{2}k_{y})+...]$ have been
observed. It is worth to mention that the previous experiments did not show
this splitting what was one of the reasons for various speculations on some
exotic electronic scattering and non-Fermi liquid scenarios. ($\mathbf{N4}$)
In the under-doped cuprates and at temperatures $T_{c}<T<T^{\ast }$ there is
a d-wave like pseudogap $\Delta _{pg}(\mathbf{k})\sim \Delta _{pg,0}(\cos
k_{x}-\cos k_{y})$ in the quasi-particle spectrum where $\Delta _{pg,0}$
increases by lowering doping. We stress that the pseudogap phenomenon is not
well understood at present and since we are interested in systems near
optimal doping where the pseudogap phenomena are absent or much less
pronounced we shall not discuss this problem here. Its origin can be due to
a precursor superconductivity or due to a competing order, such as spin- or
charge-density wave, strong correlations, etc. ($\mathbf{N5}$) The ARPES
self-energy gives evidence that EPI interaction is rather strong. The
arguments for latter statement are the following: (\textit{i}) at $T>T_{c}$
there are \textit{kinks} in the quasi-particle dispersion $\omega (\xi _{%
\mathbf{k}})$ in the \textit{nodal} direction (along the $(0,0)-(\pi ,\pi )$
line) at the characteristic phonon energy $\omega _{ph}^{(70)}\sim (60-70)$ $%
meV$ \citep{Lanzara}, see Fig.~\ref{ARPESLanzFig}, and near the \textit{%
anti-nodal point} $(\pi ,0)$ at $40$ $meV$ \citep{Cuk} - see Fig.~\ref%
{ARPESLanzFig}.

\begin{figure}[tbp]
\resizebox{.5\textwidth}{!} {\includegraphics*[
width=6cm]{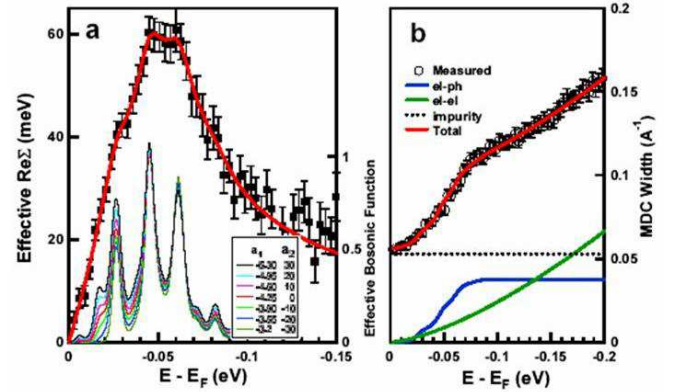}}
\caption{(a) Effective real self-energy for the non-superconducting $%
La_{2-x}Sr_{x}CuO_{4}$, $x=0.03$. Extracted $\protect\alpha _{eff}^{2}(%
\protect\omega )F(\protect\omega )$ is in the inset. (b) Top: the total MDC
width - open circles. Bottom: the EPI contribution shows saturation,
impurity contribution - dotted black line. The residual part is growing $%
\sim \protect\omega ^{1.3}$. From \protect\citep{ZhouPRL}.}
\label{Phonons-Sigma}
\end{figure}

\textit{(ii}) This kink structure is observed in a variety of the hole-doped
cuprates such as $LSCO$, $Bi-2212$, $Bi2201$, $Tl2201$ ($Tl_{2}Ba_{2}CuO_{6}$%
), $Na-CCOC$ ($Ca_{2-x}Na_{x}CuO_{2}Cl_{2}$). These kinks exist also above $%
T_{c}$ what excludes the scenario with the magnetic resonance peak in $%
Im\chi _{s}(\mathbf{Q},\omega )$. Moreover, since the tunnelling and
magnetic neutron scattering measurements give small SFI coupling constant $%
g_{sf}<0.2$ $eV$ then the kinks can not be due to SFI. (\textit{iii}) The
position of the nodal kink is practically doping independent which points
towards phonons as the scattering and pairing boson. ($\mathbf{N6}$) The
quasi-particles (holes) at and near the nodal-point $\mathbf{k}_{N}$ couple
practically to a rather broad spectrum of phonons since at least three group
of phonons were extracted in the bosonic spectral function $\alpha ^{2}F(%
\mathbf{k}_{N},\omega )$ from the ARPES effective self-energy in $%
La_{2-x}Sr_{x}CuO_{4}$ \citep{ZhouPRL} - Fig.~\ref{Phonons-Sigma}.

\begin{figure}[!tbp]
\begin{center}
\resizebox{.5 \textwidth}{!} {
\includegraphics*[width=6cm]{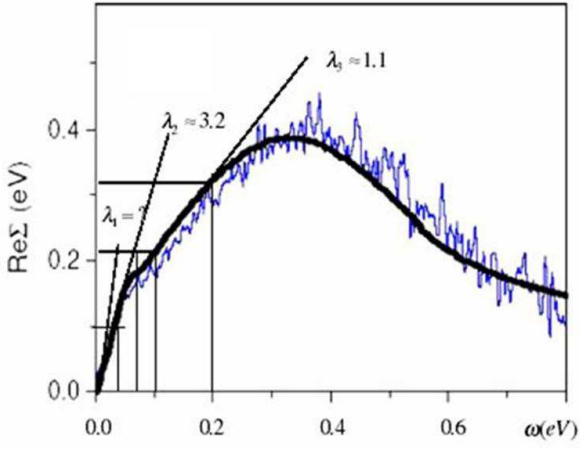}}
\end{center}
\caption{ Fig.4b from \protect\citep{Valla2006}: $Re\Sigma
(\protect\omega ) $ measured in Bi2212 (thin line) and model
$Re\Sigma (\protect\omega )$ (bold
line) obtained in \protect\citep{Valla2006}. The three thin lines ($\protect%
\lambda _{1},\protect\lambda _{2},\protect\lambda _{3}$) are the slopes of $%
Re\Sigma (\protect\omega )$ in different energy regions - see the text.}
\label{Valla-Sigma}
\end{figure}

This result is in a qualitative agreement with numerous tunnelling
measurements \citep{TunnelingVedeneev}-\citep{Tsuda} which
apparently demonstrate that the very broad spectrum of phonons
couples with holes
without preferring any particular phonons - see discussion below. ($\mathbf{%
N7}$) Recent ARPES measurements in B2212 \citep{Valla} show very
different slope $d\omega /d\xi _{\mathbf{k}}$ of the
quasi-particle energy $\omega (\xi _{\mathbf{k}})$ at small $\mid
\xi _{\mathbf{k}}\mid \ll \omega _{ph}$ and at large energies
$\mid \xi _{\mathbf{k}}\mid \gg \omega _{ph}$ - see
Fig.~\ref{Valla-Sigma}. The theoretical analysis
\citep{KulicDolgovLambda} of these results gives the total
coupling constant $\lambda ^{Z}=\lambda _{ep}^{Z}+\lambda
_{c}^{Z}\approx 3$, and for the EPI coupling $\lambda
_{ep}^{Z}\approx 2$ while the Coulomb coupling (SFI is a part of it) is $%
\lambda _{c}^{Z}\approx 1$ \citep{KulicDolgovLambda} - see Fig.~\ref%
{Valla-Sigma}. (Note, that the upper index Z in the coupling constants means
the quasi-particle renormalization in the normal part of the self-energy.)
To this end let us mention some confusion which is related to the value of
the EPI coupling constant extracted from ARPES. Namely, in \citep{ShenReview}%
, \citep{ShenCukReview}, \citep{LanzaraIsotope} the EPI
self-energy was obtained by subtracting the high energy slope of
the quasi-particle spectrum $\omega (\xi _{k})$ at $\omega \sim
0.3$ $eV$. The latter is apparently due to the Coulomb
interaction. Although the position of the low-energy kink is not
affected by this procedure (if $\omega _{ph}^{\max }\ll \omega
_{c}$), this subtraction procedure gives in fact an
\textit{effective EPI self-energy }$\Sigma
_{eff}^{ep}(\mathbf{k},\omega )$ and\textit{\ the effective
coupling constant} $\lambda _{ep,eff}^{Z}(\mathbf{k})$ only. We
demonstrate below that the $\lambda _{ep,eff}^{Z}(\mathbf{k})$ is
smaller than the real EPI coupling constant $\lambda
_{ep}^{Z}(\mathbf{k})$. The
total self-energy is $\Sigma (\mathbf{k},\omega )=\Sigma ^{ep}(\mathbf{k}%
,\omega )+\Sigma ^{c}(\mathbf{k},\omega )$ where $\Sigma ^{c}$ is the
contribution due to the Coulomb interaction. At very low energies $\omega
\ll \omega _{c}$ one has usually $\Sigma ^{c}(\mathbf{k},\omega )=-\lambda
_{c}^{Z}(\mathbf{k})\omega $, where $\omega _{c}(\sim 1$ $eV)$ is the
characteristic Coulomb energies and $\lambda _{c}^{Z}$ is the Coulomb
coupling constant. The quasi-particle spectrum $\omega (\mathbf{k})$ is
determined from the condition $\omega -\xi (\mathbf{k})-Re[\Sigma ^{ep}(%
\mathbf{k},\omega )+\Sigma ^{c}(\mathbf{k},\omega )]=0$ where $\xi (\mathbf{k%
})$ is the bare band structure energy. At low energies $\omega <\omega
_{ph}^{\max }\ll \omega _{c}$ it can be rewritten in the form
\begin{equation}
\omega -\xi ^{ren}(\mathbf{k})-Re\Sigma _{eff}^{ep}(\mathbf{k},\omega )=0,
\label{ren-spec}
\end{equation}%
with $\xi ^{ren}(\mathbf{k})=[1+\lambda _{c}^{Z}(\mathbf{k})]^{-1}\xi (%
\mathbf{k})$ and%
\begin{equation}
Re\Sigma _{eff}^{ep}(\mathbf{k},\omega )=\frac{Re\Sigma _{eff}^{ep}(\mathbf{k%
},\omega )}{1+\lambda _{c}^{Z}(\mathbf{k})}.  \label{eff-SE}
\end{equation}%
Since at very low energies $\omega \ll \omega _{ph}^{\max }$, one has $%
Re\Sigma ^{ep}(\mathbf{k},\omega )=-\lambda _{ep}^{Z}(\mathbf{k})\omega $
and $Re\Sigma _{eff}^{ep}(\mathbf{k},\omega )=-\lambda _{ep,eff}^{Z}(\mathbf{%
k})\omega $, then the real coupling constant is related to the effective one
by $\lambda _{ep}^{Z}(\mathbf{k})=[1+\lambda _{c}^{Z}(\mathbf{k})]\lambda
_{ep,eff}^{Z}(\mathbf{k})>\lambda _{ep,eff}^{Z}(\mathbf{k})$. At higher
energies $\omega _{ph}^{\max }<\omega <\omega _{c}$ the EPI effects are
suppressed and $\Sigma ^{ep}(\mathbf{k},\omega )$ stops growing, one has $%
Re\Sigma (\mathbf{k},\omega )\approx Re\Sigma ^{ep}(\mathbf{k},\omega
)-\lambda _{c}^{Z}(\mathbf{k})\omega $. The measured $Re\Sigma ^{\exp }(%
\mathbf{k},\omega )$ at $T=10$ $K$ near and slightly away from the \textit{%
nodal point} in the optimally doped Bi-2212 with $T_{c}=91$ $K$ \citep%
{Valla2006} is shown in Fig.~\ref{Valla-Sigma}.

\begin{figure}[!tbp]
\resizebox{.5 \textwidth}{!} {
\includegraphics*[width=6cm]{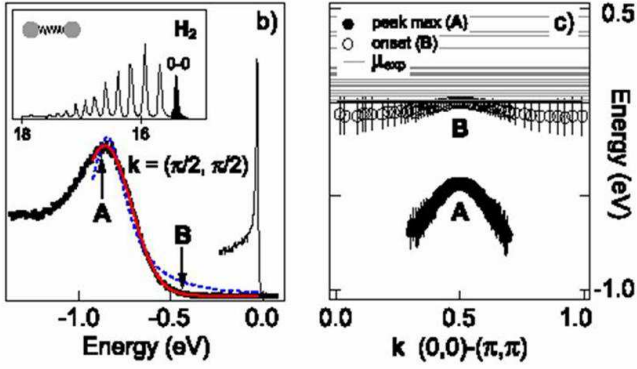}}
\caption{(left-b) The ARPES spectrum of undoped $Ca_{2}CuO_{2}Cl_{2}$ at $%
\mathbf{k}=(\protect\pi/2,\protect\pi/2)$. Gaussian shape - solid
line, Lorentzian shape - dashed line. (right-c) Dispersion of the
polaronic band - A and of the quasi-particle band - B along the
nodal direction. Horizontal lines are the chemical potentials for
a large number of samples. From \protect\citep{ShenPolarons}.}
\label{ARPES-polaronE}
\end{figure}

It is seen that $Re\Sigma ^{\exp }(\mathbf{k},\omega )$ has \textit{two kinks%
} - the first one at \textit{low energy} $\omega _{1}\approx
\omega _{ph}^{high}\approx 50-70$ $meV$ which is (as we already
argued) most probably of the phononic origin \citep{ShenReview},
\citep{ShenCukReview},
\citep{LanzaraIsotope}, while the second kink at \textit{higher energy} $%
\omega _{2}\approx \omega _{c}\approx 350$ $meV$ is due to the
Coulomb interaction. However, the important results in Ref.
\citep{Valla2006} is that the slopes of $Re\Sigma ^{\exp
}(\mathbf{k},\omega )$ at low ($\omega <\omega _{ph}^{high}$) and
high energies ($\omega _{ph}^{high}<\omega <\omega _{c}$) are very
\textit{different}. The low-energy and high-energy slope
\textit{near the nodal point} are shown in Fig.~\ref{Valla-Sigma}
schematically (thin lines). From Fig.~\ref{Valla-Sigma} it is
obvious that EPI prevails at low energies $\omega <\omega
_{ph}^{high}$. More precisely
digitalization of $Re\Sigma ^{\exp }(\mathbf{k},\omega )$ in the interval $%
\omega _{ph}^{high}<\omega <0.4$ $eV$ gives the Coulomb coupling $\lambda
_{c}^{Z}\approx 1.1$ while the same procedure at $20$ $meV\approx \omega
_{ph}^{low}<\omega <\omega _{ph}^{high}\approx 50-70meV$ gives the total
coupling constant $(\lambda _{2}\equiv )\lambda ^{Z}=\lambda
_{ep}^{Z}+\lambda _{c}^{Z}\approx 3.2$ and the EPI coupling constant $%
\lambda _{ep}^{Z}(\equiv \lambda _{ep,high}^{Z})\approx 2.1>2\lambda
_{ep,eff}^{Z}(\mathbf{k})$, i.e. the EPI coupling is at least twice larger
than the effective EPI coupling constant obtained in the previous analysis
of ARPES results \citep{ShenReview}, \citep{ShenCukReview}, \citep%
{LanzaraIsotope}. This estimation tells us that at (and near) the nodal
point, \textit{the EPI interaction dominates }in the quasi-particle
scattering at low energies since $\lambda _{ep}^{Z}(\approx 2.1)\approx
2\lambda _{z}^{c}>2\lambda _{sf}^{Z}$, while at large energies (compared to $%
\omega _{ph}$), the Coulomb interaction with $\lambda _{c}^{Z}\approx 1.1$
dominates. We point out that EPI near the anti-nodal point can be even
larger than in the nodal point, mostly due to the higher density of states
near the anti-nodal point. ($\mathbf{N8}$) Recent ARPES spectra in the
optimally doped \ $Bi-2212$ near the nodal and anti-nodal point \citep%
{LanzaraIsotope} show a low energy isotope effect in $Re\Sigma ^{\exp }(%
\mathbf{k},\omega )$,\ which can be well described in the
framework of the Migdal-Eliashberg theory for EPI \citep{MaKuDo}.
At higher energies $\omega
>\omega _{ph}$ the obtained in \citep{LanzaraIsotope} very pronounced isotope
effect cannot be explained by the simple Migdal-Eliashberg theory \citep%
{MaKuDo}. However, there are controversies with the strength of
the high-energy isotope effect since it was not confirmed in other
measurements \citep{DouglasIsotop}, \citep{IwasawaIsotop} - see
the discussion in Section F.2 related to the isotope effects in
HTSC cuprates. ($\mathbf{N9}$) ARPES experiments on
$Ca_{2}CuO_{2}Cl_{2}$ give strong evidence for the formation of
\textit{small polarons in undoped cuprates} which are due to
phonons and strong EPI, while by doping quasi-particles appear and
there are no small polarons \citep{ShenPolarons}. Namely, in
\citep{ShenPolarons} a broad peak around $-0.8$ $eV$ is observed
at the top of the band ($\mathbf{k}=(\pi /2,\pi /2)$) with the
dispersion similar to that predicted by the $t-J$ model - see Fig.
\ref{ARPES-polaronE}.

\begin{figure}[!tbp]
\begin{center}
\resizebox{.5 \textwidth}{!} {
\includegraphics*[width=6cm]{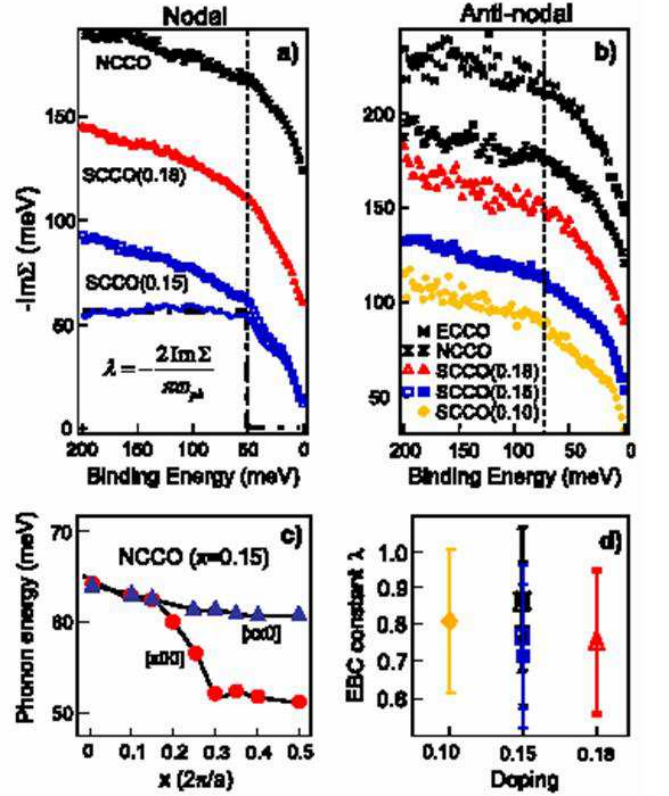}}
\end{center}
\caption{NCCO electron-doped: (a) $Im\Sigma (\protect\omega )$ measured in
the nodal point. Curves are offsets by 50 meV for clarity. The change of the
slope in the last bottom curve is at the phonon energy. (b) $Im\Sigma (%
\protect\omega )$ for the antinodal direction with $30$ $meV$ offset. (c)
Experimental phonon dispersion of the bond stretching modes. (d) Estimated $%
\protect\lambda _{eff}^{ep}$ from $Im\Sigma (\protect\omega )$.
From \protect\citep{Eisaki}.} \label{Eisaki-kink-el}
\end{figure}

However, the peak in Fig.~\ref{ARPES-polaronE} (left) is of
Gaussian shape and can be described only by coupling to bosons,
i.e. this peak is a boson side band - see more in
\citep{GunnarssonReview2008} and references therein. The theory
based on the t-J model (in the antiferromagnetic state of the
undoped compound) by including coupling to several
(half-breathing, apical oxygen, low-lying) phonons, which is given
in \citep{GunnarNagaosaCiuchi}, explains successfully this broad
peak of the boson side band by the formation of small polarons due
to the EPI coupling ($\lambda _{ep}\approx 1.2$). Note that this
value of $\lambda _{ep}$ is for the polaron at the bottom of the
band while in the case where the Fermi surface exists (in doped
systems) this coupling is even larger due to the larger density of
states at the Fermi surface \citep{GunnarNagaosaCiuchi}. In \citep%
{GunnarNagaosaCiuchi} it was stressed that even when the electron-magnon
interaction is stronger than EPI the polarons in the undoped systems are
formed due to EPI. The latter mechanism involves excitation of many phonons
at the lattice site (where the hole is seating), while it is possible to
excite only one magnon at the given site. ($\mathbf{N10}$) Recent soft x-ray
ARPES measurements on the \textit{electron doped} HTSC $%
Nd_{1.85}Ce_{0.15}CuO_{4}$ \citep{Tsunekawa}, and $Sm_{(2-x)}Ce_{x}CuO_{4}$ ($%
x=0.1,$ $0.15,$ $0.18$), $Nd_{1.85}Ce_{0.15}CuO_{4}$, $%
Eu_{1.85}Ce_{0.15}CuO_{4}$ \citep{Eisaki} show kink at energies
$50-70$ $meV$ in the quasi-particle dispersion relation along both
nodal and antinodal directions as it is shown in Fig.
\ref{Eisaki-kink-el}.

It is seen from this figure that the effective EPI coupling constant $%
\lambda _{ep,eff}(<\lambda _{ep})$ is isotropic and $\lambda
_{ep,eff}\approx 0.8-1$. It seems that the kink in the electron-doped
cuprates is due solely to EPI and in that respect the situation is similar
to the one in the hole-doped cuprates.

\textit{ARPES results in the superconducting state}

\begin{figure}[!tbp]
\begin{center}
\resizebox{.5 \textwidth}{!} {\includegraphics*[
width=6cm]{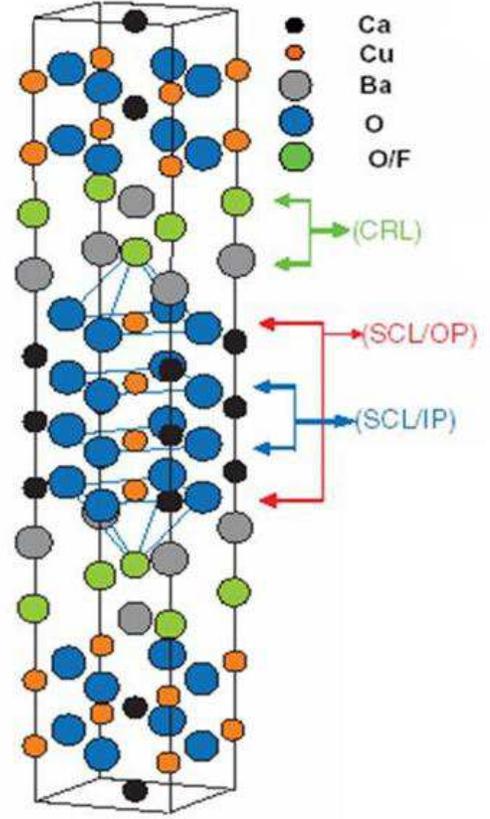}}
\end{center}
\caption{Crystal structure of $Ba_{2}Ca_{3}Cu_{4}O_{8}(O_{\protect\delta %
}F_{1-\protect\delta })_{2}$. There are four $CuO_{2}$ layers in a
unit cell with the outer having apical F atoms. CRL - charge
reservoir layer; SCL/OP - superconducting layer/outer plane;
SCL/IP - superconducting layer/inner plane. From
\protect\citep{ChenShen4layered}.} \label{4lay}
\end{figure}

($\mathbf{S1}$) There is an anisotropic superconducting gap in most HTSC
compounds \citep{ShenReview}, which is predominately d-wave like, i.e. $%
\Delta (\mathbf{k})\approx \Delta _{0}(\cos k_{x}-\cos k_{y})$
with $2\Delta _{0}/T_{c}\approx 5-6$. ($\mathbf{S2}$) The
particle-hole coherence in the superconducting state which is
expected for the BCS-like theory of superconductivity has been
observed first in \citep{CampuzanoBCSCoh} and confirmed with
better resolution in \citep{MatsuiCoh}, where the particle-hole
mixing is clearly seen in the electron and hole quasi-particle
dispersion. To remaind the reader, the excited Bogoliubov-Valatin
quasi-particles ($\hat{\alpha}_{\mathbf{k},\pm }$) with energies $E_{\mathbf{%
k}}^{\alpha _{\pm }}=\sqrt{\xi _{\mathbf{k}}^{2}+\left\vert \Delta _{\mathbf{%
k}}\right\vert ^{2}}$ are a mixture of electron ($\hat{c}_{\mathbf{k},\sigma
}$) and hole ($\hat{c}_{-\mathbf{k},-\sigma }^{\dagger }$), i.e. $\hat{\alpha%
}_{\mathbf{k},+}=u_{\mathbf{k}}\hat{c}_{\mathbf{k\uparrow }}+v_{\mathbf{k}}%
\hat{c}_{-\mathbf{k\downarrow }}^{\dagger }$, $\hat{\alpha}_{\mathbf{k}%
,-}=u_{\mathbf{k}}\hat{c}_{-\mathbf{k\downarrow }}+v_{\mathbf{k}}\hat{c}_{%
\mathbf{k\uparrow }}^{\dagger }$ where the coherence factors $u_{\mathbf{k}}$%
, $v_{\mathbf{k}}$ are given by $\left\vert u_{\mathbf{k}}\right\vert
^{2}=1-\left\vert v_{\mathbf{k}}\right\vert ^{2}=(1+\xi _{\mathbf{k}}/E_{%
\mathbf{k}})/2$. Note, that $\left\vert u_{\mathbf{k}}\right\vert
^{2}+\left\vert v_{\mathbf{k}}\right\vert ^{2}=1$ what is exactly
observed, together with d-wave pairing $\Delta (\mathbf{k})=\Delta
_{0}(\cos k_{x}-\cos k_{y})$, in experiments \citep{MatsuiCoh}.
This is very important result since it proves that the
\textit{pairing in HTSC cuprates is of the BCS type} and not
exotic one as was speculated long time after the discovery of HTSC
cuprates. ($\mathbf{S3}$) The kink at $(60-70)$ $meV$ in the
quasi-particle energy around the nodal point is \textit{not
shifted} (in energy) while the antinodal kink at $\omega
_{ph}^{(40)}\sim 40$ $meV$ is \textit{shifted} (in energy) in the
superconducting state by $\Delta _{0}(=(25-30)meV)$, i.e. $\omega
_{ph}^{(40)}\rightarrow \omega _{ph}^{(40)}+\Delta
_{0}=(65-70)meV$ \citep{ShenReview}. To remind the
reader, in the standard Eliashberg theory the kink in the normal state at $%
\omega =\omega _{ph}$ should be shifted in the superconducting state to $%
\omega _{ph}+\Delta _{0}$ at all points at the Fermi surface. This puzzling
result (that the quasi-particle energy around the nodal point is \textit{not
shifted} in the superconducting state) might be a smoking gun result since
it makes an additional constraint on the quasi-particle interaction in
cuprates. Until now there is only one plausible explanation \citep%
{KulicDolgovShift} of this \textit{non-shift puzzle} which is based on an
assumption of the forward scattering peak (FSP) in EPI - see more in Part
II. The FSP in EPI means that electrons scatter into a narrow region ($%
q<q_{c}\ll k_{F}$)\ around the initial point in the $k$-space, so that at
the most part of the Fermi surface there is practically no mixing of states
with different signs of the order parameter $\Delta (\mathbf{k})$. In that
case the EPI bosonic spectral function (which is defined in Appendix A) $%
\alpha ^{2}F(\mathbf{k},\mathbf{k}^{\prime },\Omega )\approx
\alpha ^{2}F(\varphi ,\varphi ^{\prime },\Omega )$ ($\varphi $ is
the angle on the Fermi surface) has a pronounced forward
scattering peak (at $\delta \varphi =\varphi -\varphi ^{\prime
}=0$) due to strong correlations - see Part II. Its width $\delta
\varphi _{c}$ is narrow, i.e $\delta \varphi _{c}\ll 2\pi $ and
the angle integration goes over the region $\delta \varphi _{c}$
around the point $\varphi $. In that case the kink is shifted
(approximately) by the local gap $\Delta (\varphi )=\Delta _{\max
}\cos 2\varphi $ - for more details see \citep{KulicDolgovShift}.
As the consequence, the anti-nodal kink is shifted by the maximal
gap, i.e. $\left\vert \Delta (\varphi _{AN}\approx \pi
/2)\right\vert =\Delta _{\max }$ while the nodal gap is
practically unshifted since $\left\vert \Delta (\varphi
_{AN}\approx \pi /4)\right\vert \approx 0$. ($\mathbf{S4}$) The
recent ARPES spectra \citep{ChenShen4layered} on an undoped single
crystalline 4-layered cuprate with apical fluorine (F),
$Ba_{2}Ca_{3}Cu_{4}O_{8}F_{2}$ (F0234) give rather convincing
evidence against the SFI mechanism of pairing - see
Fig.~\ref{4lay}.

\textit{First}, F0234 is not a Mott insulator - as expected from valence
charge counting which puts $Cu$ valence as $2^{+}$, but it is a
superconductor with $T_{c}=60$ $K$. Moreover, the ARPES data \citep%
{ChenShen4layered} reveal at least two metallic Fermi-surface sheets with
corresponding volumes equally below and above half-filling - see Fig.~\ref%
{Fermi4lay}.

\begin{figure}[!tbp]
\begin{center}
\resizebox{.5 \textwidth}{!} {\includegraphics*[
width=6cm]{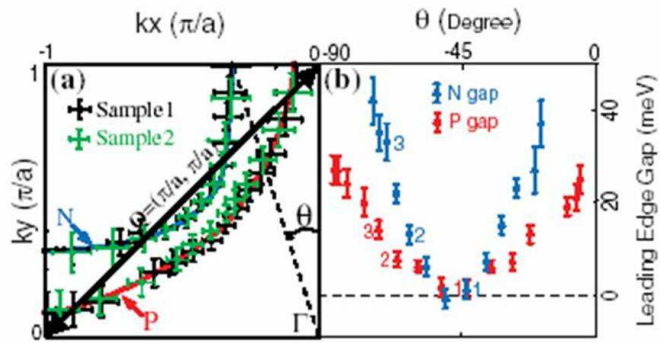}}
\end{center}
\caption{(a) Fermi surface (FS) contours from two samples of
F0234. $N$ - electron-like; $P$ - hole-like. Bold arrow is
$(\protect\pi ,\protect\pi )$ scattering vector. Angle
$\protect\theta $ defines the horizontal axis in (b). (b) Leading
gap edge along k-space angle from the two FS contours. From
\protect\citep{ChenShen4layered}.} \label{Fermi4lay}
\end{figure}
\textit{Second}, one of the Fermi-surfaces is due to the electron-like ($N$)
band (with $20\pm 6$\% electron-doping) and the other one due to the
hole-like ($P$) band (with $20\pm 8$ $\%$ hole-doping) and their splitting
along the nodal direction is significant and cannot be explained by the
LDA-DFT calculations \citep{ShenAnders4layer}. This electron and hole \textit{%
self-doping of inner and outer layers} is in an appreciable contrast to
other multilayered cuprates where there is only hole self-doping. For
instance, in HgBa$_{2}$Ca$_{n}$Cu$_{n+1}$O$_{2n+2}$ ($n=2,3$) and (Cu,C)Ba$%
_{2}$Ca$_{n}$Cu$_{n+1}$O$_{3n+2}$ ($n=2,3,4$), the inner CuO$_{2}$ layers
are less hole-doped than outer layers. It turns out, unexpectedly, that the
\textit{superconducting gap on the }$N$\textit{-band Fermi-surface is
significantly larger} than on the $P$-one, where in $%
Ba_{2}Ca_{3}Cu_{4}O_{8}F_{2}$ the ratio is anomalous $(\Delta _{N}/\Delta
_{P})\approx 2$ and $\Delta _{N}$ is an order of magnitude larger than in
the electron-doped cuprate $Nd_{2-x}Ce_{x}CuO_{4}$. Third, the $N$\textit{%
-band Fermi-surface} \textit{is rather far from the antinodal point} at ($%
\pi ,0$). This is very important result which means that the
antiferromagnetic spin fluctuations with the AF wave-vector $\mathbf{Q}=(\pi
,\pi )$, as well as the van Hove singularity, are not dominant in the
pairing in the $N$-band. To remind the reader, the SFI scenario assumes that
the pairing is due to spin fluctuations with the wave-vector $\mathbf{Q}$
(and near it) which connects two anti-nodal points which are near the van
Hove singularity at the hole-surface (at ($\pi ,0$) and ($0,\pi $)) giving
rise to large density of states. This is apparently not the case for the $N$%
-band Fermi-surface - see Fig.~\ref{Fermi4lay}. The ARPES data give further
that there is a kink at $\sim 85$ $meV$ in the quasi-particle dispersion of
both bands, while the kink in the $N$-band is stronger than that in the $P$%
-band. This result, together with the anomalous ratio $(\Delta _{N}/\Delta
_{P})\approx 2$, strongly disfavors SFI as a pairing mechanism. ($\mathbf{S5}
$) Despite the presence of significant elastic quasi-particle scattering in
a number of samples of optimally doped Bi-2212, there are dramatic
sharpening of the spectral function near the anti-nodal point $(\pi ,0)$ at $%
T<T_{c}$ (in the superconducting state) \citep{Zhu}. This effect
can be explained by assuming that the small $q$-scattering (the
forward scattering
peak) dominates in the elastic impurity scattering as it is pointed in \citep%
{Kulic1}, \citep{Kulic2}, \citep{KuOudo}, \citep{KulicDolgovImp}.
As a result, one finds that the impurity scattering rate in the
superconducting state is
almost zero, i.e. $\gamma _{imp}(\mathbf{k},\omega )=\gamma _{n}(\mathbf{k}%
,\omega )+\gamma _{a}(\mathbf{k},\omega )\approx 0$ for $\mid
\omega \mid <\Delta _{0}$ for any kind of pairing ($s$- $p$-
$d$-wave etc.) since the normal ($\gamma _{n}$)\ and the anomalous
($\gamma _{a}$) scattering rates compensate each other. This
\textit{collapse of the elastic scattering rate} is elaborated in
details in\textit{\ }\citep{Zhu} and it is a consequence of the
Anderson-like theorem for unconventional superconductors which is
due to
the dominance of the small $q$-scattering \citep{Kulic1}, \citep{Kulic2}, \citep%
{KuOudo}-\citep{KulicDolgovImp}. In such a case $d$-wave pairing
is weakly
unaffected by impurities and there is small reduction in $T_{c}$ \citep%
{KulicDolgovImp}, \citep{KeeTc}. The physics behind this result is
rather simple. The small $q$-scattering (usually called forward
scattering) means that electrons scatter into a small region in
the k-space, so that at the most part of the Fermi surface there
is no mixing of states with different signs of the order parameter
$\Delta (\mathbf{k})$. In such a way the detrimental effect of
impurities on $d$-wave pairing is significantly reduced. This
result points to the importance of strong correlations in the
renormalization of the impurity scattering too - see discussion in
Part II.

In conclusion, in order to explain ARPES results in cuprates it is necessary
to take into account: (1) the electron-phonon interaction (EPI) since it
dominates in the quasi-particle scattering in the energy region important
for pairing; (2) effects of elastic nonmagnetic impurities with the forward
scattering peak (FSP) due to strong correlations; (3) the Coulomb
interaction which dominates at higher energies $\omega >\omega _{ph}$. In
this respect, the presence of ARPES kinks and the knee-like shape of the
T-dependence of the spectral width are important constraints on the
scattering and pairing mechanism in HTSC cuprates.

\subsection{Tunnelling spectroscopy and spectral function $\protect\alpha %
^{2}F(\protect\omega )$}

By measuring current-voltage $I-V$ characteristics in $NIS$ (normal
metal-insulator-superconductor) tunnelling junctions with large tunnelling
barrier one obtains from tunnelling conductance $G_{NS}(V)=dI/dV$ the so
called tunnelling density of states in superconductors $N_{T}(\omega )$.
Moreover, by measuring of $G_{NS}(V)$ at voltages $eV>\Delta $ it is
possible to determine the Eliashberg spectral function $\alpha ^{2}F(\omega
) $ and finally to confirm the phonon mechanism of pairing in $LTSC$
materials. Four tunnelling techniques were used in the study of $HTSC$
cuprates: $\mathbf{(}1\mathbf{)}$ vacuum tunnelling by using the $STM$
technique - scanning tunnelling microscope; $\mathbf{(}2\mathbf{)}$
point-contact tunnelling; $\mathbf{(}3\mathbf{)}$ break-junction tunnelling;
$\mathbf{(}4\mathbf{)}$ planar-junction tunnelling. Each of these techniques
has some advantages although in principle the most potential one is the STM
technique since it measures superconducting properties locally \citep{Kirtley}%
. Since tunnelling measurements probe a surface region of the order of
superconducting coherence length $\xi _{0}$, then this kind of measurements
in $HTSC$ materials with small coherence length $\xi _{0}$ ($\xi _{ab}\sim
20 $ \AA\ in the $a-b$ plane and $\xi _{c}\sim 1-3$ \AA\ along the $c-axis$)
depends strongly on the surface quality and sample preparation. Nowadays,
many of the material problems in $HTSC$ cuprates are understood and as a
result consistent picture of tunnelling features is starting to emerge.

From tunnelling experiments one obtains the (energy dependent) gap function $%
\Delta (\omega )$ in the superconducting state. Since we have
already discussed this problem in \citep{KulicReview}, we will
only briefly mention some important result, that in most cases
$G_{NS}(V)$ has $V$-shape in all families of HTSC hole and
electron doped cuprates. The $V$-shape is characteristic for
$d-wave$ pairing with gapless spectrum, which is also confirmed in
the interference experiments on hole and electron doped
cuprates \citep{TsuiKirtley}. Some experiments give an $U$-shape of $%
G_{NS}(V) $ which resembles $s$-wave pairing. This controversy is explained
to be the property of the tunnelling matrix element which filters out states
with the maximal gap.

\begin{figure}[!tbp]
\begin{center}
\resizebox{.5 \textwidth}{!} {\includegraphics*[
width=6cm]{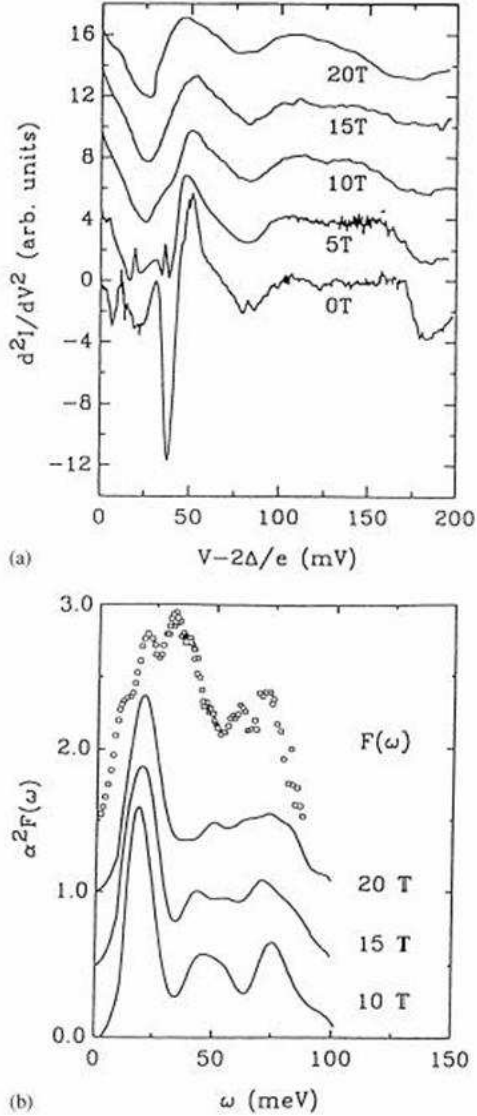}}
\end{center}
\caption{(a) Second derivative of $I(V)$ for a $Bi2212$ break junction in
various magnetic fields (from $0-20$ $T$). The structure of dips (minima) in
$d^{2}I/dV^{2}$ can be compared with the phonon density of states $F(\protect%
\omega )$; (b) the spectral functions $\protect\alpha
^{2}F(\protect\omega )$ in various magnetic fields. From
\protect\citep{TunnelingVedeneev}.} \label{Veden-Break-JJ}
\end{figure}

Here we are interested in the \textit{bosonic spectral function} $\alpha
^{2}F(\omega )$ of HTSC cuprates \textit{near optimal doping} which can be
extracted by using tunnelling spectroscopy. We inform the reader in advance,
that the shape and the energy width of $\alpha ^{2}F(\omega )$, which are
extracted from the second derivative $d^{2}I/dV^{2}$ at voltages above the
superconducting gap, in most HTSC cuprates resembles the phonon density of
states $F_{ph}(\omega )$. This result is strong evidence for the importance
of $EPI$ in the pairing potential of $HTSC$ cuprates. For instance, plenty
of break-junctions made from $Bi-2212$ single crystals \citep%
{TunnelingVedeneev} show that the peaks (and shoulders) in $-d^{2}I/dV^{2}$
(or dips-negative peaks in $d^{2}I/dV^{2}$) coincide with the peaks (and
shoulders) in the phonon density of states $F_{ph}(\omega )$ measured by
neutron scattering - see Fig.~\ref{Veden-Break-JJ}.

\begin{figure}[!tbp]
\begin{center}
\resizebox{.5 \textwidth}{!} {\includegraphics*[
width=6cm]{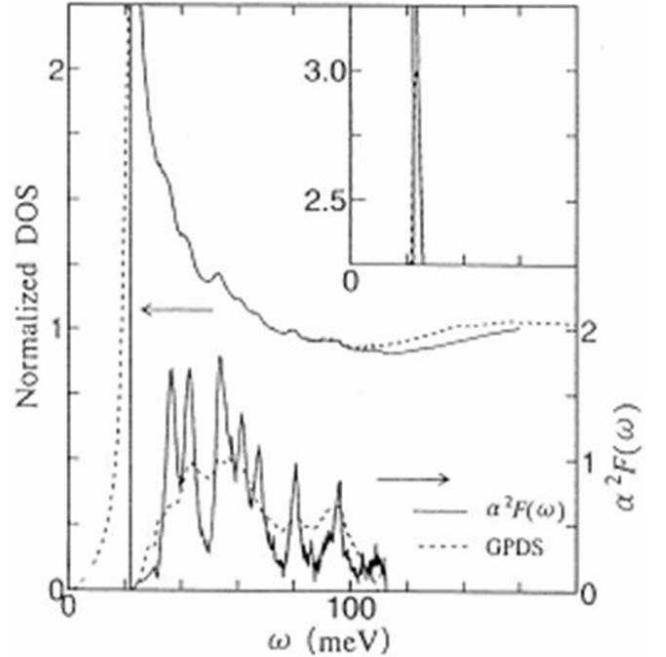}}
\end{center}
\caption{The spectral functions
$\protect\alpha^{2}F(\protect\omega )$ and the calculated density
of states at $0 K$ (upper solid line) obtained from the
conductance measurements the $Bi(2212)-Au$ planar junctions. GPDS
- generalized phonon density of states. From
\protect\citep{Tun2}.} \label{Tun-Shimada}
\end{figure}

The tunnelling spectra in Bi-2212 break junctions
\citep{TunnelingVedeneev}, which are shown in
Fig.~\ref{Veden-Break-JJ}, indicates that the spectral function
$\alpha ^{2}F(\omega )$ is independent of magnetic field, which is
in contradiction with the theoretical prediction based on the SFI
pairing mechanism where this function should be sensitive to
magnetic field. The reported broadening of the peaks in $\alpha
^{2}F(\omega )$ are partly due to the gapless spectrum of $d-wave$
pairing in $HTSC$ cuprates.
Additionally, the tunnelling density of states $N_{T}(\omega )$ at very low $%
T$ and for $\omega >\Delta $ show a pronounced gap structure. It was found
that $2\bar{\Delta}/T_{c}=6.2-6.5$, where $T_{c}=74-85$ $K$ $\ $and $\bar{%
\Delta}$ is some average value of the gap. In order to obtain
$\alpha ^{2}F(\omega )$ the inverse procedure was used by assuming
$s-wave$ superconductivity and the effective Coulomb parameter
$\mu ^{\ast }\approx 0.1$ \citep{TunnelingVedeneev}. The obtained
$\alpha ^{2}F(\omega )$ gives large $EPI$ coupling constant
$\lambda _{ep}\approx 2.3$. Although this analysis
\citep{TunnelingVedeneev} was done in terms of $s-wave$ pairing,
it
mimics qualitatively the case of $d-wave$ pairing, since one expects that $%
d-wave$ pairing does not change significantly the global structure of $%
d^{2}I/dV^{2}$ at $eV>\Delta $ albeit introducing a broadening in
it - see the physical meaning in Appendix A. We point out that the
results obtained in \citep{TunnelingVedeneev} were
\textit{reproducible} on more than $30$
junctions. In that respect very important results on slightly overdoped $%
Bi-2212-GaAs$ and on $Bi-2212-Au$ planar tunnelling junctions are
obtained in \citep{Tun2} - see Fig.~\ref{Tun-Shimada}.

\begin{figure}[!tbp]
\begin{center}
\resizebox{.4 \textwidth}{!} {\includegraphics*[
width=6cm]{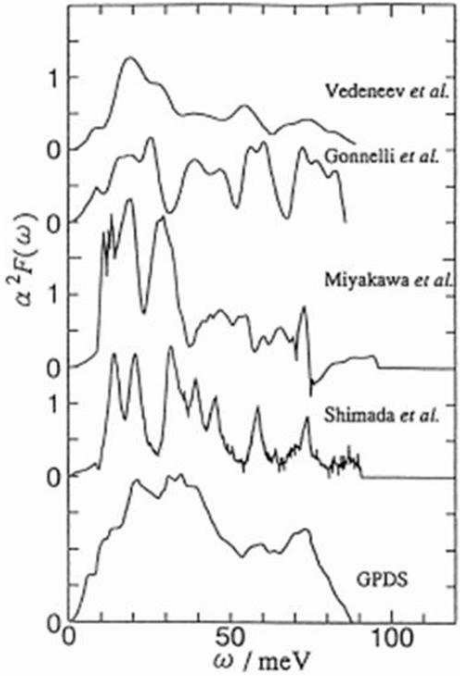}}
\end{center}
\caption{The spectral functions $\protect\alpha ^{2}F(\protect\omega )$ from
measurements of various groups: Vedeneev et al. \protect\citep%
{TunnelingVedeneev}, Gonnelli et al.\protect\citep{Gonnelli},
Miyakawa et al. \protect\citep{Tun3}, Shimada et al.
\protect\citep{Tun2}. The generalized
density of states GPDS for Bi2212 is at the bottom. From \protect\citep{Tun2}%
. }
\label{Tun-All}
\end{figure}

These results show very similar features to those obtained in \citep%
{TunnelingVedeneev} on break-junctions. It is worth mentioning
that several groups \citep{Tun3}, \citep{Tun4}, \citep{Gonnelli}
have obtained similar results for the shape of the spectral
function $\alpha ^{2}F(\omega )$ from the $I-V$ measurements on
various $HTSC$ cuprates - see the comparison in
Fig.~\ref{Tun-All}. These facts leave no much doubts about the
importance of the $EPI$ in pairing mechanism of $HTSC$ cuprates.

\begin{figure}[!tbp]
\begin{center}
\resizebox{.4 \textwidth}{!} {\includegraphics*[
width=6cm]{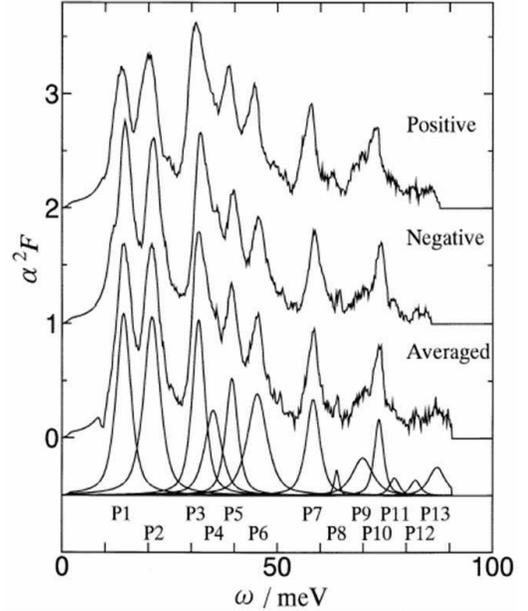}}
\end{center}
\caption{The spectral functions $\protect\alpha
^{2}F(\protect\omega )$ from the tunnelling conductance of
$Bi_{2}Sr_{2}CaCu_{2}O_{8}$ for the positive and the negative bias
voltages, and the averaged one \protect\citep{Tun2}. The averaged
one is divided into $13$ components. The origin of the ordinate is
$2,1,0$ and $-0.5$ from the top down. From \protect\citep{Tun2},
\protect\citep{Tsuda}.} \label{Shimada-Wette-SF}
\end{figure}

\begin{figure}[!tbp]
\begin{center}
\resizebox{.4 \textwidth}{!} {\includegraphics*[
width=6cm]{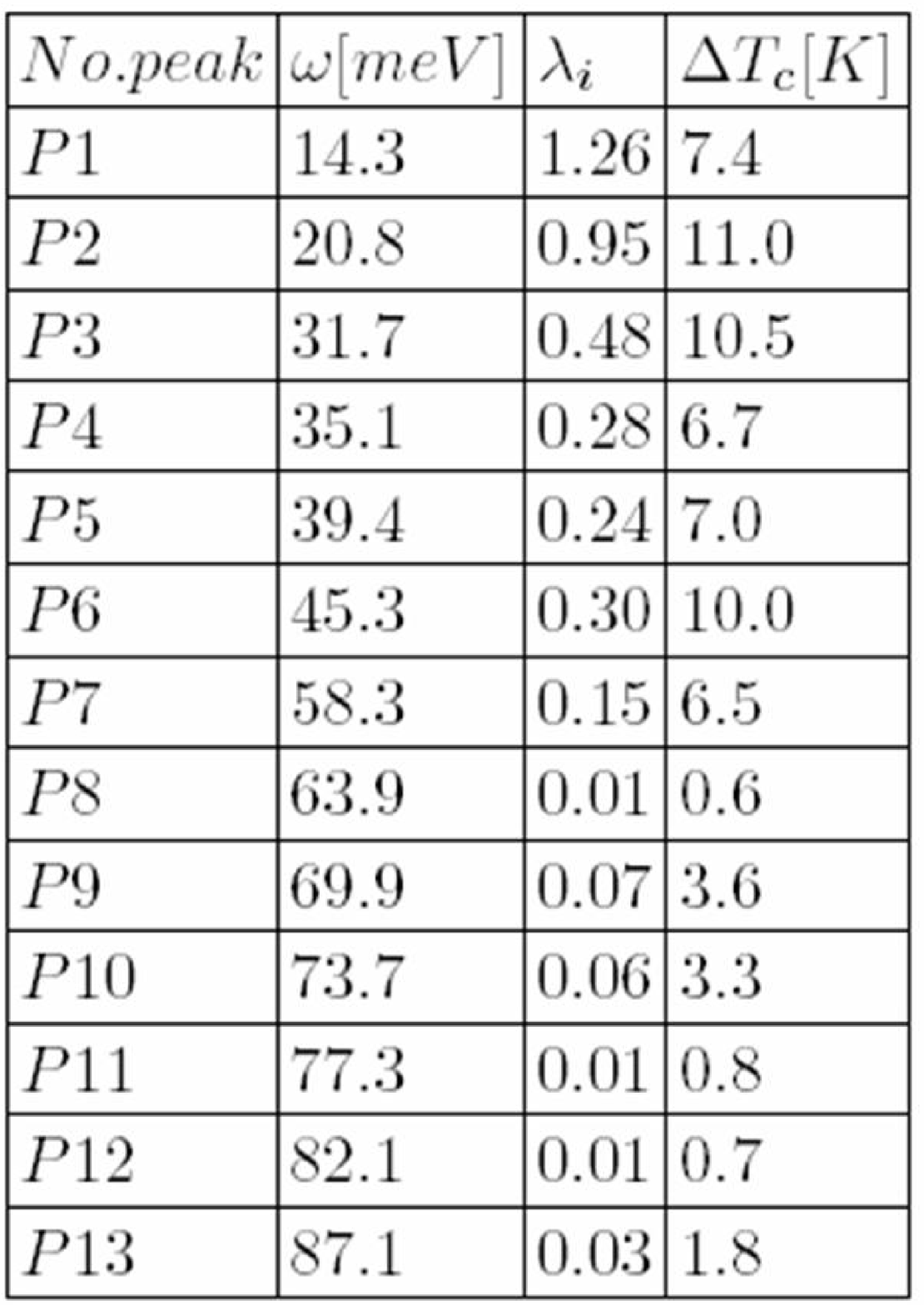}}
\end{center}
\caption{Table I - Phonon frequency $\protect\omega$, EPI coupling constant $%
\protect\lambda_{i}$ of the peaks $P1-P13$ and contribution
$\Delta T_{c}$ to $T_{c}$ of each peak in $\protect\alpha
^{2}F(\protect\omega )$-shown in
Fig.\protect\ref{Shimada-Wette-SF}, obtained from the tunnelling
conductance of $Bi_{2}Sr_{2}CaCu_{2}O_{8}$. $\Delta T_{c}$ is the
decrease in $T_{c}$ when the peak in $\protect\alpha
^{2}F(\protect\omega )$ is eliminated. From
\protect\citep{Tsuda}.} \label{TableI}
\end{figure}

In that respect, tunnelling measurements on \textit{slightly overdoped} $%
Bi_{2}Sr_{2}CaCu_{2}O_{8}$ \citep{Tun2}, \citep{Tsuda} give
impressive results. The Eliashberg spectral function $\alpha
^{2}F(\omega )$ of this compound was extracted from the
measurements of $d^{2}I/dV^{2}$ and by solving the inverse problem
- see Appendix A. The extracted $\alpha ^{2}F(\omega )$ has
several peaks in broad energy region up to $80$ $meV$ as it is
seen in Figs.~\ref{Tun-Shimada}-\ref{Tun-All}, which coincide
rather well with the peaks in the phonon density of states
$F_{ph}(\omega )$ - more precisely the generalized phonon density
of states $GPDS(\omega )$ defined in Appendix A. In \citep{Tsuda}
numerous peaks, from $P1-P13$, in $\alpha ^{2}F(\omega )$ are
discerned as shown in Fig. \ref{Shimada-Wette-SF}, which
correspond to various groups of phonon modes - laying in (and
around) these peaks. Moreover, in \citep{Tun2}, \citep{Tsuda} the
coupling constants for
these modes are extracted as well as their contribution ($\Delta T_{c}$) to $%
T_{c}$ as it is seen in Fig.~\ref{TableI}. Note, that due to the
nonlinearity of the problem the sum of $\Delta T_{c}$ is not equal to $T_{c}$%
.

The next remarkable result is that the extracted $EPI$ coupling constant is
very large, i.e. $\lambda _{ep}(=2\int d\omega \alpha ^{2}F(\omega )/\omega
)=\sum_{i}\lambda _{i}\approx 3.5$ - see Fig.~\ref{TableI}. It is obvious
from Figs.~(\ref{Shimada-Wette-SF}-\ref{TableI}) that almost \textit{all
phonon modes contribute} to $\lambda _{ep}$ and $T_{c}$, which means that on
the average each particular phonon mode is not too strongly coupled to
electrons since $\lambda _{i}<1.3$, thus \textit{keeping the lattice stable}%
. \newline
\newline

\begin{figure}[!tbp]
\begin{center}
\resizebox{.4 \textwidth}{!} {\includegraphics*[
width=6cm]{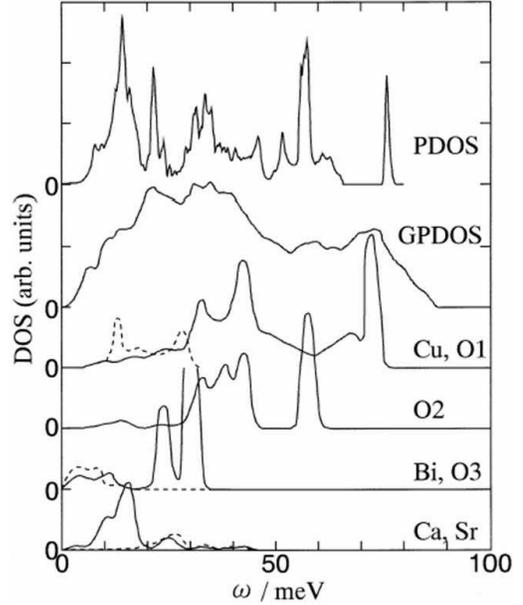}}
\end{center}
\caption{The phonon density of states $F(\protect\omega )$ (PDOS) of $%
Bi_{2}Sr_{2}CaCu_{2}O_{8}$ compared with the generalized density
of states (GPDOS) \protect\citep{Renker}. Atomic vibrations: O1 -
O in the $CuO_{2}$ plane; O2 - apical O; O3 - O in the BiO plane.
From \protect\citep{Tun2}.} \label{Shimada-Wette-DOS}
\end{figure}

\begin{figure}[!tbp]
\begin{center}
\resizebox{.5 \textwidth}{!} {\includegraphics*[
width=6cm]{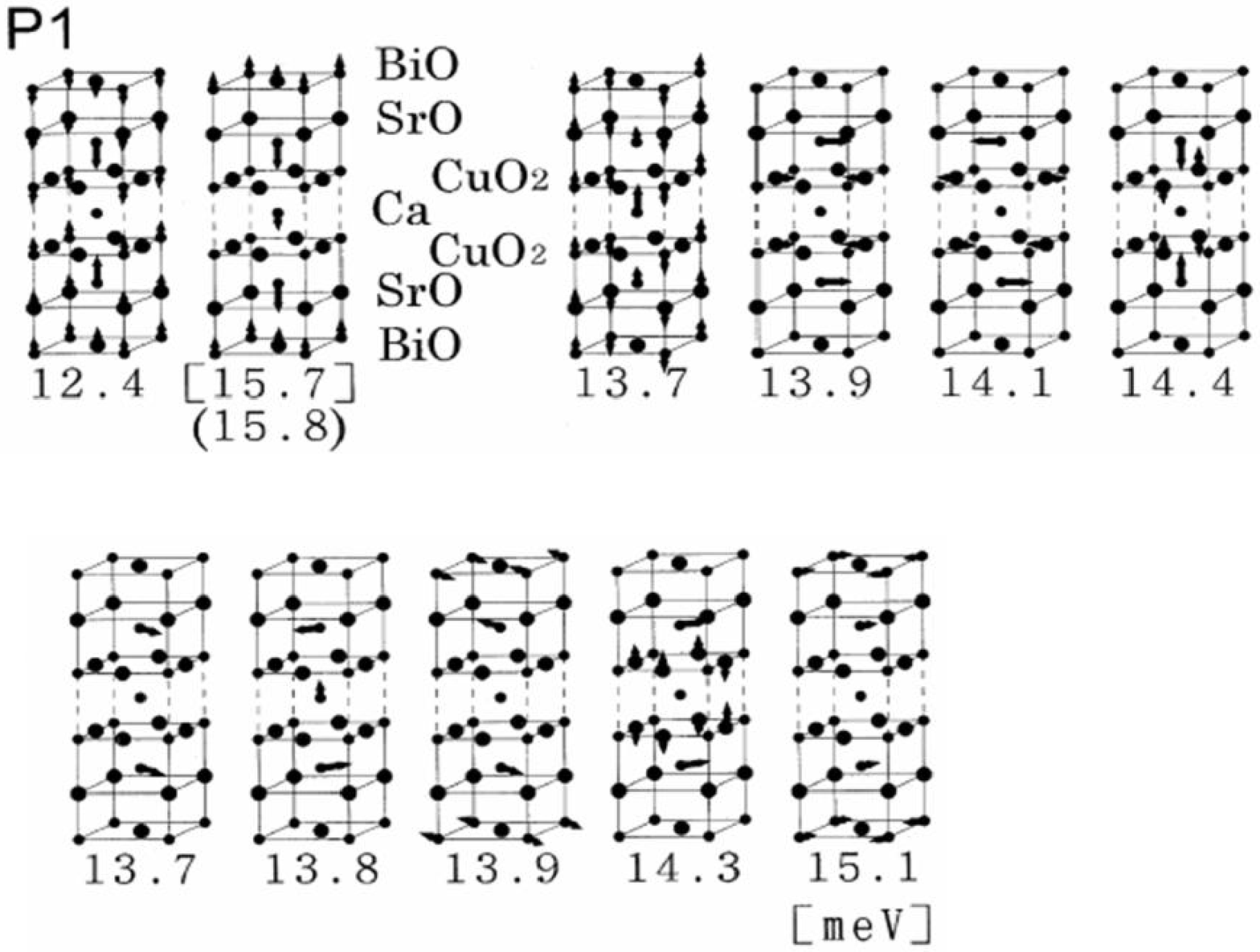}} {\includegraphics*[
width=9cm]{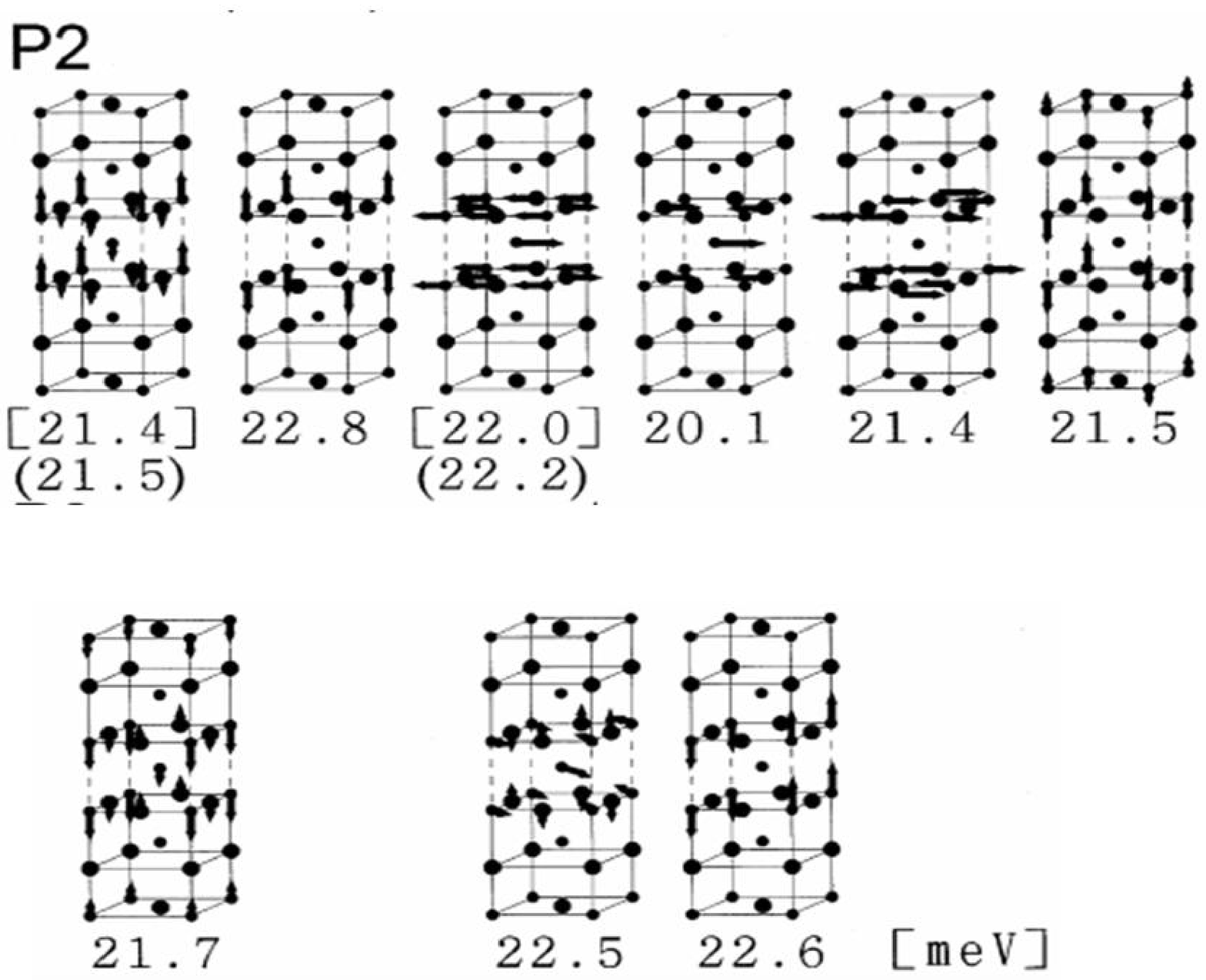}}
\end{center}
\caption{Atomic polarization vectors and their frequencies (in
$meV$) at special points in the Brillouin zone. The larger closed
circles in the lattice are O-ions. $\Gamma - X$ is along the
Cu-O-Cu direction. Arrows indicate displacements. The modes in
square and round brackets are the transverse and longitudinal
optical modes respectively. (top) - modes of the P1 peak. (bottom)
- modes of the P2 peak. From \protect\citep{Tun2},
\protect\citep{Tsuda}.} \label{Shimada-Wette-Phonons12}
\end{figure}

For a better understanding of the the EPI coupling in these systems we show
in Fig. \ref{Shimada-Wette-DOS} the total and partial density of phononic
states. Let us discuss the content of the Table in Fig.~\ref{TableI} in more
details where it is shown the strength of the EPI coupling and the relative
contribution of different phononic modes to $T_{c}$.
In Fig.~\ref{TableI} it is seen that lower frequency modes from $P1-P3$,
corresponding to $Cu,Sr$ and $Ca$ vibrations, are rather strongly coupled to
electrons (with $\lambda _{\kappa }\sim 1$) which give appreciable
contributions to $T_{c}$. It is also seen in Fig.~\ref{TableI} that the
coupling constants $\lambda _{i}$ of the high-energy phonons ($P9-P13$ with $%
\omega \geq 70$ $meV$) have $\lambda _{i}\ll 1$ and give moderate
contribution to $T_{c}$ - around $10$ $\%$. These results give
solid evidence for the \textit{importance of the low-energy modes
related to the change of the Madelung energy} in the
ionic-metallic structure of HTSC cuprates - the idea advocated in
\citep{MaksimovReview}, \citep{KulicReview} and discussed in Part
II. If confirmed in other HTSC\ families, these results are in
favor of the \textit{moderate oxygen isotope} effect in cuprates
near the optimal doping since the oxygen modes are higher energy
modes and gives smaller contribution to $T_{c}$. We stress that each peak $%
P1-P13$ in $\alpha ^{2}F(\omega )$ corresponds to many modes. In order to
obtain information on the structure of vibrations which are strongly
involved in pairing, we show in Figs. \ref{Shimada-Wette-Phonons12}-\ref%
{Shimada-Wette-Phonons34} the structure of these vibrations at special
points in the Brillouin zone. It is seen in Fig. \ref%
{Shimada-Wette-Phonons12} that the low-frequency phonons $P1-P2$ are
dominated by $Cu$, $Sr$, $Ca$ vibrations.

\begin{figure}[!tbp]
\begin{center}
\resizebox{.5 \textwidth}{!} {\includegraphics*[
width=6cm]{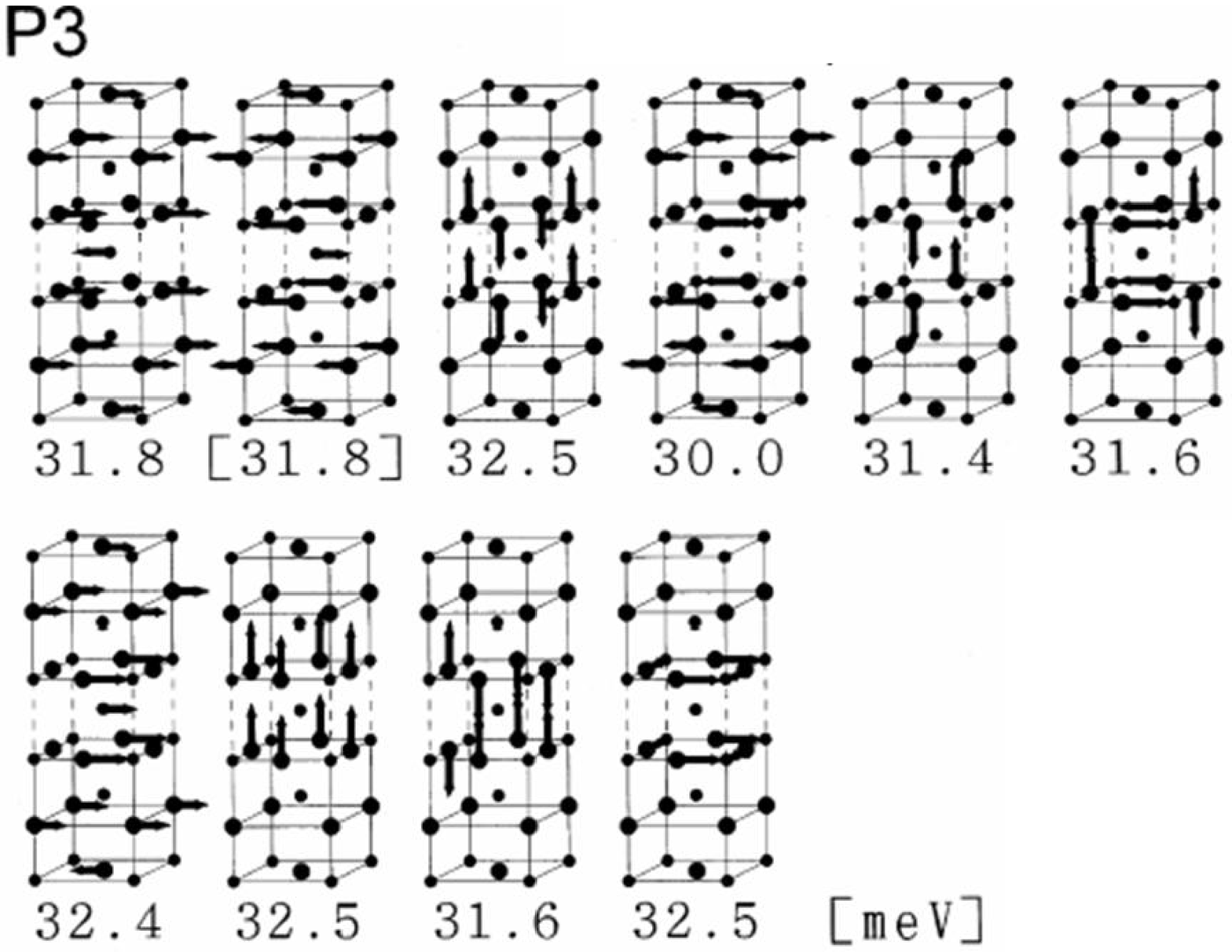}} {\includegraphics*[
width=9cm]{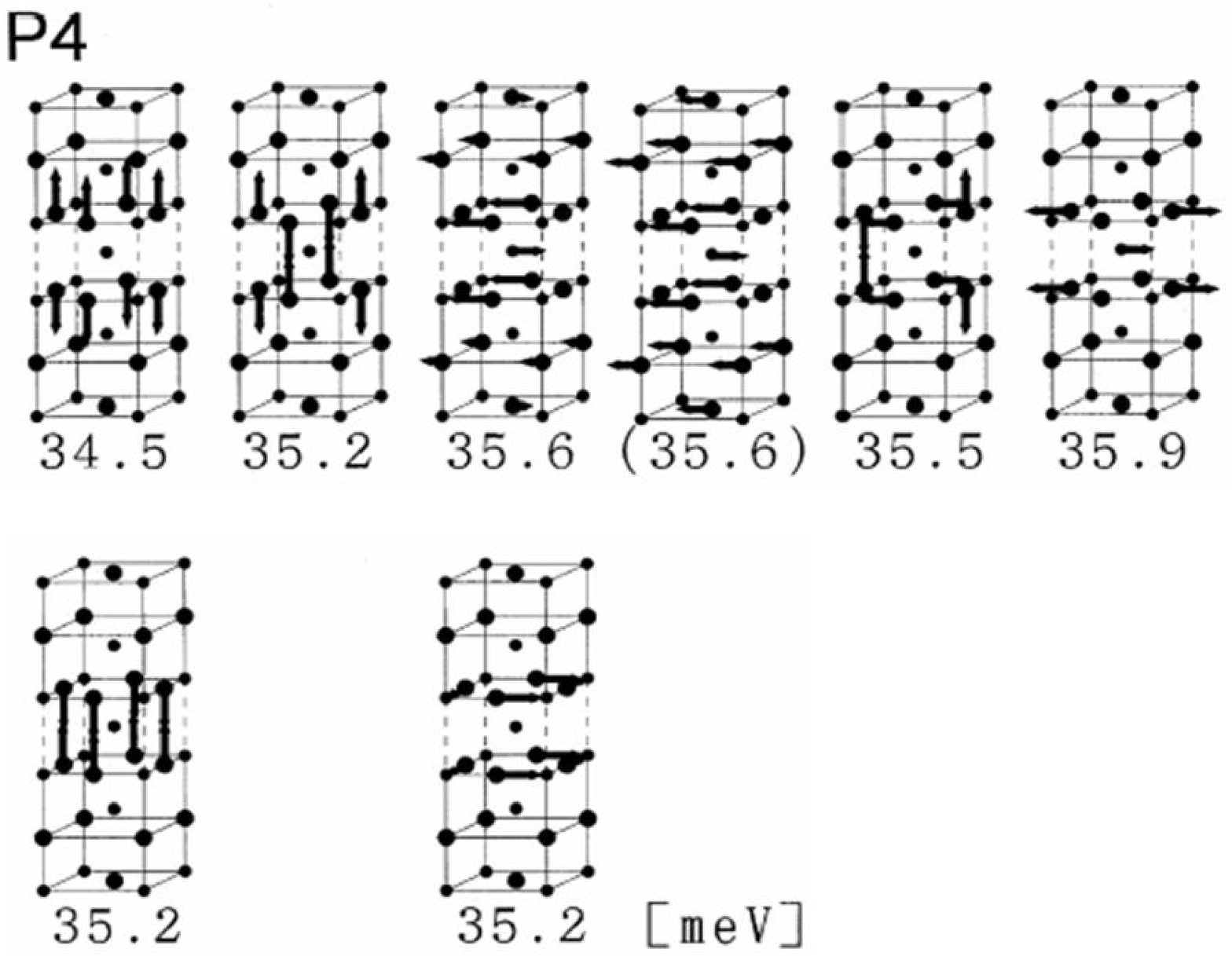}}
\end{center}
\caption{Atomic polarization vectors and their frequencies (in
$meV$) at special points in the Brillouin zone. The larger closed
circles in the lattice are O-ions.$\Gamma - X$ is along the
Cu-O-Cu direction. Arrows indicate displacements. The modes in
square and round brackets are the transverse and longitudinal
optical modes, respectively.(top) - modes of the P3 peak. (bottom)
- modes of the P4 peak. From \protect\citep{Tun2},
\protect\citep{Tsuda}.} \label{Shimada-Wette-Phonons34}
\end{figure}

Further, based on the Table in Fig.~\ref{TableI} one can conclude that the $%
P3$ modes are stronger coupled to electrons than $P4$ ones,
although the density of state for the $P4$ modes is larger. The
reason for such an anomalous behavior might be due to symmetries
of the corresponding phonons as it is seen in Fig.
\ref{Shimada-Wette-Phonons34}. Namely, to the $P3$ peak contribute
\textit{axial vibrations} of $O(1)$ in the $Cu_{2}$ plane which
are odd under inversion, while in the $P4$ peak these modes are
even. The in-plane modes of $Ca$ and $O(1)$ are present in $P3$
which are in-phase and out-of-phase modes, while in $P4$ they are
all out-of-phase modes. For more information on other modes
$P5-P13$ see \citep{Tsuda}. We stress that the Eliashberg
equations based on the extracted $\alpha ^{2}F(\omega )$ of the
slightly overdoped $Bi_{2}Sr_{2}CaCu_{2}O_{8}$ with the ratio
$(2\Delta /T_{c})\approx 6.5$ describe rather well numerous
optical, transport and
thermodynamic properties \citep{Tsuda}. However, in \textit{underdoped systems%
} with $(2\Delta /T_{c})\approx 10$, where the pseudogap phenomena
are pronounced, there are \textit{serious disagreements} between
experiments and the Migdal-Eliashberg theory \citep{Tsuda}. We
would like to stress that the
contribution of the high frequency modes (mostly the oxygen modes) to $%
\alpha ^{2}F(\omega )$ may be underestimated in tunnelling measurements due
to their sensitivity to surface contamination and defects. Namely, the
tunnelling current probes a superconductor to a depth of order of the
quasi-particle mean-free path $l(\omega )=v_{F}\gamma ^{-1}(\omega )$. Since
$\gamma ^{-1}(\omega )$ decreases with increasing $\omega $ the mean free
path can be rather small and the effects of the high energy phonons are
sensitive to the surface contamination.

\begin{figure}[!tbp]
\begin{center}
\resizebox{.5 \textwidth}{!} {\includegraphics*[
width=6cm]{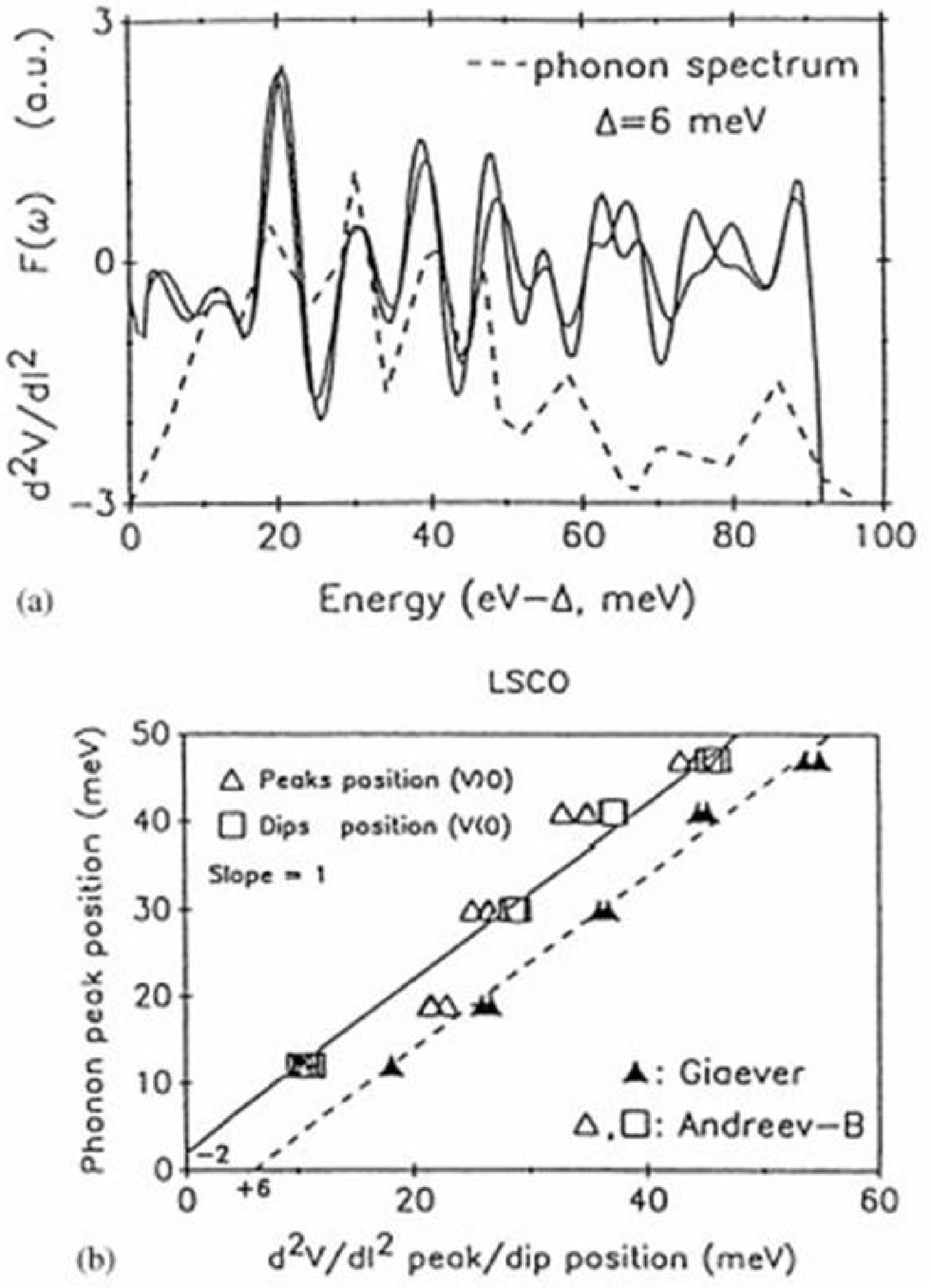}}
\end{center}
\caption{(a) $d^{2}I/dV^{2}$ of a Giaver-like contact in $La$ $%
_{2-x}Sr_{x}CuO_{4}$ - note the large structure below $50meV$; (b) $%
d^{2}I/dV^{2}$ of an Andreev- and Giaver-like contact compared to
the peaks in the phonon density of states. From
\protect\citep{Deutscher}.} \label{Tun-Deutsch}
\end{figure}

Similar conclusion regarding the structure of the $EPI$ spectral function $%
\alpha ^{2}F(\omega )$ in $HTSC$ cuprates comes out from
tunnelling measurements on Andreev junctions ($Z\ll 1$ - low
barrier) and Giaver junctions ($Z\gg 1$ - high barrier) in
$La_{2-x}Sr_{x}CuO_{4}$ and $YBCO$ compounds \citep{Deutscher},
where the extracted $\alpha ^{2}F(\omega )$ is in good accordance
with the phonon density of states $F_{ph}(\omega )$ - see Fig.
\ref{Tun-Deutsch}.

Note, that the BTK parameter $Z$ is related to the transmission and
reflection coefficients for the normal metal $(1+Z^{2})^{-1}$ and $%
Z^{2}(1+Z^{2})^{-1}$ respectively.

\begin{figure}[!tbp]
\begin{center}
\resizebox{.45 \textwidth}{!} {\includegraphics*[
width=7cm]{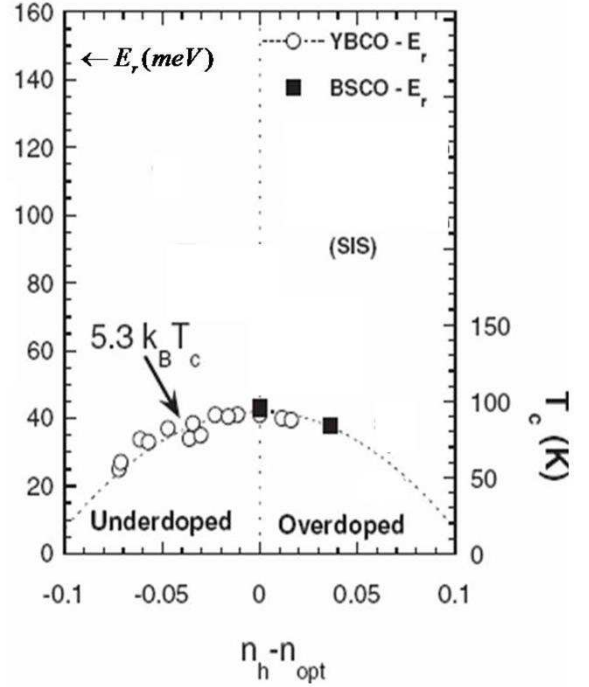}}
\end{center}
\caption{Doping dependence of the energy $E_{r}$ of the magnetic
resonance peak at $\protect\pi ,\protect\pi $ in YBCO and Bi2212
measured at low temperatures by inelastic neutron scattering. From
\protect\citep{KeimerPss}. } \label{Keimer-reson}
\end{figure}

\begin{figure}[!tbp]
\begin{center}
\resizebox{.45 \textwidth}{!} {\includegraphics*[
width=6cm]{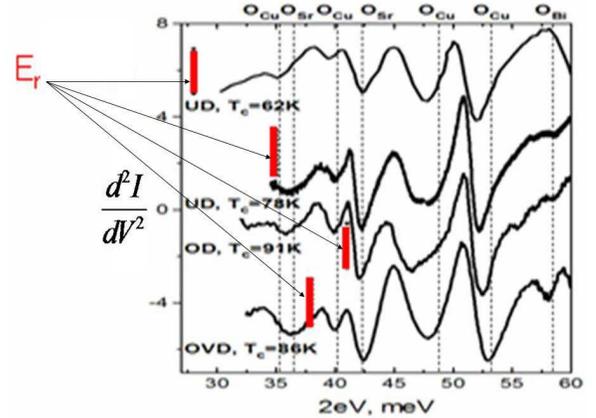}}
\end{center}
\caption{Second derivative of $I(V)$ for a $Bi2212$ tunnelling junctions for
various doping: UD-underdoped; OD-optimally doped; OVD-overdoped system. The
structure of minima in $d^{2}I/dV^{2}$ can be compared with the phonon
density of states $F(\protect\omega )$. The full and vertical lines mark the
positions of the magnetic resonance energy $E_{r}\approx 5.4T_{c}$ for
various doping taken from Fig.\protect\ref{Keimer-reson}. Red tiny arrows
mark positions of the magnetic resonance $E_{r}$ in various doped systems.
Dotted vertical lines mark various phonon modes. From \protect\citep%
{PonomarevTunnel}.}
\label{Ponomarev-tunnel}
\end{figure}

Although most of the peaks in $\alpha ^{2}F(\omega )$ of $HTSC$
cuprates coincide with the peaks in the phonon density of states
it is legitimate to put the question - can the magnetic resonance
in the superconducting state give significant contribution to
$\alpha ^{2}F(\omega )$? In that respect the inelastic magnetic
neutron scattering measurements of the magnetic resonance as a
function of doping \citep{KeimerPss} give that the resonance
energy $E_{r}$ scales with $T_{c}$, i.e. $E_{r}=(5-6)T_{c}$ as shown in Fig.%
\ref{Keimer-reson}.
This means that if one of the peaks in $\alpha ^{2}F(\omega )$ is
due to the magnetic resonance at $\omega =E_{r}$, then it should
shift strongly with doping as it is observed in \citep{KeimerPss}.
This is contrary to phonon peaks (energies) whose positions are
practically doping independent. To this end, recent tunnelling
experiments on Bi-2212 \citep{PonomarevTunnel} show clear
\textit{doping independence }of $\alpha ^{2}F(\omega )$ as it is
seen in Fig. \ref{Ponomarev-tunnel}. This remarkable result is an
additional and strong evidence in favor of EPI and against the SFI
mechanism of pairing in HTSC cuprates which is based on the
magnetic resonance peak in the superconducting state. In that
respect the analysis in \citep{GuoMeng} of tunneling spectra of
electron-doped cuprates Pr$_{0.88}Ce_{0.12}CuO_{4}$ with
$T_{c}=24$ $K$ shows the existence of the bosonic mode at $\omega
_{B}=16$ $meV$ which is significantly larger than the
magnetic-resonance mode with $\omega _{r}=(10-11)$ $meV$. This
result excludes the magnetic-resonance mode as an important factor
which modifies superconductivity.

\begin{figure}[!tbp]
\begin{center}
\resizebox{.5 \textwidth}{!} {\includegraphics*[
width=6cm]{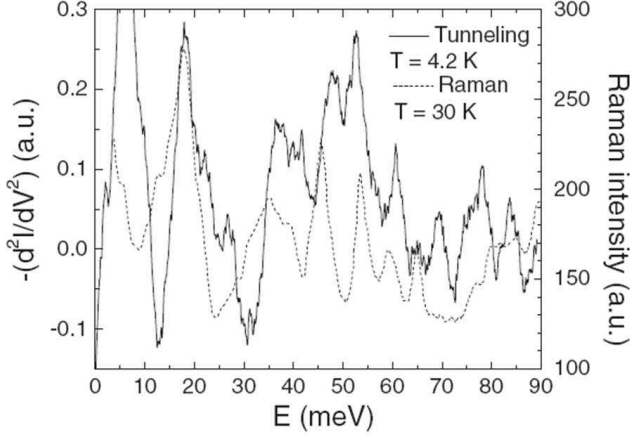}}
\end{center}
\caption{Second derivative data $d^2I(V)/dV^2$ of the tunnelling
spectra on thin films of $La_{1.85}Sr_{0.15}CuO_{4}$ are shown
along with phonon Raman scattering data on single crystals of LSCO
with $20$ $\%$ Sr. The polarization of the incident and scattered
light in the Raman spectra is parallel to the $CuO_{2}$ planes.
From \protect\citep{Chaudhari}.} \label{Chaudhari-tunnel}
\end{figure}

The presence of pronounced phononic structures (and the importance
of EPI)\ in the $I(V)$ characteristics was quite recently
demonstrated by the tunnelling measurements on the very good
$La_{1.85}Sr_{0.15}CuO_{4}$ films prepared by the molecular beam
epitaxy on the [001]-symmetric $SrTiO_{3}$ bi-crystal substrates
\citep{Chaudhari}. They give unique evidence for eleven peaks in
the (negative) second derivative, i.e. $-d^{2}I/dV^{2}$.
Furthermore, \textit{these peaks coincides with the peaks in the
intensities of the phonon Raman scattering} data measured at $30$
$K$ in single crystals of LSCO with $20$ $\%$ of Sr
\citep{SugaiRaman}. These results are shown in Fig.
\ref{Chaudhari-tunnel}. In spite of the lack of a quantitative
analysis of the data in the framework of the Eliashberg equations
the results in Ref. \citep{Chaudhari} are important evidence that
phonons are relevant pairing bosons in HTSC cuprates.

\begin{figure}[!tbp]
\begin{center}
\resizebox{.5 \textwidth}{!} {\includegraphics*[
width=6cm]{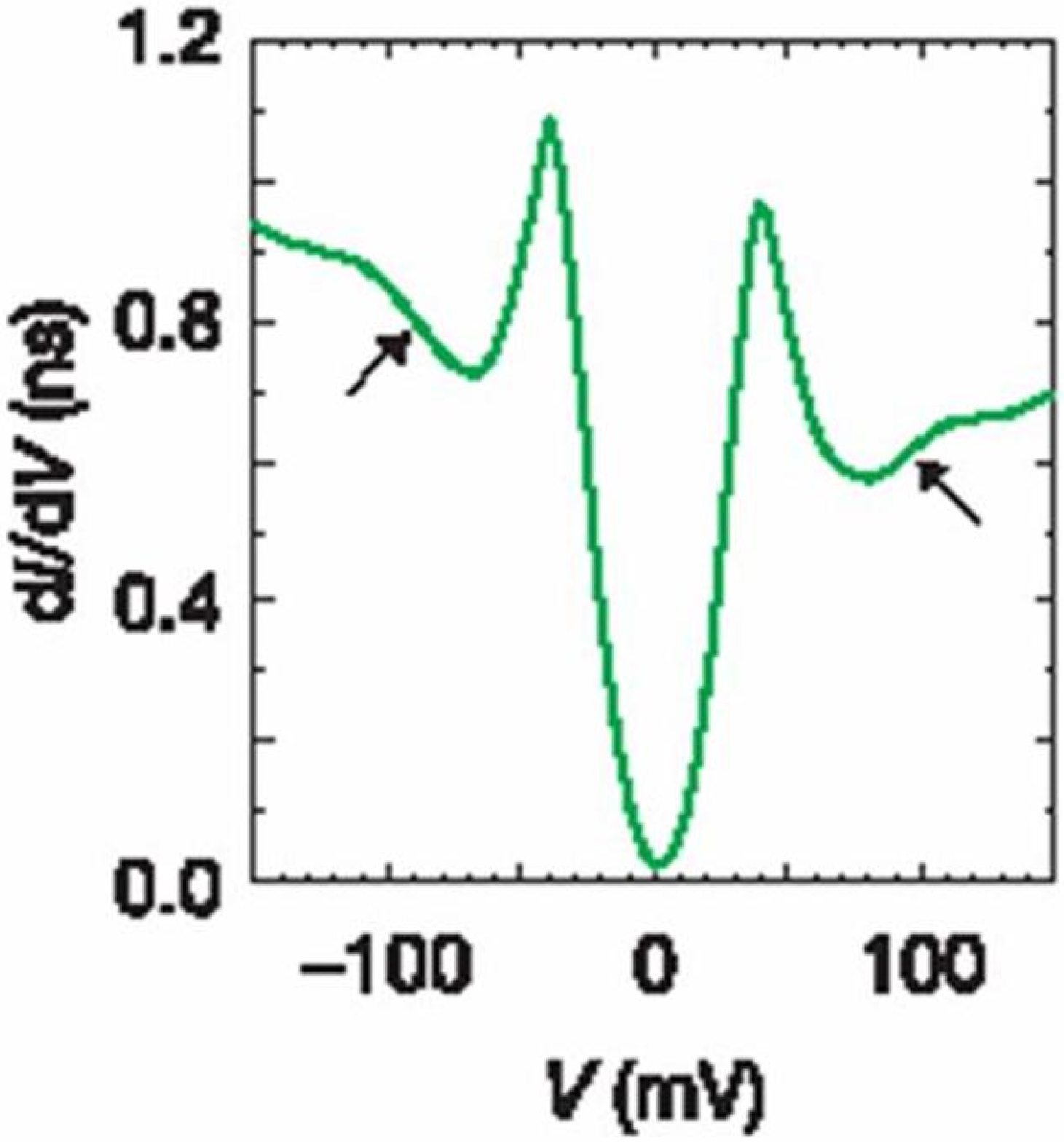} \includegraphics*[
width=8cm]{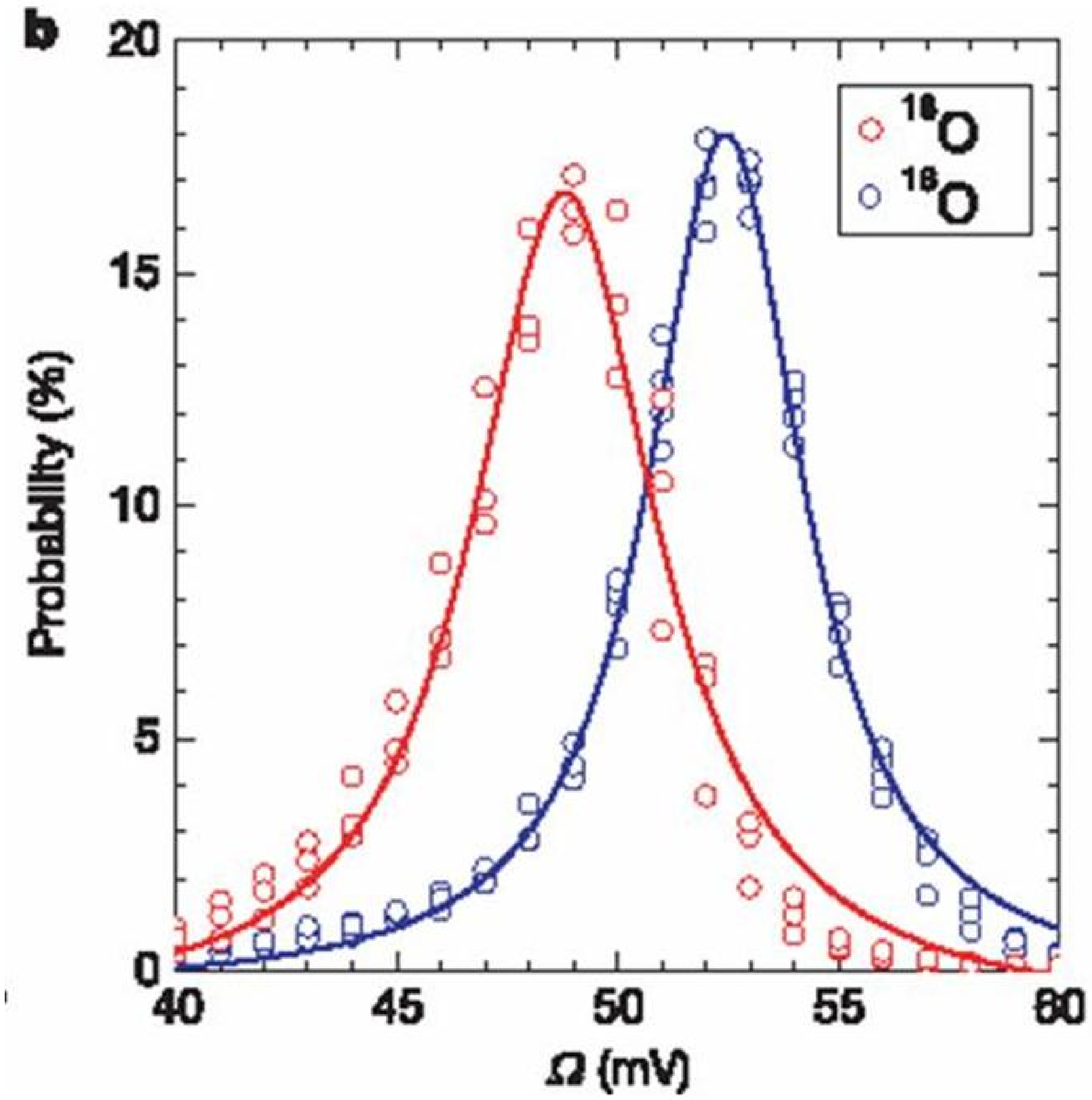}}
\end{center}
\caption{(left) Typical conductance $dI/dV(\mathbf{r},E)$. The ubiquitous
feature at $eV>\Delta$(gap) with maximal slopes, which give peaks in $%
d^{2}I/dV^{2}(\mathbf{r},E)$ are indicated by arrows. (right-b)
The histograms of all values of $\Omega (\mathbf{r})$ for samples
with $O^{16}$ - right curve and with $O^{18}$ - left curve. From
\protect\citep{Davis}.} \label{Tun-Davis}
\end{figure}

It is interesting that in the $c$-axis vacuum tunnelling $STM$
measurements \citep{FischerRennerRM} the fine structure in
$d^{2}I/dV^{2}$ at $eV>\Delta $ was not seen below $T_{c}$, while
the pseudogap structure is observed at temperatures near and above
$T_{c}$. This result could mean that the $STM$ tunnelling is
likely dominated by the nontrivial structure of the tunnelling
matrix element (along the $c$-axis), which is derived from the
band
structure calculations \citep{AJLM}. However, recent $STM$ experiments on $%
Bi-2212$ \citep{Davis} give information on possible nature of the
bosonic
mode which couples with electrons. In \citep{Davis} the local conductance $%
dI/dV(\mathbf{r},E)$ is measured where it is found that $d^{2}I/dV^{2}(%
\mathbf{r},E)$ has peak at $E(\mathbf{r})=\Delta (\mathbf{r})$ $+\Omega (%
\mathbf{r})$ where $dI/dV(\mathbf{r},E)$ has the maximal slope - see Fig.~ %
\ref{Tun-Davis}(a).

It turns out that the corresponding average phonon energy $\bar{\Omega}$
depends on the oxygen mass, i.e. $\bar{\Omega}\sim M_{O}^{-1/2}$, with $\bar{%
\Omega}_{16}=52$ $meV$ and $\bar{\Omega}_{18}\approx 48$ $meV$ -
as it is seen in Fig.~\ref{Tun-Davis}(b). This result is
interpreted in \citep{Davis} as an evidence that oxygen phonons
are strongly involved in the
quasi-particle scattering. A possible explanation is put forward in \citep%
{Davis} by assuming that this isotope effect is due to the
B$_{1g}$ phonon which interacts with the anti-nodal
quasi-particles. However, this result requires a reanalysis since
the energy of the bosonic mode in fact coincides with the dip and
not with the peak of $d^{2}I/dV^{2}(\mathbf{r},E)$ - as reported
in \citep{Davis}.

The important message of most tunnelling experiments in $HTSC$ cuprates
\textit{near and at the optimal doping} is that there is strong evidence for
the importance of $EPI$ in the quasi-particle scattering and that no
particular phonon mode can be singled out in the spectral function $\alpha
^{2}F(\omega )$ as being the only one which dominates in pairing mechanism.
This important result means that the high $T_{c}$ is not attributable to a
particular phonon mode in the $EPI$ mechanism but all phonon modes
contribute to $\lambda _{ep}$. Having in mind that the phonon spectrum in $%
HTSC$ cuprates is very broad (up to $80$ $meV$ ), then the large $EPI$
constant ($\lambda _{ep}\gtrsim 2$) obtained in tunnelling experiments is
not surprising at all. Note, that similar conclusion holds for some other
oxide superconductors such as $Ba_{9.6}K_{0.4}BiO_{3}$ with $T_{c}=30$ $K$
where the peaks in the bosonic spectral function /extracted from tunnelling
measurements) coincide with the peaks in the phononic density of states \citep%
{Huang}, \citep{Jensen}).

\subsection{Phonon spectra and EPI}

Although experiments related to phonon spectra and their renormalization by $%
EPI$, such as inelastic neutron, inelastic x-ray and Raman
scattering, do not give the spectral function $\alpha ^{2}F(\omega
)$ they nevertheless can give useful, but indirect, information on
the strength of EPI for some particular phonons. We stress in
advance that the interpretation of experimental results in HTSC
cuprates by the theory of EPI for weakly correlated electrons is
risky since in strongly correlated systems, such as HTSC cuprates,
the phonon renormalization due to EPI is different than in weakly
correlated metals \citep{GunnarssonRoschEpi}. Since these
questions are reviewed in \citep{GunnarssonRoschEpi} we shall
briefly enumerate the main points: (\textit{1}) in strongly
correlated systems the EPI coupling for a number of phononic modes
can be significantly larger than the LDA-DFT and Hartree-Fock
methods predict. This is due to many-body effects not contained in
LDA-DFT \citep{GunnarssonRoschEpi}, \citep{Becker}. The lack of
the LDA-DFT calculations in obtaining phonon line-widths is
clearly demonstrated, for instance in experiments on
L$_{2-x}$Sr$_{x}$CuO$_{4}$ - see review \citep{Reznik} and
References therein, where the bond-stretching phonons at
$\mathbf{q}=(0.3,0,0)$ are \textit{softer and much broader} than
the LDA-DFT calculations predict. (Note, the wave vector
$\mathbf{q}$ is in units ($2\pi /a,2\pi /b,2\pi /c$) - for
instance in these units $(\pi ,\pi )$ corresponds to $(0.5,0.5)$.)
(\textit{2}) The calculation of phonon spectra is in principle
very difficult problem since besides the complexity of structural
properties in a given material one should take into account
appropriately the long-range Coulomb interaction of electrons as
well as strong short-range repulsion. Our intention is not to
discuss this complexity here - for that see for instance in
\citep{Ginzburg}, but we only stress some important points which
will help to understand problems with which is confronted the
theory of phonons in cuprates.

The phonon Green's function $D\mathbf{(q},\omega )$ depends on the phonon
self-energy $\Pi \mathbf{(q},\omega )$ which takes into account all the
enumerated properties (note $D^{-1}\mathbf{(q},\omega )=D_{0}^{-1}\mathbf{(q}%
,\omega )-\Pi \mathbf{(q},\omega )$). In cases when the EPI coupling
constant $g(\mathbf{k},\mathbf{k}^{\prime })$ is a function on the transfer
momentum $\mathbf{q}=\mathbf{k}-\mathbf{k}^{\prime }$ only, then $\Pi
\mathbf{(}q)$ ($q=(\mathbf{q},i\omega _{n})$) depends on the quasi-particle
charge susceptibility $\chi _{c}\mathbf{(}q)=P\mathbf{(}q)/\varepsilon _{e}%
\mathbf{(}q)$
\begin{equation}
\Pi \mathbf{(}q)=\left\vert g_{\mathbf{q}}\right\vert ^{2}\chi _{c}\mathbf{(}%
q),  \label{phonon-self}
\end{equation}%
and $P\mathbf{(}q)$ is the \textit{irreducible electronic polarization}
given by
\begin{equation}
P\mathbf{(}q)=-\sum_{p}G(p+q)\Gamma _{c}(p,q)G(p).  \label{el-pol}
\end{equation}%
The screening due to the long-range Coulomb interaction is contained in the
electronic dielectric function $\varepsilon _{e}\mathbf{(}q)$ while
the"screening" due to (strong) correlations is described by the charge
vertex function $\Gamma _{c}(p,q)$. Due to complexity of the physics of
strong correlations the phonon dynamics was studied in the $t-J$ model but
without the long-range Coulomb interaction \citep{GunnarssonRoschEpi}, \citep%
{Becker}, \citep{Khaliullin}, in which case one has $\varepsilon _{e}=1$ and $%
\chi _{c}\mathbf{(}q)=P\mathbf{(}q)$. However, in studying the
phonon spectra in cuprates it is expected that this deficiency is
partly compensated by choosing the bare phonon frequency $\omega
_{0}(\mathbf{q})$ (contained in $D_{0}^{-1}\mathbf{(q},\omega )$)
to correspond to undoped compounds \citep{GunnarssonRoschEpi},
\citep{Khaliullin}. It is a matter of future investigations to
incorporate all relevant interactions in order to obtain a fully
microscopic and reliable theory of phonons in cuprates.
Additionally, the electron-phonon interaction (with the bare
coupling constant $g_{\mathbf{q}}$) is dominated by the change of
the energy of the
Zhang-Rice singlet - see more Part II Section VII, and in that Eq.(\ref%
{phonon-self}) for $\Pi \mathbf{(}q)$ is adequate one
\citep{KulicReview}, \citep{GunnarssonRoschEpi}, \citep{Becker}.
Since the charge fluctuations in HTSC cuprates are strongly
suppressed (no doubly occupancy of the Cu 3d$^{9}$ state) due to
strong correlations and since the suppressed value of $\chi
_{c}\mathbf{(}q)$ can not be obtained by the band-structure
calculations this means that \textit{LDA-DFT underestimates the
EPI coupling constant significantly}. In the following we discuss
this important result briefly.

\subsubsection{\textit{Inelastic neutron and x-ray scattering - phonon
softening and width due to EPI }}

\begin{figure}[!tbp]
\begin{center}
\resizebox{.55 \textwidth}{!} {\includegraphics*[
width=6cm]{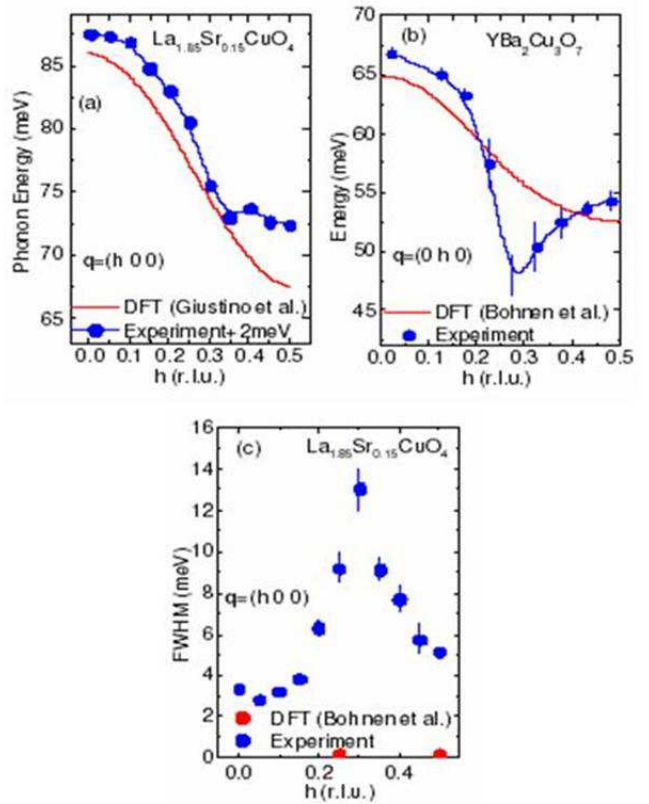}}
\end{center}
\caption{Comparison of DFT calculations with experimental results of
inelastic x-ray scattering:(a) phonon energies in $La_{1.85}Sr_{0.15}CuO_{4}$
and (b) in $YBa_{2}Cu_{3}O_{7}$; (c) phonon line widths in $%
La_{1.85}Sr_{0.15}CuO_{4}$. DFT calculations
\protect\citep{BohnenCohen} gives much smaller width than
experiments \protect\citep{PhononExp}. From
\protect\citep{ReSaGuDe}.} \label{LDAvsExp}
\end{figure}

The softening and broadening of numerous phonon modes has been observed in
the normal state of cuprates giving evidence for pronounced EPI effects and
for inadequacy of the LDA-DFT calculations in treating strong correlations
and suppression of the charge susceptibility \citep{KulicReview}, \citep%
{GunnarssonReview2008}, \citep{GunnarssonRoschEpi},
\citep{Khaliullin}. There are several relevant reviews on this
subject \citep{GunnarssonReview2008} , \citep{Reznik},
\citep{GunnarssonRoschEpi}, \citep{Pintschovius} and here we
discuss briefly two important examples which demonstrate the
inefficiency of the LDA-DFT band structure calculations to treat
quantitatively EPI in HTSC
cuprates. For instance, the Cu-O bond-stretching phonon mode shows a \textit{%
substantial softening} at $\mathbf{q}_{hb}=(0.3,0,0)$ by doping \ of $%
La_{1.85}Sr_{0.15}CuO_{4}$ and $YBa_{2}Cu_{3}O_{7\text{ }}$
\citep{Reznik}, \citep{Pintschovius} - called the
\textit{half-breathing phonon}, and a \textit{large broadening} by
$5$ $meV$ at $15$ $\%$ doping \citep{PhononExp} as it is seen in
Fig.~\ref{LDAvsExp}. While the softening can be partly described
by the LDA-DFT method \citep{BohnenHeid2003}, the latter theory
\textit{\ predicts an order of magnitude smaller broadening} than
the experimental one. This failure of LDA-DFT is due to incorrect
treatment of the ffects of strong correlations on the charge
susceptibility $\chi _{c}(q)$ and due to absence of many body
effects which can increase the coupling constant $g_{\mathbf{q}}$
- see more in \textit{Part II}. The neutron scattering
measurements in $La_{1.85}Sr_{0.15}CuO_{4}$ give evidence for
large ($30$ $\%$) softening of the $O_{Z}^{Z}$ with $\Lambda _{1}$
symmetry with the energy $\omega \approx 60$ $meV$, which is
theoretically predicted in \citep{FalterOPhonon}, and for the
large line-width about $17$ $meV$ which also suggests strong EPI.
These apex modes are favorable for $d$-wave
pairing since their coupling constants are peaked at small momentum $q$ \citep%
{GunnarssonReview2008}. Having in mind the above results, there is not
surprising at all that the recent calculations of the EPI coupling constant $%
\lambda _{ep}$ in the framework of LDA-DFT give very small EPI
coupling constant $\lambda _{ep}\approx 0.3$ \citep{BohnenCohen},
\citep{Giuistino}. The critical analysis of the LDA-DFT results in
HTSC cuprates is done in
\citep{KulicReview} and recently additionally argued in \citep%
{GunnarssonReview2008}, \citep{ReSaGuDe} by their disagreement
with inelastic
neutron and x-ray scattering measurements - as it is shown in Fig.~\ref%
{LDAvsExp}.

In \textit{Part II} we shall discuss some theoretical approaches
related to EPI in strongly correlated systems but we shall not
study the phonon renormalization. The latter problem was studied
more detailed and thoroughly in the review article
\citep{GunnarssonReview2008}, \citep{Reznik}. Here, we point out
only three important results. \textit{First}, there is an
appreciable difference in the \textit{phonon renormalization} in
strongly and weakly correlated systems. Namely, the change of
phonon frequencies in the presence of conduction electrons is
proportional to the squared coupling
constant $\left\vert g_{\mathbf{q}}\right\vert $ and charge susceptibility $%
\chi _{c}$, i.e. $\delta \omega (\mathbf{q})\sim \left\vert g_{\mathbf{q}%
}\right\vert ^{2}Re\chi _{c}$, while the \textit{line width} is given by $%
\Gamma _{\omega (\mathbf{q})}\sim \left\vert g_{\mathbf{q}}\right\vert
^{2}\left\vert Im\chi _{c}\right\vert $. All these quantities can be
calculated in LDA-DFT and as we discussed above where\ for some modes one
obtains that $\Gamma _{\omega (\mathbf{q})}^{(LDA)}\ll \Gamma _{\omega (%
\mathbf{q})}^{(\exp )}$. However, it turns out that in strongly correlated
systems doped by holes (with the concentration $\delta \ll 1$) the charge
fluctuations are suppressed in which case the following sum-rule holds \citep%
{GunnarssonReview2008}, \citep{Khaliullin}
\begin{equation}
\frac{1}{\pi N}\sum_{\mathbf{q}\neq 0}\int_{-\infty }^{\infty }\left\vert
Im\chi _{c}(\mathbf{q},\omega )\right\vert d\omega =2\delta (1-\delta )N,
\label{chi-c}
\end{equation}%
while in the LDA-DFT method one has
\begin{equation}
\frac{1}{\pi N}\sum_{\mathbf{q}\neq 0}\int_{-\infty }^{\infty }\left\vert
Im\chi _{c}(\mathbf{q},\omega )\right\vert ^{LDA}d\omega =(1-\delta )N.
\label{chi-c-LDA}
\end{equation}%
The inequality $\Gamma _{\omega (\mathbf{q})}^{(LDA)}\ll \Gamma _{\omega (%
\mathbf{q})}^{(\exp )}$ (for some phonon modes) together with Eqs.(\ref%
{chi-c}-\ref{chi-c-LDA}) means that for low doping $\delta \ll 1$ the LDA
\textit{strongly underestimates the EPI\ coupling constant }in the large
portion of the Brillouin zone, i.e. one has $\left\vert g_{\mathbf{q}%
}^{(LDA)}\right\vert \ll \left\vert g_{\mathbf{q}}^{(\exp )}\right\vert $.
The large softening and the large line width of the half-breathing mode at $%
q=(0.5,0)$ but very moderate effects for the breathing mode at
$q=(0.5,0.5)$ is explained in the framework of the one
\textit{slave-boson (SB) theory} (for $U=\infty $) in
\citep{Khaliullin}, where $\chi _{c}(\mathbf{q},\omega )$ (i.e.
$\Gamma _{c}(p,q)=\Gamma _{c}(\mathbf{p},q)$) is calculated in
leading $O(1/N)$ order. We stress that there is another method for
studying strong correlations - the \textit{X-method}, where the
controllable $1/N$ expansion
is performed in terms of Hubbard operators and where the charge vertex $%
\Gamma _{c}(\mathbf{p},q)$ is calculated \citep{KulicReview},
\citep{Kulic1},
\citep{Kulic2}, \citep{Kulic3}. It turns out that in the adiabatic limit ($%
i\omega _{n}=0$) the vertex functions $\Gamma _{c}(\mathbf{p}_{F},\mathbf{q}%
) $ in two methods have some important differences. For instance,
$\Gamma _{c}^{(X)}(\mathbf{p}_{F},q)$ (in the X-method) is peaked
at $q=0$ - the so called forward scattering peak, while $\Gamma
_{c}^{(SB)}(\mathbf{p}_{F},q)$ has maximum at finite $q\neq 0$
\citep{KuAk2} - see Part II, Section VI.E. The properties of
$\Gamma _{c}^{(X)}(\mathbf{p}_{F},q)$ are confirmed by
numerical Monte Carlo calculations in the finite-U Hubbard model \citep%
{ScalapinoHanke}, where it is found that FSP exists for all $U$,
but it is especially pronounced in the limit $U\gg t$. These
results are also confirmed in \citep{Cerruti4SB} where the
calculations are performed in the four slave-bosons technique -
see more Part II Section VII.E. Having in mind
this difference it would be useful to have calculations of $\chi _{c}(%
\mathbf{q},\omega )$ in the framework of the X-method which are
unfortunately not done yet. \textit{Second}, the many body theory gives that
for coupling to some modes the coupling constant $\left\vert g_{\mathbf{q}%
}\right\vert $ in HTSC cuprates can be significantly larger than
LDA-DFT predicts \citep{GunnarssonReview2008}, which is due to
some many-body effects not present in the latter \citep{Becker}.
In Part II it will be argued that for some phonon modes one has
$\left\vert g_{\mathbf{q}}\right\vert ^{2}\gg $ $\left\vert
g_{\mathbf{q}}^{(LDA)}\right\vert ^{2}$. For instance, for the
half-breathing mode, one has $\left\vert g_{\mathbf{q}}\right\vert
^{2}\approx 3\left\vert g_{\mathbf{q}}\right\vert _{LDA}^{2}$ \citep%
{GunnarssonReview2008}, \citep{Becker} - see Part II. These two
results point to an inadequacy of LDA-DFT in calculations of EPI
effects in HTSC cuprates. \textit{Third}, the phonon self-energy
($\Pi \mathbf{(}q)$) and quasi-particle self-energy $\Sigma (k)$
are differently renormalized by
strong correlations \citep{KulicReview}, \citep{GunnarssonReview2008}, \citep%
{Kulic1}, \citep{Kulic2}, \citep{Kulic3}, which is the reason that
$\Pi \mathbf{(}q)$ is much more suppressed than $\Sigma (k)$ - see
Part II. The effects of the charge vertex on $\Pi (q)$ and $\Sigma
(k)$ are differently
manifested. Namely, the vertex function enters \textit{quadratically} in $%
\Sigma (k)$ and the presence of the forward scattering peak in the charge
vertex \textit{strongly affects} the EPI coupling constant $g_{\mathbf{q}}$
in $\Sigma (k)$%
\begin{equation}
\Sigma (k)=-\sum_{q}\left\vert g_{\mathbf{q}}\cdot \gamma _{c}(\mathbf{k}%
,q)\right\vert ^{2}D(q)g(k+q),  \label{elec-self}
\end{equation}%
where $g(k)(\equiv G(k)/Q)$ is the quasi-particle Green's function, $\gamma
_{c}(\mathbf{k},q)=\Gamma _{c}(\mathbf{k},q)/Q$ is the quasi-particle vertex
and $Q(\sim \delta )$ is the Hubbard quasi-particle spectral weight - see
Part II, Section VII. In the adiabatic limit $\left\vert \mathbf{q}%
\right\vert >q_{\omega }=\omega _{ph}/v_{F}$ one has $\gamma _{c}(\mathbf{k}%
,q)\approx \gamma _{c}(\mathbf{k},\mathbf{q})$ and for $q\gg q_{c}(\approx
\delta \cdot \pi /a)$ the charge vertex is strongly suppressed ($\gamma _{c}(%
\mathbf{k},\mathbf{q})\ll 1$) making the effective EPI coupling (which also
enters the pairing potential) small at large (transfer) momenta $\mathbf{q}$%
. This has strong repercussion on the pairing due to EPI since for small
doping it makes the d-wave pairing coupling constant to be of the order of
the s-one ($\lambda _{d}\approx \lambda _{s}$). Then in the presence of the
residual Coulomb interaction EPI gives rise to d-wave pairing. On the other
side the charge vertex $\Gamma _{c}(\mathbf{k},q)$ enters $\Pi \mathbf{(}q)$
\textit{linearly }and it is \textit{additionally integrated} over the
quasi-particle momentum $\mathbf{k}\mathbb{\ }$- see Eq.(\ref{el-pol}).
Therefore, one expects that the effects of the forward scattering peak on $%
\Pi \mathbf{(}q)$ are less pronounced than on $\Sigma (k)$. Nevertheless,
the peak of $\Gamma _{c}(\mathbf{k},q)$ at $q=0$ may be (partly) responsible
that the maximal experimental softening and broadening of the stretching
(half-breading) mode in $La_{1.85}Sr_{0.15}CuO_{4}$ and $YBa_{2}Cu_{3}O_{7%
\text{ }}$is at $\mathbf{q}_{hb}=(0.3,0,0)$ \citep{Reznik} and not at $%
\mathbf{q}_{hb}=(0.5,0)$ for which $g_{\mathbf{q}_{b}}$ reaches maximum.
This means that the charge vertex function pushes the renormalized EPI
coupling constant to smaller $\mathbf{q}$. It would be very interesting to
have calculations for other phonons by including the vertex function
obtained by the X-method - see Section.VII

\subsubsection{\textit{The phonon Raman scattering }}

The \textit{phonon} Raman scattering gives an indirect evidence for
importance of EPI in cuprates \citep{Cardona1}, \citep{Cardona2}, \citep%
{Hadjiev}. We enumerate some of them - see more in
\citep{KulicReview} and References therein. (i) There is a
pronounced \textit{asymmetric line-shape}
(of the Fano resonance) in the metallic state. For instance, in $%
YBa_{2}Cu_{3}O_{7}$ two Raman modes at $115$ $cm^{-1}$ ($Ba$ dominated mode)
and at $340$ $cm^{-1}$ ($O$ dominated mode in the $CuO_{2}$ planes)\ show
pronounced asymmetry which is absent in $YBa_{2}Cu_{3}O_{6}$. This asymmetry
means that there is an appreciable interaction of Raman active phonons with
continuum states (quasi-particles). (ii) The phonon frequencies for some $%
A_{1g}$ and $B_{1g}$ are strongly renormalized in the
superconducting state, between $(6-10)$ $\%$, pointing again to
the importance of EPI \citep{Hadjiev} - see also in
\citep{KulicReview}, \citep{KulicAIP}. To this point we mention
that there is a remarkable correlation between the electronic
Raman cross-section $\tilde{S}_{\exp }(\omega )$ and the optical
conductivity in the $a-b$ plane $\sigma (\omega )$, i.e.
$\tilde{S}_{\exp }(\omega )\sim \sigma _{ab}(\omega )$
\citep{KulicReview}. In previous Subsections it is argued that EPI
with the very broad spectral function $\alpha ^{2}F(\omega )$
($0<\omega \lesssim 80$ $meV$) explains in a natural way the $\omega $ and $%
T $ dependence of $\sigma _{ab}(\omega )$. This means that the
electronic Raman spectra in cuprates can be explained by EPI in
conjunction with strong correlations. This conclusion is supported
by calculations of the Raman cross-section \citep{Rashkeev} which
take into account EPI with $\alpha
^{2}F(\omega )$ extracted from tunnelling measurements in $%
YBa_{2}Cu_{3}O_{6+x}$ and $Bi_{2}Sr_{2}CaCu_{2}O_{8+x}$
\citep{KulicReview}, \citep{TunnelingVedeneev}-\citep{Tsuda}.
Quite similar properties (to cuprates) of the electronic Raman
scattering, as well as of $\sigma (\omega ) $, $R(\omega )$ and
$\rho (T)$, were observed in experiments \citep{Bozovic}
on isotropic 3D metallic oxides $La_{0.5}Sr_{0.5}CoO_{3}$ and $%
Ca_{0.5}Sr_{0.5}RuO_{3}$ where there are no signs of antiferromagnetic
fluctuations. This means that low-dimensionality and antiferromagnetic spin
fluctuations cannot be a prerequisite for anomalous scattering of
quasi-particles and EPI must be inevitably taken into account since it is
present in all these compounds.

\subsection{Isotope effect in $T_{c}$ and $\Sigma (\mathbf{k},\protect\omega %
)$}

The isotope effect $\alpha _{T_{C}}$ in the critical temperature $T_{c}$ was
one of the very important proof for the EPI pairing in low-temperature
superconductors (LTSC). As a curiosity the isotope effect in $LTSC$ systems
was measured almost exclusively in mono-atomic systems and in few polyatomic
systems: the hydrogen isotope effect in $PdH$, the $Mo$ and $Se$ isotope
shift of $T_{c}$ in $Mo_{6}Se_{8}$, and the isotope effect in $Nb_{3}Sn$ and
$MgB_{2}$. We point out that very small ($\alpha _{T_{C}}\approx 0$ in $Zr$
and $Ru$) and even negative (in $PdH$) isotope effect in some polyatomic
systems of $LTSC$ materials are compatible with the $EPI$ pairing mechanism
but in the presence of substantial Coulomb interaction or lattice
anharmonicity. The isotope effect $\alpha _{T_{C}}$cannot be considered as
the smoking gun effect since it is sensitive to numerous influences. For
instance, in $MgB_{2}$ it is with certainty proved that the pairing is due
to EPI and strongly dominated by the boron vibrations, but the boron isotope
effect is significantly reduced, i.e. $\alpha _{T_{C}}\approx 0.3$ and the
origin for this smaller value is still unexplained. The situation in HTSC
cuprates is much more complicated because they are \textit{strongly
correlated systems and contain many-atoms in unit cell}. Additionally, the
situation is complicated with the presence of \textit{intrinsic and
extrinsic inhomogeneities, low dimensionality} which can mask the isotope
effects. On the other hand new techniques such as ARPES, STM, $\mu SR$ allow
studies of the isotope effects in quasi-particle self-energies, i.e. $\alpha
_{\Sigma }$, which will be discussed below.

\subsubsection{Isotope effect $\protect\alpha _{T_{C}}$ in $T_{c}$}

This problem will be discussed only briefly since more extensive
discussion can be found in \citep{KulicReview}. It is well known
that in the pure EPI
pairing mechanism, the total isotope coefficient $\alpha $ is given by $%
\alpha _{T_{C}}=\sum_{i,p}\alpha _{i}^{(p)}=-\sum_{i,p}d\ln T_{c}/d\ln
M_{i}^{(p)}$, where $M_{i}^{(p)}$ is the mass of the i-th element in the $p$%
-th crystallographic position. We stress that the total isotope effect is
not measured in HTSC cuprates but only some partial ones. Note, that in the
case when the screened Coulomb interaction is negligible, i.e. $\mu
_{c}^{\ast }=0$, the theory predicts $\alpha _{T_{C}}=1/2$. From this
formula one can deduce that the relative change of $T_{c}$, $\delta
T_{c}/T_{c}$, for heavier elements should be rather small - for instance it
is $0.02$ for $^{135}Ba\rightarrow ^{138}Ba$, $0.03$ for $^{63}Cu\rightarrow
^{65}Cu$ and $0.07$ for $^{138}La\rightarrow ^{139}La$. This means that
measurements of $\alpha _{i}$ for heavier elements are at/or beyond the
ability of present day experimental techniques. Therefore most isotope
effect measurements were done by substituting light atoms $^{16}O$ by $%
^{18}O $ only. It turns out that in most optimally doped HTSC cuprates $%
\alpha _{O}$ is rather small. For instance $\alpha _{O}\approx 0.02-0.05$ in
$YBa_{2}Cu_{3}O_{7}$ with $T_{c,\max }\approx 91$ $K$, but it is appreciable
in $La_{1.85}Sr_{0.15}CuO_{4}$ with $T_{c,\max }\approx 35$ $K$ where$\
\alpha _{O}\approx 0.1-0.2$. In $Bi_{2}Sr_{2}CaCu_{2}O_{8}$ with $T_{c,\max
}\approx 76$ $K$ one has $\alpha _{O}\approx 0.03-0.05$ while $\alpha
_{O}\approx 0.03$ and even negative ($-0.013$) in $%
Bi_{2}Sr_{2}Ca_{2}Cu_{2}O_{10}$ with $T_{c,\max }\approx 110$ $K$. The
experiments on $Tl_{2}Ca_{n-1}BaCu_{n}O_{2n+4}$ ($n=2,3$) with $T_{c,\max
}\approx 121$ $K$ are still unreliable and $\alpha _{O}$ is unknown. In the
electron-doped $(Nd_{1-x}Ce_{x})_{2}CuO_{4}$ with $T_{c,\max }\approx 24$ $K$
one has $\alpha _{O}<0.05$ while in the underdoped materials $\alpha _{O}$
increases. The largest $\alpha _{O}$ is obtained even in the optimally doped
compounds like in systems with substitution, such as $%
La_{1.85}Sr_{0.15}Cu_{1-x}M_{x}O_{4}$, $M=Fe,Co$, where $\alpha _{O}\approx
1.3$ for $x\approx 0.4$ $\%$. In $La_{2-x}M_{x}CuO_{4}$ there is a $Cu$
isotope effect which is of the order of the oxygen one, i.e. $\alpha
_{Cu}\approx \alpha _{O}$ giving $\alpha _{Cu}+\alpha _{O}\approx 0.25-0.35$
for optimally doped systems ($x=0.15$). In case when $x=0.125$ with $%
T_{c}\ll T_{c,\max }$ one has$\ \alpha _{Cu}\approx 0.8-1$ with
$\alpha _{Cu}+\alpha _{O}\approx 1.8$ \citep{Franck}. The
appreciation of copper isotope effect in $La_{2-x}M_{x}CuO_{4}$
tells us that vibrations other than oxygen ions are important in
giving high T$_{c}$. In that sense one should have in mind the
tunnelling experiments discussed above, which tell us that
all phonons contribute to the Eliashberg pairing function $\alpha ^{2}F(%
\mathbf{k},\omega )$ and according to these results the oxygen
modes give moderate contribution to $T_{c}$ \citep{Tsuda}. Hence
the small oxygen isotope effect $\alpha _{T_{c}}^{(O)}$ in
optimally doped cuprates, if it is an intrinsic property at all
(due to pronounced local inhomogeneities of samples and quasi-two
dimensionality of the system), does not exclude the EPI mechanism
of pairing.

\subsubsection{Isotope effect $\protect\alpha_{\Sigma }$ in the self-energy}

\begin{figure}[!tbp]
\begin{center}
\resizebox{.45 \textwidth}{!} {\includegraphics*
[width=6cm]{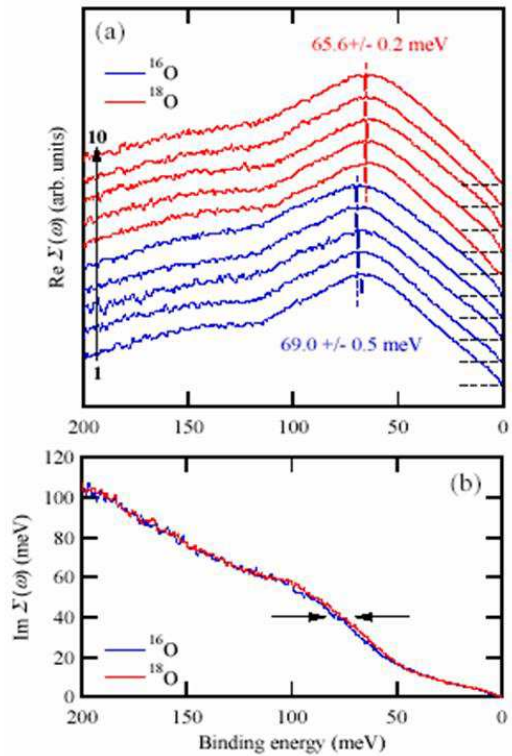}}
\end{center}
\caption{(a) Effective $Re\Sigma $ for five samples for $O^{16}$ (blue) and $%
O^{18}$ (red) along the nodal direction. (b) Effective $Im\Sigma $
determined from MDC full widths. An impurity term is subtracted at $\protect%
\omega =0$. From \protect\citep{IwasawaIsotop2}.}
\label{Iwasawa-isotop}
\end{figure}

The fine structure of the quasi-particle self-energy $\Sigma (\mathbf{k}%
,\omega )$, such as kinks and slopes, can be resolved in ARPES
measurements and in some respect in STM. It turns out that there
is isotope effect in the self-energy in the optimally doped
$Bi-2212$ samples \citep{LanzaraIsotope}, \citep{DouglasIsotop},
\citep{IwasawaIsotop}. In the first paper on this subject
\citep{LanzaraIsotope} it is reported a \textit{red shift} $\delta
\omega _{k,70}\sim -(10-15)$ $meV$ of the nodal kink at $\omega
_{k,70}\simeq 70$ $meV$ for the $^{16}O\rightarrow O^{18}$
substitution. In \citep{LanzaraIsotope} it is reported that the
isotope shift of the self-energy $\delta \Sigma =\Sigma
_{16}-\Sigma _{18}\sim 10$ $meV$ is very pronounced at large
energies $\omega =100-300$ $meV$. Concerning the latter
result, there is a dispute since it is not confirmed experimentally in \citep%
{DouglasIsotop}, \citep{IwasawaIsotop}. However, the isotope effect in $%
Re\Sigma (\mathbf{k},\omega )$ at low energies \citep{DouglasIsotop}, \citep%
{IwasawaIsotop} is well described in the framework of the
Migdal-Eliashberg theory for EPI \citep{MaKuDo} which is in
accordance with the recent ARPES measurements with low-energy
photons $\sim 7$ $eV$ \citep{IwasawaIsotop2}. The latter allowed
very good precision in measuring the isotope effect in
the nodal point of Bi-2212 with $T_{c}^{16}=92.1$ $K$ and $T_{c}^{18}=91.1$ $%
K$ \citep{IwasawaIsotop2}. They observed shift in the maximum of $Re\Sigma (%
\mathbf{k}_{N},\omega )$ - at $\omega _{k,70}\approx 70$ $meV$ (it
corresponds to the half-breathing or to the breathing phonon) by $\delta
\omega _{k,70}\approx 3.4\pm 0.5$ $meV$ as shown in Fig. \ref{Iwasawa-isotop}%
.

By analyzing the shift in Im$\Sigma (\mathbf{k}_{N},\omega )$ -
shown in Fig. \ref{Iwasawa-isotop}, one finds similar result for
$\delta \omega _{k,70}\approx 3.2\pm 0.6$ $meV$. The similar shift
was obtained in STM measurements \citep{Davis} which is shown in
Fig.~\ref{Tun-Davis}(b) and can
have its origin in different phonons. We would like to stress two points: ($%
i $) in compounds with $T_{c}\sim 100$ $K$ the oxygen isotope effect in $%
T_{c}$ is moderate, i.e. $\alpha _{T_{c}}^{(O)}<0.1$,
\citep{IwasawaIsotop2}. If we consider this value to be intrinsic
then even in this case it is not in conflict with the tunnelling
experiments \citep{Tsuda} since the latter
give evidence that vibrations of heavier ions contribute significantly to $%
T_{c}$ - see the discussion in Subsection D on the tunnelling spectroscopy; (%
$ii$) in ARPES measurements of \citep{IwasawaIsotop2} the
effective EPI coupling constant $\lambda _{ep,eff}$ $\gtrsim 0.6$
is extracted, while the theory in Subsection C gives that the real
coupling constant is larger, i.e. $\lambda _{ep}>1.2$. This value
is significantly larger than the LDA-DFT
theory predicts $\lambda _{ep,LDA}<0.3$ \citep{BohnenCohen}, \citep{Giuistino}%
. This again points that the \textit{LDA-DFT method does not pick up the
many-body effects} due to strong correlations- see Part II.

\section{Summary of \textit{Part I}}

The analysis of experimental data in HTSC cuprates which are related to
optics, tunnelling and ARPES measurements \textit{near and at the optimal
doping} give evidence for the large electron-phonon interaction (EPI) with
the coupling constant $1<\lambda _{ep}<3.5$. We stress that this analysis is
done in the framework of the Migdal-Eliashberg theory for EPI which is a
reliable approach for systems near the optimal doping. The spectral function
$\alpha ^{2}F(\omega )$, averaged over the Fermi surface, is extracted from
various tunnelling measurements on bulk materials and tin films. It contains
peaks at the same energies as the phonon density of states $F_{ph}(\omega )$%
. So obtained spectral function when inserted in the Eliashberg equations
provides sufficient strength for obtaining high critical temperature T$%
_{c}\sim 100$ $K$. These facts are a solid proof for the important role of
EPI in the normal state scattering and pairing mechanism of cuprates. Such a
large (experimental) value of the EPI coupling constant and the robustness
of the d-wave superconductivity in the presence of impurities imply that the
EPI and impurity scattering amplitude must be strongly momentum dependent.
The IR optical reflectivity data provide additional support for the
importance of EPI since by using the spectral function (extracted from
tunnelling measurements) one can quantitatively explain frequency dependence
of the dynamical conductivity, optical relaxation rate and optical mass.
These findings related to EPI are additionally supported by ARPES
measurements on BISCO compounds. The ARPES kinks, the phononic features and
isotope effect in the quasi-particle self-energy in the nodal and anti-nodal
points at low energies ($\omega \ll \omega _{c}$) persist in the normal and
superconducting state. They are much more in favor of EPI than for the spin
fluctuation (SFI)\ scattering mechanism. The transport EPI coupling constant
in HTSC cuprates is much smaller than $\lambda _{ep}$, i.e. $\lambda
_{tr}\sim \lambda _{ep}/3$, which points to some peculiar scattering
mechanism not met in low-temperature superconductors. The different
renormalization of the quasi-particle and transport self-energies by the
Coulomb interaction (strong correlations) hints to the importance of the
small-momentum scattering in EPI. This will be discussed in \textit{Part II}.

The ineffectiveness of SFI to solely provide pairing mechanism in cuprates
comes out also from magnetic neutron scattering on YBCO, BISCO. As a result,
the imaginary part of the susceptibility is drastically reduced in the low
energy region by going from slightly underdoped toward optimally doped
systems, while $T_{c}$ is practically unchanged. This implies that the real
SFI coupling constant $\lambda _{sf}(\sim g_{sf}^{2})$ is small since the
experimental value $g_{sf}^{(\exp )}<0.2$ $eV$ is much smaller than the
assumed theoretical value $g_{sf}^{(th)}\approx (0.7-1.5)$ $eV$.

Inelastic neutron and x-ray scattering measurements in cuprates show that
the broadening of some phonon lines is by an order of magnitude larger than
the LDA-DFA methods predict. Since the phonon line-widths depend on the EPI
coupling and the charge susceptibility it is evident that calculations of
both quantities are beyond the range of applicability of LDA-DFT. As a
consequence, LDA-DFT overestimates electronic screening and thus
underestimates the EPI coupling, since many-body effects due to strong
correlations are not contained in this mean-field type theory. However, in
spite of the promising and encouraging experimental results about the
dominance of EPI in cuprates the theory is still confronted with
difficulties in explaining sufficiently large coupling constant in the
d-channel. At present there is no such a satisfactory microscopic theory
although some concepts, such as the the dominant EPI scattering at small
transfer momenta, are understood at least qualitatively. These set of
problems and questions will be discussed in \textit{Part II}.

\part{Theory of EPI in HTSC}

The experimental results in Part I give evidence that the
electron-phonon interaction (EPI) in HTSC cuprates is strong and
in order to be conform with $d$-wave pairing \textit{EPI must be
peaked at small transfer momenta}. A number of other experiments
in HTSC cuprates give evidence that these are \textit{strongly
correlated systems }with large on-site Coulomb repulsion of
electrons on the Cu-ions\textit{. }However, at present there is no
satisfactory microscopic theory of pairing in HTSC\ cuprates which
is capable to calculate $T_{c}$ and the order parameter. This is
due to mathematical difficulties in obtaining a solution of the
formally exact \textit{ab initio many-body equations} which take
into account two important ingredients - EPI and strong
correlations \citep{KulicReview}. In Section V we discuss first
the ab initio many body theory of superconductivity in order to
point places which are most difficult to be solved. Since the
superconductivity is low energy phenomenon (also in HTSC cuprates)
one can simplify the structure of the ab initio equations in the
low-energy sector (the Migdal-Eliashberg theory) where the
high-energy processes are incorporated in the (so called)
\textit{ideal band structure} (non-local)
potential $V_{IBS}(\mathbf{x},\mathbf{y})$ and \textit{vertex functions }$%
\Gamma $. Due to its complexity this program of calculation of $V_{IBS}(%
\mathbf{x},\mathbf{y})$, $\Gamma $, EPI couplings $g_{ep}(\mathbf{x},\mathbf{%
y})$ is not realized in HTSC superconductors. However, one pragmatical way
out is to calculate $g_{ep}$ in the framework of the LDA-DFT method which is
at present stage unable to treat strong correlations in a satisfactory
manner. Some achievements and results of LDA-DFT which are related to HTSC
cuprates are discussed in Section VI.

In the case of very complicated systems, such as the HTSC cuprates, the
standard (pragmatical) procedure in physics is to formulate a minimal
theoretical model - sometimes called toy model, which includes minimal set
of important ingredients necessary for qualitative and and semi-quantitative
study of a phenomenon. As the consequence of the experimental results\textit{%
, }the \textit{minimal theoretical model} for cuprates must comprise \textit{%
two important ingredients}: 1. \textit{EPI and 2. strong correlations}. In
Section VII we shall introduce such a \textit{minimal theoretical }$t-J$%
\textit{\ model which includes EPI }and discuss the renormalization of EPI
by strong correlations. In recent years the interest in this set of problems
is increased and numerous numerical calculations were done mostly on small
clusters with $n\times n$ atoms ($n<8$). We shall not discuss this subject
which is fortunately covered in the recent comprehensive review \citep%
{GunnarssonReview2008}. The analytical approaches in studying the
renormalization of EPI by strong correlations, which are based on a
controllable and systematic theory, are rather scarce. We shall discuss such
a systematic and controllable theory in the framework of \ the $t-J$ model
with $EPI$, which is formulated and solved in terms of Hubbard operators.
The theory of this (toy) model predicts some interesting effects which might
be important for understanding the physics of HTSC cuprates. It predicts
that the high-energy processes (due to the suppression of doubly occupancy
for $U\gg W_{b}$) give rise to a nonlocal contribution to the band structure
potential (self-energy $\Sigma (\mathbf{x},\mathbf{y},\omega =0)$) as well
as to EPI. This non-locality in EPI is responsible for the peak in the
effective pairing potential ($V_{ep,eff}(\mathbf{q},\omega )$) at small
transfer momenta $q$($\leqslant q_{c}\ll k_{F}$) \citep{KulicReview}, \citep%
{Kulic1}, \citep{Kulic2}. The latter property allows that the
(strong) EPI is conform with d-wave pairing in HTSC cuprates.
Furthermore, the peculiar structural properties of HTSC cuprates
and corresponding electronic quasi-two-dimensionality give an
additional non-locality in EPI. The latter is due to the weakly
screened Madelung energy for a number of lattice vibrations along
the c-axis. Since at present there is no quantitative theory for
the latter effect we tackle this problem here only briefly. The
next task for the future studies of the physics of HTSC cuprates
is to incorporate these structural properties in the minimal
theoretical model problem.

Finally, by writing this chapter our intention is not to overview the
theoretical studies of EPI in HTSC cuprates - an impossible task, but more
(i) to elucidate the descending way from the (old) well-defined ab initio
microscopic theory of superconductivity to the one of the minimal model
which treats the interplay of EPI and strong correlations, (ii) to encourage
the reader to further develop the theory of HTSC cuprates.

\section{Microscopic theory of superconductivity}

\subsection{Ab iniitio many-body theory}

The many-body theory of superconductivity is based on the fully
microscopic electron-ion Hamiltonian for electrons and ions in the
crystal - see for instance \citep{ScalapRev},
\citep{DierkLowTemp}. It comprises mutually interacting electrons
which also interact with ions and ionic vibrations. In order to
pass continually to the problem of interplay of EPI and strong
correlations and also to explain why the LDA-DFT method is
inadequate for HTSC cuprates we discuss this problem here with
restricted details - more extended discussion can be found in
\citep{KulicReview}, \citep{DierkLowTemp}.
In order to describe superconductivity the Nambu-spinor $\hat{\psi}^{\dag }(%
\mathbf{r})=(\hat{\psi}_{\uparrow }^{\dagger }(\mathbf{r}) $ $\hat{\psi}%
_{\downarrow }(\mathbf{r}))$ is introduced which operates in the
electron-hole space ($\hat{\psi}(\mathbf{r})=($\ $\hat{\psi}^{\dagger }(%
\mathbf{r}))^{\dag }$) where $\hat{\psi}_{\uparrow }(\mathbf{r})$, $\hat{\psi%
}_{\uparrow }^{\dagger }(\mathbf{r})$ are annihilation and creation
operators for spin up, respectively etc. The microscopic Hamiltonian which
in principle describes normal and superconducting states of the system under
consideration contains three parts $\hat{H}=\hat{H}_{e}+\hat{H}_{i}+\hat{H}%
_{e-i}$. The \textit{electronic Hamiltonian} $\hat{H}_{e}$, which describes
the kinetic energy and the Coulomb interactions of electrons is given by
\begin{equation*}
\hat{H}_{e}=\int d^{3}r\hat{\psi}^{\dagger }(\mathbf{r})\hat{\tau}%
_{3}\epsilon _{0}(\hat{p})\hat{\psi}(\mathbf{r})+
\end{equation*}%
\begin{equation}
+\frac{1}{2}\int d^{3}rd^{3}r^{\prime }\hat{\psi}^{\dagger }(\mathbf{r})\hat{%
\tau}_{3}\hat{\psi}(\mathbf{r})V_{c}(\mathbf{r-r}^{\prime })\hat{\psi}%
^{\dagger }(\mathbf{r}^{\prime })\hat{\tau}_{3}\hat{\psi}(\mathbf{r}^{\prime
}),  \label{Ab-3.3}
\end{equation}%
where $\epsilon _{0}(\hat{p})=\hat{p}^{2}/2m$ is the kinetic energy of
electron and $V_{c}(\mathbf{r-r}^{\prime })=e^{2}/\mid \mathbf{r-r}^{\prime
}\mid $ is the electron-electron Coulomb interaction. Note, in the
electron-hole space the pseudo-spin (Nambu) matrices $\hat{\tau}_{i}$, $%
i=0,1,2,3$ are Pauli matrices. Since we shall discuss only the electronic
properties the explicit form of the \textit{lattice Hamiltonian}\textbf{\ }$%
\hat{H}_{i}$ \citep{KulicReview}, \citep{DierkLowTemp} is omitted
here. The \textit{electron-ion Hamiltonian} describes the
interaction of electrons with the equilibrium lattice and with its
vibrations, respectively

\begin{equation*}
\hat{H}_{e-i}=\sum_{n}\int d^{3}rV_{e-i}(\mathbf{r}-\mathbf{R}_{n}^{0})\hat{%
\psi}^{\dagger }(\mathbf{r})\hat{\tau}_{3}\hat{\psi}(\mathbf{r})
\end{equation*}%
\begin{equation}
+\int d^{3}r\hat{\Phi}(\mathbf{r})\hat{\psi}^{\dagger }(\mathbf{r})\hat{\tau}%
_{3}\hat{\psi}(\mathbf{r}).  \label{Ab-3.5}
\end{equation}%
Here, $V_{e-i}(\mathbf{r}-\mathbf{R}_{n}^{0})$ is the electron-ion potential
where $Ze$ is the ionic charge and in the all-electron calculations it is
given by $V_{e-i}(\mathbf{r}-\mathbf{R}_{n}^{0})=-$ $Ze^{2}/\mid \mathbf{r-R}%
_{n}^{0}\mid $. The second term which is proportional to the lattice
distortion operator $\hat{\Phi}(\mathbf{r})=-\sum_{n,\alpha }\hat{u}_{\alpha
n}\nabla _{\alpha }V_{e-i}(\mathbf{r}-\mathbf{R}_{n}^{0})+\hat{\Phi}_{anh}(%
\mathbf{r})$ (because of convenience it includes also the $EPI$ coupling $%
\nabla _{\alpha }V_{e-i}$) describes the interaction of electrons with
harmonic ($\sim \hat{u}_{\alpha n}$) (or anharmonic $\sim \hat{\Phi}_{anh}(%
\mathbf{r})$) lattice vibrations.

The Dyson's equations for the electron and phonon Green's functions $\hat{G}%
(1,2)=-\langle T\hat{\psi}(1)\hat{\psi}^{\dagger }(2)\rangle $, $\tilde{D}%
(1-2)=-\langle T\hat{\Phi}(1)\hat{\Phi}(2)\rangle $ are $\hat{G}^{-1}(1,2)=%
\hat{G}_{0}^{-1}(1,2)-\hat{\Sigma}(1,2)$ and $\tilde{D}^{-1}(1,2)=\tilde{D}%
_{0}^{-1}(1,2)-\tilde{\Pi}(1,2)$, where the $\hat{G}_{0}^{-1}(1,2)=[(-\frac{%
\partial }{\partial \tau _{1}}-\epsilon _{0}(\mathbf{p}_{1})+\mu )\hat{\tau}%
_{0}-u_{eff}(1)\hat{\tau}_{3}]\delta (1-2)$ is the bare inverse electronic
Green's function. Here, $1=(\mathbf{r}_{1},\tau _{1})$, where $\tau _{1}$ is
the imaginary time in the Matsubara technique and the effective one-body
potential $u_{eff}(1)=V_{e-i}(1)+V_{H}++\langle \hat{\Phi}(1)\rangle $, $%
V_{H}$ is the Hartree potential. The electron and phonon self-energies $\hat{%
\Sigma}(1,2)$ and $\tilde{\Pi}(1,2)$ take into account many-body dynamics of
the interacting system. The \textit{electronic self-energy} $\hat{\Sigma}%
(1,2)=\hat{\Sigma}_{c}(1,2)+\hat{\Sigma}_{ep}(1,2)$ is obtained in the form
\begin{equation}
\hat{\Sigma}(1,2)=-V_{eff}(1,\bar{1})\hat{\tau}_{3}\hat{G}(1,\bar{2})\hat{%
\Gamma}_{eff}(\bar{2},2;\bar{1}),  \label{Ab-3.15}
\end{equation}%
where integration (summation) over the bar indices is understood. The
effective retarded potential $V_{eff}(1,\bar{1})$ in Eq.(\ref{Ab-3.15})
contains the screened (by the electron dielectric function\textbf{\ }$%
\varepsilon _{e}(1,2)$) Coulomb and $EPI$ interactions

\begin{equation}
V_{eff}(1,2)=V_{c}(1-\bar{1})\varepsilon _{e}^{-1}(\bar{1},2)+\varepsilon
_{e}^{-1}(1,\bar{1})\tilde{D}(\bar{1},\bar{2})\varepsilon _{e}^{-1}(\bar{2}%
,2).  \label{Ab-3.16}
\end{equation}%
The inverse electronic dielectric permeability is $\varepsilon
_{e}^{-1}(1,2)=\delta (1-2)+V_{c}(1-\bar{1})P(\bar{1},\bar{2})\varepsilon
_{e}^{-1}(\bar{2},2)$ is defined via the irreducible electronic polarization
operator $P(1,2)=-Sp\{\hat{\tau}_{3}\hat{G}(1,\bar{2})\hat{\Gamma}_{eff}(%
\bar{2},\bar{3};2)\hat{G}(\bar{3},1^{+})\}$. The \textit{vertex function} $%
\hat{\Gamma}_{eff}(1,2;3)=-\delta \hat{G}(1,2)/\delta u_{eff}(3)$ in Eq.(\ref%
{Ab-3.15}) is the solution of the complicated (and practically unsolvable)
integro-differential functional equation $\hat{\Gamma}_{eff}(1,2;3)=\hat{\tau%
}_{3}\delta (1-2)\delta (1-3)+\{\delta \hat{\Sigma}(1,2)/\delta \hat{G}(\bar{%
1},\bar{2})\}\hat{G}(\bar{1},\bar{3})\hat{\Gamma}_{eff}(\bar{3},\bar{4};3)%
\hat{G}(\bar{4},\bar{2})$. Note, that the effective vertex function $\hat{%
\Gamma}_{eff}(1,2;3)$, which takes into account all renormalizations going
beyond the simple Coulomb ($RPA$) screening, is the functional of both the
electronic and phononic Green's functions $\hat{G}$ and $\tilde{D}$, thus
making at present the ab initio microscopic equations practically unsolvable.

\subsection{Low-energy Migdal-Eliashberg\ theory}

If the vertex function $\hat{\Gamma}_{eff}$ would be known we would have
closed set of equations for Green's functions which describe dynamics of
interacting electrons and lattice vibrations (phonons) in the normal and
superconducting state. However, this is a formidable task and at present far
from any practical realization. Fortunately, we are mostly interested in%
\textit{\ low-energy phenomena }\ (with energies $\left\vert \omega
_{n}\right\vert ,\xi \ll \omega _{c}$ and for momenta $k=k_{F}+\delta k$ in
the shall $\delta k\ll \delta k_{c}$ near the Fermi momentum $k_{F}$; $%
\omega _{c}$ and $\delta k_{c}$ are some cutoffs which disappear after
calculations) which allows us further simplification of equations \citep%
{Migdal}. Therefore, the strategy is to \textit{integrate
high-energy processes} - see more in \citep{DierkLowTemp}, and we
sketch it only briefly.
Namely, the Green's function $\hat{G}(\mathbf{k},\omega _{n})=[i\omega _{n}-(%
\mathbf{k}^{2}/2m-\mu )\hat{\tau}_{3}-\hat{\Sigma}(\mathbf{k},\omega
_{n})]^{-1}$ can be formally written in the form $\hat{G}(\mathbf{k},\omega
_{n})=\hat{G}^{low}(\mathbf{k},\omega _{n})+\hat{G}^{high}(\mathbf{k},\omega
_{n})$, where $\hat{G}^{low}(\mathbf{k},\omega _{n})=\hat{G}(\mathbf{k}%
,\omega _{n})\Theta (\omega _{c}-\left\vert \omega _{n}\right\vert )\Theta
(\delta k_{c}-\delta k)$ is the low-energy Green's function and $\hat{G}%
^{high}(\mathbf{k},\omega _{n})=\hat{G}(\mathbf{k},\omega _{n})\Theta
(\left\vert \omega _{n}\right\vert -\omega _{c})\Theta (\delta k-\delta
k_{c})$ the high-energy one\ and analogously $D=D^{low}(\mathbf{k},\omega
_{n})+D^{high}(\mathbf{k},\omega _{n})$. By introducing the \textit{small
parameter} of the theory $s\sim (\omega /\omega _{c})\sim (\delta k/\delta
k_{c})\ll 1$ one has in leading order $\hat{G}^{low}(\mathbf{k},\omega
_{n})\sim s^{-1}$, \ $\hat{G}^{high}(\mathbf{k},\omega _{n})\lesssim 1$ and $%
D^{low}(\mathbf{k},\omega _{n})\sim s^{0}$, \ $D^{high}(\mathbf{k},\omega
_{n})\sim s^{2}$. Note, that the coupling constants ($V_{ei},\nabla
V_{ei},V_{ii}$ etc.) are of the order $s^{0}=1$.

This procedure of separating low-energy and high-energy processes lies also
behind the \textit{adiabatic approximation} since in most materials the
characteristic phonon (Debye) energy of lattice vibrations $\omega _{D\text{
}}$ is much smaller than the characteristic electronic Fermi energy $E_{F}$ (%
$\omega _{D\text{ }}\ll E_{F}$). In the small $s(\ll 1)$ limit the
Migdal theory \citep{Migdal} keeps in the total self-energy
$\Sigma $ linear terms in the phonon propagator $\tilde{D}$ ($D$)
only. In that case the effective vertex function can be written in
the form $\hat{\Gamma}_{eff}(1,2;3)\cong
\hat{\Gamma}_{c}(1,2;3)+\delta \hat{\Gamma}_{ep}(1,2;3)$
\citep{Migdal},
where the Coulomb charge vertex $\hat{\Gamma}_{c}(1,2;3)=\hat{\tau}%
_{3}\delta (1-2)\delta (1-3)+\delta \hat{\Sigma}_{c}(1,2)/\delta u_{eff}(3)$
contains correlations due to the Coulomb interaction only but does not
contain $EPI$ \ and phonon propagator $\tilde{D}$ explicitly. The part $%
\delta \hat{\Gamma}_{ep}(1,2;3)=\delta \hat{\Sigma}_{ep}(1,2)/\delta
u_{eff}(3)$ contains all linear terms with respect to $EPI$. Note, that in
these diagrams enters the dressed Green's function which contains implicitly
$EPI$ up to infinite order. By careful inspection of all (explicit)
contributions to $\delta \hat{\Gamma}_{ep}(1,2;3)$ which is linear in $%
\tilde{D}$ one can express the self-energy in terms of the charge (Coulomb)
vertex $\hat{\Gamma}_{c}(1,2;3)$ only. As a result of this approximation,
the part of the self-energy due to Coulomb interaction is given by

\begin{equation}
\hat{\Sigma}_{c}(1,2)=-V_{c}^{sc}(1,\bar{1})\hat{\tau}_{3}\hat{G}(1,\bar{2})%
\hat{\Gamma}_{c}(\bar{2},2;\bar{1}),  \label{Ab-3.29a}
\end{equation}%
where $V_{c}^{sc}(1,2)=V_{c}(1,\bar{2})\varepsilon _{e}^{-1}(\bar{2},2)$ is
the screened Coulomb interaction. The part which is due to $EPI$ has the
following form
\begin{equation}
\hat{\Sigma}_{ep}(1,2)=-V_{ep}(\bar{1},\bar{2})\hat{\Gamma}_{c}(1,\bar{3};%
\bar{1})\hat{G}(\bar{3},\bar{4})\hat{\Gamma}_{c}(\bar{4},2;\bar{2}),
\label{Ab-3.30a}
\end{equation}%
where $V_{ep}(1,2)=\varepsilon _{e}^{-1}(1,\bar{1})\tilde{D}(\bar{1},\bar{2}%
)\varepsilon _{e}^{-1}(\bar{2},2)$ is the screened $EPI$ potential. Note,
that $\hat{\Sigma}_{ep}(1,2)$ depends now quadratically on the charge vertex
$\hat{\Gamma}_{c}$, which is due to the adiabatic theorem.

It is well known that the Coulomb self-energy $\hat{\Sigma}_{c}(1,2)$ is the
most complicating part of the electronic dynamics but since we are
interested in low-energy physics when $s\ll 1$ then the term $\hat{\Sigma}%
_{c}(1,2)$ can be further simplified by separating it in two parts%
\begin{equation}
\hat{\Sigma}_{c}(1,2)=\hat{\Sigma}_{c}^{(h)}(1,2)+\hat{\Sigma}%
_{c}^{(l)}(1,2).  \label{Ab-3.42}
\end{equation}%
The term $\hat{\Sigma}_{c}^{(h)}(1,2)$ is \textit{due to high-energy
processes} contained in the product $\hat{G}^{high}(1,\bar{2})\hat{\Gamma}%
_{c}^{high}(\bar{2},2;\bar{1})$ (for instance due to the large Hubbard $U$
in strongly correlated systems) and $\hat{\Sigma}_{c}^{(l)}(1,2)$ is due to
\textit{low energy processes}. The leading part part of $\hat{\Sigma}%
_{c}^{(h)}(1,2)$ is $1$, i.e. $\hat{\Sigma}_{c}^{(h)}(1,2)\sim s^{0}$, while
$\hat{\Sigma}_{c}^{(l)}(1,2)$ is small of order $1$, i.e. $\hat{\Sigma}%
_{c}^{(l)}(1,2)\sim s^{1}$. For further purposes we define the quantity $%
\hat{V}_{0}$
\begin{equation}
\hat{V}_{0}(1,2)=\{V_{e-i}(1)+V_{H}(1)\}\tau _{3}\delta (1-2)+\hat{\Sigma}%
_{c}^{(h)}(1,2).  \label{Ab-3.43}
\end{equation}%
where $V_{e-i}$, $V_{H}$ are also of order $s^{0}$. After the Fourier
transform with respect to time (and for small $\left\vert \omega
_{n}\right\vert \ll \omega _{c}$) $\hat{\Sigma}_{c}^{(h)}$ is given by
\begin{equation}
\hat{\Sigma}_{c}^{(h)}(\mathbf{x}_{1},\mathbf{x}_{2},\omega _{n})\simeq
\Sigma _{c0}^{(h)}(\mathbf{x}_{1},\mathbf{x}_{2},0)\hat{\tau}_{3}+(\hat{%
\Sigma}_{c0}^{(h)})^{\prime }(\mathbf{x}_{1},\mathbf{x}_{2},0)\cdot i\omega
_{n}.  \label{Ab-3.44}
\end{equation}%
As we said $\Sigma _{c0}^{(h)}\sim s^{0}$ while $(\hat{\Sigma}%
_{c0}^{(h)})^{\prime }\cdot \omega _{n}\sim s^{1}$ because $\omega _{n}\sim
s^{1}$. From Eq.(\ref{Ab-3.29a}) it is seen that the part $\hat{\Sigma}%
_{c}^{(l)}(1,2)$ contains the low-energy Green's function $\hat{G}^{low}(1,%
\bar{2})$ and this skeleton diagram is of order $s^{1}$. The similar
analysis based on Eq.(\ref{Ab-3.30a}) for $\hat{\Sigma}_{ep}(1,2)$ gives
that the leading order is $s^{1}$ which describes the low-energy part of $%
EPI $. After the separations of terms (of $s^{0}$ and $s^{1}$ orders) the
\textit{Dyson equation in the low-energy region} has the form
\begin{equation*}
\lbrack i\omega _{n}Z_{c}(\mathbf{x},\mathbf{\bar{x}})-\hat{H}_{0}(\mathbf{x}%
,\mathbf{\bar{x}})-\hat{\Sigma}_{c}^{(l)}(\mathbf{x},\mathbf{\bar{x},}\omega
_{n})-\hat{\Sigma}_{ep}(\mathbf{x},\mathbf{\bar{x},}\omega _{n})]
\end{equation*}%
\begin{equation}
\times \hat{G}^{low}(\mathbf{\bar{x},y},\omega _{n})=\delta (\mathbf{x-y})%
\hat{\tau}_{0},  \label{Ab-G-low}
\end{equation}%
where $\mathbf{\bar{x}}$ means integration $\int d^{3}\bar{x}$ over the
crystal volume. The Coulomb renormalization function $Z_{c}(\mathbf{x},%
\mathbf{y})=\delta (\mathbf{x-y})-(\Sigma _{0c}^{(h)})^{\prime }(\mathbf{x},%
\mathbf{y},0)$ and the single-particle Hamiltonian $\hat{H}_{0}(\mathbf{x},%
\mathbf{y})$ collect formally all high-energy processes which are unaffected
by superconductivity (which is low-energy process) where%
\begin{equation}
\hat{H}_{0}(\mathbf{x},\mathbf{y})=\{(-\frac{1}{2m}\nabla _{\mathbf{x}%
}^{2}-\mu )\delta (\mathbf{x-y})+V_{0}^{(h)}(\mathbf{x},\mathbf{y},0)\}\hat{%
\tau}_{3}  \label{H0-he}
\end{equation}%
with
\begin{equation}
V_{0}^{(h)}(\mathbf{x},\mathbf{y},0)=\{V_{e-i}(\mathbf{x})+V_{H}(\mathbf{x}%
)\}\delta (\mathbf{x}-\mathbf{y})+\Sigma _{c0}^{(h)}(\mathbf{x},\mathbf{y}%
,0).  \label{V0-he}
\end{equation}%
One can further absorb $Z_{c}(\mathbf{x},\mathbf{y})$ into the \textit{%
renormalized} Green's function%
\begin{equation}
\hat{G}_{r}(x\mathbf{,y},\omega _{n})=Z_{c}^{1/2}(\mathbf{x},\mathbf{\bar{x}}%
)\hat{G}^{low}(\bar{x}\mathbf{,\bar{y}},\omega _{n})Z_{c}^{1/2}(\mathbf{\bar{%
y}},\mathbf{y}),  \label{G-ren}
\end{equation}
the renormalized vertex function $\hat{\Gamma}_{ren}(1,2;3)=Z_{c}^{-1/2}\hat{%
\Gamma}_{c}Z_{c}^{-1/2}$, the renormalized self-energies $\hat{\Sigma}%
_{r;c,ep}^{(l)}(x\mathbf{,y},\omega _{n})=Z_{c}^{-1/2}(\mathbf{x},\mathbf{%
\bar{x}})\hat{\Sigma}_{c,ep}^{(l)}(\bar{x}\mathbf{,\bar{y}},\omega
_{n})Z_{c}^{-1/2}(\mathbf{\bar{y}},\mathbf{y})$ and introduce the \textit{%
ideal band-structure Hamiltonian }$\hat{h}_{0}(x\mathbf{,y})=Z_{c}^{-1/2}(%
\mathbf{x},\mathbf{\bar{x}})\hat{H}_{0}(\bar{x}\mathbf{,\bar{y}}%
)Z_{c}^{-1/2}(\mathbf{\bar{y}},\mathbf{y})$ given by
\begin{equation}
\hat{h}_{0}(\mathbf{x,y})=\{(-\frac{1}{2m}\nabla _{\mathbf{x}}^{2}-\mu
)\delta (\mathbf{x-y})+V_{IBS}(\mathbf{x},\mathbf{y})\}\hat{\tau}_{3}.
\label{Ab-3.51}
\end{equation}%
Here,
\begin{equation}
V_{IBS}(\mathbf{x},\mathbf{y})=Z_{c}^{-1/2}(\mathbf{x},\mathbf{\bar{x}}%
)V_{0}^{(h)}(\bar{x}\mathbf{,\bar{y}})Z_{c}^{-1/2}(\mathbf{\bar{y}},\mathbf{y%
})  \label{Vibs}
\end{equation}%
is the \textit{ideal band structure potential} (sometimes called the
excitation potential) and apparently nonlocal quantity, which is contrary to
the standard local potential $V_{g}(\mathbf{x})$ in the LDA-DFT theories -
see Section VI. The static potential $V_{IBS}(\mathbf{x},\mathbf{y})$ is of
order $s^{0}$ and includes high-energy processes

Finally, we obtain the matrix Dyson equation for the renormalized Green's
function $\hat{G}_{r}(x\mathbf{,y},\omega _{n})$ which is the basis for the
(strong-coupling) Migdal-Eliashberg theory in the low-energy region
\begin{equation*}
\lbrack i\omega _{n}\delta (\mathbf{x}-\mathbf{\bar{x}})-\hat{h}_{0}(\mathbf{%
x},\mathbf{\bar{x}})-\hat{\Sigma}_{c,r}^{(l)}(\mathbf{x},\mathbf{\bar{x},}%
\omega _{n})-\hat{\Sigma}_{ep,r}(\mathbf{x},\mathbf{\bar{x},}\omega _{n})]
\end{equation*}%
\begin{equation}
\times \hat{G}_{r}(\mathbf{\bar{x},y},\omega _{n})=\delta (\mathbf{x-y})\hat{%
\tau}_{0},  \label{Ab-3.53}
\end{equation}%
where $\hat{\Sigma}_{c,r}^{(l)}$ and $\hat{\Sigma}_{ep,r}$ have the same
form as Eqs.(\ref{Ab-3.29a}-\ref{Ab-3.30a}) but with renormalized Green's
and vertex functions $\hat{G}_{r}$, $\hat{\Gamma}_{r}$ instead of $\hat{G}$,
$\hat{\Gamma}$. We stress that Eq.(\ref{Ab-3.53}) holds in the low-energy
region only. In the superconducting state the set of Eliashberg equations in
Eq. (\ref{Ab-3.53}) are writen explicitely in Appendix A, where it is seen
that the superconducting properties depend on the Eliashberg spectral
function $\alpha _{\mathbf{kp}}^{2}F(\omega )$. The latter function is
defined also in Appendix A, Eq.(\ref{alpha-kp}), and it depends on material
properties of the system.

The important ingredients of the low-energy Migdal-Eliashberg theory is the
ideal band-structure Hamiltonian\textbf{\ }$\hat{h}_{0}(\mathbf{x},\mathbf{y}%
)$ - given by Eq.(\ref{Ab-3.51}), which contains many-body (excitation)
ideal band-structure non-local periodic crystal potential\textbf{\ }$V_{IBS}(%
\mathbf{x},\mathbf{y})$. The Hamiltonian\textbf{\ }$\hat{h}_{0}(\mathbf{x},%
\mathbf{y})$ determines the \textit{ideal energy spectrum} $\epsilon (%
\mathbf{k})$ of the conduction electrons and the wave function $\psi _{i,%
\mathbf{p}}(\mathbf{x})$ through
\begin{equation}
\hat{h}_{0}(\mathbf{x}\mathbf{,\bar{y}})\psi _{i,\mathbf{k}}(\mathbf{\bar{y}}%
)=[\epsilon _{i}(\mathbf{k})-\mu ]\psi _{i,\mathbf{k}}(\mathbf{x}),
\label{Ab-3.54}
\end{equation}%
where $\mu $ is the chemical potential. We stress that the Hamiltonian $\hat{%
h}_{0}(x\mathbf{,y})$ also governs transport properties of metals in
low-energy region.

After solving Eq.(\ref{Ab-3.54}) the next step is to expand all
renormalized Green's function, self-energies, vertices and the
renormalized $EPI$ constant (symbolically $g_{ep,r}=g_{ep,0}\Gamma
_{ren}/\varepsilon _{e}$) in the basis of $\psi
_{i,\mathbf{p}}(\mathbf{x})$ and write down the Eliashberg
equations in this basis. We shall not elaborate further this
program and we refer the reader to the relevant literature
\citep{ScalapRev}, \citep{DierkLowTemp}. We point out, that even
such simplified program of the low-energy Migdal-Eliashberg theory
was never fully realized in
low-temperature superconductors, because the non-local potential $V_{IBS}(%
\mathbf{x},\mathbf{y})$ (enters the ideal band-structure Hamiltonian\textbf{%
\ }$h_{0}(x\mathbf{,y})$) and the renormalized vertex function (entering the
$EPI$ coupling constant $g_{ep,r}$) which include electronic correlations
are difficult to calculate especially in strongly correlated metals.
Therefore, it is not surprising at all, that the situation is even worse in $%
HTS$ materials which are strongly correlated systems with complex structural
and material properties. Due to these difficulties the calculations of the
electronic band structure and the $EPI$ coupling is usually done in the
framework of LDA-DFT where the many-body excitation potential $V_{IBS}(%
\mathbf{x},\mathbf{y})$ is replaced by some (usually local) potential $%
V_{LDA}(\mathbf{x})$ which in fact determines ground state properties of the
crystal. In the next section we briefly describe: (\textit{i}) the LDA-DFT
procedure in calculating the EPI\ coupling constant and (\textit{ii}) some
results related to HTSC cuprates. We shall also discuss why this
approximation is a serious drawback if applied to $HTS$ materials.

\section{ LDA-DFT calculations of the EPI coupling}

We point out again two results which are important for the future
microscopic theory of pairing in HTSC cuprates. \textit{First}, numerous
experiment (discussed in Part I) give evidence that the EPI coupling
constant which enters the normal part of the quasi-particle self-energy $%
\lambda _{ep}^{Z}=$ $\lambda _{s}+\lambda _{d}+...$ is rather large, i.e. $%
1<\lambda _{ep}^{Z}<3.5$. In order to be conform with d-wave pairing the
effective EPI potential must be rather nonlocal (peaked at small transfer
momenta $q$) which implies that the s-wave and d-wave coupling constants are
of the same order, i.e. $\lambda _{d}\approx \lambda _{s}$. \textit{Second},
the theory based on the minimal $t-J$ model, which will be discussed in
Section VII, gives that strong electronic correlations produce a peak at
small transfer momenta in the effective EPI pairing potential thus giving
rise to $\lambda _{d}\approx \lambda _{s}$. This is a striking property
which allows that EPI is conform with d-wave pairing. However, the theory is
seriously confronted with the problem of calculation of the coupling
constants $\lambda _{ep}^{Z}$. It turns out, that at present it is an
illusory task to calculate $\lambda _{ep}^{Z}$ and $\lambda _{d}$ since it
is extremely difficult (if possible at all) to incorporate the peculiar
structural properties of HTSC cuprates (layered structure, ionic-metallic
system, etc.) and strong correlations effects in a consistent and reliable
microscopic theory. As it is stressed several times, the LDA-DFT methods
miss some important many body effects (especially in the band structure
potential) and therefore fail to describe correctly screening properties of
HTSC cuprates and the strength of EPI. However, the LDA-DFT methods are
capable to incorporate structural properties much better than the simplified
minimal $t-J$ (toy) model. Here, we discuss briefly some achievements of the
improved LDA-DFT calculations which are able to take partially into account
some nonlocal effects in the EPI. The latter are mainly due to the almost
\textit{ionic structure along the c-axis} which is reflected in the very
small c-axis plasma frequency $\omega _{c}\ll \omega _{ab}$).

The main task of the LDA-DFT theory in obtaining the EPI coupling is to
calculate the change of the ground state (self-consistent) potential $\delta
V_{g}(\mathbf{r})/\delta R_{\alpha }$ and the EPI coupling constant (\textit{%
matrix element}) $g_{\alpha }^{LDA}(\mathbf{k},\mathbf{k}^{\prime
})$ (see its definition below) what is the most difficult part of
calculations. Since in LDA-DFT there is no EPI coupling, the
recipe is that the calculated "EPI" coupling is inserted into the
Eliashberg equations. By knowing $g_{\alpha
}^{LDA}(\mathbf{k},\mathbf{k}^{\prime })$ one can define the total
($\lambda $) and partial ($\lambda _{\mathbf{q},\nu }$) $EPI$
coupling constants for the $\nu $-th mode, respectively
\citep{Krakauer} \
\begin{equation}
\lambda =\frac{1}{Np}\sum_{\mathbf{q},\nu }\lambda _{\mathbf{q},\nu },\text{
\ \ \ \ }\lambda _{\mathbf{q},\nu }=\frac{p\gamma _{\mathbf{q},\nu }}{\pi
N(0)\omega _{\mathbf{q,}\nu }},  \label{lamda-lda}
\end{equation}%
where $p=3\kappa $ is the number of phonon branches ($\kappa $ is the number
of atoms in the unit cell) and $N(0)$ is the density of states at the Fermi
energy (per spin and unit cell). The \textit{phonon line-width} $\gamma _{%
\mathbf{q},\nu }$ is defined in the Migdal-Eliashberg theory by
\begin{equation*}
\gamma _{\mathbf{q},\nu }=2\pi \omega _{\mathbf{q,}\nu }\frac{1}{N}%
\sum_{ll^{\prime }\mathbf{k}}\frac{1}{2M\omega _{\mathbf{k-q,}\nu }}\mid
e_{\nu }^{\alpha }(\mathbf{q})\cdot g_{\alpha ,ll^{\prime }}(\mathbf{k},%
\mathbf{k}-\mathbf{q})\mid ^{2}
\end{equation*}%
\begin{equation}
\times \lbrack \frac{n_{F}(\xi _{l,\mathbf{k}})-n_{F}(\xi _{l,\mathbf{k}%
}+\hbar \omega _{\mathbf{q,}\nu })}{\hbar \omega _{\mathbf{q,}\nu }}]\delta
(\xi _{l^{\prime },\mathbf{k-q}}-\xi _{l,\mathbf{k}}-\hbar \omega _{\mathbf{%
q,}\nu }).  \label{gamma-lda}
\end{equation}%
Here, $e_{\nu }^{\alpha }(\mathbf{q})$ is the phonon polarization vectors, $%
n_{F}$ is the Fermi function. Since the ideal energy spectrum from Eq. (\ref%
{Ab-3.54}) $\xi _{l,\mathbf{k}}=E_{l,\mathbf{k}}-\mu $ and the corresponding
eigenfunctions $\psi _{\mathbf{k}l}$ are unknown then instead of these one
sets in Eq. (\ref{gamma-lda}) the $LDA-DFT$ eigenvalues for the $l-th$ band $%
\xi _{l,\mathbf{k}}^{(LDA)}$ and $\psi _{\mathbf{k}l}^{(LDA)}$. In the
LDA-DFT method the EPI\ coupling constant (\textit{matrix element}) $%
g_{\alpha ,ll^{\prime }}^{(LDA)}$ is defined which by the change of the
ground-state potential $\delta V_{g}(\mathbf{r})/\delta R_{\alpha }$
\begin{equation}
g_{\alpha ,ll^{\prime }}^{(LDA)}(\mathbf{k},\mathbf{k}^{\prime })=\langle
\psi _{\mathbf{k}l}^{(LDA)}\mid \sum_{n}\frac{\delta V_{g}(\mathbf{r})}{%
\delta R_{n\alpha }}\mid \psi _{\mathbf{k}^{\prime }l^{\prime
}}^{(LDA)}\rangle .  \label{g-lda}
\end{equation}%
The index $n$ means summation over the lattice sites, $\alpha
=x,y,z$ and the wave function $\psi _{\mathbf{k}l}^{(LDA)}$ are
the solutions of the Kohn-Sham equation - see \citep{KulicReview}.
In the past various
approximations within the $LDA-DFT$ method have been used in calculating $%
\delta V_{g}(\mathbf{r})/\delta R_{\alpha }$ and $\lambda $ while here we
comment some of them only: $\mathbf{(i)}$ In most calculations in $LTS$
systems and in $HTSC$ cuprates the \textit{rigid-ion}\textbf{\ (}$RI$\textbf{%
) }approximation was used as well as its further simplifications which
inevitable (due to its shortcomings and obtained small $\lambda $) deserves
to be commented. The $RI$ approximation is based on the very specific
assumption that the ground-state (crystal) potential $V_{g}(\mathbf{r})$ can
be considered as a sum of ionic potentials $V_{g}(\mathbf{r})=\sum_{n}V_{g}(%
\mathbf{r-R}_{n})$ where the ion potential $V_{g}(\mathbf{r-R}_{n})$ and the
electron density $\rho _{e}(\mathbf{r})$ are carried rigidly with the ion at
$\mathbf{R}_{n}$ during the ion displacement ($\mathbf{R}_{n}=\mathbf{R}%
_{n}^{0}+\hat{u}_{\alpha n}$). In the $RI$ approximation the change of $%
V_{g}(\mathbf{r})$ is given by
\begin{equation}
\delta V_{g}(\mathbf{r})=\sum_{n}\nabla _{\alpha }V_{g}(\mathbf{r-R}%
_{n}^{0})u_{\alpha n}  \label{Vg}
\end{equation}%
\
\begin{equation}
\frac{\delta V_{g}(\mathbf{r})}{\delta R_{n\alpha }}=\nabla _{\alpha }V_{g}(%
\mathbf{r-R}_{n}^{0}),  \label{Vg-prime}
\end{equation}%
which means that $RI$ does not take into account changes of the electron
density during the ion displacements. In numerous calculations applied to $%
HTSC$ cuprates the rigid-ion model is even further simplified by using the
\textit{rigid muffin-tin approximation} ($RMTA$) (or similar version with
the rigid-atomic sphere) - see discussions in \citep{Krakauer}, \citep%
{Falter97}, \citep{Falter98}. The $RMTA$ assumes that the
ground-state potential and the electron density follow ion
displacements rigidly inside the Wigner-Seitz cell but outside it
$V_{g}(\mathbf{r})$ is not changed because of the assumed very
good metallic screening (for instance in simple metals)
\begin{equation}
\nabla _{\alpha }V_{g}(\mathbf{r-R}_{n})=\{%
\begin{tabular}{l}
$\nabla _{\alpha }V_{g}(\mathbf{r-R}_{n}),$ \ \ $\mathbf{r}$\ in cell $n$ \\
$0,$ \ \ \ \ \ \ \ \ \ \ \ \ \ \ \ \ \ \ \ \ \ \ \ outside%
\end{tabular}%
\ .  \label{Grad-vg}
\end{equation}%
This means that the dominant $EPI$ scattering is due to the nearby atoms
only and that the scattering potential is isotropic. All nonlocal effects
related to the interaction of electrons with ions far away, are neglected in
the $RMTA$. In this case $g_{\alpha ,n}^{LDA}(\mathbf{k},\mathbf{k}^{\prime
})$ is calculated by the wave function centered at the given ion $\mathbf{R}%
_{n}^{0}$ which can be expanded inside the muffin-tin sphere (outside it the
potential is assumed to be constant) in the angular momentum basis $\{l,m\}$%
, i.e.
\begin{equation}
\langle \mathbf{r}\mid \psi _{\mathbf{k}}^{(RMTA)}\rangle =\sum_{lm}C_{lm}(k,%
\mathbf{R}_{n}^{0})\rho _{l}(\mid \mathbf{r-R}_{n}^{0}\mid )Y_{lm}(\phi
,\theta )  \label{psi-lda}
\end{equation}%
(the angles $\phi ,\theta $ are related to the vector $\hat{r}=(\mathbf{r-R}%
_{n}^{0})/\mid \mathbf{r-R}_{n}^{0}\mid $). The radial function $\rho
_{l}(\mid \mathbf{r-R}_{n}^{0}\mid )$ is zero outside the muffin-tin sphere.
In that case the $EPI$ matrix element is given by $g_{\alpha ,n}^{RMTA}(%
\mathbf{k},\mathbf{k}^{\prime })\sim \langle Y_{lm}\mid \hat{r}\mid
Y_{l^{\prime }m^{\prime }}\rangle $ and because $\hat{r}$ is a vector the
selection rule implies that only terms with $\Delta l\equiv l^{\prime
}-l=\pm 1$ contribute to the $EPI$ coupling constant in the $RMTA$. This
result is an immediate consequence of the assumed locality of the $EPI$
potential in $RMTA$. However, since \textit{nonlocal effects}, such as the
long-range Madelung-like interaction, are important in $HTSC$ cuprates then
additional terms contribute also to the coupling constant $g_{\alpha ,n}$,
i.e. $g_{\alpha ,n}(\mathbf{k},\mathbf{k}^{\prime })=g_{\alpha ,n}^{RMTA}(%
\mathbf{k},\mathbf{k}^{\prime })+g_{\alpha ,n}^{nonloc}(\mathbf{k},\mathbf{k}%
^{\prime })$, where a part ($\delta g_{\alpha ,n}^{nonloc}$) of the nonlocal
contribution to $g_{\alpha ,n}^{nonloc}$ is represented schematically
\begin{equation}
\delta g_{\alpha ,n}^{nonloc}(\mathbf{k},\mathbf{k}^{\prime })\sim \langle
Y_{lm}\mid (\mathbf{R}_{n}^{0}-\mathbf{R}_{m}^{0})_{\alpha }\mid
Y_{l^{\prime }m^{\prime }}\rangle  \label{g-nonloc}
\end{equation}%
From Eq.(\ref{g-nonloc}) comes out the selection rule\textbf{\ }$\Delta
l=l\prime -l=0$ for the nonlocal part of the $E-P$ interaction. We stress
that the $\Delta l=0$ (nonlocal) terms are omitted in the $RMTA$ approach
and therefore it is not surprising that this approximation\ works
satisfactory in elemental (isotropic) metals only. The latter are
characterized by the large density of states at the Fermi surface which
makes electronic screening very efficient. This gives rise to a local $EPI$
where an electron feels potential changes of the nearby atom only. One can
claim with certainty, that the $RMTA$\ method is not suitable for $HTSC$
cuprates which are highly anisotropic systems with pronounced ionic
character of binding and pronounced strong electronic correlations. The $%
RMTA $\ method applied to $HTSC$ cuprates misses just this
important part - the long-range part $EPI$ due to the change of
the long range Madelung energy in the almost ionic structure of
HTSC cuprates. For instance, the first calculations done in
\citep{Mazin} which are based on the $RMTA$ give very small $EPI$
coupling constant $\lambda ^{RMTA}\sim 0.1$ in $YBCO$,
which is in apparent contradiction with the experimental finding that $%
\lambda _{ep}$ is large - discussed in Part I.

However, these nonlocal effects are taken into account in
\citep{Krakauer} by using the \textit{frozen-in phonon} ($FIP$)
method in evaluating of $\lambda _{ep}$ in $La_{2-x}M_{x}CuO_{4}$.
In this method some symmetric phonons are considered and the band
structure is calculated for the system with the super-cell which
is determined by the periodicity of the phonon displacement. By
comparing the unperturbed and perturbed energies the
corresponding $EPI$ coupling $\lambda _{\nu }$ (for the considered phonon $%
\nu $-th mode) is found. More precisely speaking, in this approach the
matrix elements of $\delta V_{g}(\mathbf{r})/\delta R_{0,\alpha }^{\kappa }$
are determined from the finite difference of the ground-state potential
\begin{equation*}
\Delta V_{g,\mathbf{q},\nu }(\mathbf{r})=V_{g}(\mathbf{R}_{0,L}^{\kappa
}+\Delta \mathbf{\tau }_{\mathbf{q},\nu }^{\kappa }(L))-V_{g}(\mathbf{R}%
_{0,L}^{\kappa })
\end{equation*}%
\begin{equation}
=\sum_{L,\kappa }\Delta \mathbf{\tau }_{\mathbf{q},\nu }^{\kappa }(L)\frac{%
\partial V_{g}(\mathbf{R}_{0,L}^{\kappa })}{\partial \mathbf{R}%
_{0,L}^{\kappa }},  \label{Vg-q}
\end{equation}%
where $L$, $\kappa $ enumerate elementary lattice cells and atoms in the
unit cell, respectively. The frozen-in atomic displacements of the phonon $%
\Delta \mathbf{\tau }_{\mathbf{q},\nu }^{\kappa }(L)$ of the $\nu $-th mode
is given by $\Delta \mathbf{\tau }_{\mathbf{q},\nu }^{\kappa }(L)=\Delta u_{%
\mathbf{q},\nu }(\hbar /2M_{\kappa }\omega _{\mathbf{q},\nu })^{1/2}Re[%
\mathbf{e}_{\kappa ,\nu }(\mathbf{q})e^{i\mathbf{q\cdot R}}]$ where $\Delta
u_{\mathbf{q},\nu }$ is the dimensionless phonon amplitude and the phonon
polarization (eigen)vector $\mathbf{e}_{\kappa ,\nu }(\mathbf{q})$ fulfills
the condition $\sum_{\kappa }\mathbf{e}_{\kappa ,\nu }^{\ast }(\mathbf{q}%
)\cdot \mathbf{e}_{\kappa ,\nu ^{\prime }}(\mathbf{q})=\delta _{\nu ,\nu
^{\prime }}$. Based on this approach various symmetric $A_{g}$ (and some $%
B_{3g}$) modes of $La_{2-x}M_{x}CuO_{4}$ were studied
\citep{Krakauer}, where it was found that the large matrix
elements are due to unusually long-range
Madelung-like, especial for the $c$-axis phonon modes. The obtained large $%
\lambda _{ep}\approx 1.37$ is the consequence of the following three main
facts: $\mathbf{(}i\mathbf{)}$ The electronic spectrum in $HTSC$ cuprates is
highly anisotropic, i.e. it is \textit{quasi-two-dimensional}. This is an
important fact for pairing because if the conduction electrons would be
uniformly spread over the whole unit cell then due to the rather low
electron density ($n\sim 10^{21}cm^{-3}$) the density of states on the $Cu$
and $O$ in-plane atoms would be an order of magnitude smaller than the real
value. This would further give an order of magnitude smaller $EPI$ coupling
constant $\lambda _{ep}$. Note, that the calculated density of states on the
(heavy) $Cu$ and (light) $O$ in-plane atoms, $N^{Cu}(0)\cong 0.54$ $%
states/eV $ \ and $N^{O_{xy}}(0)\cong 0.35$ $states/eV$, are of same order
of magnitude as in some $LTS$ materials. For instance, in $NbC$ where $%
T_{c}\approx 11$ $K$ one has on (the heavy) $Nb$ atom $N^{Nb}(0)\cong 0.58$ $%
states/eV$ and on (the light) $C$ atom $N^{C}(0)\cong 0.25$ $states/eV$. So,
the quasi-two-dimensional character of the spectrum is crucial in obtaining
appreciable density of states on the light $O$ atoms in the $CuO_{2}$
planes. $\mathbf{(}ii\mathbf{)}$\textbf{\ }In $HTSC$ cuprates there is
\textit{strong }$Cu-O$\textit{\ hybridization} leading to good in-plane
metallic properties. This \textit{large covalency in the plane} is due to
the (fortunately) small energy separation of the electron levels on $Cu$ and
$O_{xy}$ atoms which comes out from the band structure calculations \citep%
{Mattheiss}, i.e. $\Delta =\mid \epsilon _{Cu}-\epsilon _{O_{xy}}\mid
\approx 3$ $eV$ . The latter value gives rise to strong covalent mixing (the
hybridization parameter $t_{pd}$) of the $Cu_{d_{x^{2}-y^{2}}}$ and $%
O_{p_{x,y}}$ states, i.e. $t_{pd}=-1.85$ $eV$. It is interesting, that the
small value of $\Delta $ is not due to the ionic structure (crystal field
effect) of the system but it is mainly due to the \textit{natural falling}
of the $Cu_{d_{x^{2}-y^{2}}}$ states across the transition-metal series. So,
the natural closeness of the energy levels of the $Cu_{d_{x^{2}-y^{2}}}$ and
$O_{p_{x,y}}$ states is this distinctive feature of $HTSC$ cuprates which
basically allows achievement of high $T_{c}$. $\mathbf{(}iii\mathbf{)}$ The
\textit{ionic structure of HTSC} \textit{cuprates} which is very pronounced
along the $c$-axis is responsible for the weak electronic screening along
this axis and according to that for the significant contribution of the
nonlocal (long-range) Madelung-like interaction to $EPI$. It turns out that
because of the ionicity of the structure the $La$ and $O_{z}$ \textit{axial
modes are strongly coupled with charge carriers} in the $CuO_{2}$ planes
despite the fact that the local density of states on these atoms is very
small \citep{Krakauer}, i.e. $N^{La}(0)=0.13$ $states/eV$ and $%
N^{O_{z}}(0)=0.016$ $states/eV$. For comparison, on planar atoms $Cu$ and $%
O_{xy}$ one has $N^{Cu}(0)=0.54$ $states/eV$ \ and $N^{O_{xy}}(0)=0.35$ $%
states/eV$. These calculations show that the lanthanum mode (with $\omega _{%
\mathbf{q,}\nu }=202$ $cm^{-1}$) at the $q=(0,0.2\pi /c)$ zone boundary
(fully symmetric $Z$-point) has ten times larger coupling constant $\lambda
_{\mathbf{q},\nu }^{La}(FIP)=4.8$ than it is predicted in the $RMT$
approximation $\lambda _{\mathbf{q},\nu }^{La}(RMT)=0.48$. The similar
increase holds for the average coupling constant, where $\lambda _{\nu
,average}^{La}(FIP)=1.0$ but $\lambda _{\nu ,average}^{La}(RMT)=0.1$. Note,
that for the $\mathbf{q}\approx 0$ La-mode one obtains $\lambda _{\nu
}^{La}(FIP)=4.54$ compared \ to $\lambda _{\nu }^{La}(RMT)=0.12$. Similar
results hold for the axial apex-oxygen $q=(0,0.2\pi /c)$ mode ($O_{z}$) with
$\omega _{\mathbf{q,}\nu }=396$ $cm^{-1}$ where the large (compared to the $%
RMT$ method) coupling constant is obtained: $\lambda
_{\mathbf{q},\nu }^{O_{z}}=14$ and $\lambda _{\nu
,average}^{O_{z}}=4.7$, while for $q\approx 0$ axial apex-oxygen
modes with $\omega _{\mathbf{q,}\nu }=415$ $cm^{-1}$ one has
$\lambda _{\nu ,average}^{O_{z}}=11.7$. After averaging over all
calculated modes it was estimated $\lambda =1.37$ and $\omega
_{\log }\approx 400$ $K$. By assuming that $\mu ^{\ast }=0.1$ one
obtains $T_{c}=49$ $K$ by using Allen-Dynes formula for
$T_{c}\approx 0.83\omega _{\log }\exp \{-1.04(1+\lambda )/[\lambda
-\mu ^{\ast }(1+0.62\lambda )]\}$ with $\omega _{\log }=2\int
d\omega d\omega \alpha ^{2}(\omega )F(\omega )\ln \omega /\lambda
\omega $. \ For $\mu ^{\ast }=0.15$ and $0.2$ one obtains
$T_{c}=41$ and $32$ $K$, respectively. We stress that the rather
large $\lambda _{ep}$ (and $T_{c}$) are due to the nonlocal (long
range) effects of the metallic-ionic structure of HTSC cuprates
and non-muffin-tin corrections in EPI, as was first proposed in
\citep{Jarlborg}. However, we would like to
stress that the optimistic results for $\lambda _{ep}$ obtained in \citep%
{Krakauer} is in fact based on the calculation of the EPI coupling for some
wave vectors $\mathbf{q}$ with symmetric vibration patterns and in fact the
obtained $\lambda _{ep}$ is an extrapolated value. The all-$\mathbf{q}$
calculations of $\lambda _{ep,\mathbf{q}}$ which take into account
long-range effects is a real challenge for the LDA-DFT calculations and are
still awaiting.

Finally, it is worth to mention important calculations of the EPI coupling
constant in the framework of the\textit{\ linear-response full-potential
linear-muffin-tin-orbital} \textit{method} ($LRFP-LMTO$) invented in \citep%
{Savrasovi} and applied to the doped $HTSC$ cuprate $%
(Ca_{1-x}Sr_{x})_{1-y}CuO_{2}$ for $x\sim 0.7$ and $y\sim 0.1$ with $%
T_{c}=110$ $K$ \citep{SavAnd}. Namely, these calculations give
strong evidence that structural properties of HTSC cuprates
already alone make the dominance of small-q scattering in EPI,
which effect is additionally
increased by strong correlations. In order to analyze this compound in \citep%
{SavAnd} calculations are performed for $CaCuO_{2}$ doped by holes in an
uniform, neutralizing back-ground charge. The momentum ($\mathbf{q}=(\mathbf{%
q}_{\parallel },q_{\perp })$) dependent EPI coupling constant (summed over
all phonon branches $\nu $) in different $L$ channels ($s,p,d..$) is
calculated by using a standard expression \newline
\begin{equation}
\lambda _{L}(\mathbf{q})=M\sum_{\mathbf{k},\nu }Y_{L}(\mathbf{k}+\mathbf{q}%
)\left\vert g_{\mathbf{k},\mathbf{q},\nu }\right\vert ^{2}Y_{L}(\mathbf{k}%
)\times \delta (\xi _{\mathbf{k}+\mathbf{q}})\delta (\xi _{\mathbf{k}})
\label{lambda-L-q}
\end{equation}%
Here, $\xi _{\mathbf{k}}$ is the quasi-particle energy, $g_{\mathbf{k},%
\mathbf{q},\nu }$ is the EPI coupling constant (matrix element) with the $%
\nu $-th branch, $Y_{L}(\mathbf{k})$ is the $L$-channel wave function and
the normalization factor $M\propto N_{L}^{-1}(0)$ with the partial density
of states is $N_{L}(0)\propto \sum_{\mathbf{k}}Y_{L}^{2}(\mathbf{k})\delta
(\xi _{\mathbf{k}})$. The total coupling constant in the $L$-channel is an
average of $\lambda _{L}(\mathbf{q})$ over the whole 2D Brillouin zone (over
$\mathbf{q}_{\parallel }$), i.e. $\lambda _{L}(q_{\perp })=\left\langle
\lambda _{L}(\mathbf{q}_{\parallel })\right\rangle _{BZ}$. We stress three
important results of Ref. \citep{SavAnd}. \textit{First}, the $s$- and $d$%
-coupling constants, $\lambda _{s}(\mathbf{q})$, $\lambda
_{d}(\mathbf{q})$, are \textit{peaked at small transfer momenta}
$\mathbf{q\sim }(\pi /3a,0,0)$ as it is shown in Fig. 3 of Ref.
\citep{SavAnd}. This result is mainly caused
by the nesting properties of the Fermi surface shown in Fig. 1 of Ref. \citep%
{SavAnd}. \textit{Second}, the $\mathbf{q}$-dependence of the integrated EPI
matrix elements $\left\vert \bar{g}_{L,\mathbf{q}}\right\vert ^{2}=\lambda
_{L}(\mathbf{q})/\chi _{L}^{"}(\mathbf{q})$ (with $\chi _{L}^{"}(\mathbf{q}%
)\propto \sum_{\mathbf{k}}Y_{L}(\mathbf{k}+\mathbf{q})Y_{L}(\mathbf{k}%
)\delta (\xi _{\mathbf{k}+\mathbf{q}})\delta (\xi _{\mathbf{k}})$) for $%
L=s,d $ \ is similar to that of $\lambda _{L}(\mathbf{q})$, i.e. these are
peaked at small transfer momenta $q\ll 2k_{F}$. Both these results mean that
the structural properties of HTSC cuprates imply the \textit{dominance of
small-}$q$\textit{\ EPI scattering}. \textit{Third}, the calculations give
similar values for $\lambda _{s}(q_{\perp }=0)$ and $\lambda _{d}(q_{\perp
}=0)$, i.e. $\lambda _{s}=0.47$ for $s-wave$ and $\lambda _{d}=0.36$ for $%
d-wave$ pairing \citep{SavAnd}. This finding (that $\lambda
_{d}\approx \lambda _{s}$) is due to the dominance of the small
$q$-scattering in EPI. It means that the nonlocal effects (long
range forces) in EPI of HTSC cuprates are very important. This
result together with the finding of the
dominance of small-$q $ scattering in EPI due to strong correlations \citep%
{Kulic1}, \citep{Kulic2}, \citep{Kulic3} mean that strong
correlations and structural properties of HTSC cuprates make EPI
conform with d-wave pairing, either as its main cause or as its
supporter. We stress that the obtained coupling constant $\lambda
_{d}=0.36$ is rather small to give $d$-wave pairing with large
$T_{c}$ and on the first glance this result is against the EPI
mechanism of pairing in cuprates. However, it is argued throughout
this paper, that the LDA methods applied to strongly correlated
systems overestimate the screening effects and underestimate the
coupling constant and therefore their quantitative predictions are
not reliable.

\section{EPI and strong correlations in cuprates}

\subsection{Minimal model Hamiltonian}

The \textit{minimal microscopic model} for normal and superconducting state
of HTSC cuprates must include at least three orbitals: one $d_{x^{2}-y^{2}}$%
-orbital of the $Cu$-ion and two $p$-orbitals ($p_{x,y}$) of the
$O$-ion since they participate in transport properties of these
materials - see more in \citep{KulicReview} and References
therein. The electronic part of the Hamiltonian (of the minimal
model) is $\hat{H}=\hat{H}_{0}+\hat{H}_{int}$ - usually called the
Emery model \citep{Emery}, where the one particle tight-binding
Hamiltonian $\hat{H}_{0}$ describes kinetic energy in the
three-band (orbital) model
\begin{equation*}
\hat{H}_{0}=\sum_{i,\sigma }(\epsilon _{d}^{0}-\mu )d_{i\sigma }^{\dagger
}d_{i\sigma }+\sum_{j,\alpha ,\sigma .}(\epsilon _{p\alpha }^{0}-\mu
)p_{j\alpha \sigma }^{\dagger }p_{j\alpha \sigma }
\end{equation*}%
\begin{equation}
+\sum_{i,j,\alpha ,\sigma }t_{ij\alpha }^{pd}d_{i\sigma }^{\dagger
}p_{j\alpha \sigma }+\sum_{j,j^{\prime },\alpha ,\beta ,\sigma
}t_{jj^{\prime },\alpha \beta }^{pp}p_{j\alpha \sigma }^{\dagger
}p_{j^{\prime }\beta \sigma }  \label{3-band}
\end{equation}%
Here $t_{ij\alpha }^{pd}$ ($i,j$ enumerate Cu- and O-sites, respectively) is
the hopping integral between the $p_{\alpha }(\alpha =x,y)$- and $d$- states
and $t_{jj^{\prime }\alpha \beta }^{pp}$ between $p_{\alpha }$- and $%
p_{\beta }$-states, while $\epsilon _{d}^{0}$ and $\epsilon _{p\alpha }^{0}$
are the bare $d$- and $p$- local energy levels and $\mu $ is the chemical
potential. This tight-binding Hamiltonian is written in the \textit{%
electronic notation} where the charge-transfer energy $\Delta _{dp,0}\equiv
\epsilon _{d}^{0}-\epsilon _{p}^{0}>0$ by assuming that there is one $%
3d_{x^{2}-y^{2}}$ electron on the copper ($Cu^{2+}$) while electrons in the $%
p$-levels of the $O^{2-}$ ions occupy filled bands. $\hat{H}_{0}$ contains
the main ingredients coming from the comparison with the $LDA-DFT$ band
structure calculations. The $LDA-DFT$ results are reproduced by assuming $%
t^{pp}\ll t_{pd}$ (and $\epsilon _{p\alpha }^{0}=\epsilon _{p}^{0}$) where
the good fit to the $LDA-DFT$ band structure is found for $\Delta
_{dp,0}\equiv \epsilon _{d}^{0}-\epsilon _{p}^{0}\approx 3.2$ $eV$ and $%
t^{pd}(\equiv t_{pd})=(\sqrt{3}/2)(pd\sigma )$, $(pd\sigma )=-1.8$ $eV$. The
total LDA band-width $W_{b}=(4\sqrt{2})\mid t_{pd}\mid \cong 9$ $eV$ \citep%
{Picket}.

The electron interaction is described by $\hat{H}_{int}$
\begin{equation}
\hat{H}_{int}=U_{d}\sum_{i}n_{i\uparrow }^{d}n_{i\downarrow
}^{d}+U_{p}\sum_{j,\alpha }n_{j\alpha \uparrow }^{p}n_{j\alpha \downarrow
}^{p}+\hat{V}_{c}+\hat{V}_{ep},  \label{Hint-3b}
\end{equation}%
where $U_{d}$ and $U_{p}$ are the on-site Coulomb repulsion energies at $Cu$
and $O$ sites, respectively while $\hat{V}_{c}$ and $\hat{V}_{ep}$ describe
the long-range part of the Coulomb interaction of electrons (holes) and $EPI$%
, respectively. Note, the Hubbard repulsion $U_{d}$ on the Cu-ion
is different from its bare atomic value $U_{d0}$($\approx 16$ $eV$
for Cu) due to various kind of screening effects in solids
\citep{Joeren}. In turns out that in most transition metal oxides
one has $U_{d}\ll U_{d0}$. This problem is thoroughly studied in
\citep{Joeren} and applied to HTSC cuprates. An
estimation from numerical cluster calculations \citep{Sawatzky} gives $%
U_{d}=9-11$ $eV$ and $U_{p}=4-6$ $eV$ but because $n_{i}^{d}U_{d}$ $\gg $ $%
n_{j}^{p}U_{p}$ the on-site repulsion on oxygen is usually neglected at the
first stage of the analysis.

Note, that in the case of large $U_{d}(\gg t_{pd},\Delta _{dp,0})$ the
\textit{hole notation} is usually used where in the parent compound (and for
$\left\vert t_{pd}\right\vert \ll \Delta _{dp,0}$) one has $\langle
n_{i}^{d}\rangle =1$, i.e. one hole in the $3d$-shall (in the $%
3d_{x^{2}-y^{2}}$ state) in the ground state. In the limit of large $%
U_{d}\rightarrow \infty $ the \textit{doubly occupancy on the Cu atoms is
forbidden} and only two copper states are possible: $Cu^{2+}$ described by
the quantum state $d_{i\sigma }^{\dagger }\mid 0\rangle $ and $Cu^{1+}$
described by $\mid 0\rangle $. In this (hole) notation the oxygen $p$-level
is fully occupied by electrons, i.e. there is no holes ( $\langle
n_{j}^{p}\rangle =0$) in the occupied oxygen $2p$-shall of $O^{2-}$. In this
notation the vacuum state $\mid 0_{v}\rangle $ (not the ground state) of the
Hamiltonian $\hat{H}$ for large $U_{d}$ corresponds to the closed shell
configuration $Cu^{1+}O^{2-}$. In the hole notation the hole $p$-level $%
\epsilon _{ph}^{0}$ lies higher than the hole $d$-level $\epsilon _{dh}^{0}$%
, i.e. $\Delta _{pd,0}\equiv \epsilon _{ph}^{0}-\epsilon _{dh}^{0}>0$ (note
in the electron picture it is opposite) and $U_{d}$ means repulsion of two
holes (in the $3d_{x^{2}-y^{2}}$ orbital ) with opposite spins - $3d^{8}$
configuration of the $Cu^{3+}$ ion. Note, that $\epsilon _{ph}^{0}=-\epsilon
_{p}^{0}$, $\epsilon _{dh}^{0}=-\epsilon _{d}^{0}$ and $t_{pd,h}=-t_{pd}$.
In the following the index $h$ in $t_{pd,h}$ is omitted. The reason for $%
\epsilon _{ph}^{0}>\epsilon _{dh}^{0}$ is partly in different
energies for the hole sitting on the oxygen and copper,
respectively \citep{Picket}. From this model one can derive in the
limit $U\rightarrow \infty $ the $t-J$
model for the $2D$ lattice in the $CuO_{2}$ plane \citep{Zhang}, \citep%
{Yushankai}, where each lattice site corresponds to a $Cu$-atom.
In the presence of one hole in the $3d$-shall then in the undoped
(no oxygen holes) HTSC cuprate each lattice site is occupied by
one hole. By doping the systems with holes the additional holes go
onto $O$-sites. Furthermore, due to the strong Cu-O covalent
binding the energetics of the system implies that an $O$-hole
forms a Zhang-Rice singlet with a $Cu$-hole \citep{Zhang}. In the
$t-J$ model the \textit{Zhang-Rice singlet is described by an
empty site. }Since in the $t-J$ model the doubly occupancy is
forbidden one
introduces annihilation (Hubbard) operator of the composite fermion $\hat{X}%
_{i}^{\sigma 0}=c_{i\sigma }^{\dagger }(1-n_{i,-\sigma })$ which describes
creation of an hole (in the $3d$-shall of the Cu-atoms) on the $i$-th site
if this site is previously empty (thus excluding doubly occupancy) , i.e. $%
n_{i,\sigma }+n_{i,-\sigma }\leq 1$ must be fulfilled on each lattice site.
\ In this picture the doped hole concentration $\delta $ means at the same
time the concentration of the oxygen holes and of the Zhang-Rice singlets.

The \textit{bosonic-like operators} $\hat{X}_{i}^{\sigma _{1}\sigma _{2}}=%
\hat{X}_{i}^{\sigma _{1}0}\hat{X}_{i}^{0\sigma _{2}},$ \ $\sigma _{1}\neq
\sigma _{2}$ create a \textit{spin fluctuation} for $\sigma _{1}\neq \sigma
_{2}$ at the $i-th$ site and the spin operator is given by $\mathbf{S}=\hat{X%
}_{i}^{\bar{\sigma}_{1}0}(\vec{\sigma})_{\bar{\sigma}_{1}\bar{\sigma}_{2}}%
\hat{X}_{i}^{0\bar{\sigma}_{2}}$ where summation over bar indices is
understood. The operator $\sum_{\sigma }\hat{X}_{i}^{\sigma \sigma }$ has
the meaning of the\textit{\ hole number on the }$i$\textit{-th site}. It is
useful to introduce the operator $\hat{X}_{i}^{00}=\hat{X}_{i}^{0\sigma }%
\hat{X}_{i}^{\sigma 0}$ at a given lattice site which is the \textit{number
of Zhang-Rice singlets on the i-th site}, i.e. if $\hat{X}_{i}^{00}\mid
0\rangle =1\mid 0\rangle $ the $i$-th site is occupied by the Zhang-Rice
singlet, while for $\hat{X}_{i}^{00}\mid 1\rangle =0\mid 1\rangle $ there is
no Zhang-Rice singlet on the $i$-th site (i.e. this site is occupied only by
one $3d^{9}$ hole on the Cu site). The latter is due to the \textit{local
constraint}
\begin{equation}
\hat{X}_{i}^{00}+\sum_{\sigma =\uparrow \downarrow }\hat{X}_{i}^{\sigma
\sigma }=1,  \label{const}
\end{equation}%
which \textit{forbids doubly occupancy} of the $i$-th site by holes. By
projecting out doubly occupied (high energy) states the $t-J$ model reads%
\begin{equation*}
\hat{H}_{t-j}=\sum_{i,\sigma }\epsilon _{i}^{0}\hat{X}_{i}^{\sigma
\sigma }-\sum_{i,j,\sigma }t_{ij}\hat{X}_{i}^{\sigma
0}\hat{X}_{j}^{0\sigma }
\end{equation*}%
\begin{equation}
\;+\sum_{i,j}J_{ij}(\mathbf{S}_{i}\cdot \mathbf{S}_{j}-\frac{1}{4}\hat{n}_{i}%
\hat{n}_{j}\;)+\hat{H}_{3}.  \label{t-J}
\end{equation}%
The first term ($\sim \epsilon _{i}^{0}$) describes an effective local
energy of the hole (or the Zhang-Rice singlet), the second one ($\sim t_{ij}$%
) describes hopping of the holes, the third one ($\sim J_{ij}$) is
the Heisenberg-like exchange energy between two holes. The theory
\citep{Zhang} predicts that $\left\vert \epsilon
_{i}^{0}\right\vert \gg \left\vert
t_{ij}\right\vert $. This property is very important in the study of EPI. $%
\hat{H}_{3}$ contains three-sites term which is usually omitted believing it
is not important. For charge fluctuation processes it is plausible to omit
it, while for spin-fluctuation processes it is questionable approximation.
If one introduces the enumeration $\alpha ,\beta ,\gamma ,\lambda
=0,\uparrow ,\downarrow $ than the Hubbard operators satisfy the following
algebra
\begin{equation}
\left[ \hat{X}_{i}^{\alpha \beta },\hat{X}_{j}^{\gamma \lambda }\right]
_{\pm }=\delta _{ij}\left[ \delta _{\gamma \beta }\hat{X}_{i}^{\alpha
\lambda }\pm \delta _{\alpha \lambda }\hat{X}_{i}^{\gamma \beta }\;\right] ,
\label{commut}
\end{equation}%
where $\delta _{ij}$ is the Kronecker symbol. The (anti)commutation
relations in Eq.(\ref{commut}) are more complicated than the canonical Fermi
and Bose (anti)commutation relations which complicates the mathematical
structure of the theory. Note, that the Hubbard operators possess also
\textit{projection properties} with $\hat{X}_{i}^{\alpha \beta }\hat{X}%
_{i}^{\gamma \lambda }=\delta _{\beta \gamma }\hat{X}_{i}^{\alpha \lambda }$%
. To escape these complications some novel techniques have been used, such
as the \textit{slave boson technique}. In this technique $\hat{X}%
_{i}^{0\sigma }=f_{i\sigma }b_{i}^{\dag }$, $\hat{X}_{i}^{\sigma _{1}\sigma
_{2}}=f_{i\sigma _{1}}^{\dagger }f_{i\sigma _{2}}$ are represented in terms
of the fermion (spinon) operator $f_{i\sigma }$ which annihilates the spin
of the hole and the boson (holon) operator $b_{i}^{\dag }$ which creates the
Zhang-Rise singlet.

In the minimal theoretical model the electron-phonon interaction (EPI)
contains in principle two leading terms $\hat{H}_{ep}=\hat{H}_{ep}^{ion}+%
\hat{H}_{ep}^{cov}$, the "ionic" one ($\hat{H}_{ep}^{ion}$) and the
"covalent" one ($\hat{H}_{ep}^{cov}$). The "ionic"term describes the change
of the energy of the hole (or the Zhang-Rice singlet) at the $i$-th site due
to lattice vibrations and it reads \citep{KulicReview}, \citep{Kulic1}, \citep%
{Kulic2}

\begin{equation}
\hat{H}_{ep}^{ion}=\sum_{i,\sigma }\hat{\Phi}_{i}\hat{X}_{i}^{\sigma \sigma
},  \label{Hep-ion}
\end{equation}%
where the "displacement" operator $\hat{\Phi}_{i}=\sum_{L\kappa }[\epsilon (%
\mathbf{R}_{i}^{0}-\mathbf{R}_{L\kappa }^{0}+\mathbf{\hat{u}}_{i}-\mathbf{%
\hat{u}}_{L\kappa })-\epsilon (\mathbf{R}_{i}^{0}-\mathbf{R}_{L\kappa
}^{0})] $ (which as in Section V includes the bare coupling constant)
describes the change of the hole (or Zhang-Rice singlet) energy $\epsilon
_{a,i}^{0}$ by displacing atoms in the lattice by the vector $\mathbf{\hat{u}%
}_{L\kappa }$. In the harmonic approximation the EPI potential is given by $%
\hat{\Phi}_{i}=\sum g_{i}(\mathbf{q},\lambda )\exp \{i\mathbf{qR}_{i}\}[b_{%
\mathbf{q},\lambda }+b_{-\mathbf{q},\lambda }^{\dag }]$ where $b_{\mathbf{q}%
,\lambda }$ and $b_{\mathbf{q},\lambda }^{\dag }$ are annihilation and
creation operators of phonons with the polarization $\lambda $,
respectively. This term describes in principle the following processes: (1)
the change of the $O $-hole and $Cu$-hole bare energies $\epsilon _{ph}^{0}$%
, $\epsilon _{dh}^{0}$ in the three--band model due to lattice vibrations;
(2) the change of the long-range Madelung energy (which is due to the
ionicity of the structure) by lattice vibrations along the c-axis; (3) the
change of the $Cu-O$ hopping parameter $t_{pd}$ in the presence of
vibrations, etc. Here, $L$ and $\kappa $ enumerate unit lattice vectors and
atoms in the unit cell, respectively. Until now the phonon operator $\hat{%
\Phi}_{i}$ is calculated in the harmonic approximation for the EPI
interaction of holes with some specific phononic modes, such as the
breathing and half-breathing ones \citep{GunnarssonReview2008}, \citep{Becker}%
. The theory which includes also all other (than oxygen) vibrations in $\hat{%
\Phi}_{i}$ is still awaiting.

It is interesting to make comparison of the EPI coupling constants in the $%
t-J$ model and in the Hartree-Fock (HF) approximation (which is the
analogous of the LDA-DFT method) of the three-band model in Eqs.(\ref{3-band}%
-\ref{Hint-3b}) when the problem is projected on the single band. For
instance, the coupling constant with the\textit{\ half-breathing mode} at
the zone boundary in HF approximation (which mimics LDA-DFT) is given by

\begin{equation}
g_{hb}^{HF}=\pm 4t_{pd}\frac{\partial t_{pd}}{\partial R_{Cu-O}}\frac{1}{%
\epsilon _{d}-\epsilon _{p}}u_{0},  \label{g-LDA}
\end{equation}%
while the coupling constant \textit{in the t-J model} $g_{hb}^{t-J}(=%
\partial \epsilon ^{0}/\partial R_{Cu-O})$ is given by%
\begin{equation}
g_{hb}^{t-J}=\pm 4t_{pd}\frac{\partial t_{pd}}{\partial R_{Cu-O}}\left[
\frac{2p^{2}-1}{\epsilon _{d}-\epsilon _{p}}+\frac{2p^{2}}{U_{d}-\left\vert
\epsilon _{d}-\epsilon _{p}\right\vert }\right] u_{0},  \label{g-tJ}
\end{equation}%
where $p=0.96$ - see \citep{GunnarssonReview2008}, \citep{Becker}
and references therein. It is obvious that in the $t-J$ model the
electron-phonon coupling is different from the HF one, since the
former contains an additional term coming from the many body
effects, which are not
comprised by the HF (LDA-DFT) calculations. The first term in Eq. (\ref{g-tJ}%
) describes the hopping of a $3d$-hole into the $O$ $2p$-states and this
term exists also in the LDA-DFT coupling constant - see Eq. (\ref{g-LDA}).
However, the second term in Eq. (\ref{g-tJ}), which is due to many body
effects, describes the hopping of a $O$ $2p$-hole into the (already) single
occupied $Cu$ $3d$-state and it does not exist in the LDA-DFT approach.
Since the corresponding dimensionless coupling constant $\lambda _{hb}$ is
proportional to $\left\vert g_{hb}\right\vert ^{2}$ one obtains that \textit{%
the} \textit{bare} \textit{t-J coupling constant is almost three times
larger than the LDA-DFT one }$\lambda _{hb}^{t-J}\approx 3\lambda _{hb}^{HF}$%
. This example demonstrates clearly that LDA-DFT method is unreliable for
calculating the EPI coupling constant in HTSC\ cuprates.

Note, that there is also a covalent contribution to $EPI$ which comes from
the change of the effective hopping ($t$) in the $t-J$ model Eq.(\ref{t-J})
and the exchange energy ($J$) in the presence of atomic displacements
\begin{equation*}
\hat{H}_{ep}^{cov}=-\sum_{i,j,\sigma }\frac{\partial t_{ij}}{\partial (%
\mathbf{R}_{i}^{0}-\mathbf{R}_{j}^{0})}(\mathbf{\hat{u}}_{i}-\mathbf{\hat{u}}%
_{j})\hat{X}_{i}^{\sigma 0}\hat{X}_{j}^{0\sigma }\;+
\end{equation*}%
\begin{equation}
+\sum_{i,j,}\frac{\partial J_{ij}}{\partial (\mathbf{R}_{i}^{0}-\mathbf{R}%
_{j}^{0})}(\mathbf{\hat{u}}_{i}-\mathbf{\hat{u}}_{j})\mathbf{S}_{i}\cdot
\mathbf{S}_{j}.  \label{Hep-cov}
\end{equation}%
Here, we shall not go into details but only stress that since $\left\vert
\epsilon _{i}^{0}\right\vert \gg \left\vert t_{ij}\right\vert $ then \textit{%
the covalent term in the effective t-J model is much smaller than the ionic
term} - see more in \citep{KulicReview}, \citep{GunnarssonReview2008}, \citep%
{Becker} and References therein, and in the following only the term $\hat{H}%
_{ep}^{ion}$ will be considered \citep{KulicReview}, \citep{Kulic1}, \citep%
{Kulic2}.

\subsection{Controllable X-method for the quasi-particle dynamics}

The minimal model Hamiltonian for strongly correlated holes with EPI
(discussed above) is expressed via the Hubbard operators which obey "ugly"
non-canonical commutation relations. The latter property is rather
unpleasant for making a controllable theory in terms of Feynmann diagrams
(for these "ugly" operators) and some other approaches are required. There
is one very efficient and \textit{mathematically controllable approach} for
treating the problem with Hubbard operators without using slave boson (or
fermion) techniques. The method (we call it the \textit{X-method}) is based
on the general Baym-Kadanoff technique where the $1/N$ expansion for the
Green's functions in terms of Hubbard operators was first introduced in \citep%
{RuckensteinSchmittRink} and refined in \citep{Kulic1},
\citep{Kulic2}. In the paramagnetic and homogeneous state (with
finite doping) the Green's function $G_{\sigma _{1}\sigma
_{2}}(1-2)$ is diagonal, i.e. $G_{\sigma _{1}\sigma
_{2}}(1-2)=\delta _{\sigma _{1}\sigma _{2}}G(1-2)$ where
\begin{equation}
G(1-2)=-\left\langle T\hat{X}^{0\sigma }(1)\hat{X}^{\sigma
0}(2)\right\rangle =g(1-2)Q,  \label{G-sc}
\end{equation}%
with the Hubbard spectral weight $Q=\langle \hat{X}^{00}\rangle
+\left\langle \hat{X}^{\sigma \sigma }\right\rangle $. The function $g(1-2)$
plays the role of the quasi-particle Green's function - see more in \citep%
{KulicReview}, \citep{Kulic1}, \citep{Kulic2}, \citep{Kulic3}. It
turns out, that in order to have a controllable theory ($1/N$
expansion) one way is to increase the number of spin components
from two to $N$ by changing the
constraint Eq.(\ref{const}) into the new one%
\begin{equation}
\hat{X}_{i}^{00}+\sum_{\sigma =1}^{N}\hat{X}_{i}^{\sigma \sigma }=\frac{N}{2}%
.  \label{N-const}
\end{equation}%
In order to reach convergence of physical quantities in the limit $%
N\rightarrow \infty $ the hopping and exchange energy are also re-scaled,
i.e. $t_{ij}=t_{0,ij}/N$ and $J_{ij}=J_{0,ij}/N$. In order to eliminate a
possible misunderstanding we stress one important fact, that in the case $%
N>2 $ the constraint in Eq.(\ref{N-const}) spoils some projection properties
of the Hubbard operators. Fortunately, these (lost) projection properties
are not used at all in the refined theory. As the result one obtains the
functional integral equation for $G(1,2)$, thus allowing \textit{unambiguous
mathematical and physical treatment} of the problem. In \citep{Kulic1}, \citep%
{Kulic2}, \citep{Kulic3} it is developed a systematic $1/N$
expansion for the
quasi-particle Green's function $g(1-2)(=g_{0}+g_{1}/N+...)$, $%
Q(=Nq_{0}+q_{1}+...)$ (also for $G(1-2)$) and the self-energy. For large $%
N(\rightarrow \infty )$ the leading term is $G_{0}(1-2)=O(N)=g_{0}(1-2)Q_{0}$
with $g_{0}=O(1)$ and $Q_{0}=\left\langle \hat{X}_{i}^{00}\right\rangle
=N\delta /2$. Here, $\delta $ is the concentration of the oxygen holes (i.e.
of the Zhang-Rice singlets) which is related to the chemical potential by
the equation $1-\delta =2\sum_{\mathbf{p}}n_{F}(\mathbf{p})$ with $\;n_{F}(%
\mathbf{p})=\left( e^{\epsilon _{0}(\mathbf{k})-\mu }+1\right) ^{-1}$. The
quasi-particle Green's function $g_{0}(\mathbf{k},\omega )$ and
quasi-particle spectrum $\epsilon _{0}(\mathbf{k})$ in the leading order are
given by
\begin{equation}
g_{0}(\mathbf{k},\omega )\equiv \frac{G_{0}(\mathbf{k},\omega )}{Q_{0}}=%
\frac{1}{\omega -[\epsilon _{0}(\mathbf{k})-\mu ]},  \label{g0}
\end{equation}

\begin{equation}
\epsilon _{0}(\mathbf{k})=\epsilon _{c}-\delta \cdot t(\mathbf{k})-\sum_{%
\mathbf{p}}J_{0}(\mathbf{k}+\mathbf{p})n_{F}(\mathbf{p}).  \label{e0}
\end{equation}%
The level shift is $\epsilon _{c}=\epsilon ^{0}+2\sum_{\mathbf{p}}t(\mathbf{p%
})n_{F}(\mathbf{p})$ and $t(\mathbf{p})$ is the Fourier transform
of the hopping integral $t_{ij}$ - see more in
\citep{KulicReview}.

Let us summarize the \textit{main results of the X-method} in leading $O(1)$%
-order for the quasi-particle properties in the t-J model \citep{KulicReview}%
, \citep{Kulic1}, \citep{Kulic2}, \citep{Kulic3}: (\textbf{i}) The
Green's function $g_{0}(\mathbf{k},\omega )$ describes the
\textit{coherent motion} \textit{of quasi-particles} whose
contribution to the total spectral weight of the Green's function
$G_{0}(\mathbf{k},\omega )$ is $Q_{0}=N\delta /2$.
The coherent motion of quasi-particles is described in leading order by $%
G_{0}(\mathbf{k},\omega )=Q_{0}g_{0}(\mathbf{k},\omega )$ and \textit{the
quasi-particle residuum }$Q_{0}$ disappears in the undoped Mott insulating
state ($\delta =0$). This result is physically plausible since in the Mott
insulating state the coherent motion of quasi-particles, which is
responsible for finite conductivity, vanishes; (\textbf{ii}) The
quasi-particle spectrum $\epsilon _{0}(\mathbf{k})$ plays the same role as
the eigenvalues of the ideal band structure Hamiltonian $\hat{h}_{0}(\mathbf{%
x,y})$ (it contains the excitation potential $V_{IBS}(\mathbf{x,y})$ which
includes high-energy processes due to the Coulomb interaction). So, if we
would consider $\epsilon _{tb}(\mathbf{k})=-t(\mathbf{k})$ as the
tight-binding parametrization of the LDA-DFT band-structure spectrum which
takes int account only weak correlations (with the local potential $%
V_{xc}(x)\delta (x-y)$) then one can define a non-local \textit{excitation
potential} $V_{IBS}^{tJ}(\mathbf{x,y})=\tilde{V}_{IBS}^{tJ}(\mathbf{x,y}%
)+V_{xc}(\mathbf{x})\delta (\mathbf{x}-\mathbf{y})$ which mimics strong
correlations in the t-J model
\begin{equation}
\tilde{V}_{IBS}^{tJ}(\mathbf{x,y})\approx V_{0}\delta (\mathbf{x-y}%
)+(1-\delta )t(\mathbf{x-y})-\tilde{J}(\mathbf{x-y}).  \label{V-tJ}
\end{equation}%
Here, $V_{0}=2\sum_{\mathbf{p}}t(\mathbf{p})n_{F}(\mathbf{p})$ and $t(%
\mathbf{x-y})$ is the Fourier transform of $t(\mathbf{k})$ while $\tilde{J}(%
\mathbf{x-y})$ is the Fourier transform of the third term in Eq.(\ref{e0}).
The relative excitation potential $\tilde{V}_{IBS}^{tJ}(\mathbf{x,y})$ is
due to strong correlations (suppression of doubly occupancy on each lattice
site) and as we shall see below it is responsible for the screening of EPI
in such a way that the forward scattering peak appears in the effective EPI
interaction - see discussion below. (\textbf{iii}) For very low doping $%
\epsilon _{0}(\mathbf{k})$ is dominated by the exchange parameter if $%
J_{0}>\delta t_{0}$. However, in the case when $J_{0}\ll \delta t_{0}$ there
is a \textit{band narrowing} by lowering the hole doping $\delta $, where
the band width is proportional to the hole concentration $\delta $, i.e. $%
W_{b}=z\cdot \delta \cdot t_{0}$; (\textbf{iv}) The $O(1)$-order
quasi-particle Green's function $g_{0}(\mathbf{k},\omega )$ and the
quasi-particle spectrum $\epsilon _{0}(\mathbf{k})$ in the X-method have
similar form as the spinon Green's function $g_{0,f}(\mathbf{k},\omega
)=-\langle Tf_{\sigma }f_{\sigma }^{\dagger }\rangle _{\mathbf{k},\omega }$
and the spinon energy $\epsilon _{s}(\mathbf{k})$ in the $SB$-method.
However, in the $SB$ method there is a broken gauge symmetry in metallic
state which is characterized by $\langle \hat{b}_{i}\rangle \neq 0$. This
broken local gauge symmetry in the slave-boson method in O(1) order, which
is due to the local decoupling of spinon and holon is in fact forbidden by
to the Elitzur's theorem. On the other side the local gauge invariance is
not broken in the X-method where the Green's function $G_{0}(\mathbf{k}%
,\omega )$ describes motion of the \textit{composite object}, i.e.
simultaneous creation of the hole and annihilation of the spin at a given
lattice site, while in the $SB$ theory there is a spin-charge separation
because of the broken symmetry ($\langle \hat{b}_{i}\rangle \neq 0$). The
assumption of the broken symmetry $\langle \hat{b}_{i}\rangle \neq 0$ gives
qualitative satisfactory results for quasi-particle energy for the case $%
N=\infty $ in $D>2$ dimensions. However, the analysis of response
functions of the system and of higher order $1/N$ corrections to
the self-energy in the $SB$ theory is very delicate and special
techniques must be implemented in order to restore the gauge
invariance of the theory. On the other side \textit{the X}-method\
is intrinsically gauge invariant and free of spurious effects in
all orders of the $1/N$ expansion. Therefore, one expects that
these two methods may deliver different results in O(1) order in
response function. This difference is already manifested in the
calculation of EPI where charge vertices are peaked at different
wave vectors $\mathbf{q}$, i.e. at $q=0$ in the X-method and
$q\neq 0$ in the SB method - see Subsection E; (\textbf{v})\ In
the important paper \citep{Greco} it is shown
that in the superconducting state the anomalous self-energies (which are of $%
O(1/N)$-order in the $1/N$ expansion) of the $X$- and $SB$-methods
differ substantially. As a consequence, the $SB$-method
\citep{Kotliar2} predicts false superconductivity in the $t-J$
model (for $J=0$) with large $T_{c}$ (due to the kinematical
interaction), while the $X$-method gives extremely small
$T_{c}(\approx 0)$ \citep{Greco}. So, although the two approaches
yield some similar results in leading $O(1)$-order they are
different at least in the next leading $O(1/N)$-order.

\subsection{EPI effective potential in the t-J model}

The theory of EPI in the minimal $t-J$ model based on the X-method
predicts that leading term in the EPI self-energy $\Sigma _{ep}$
is given by the expression \citep{KulicReview}, \citep{Kulic1},
\citep{Kulic2}
\begin{equation}
\Sigma _{ep}(1,2)=-V_{ep}(\bar{1}-\bar{2})\gamma _{c}(1,\bar{3};\bar{1}%
)g_{0}(\bar{3}-\bar{4})\gamma _{c}(\bar{4},2;\bar{2}),  \label{Sigma-EP}
\end{equation}%
where the screened (by the dielectric constant) EPI potential $%
V_{ep}(1-2)=\varepsilon _{e}^{-1}(1-\bar{1})V_{ep}^{0}(\bar{1}-\bar{2}%
)\varepsilon _{e}^{-1}(\bar{2}-2)$ and $V_{ep}^{0}(1-2)=-\langle T\hat{\Phi}%
(1)\hat{\Phi}(2)\rangle $ is the "phonon" propagator which may also describe
an anharmonic EPI. It is obvious that Eq.(\ref{Sigma-EP}) is equivalent to
Eq.(\ref{Ab-3.30a}) in spite the fact that the theory is formulated in terms
of the Hubbard operators. The \textit{charge vertex} $\gamma
_{c}(1,2;3)=-\delta g_{0}^{-1}(1,2)/\delta u_{eff}(3)$ corresponds to the
the renormalized vertex $\Gamma _{c,r}$ in Eq.(\ref{Ab-3.30a}) and it
describes the "screening" by strong correlations. It depends on the relative
excitation potential $\tilde{V}_{IBS}^{tJ}(\mathbf{x,y})$. The electronic
dielectric function $\varepsilon _{e}(1-2)$ describes screening of EPI by
the long-range part of the Coulomb interaction.\ Note, that in the harmonic
approximation $\hat{\Phi}(1)$ contains the bare EPI coupling constant $%
g_{ep}^{0}$ and lattice displacement $\hat{u}$, i.e.
$\hat{\Phi}\sim g_{ep}^{0}\hat{u}$ - see more in
\citep{KulicReview}. (Note, that in the above equations summation
and integration over bar indices are understood.)
The self-energy $\Sigma _{ep}(\mathbf{k},\omega )$ due to EPI reads%
\begin{equation}
\Sigma _{ep}(\mathbf{k},\omega )=\int_{0}^{\infty }d\Omega \langle \alpha
^{2}F(\mathbf{k,k}^{\prime },\Omega )\rangle _{\mathbf{k}^{\prime }}R(\omega
,\Omega ),  \label{ep-self}
\end{equation}%
with $R(\omega ,\Omega )=-2\pi i(n_{B}(\Omega )+1/2)+\psi (1/2+i)-\psi
(1/2-i(\Omega +\omega )/2\pi T)$ where $n_{B}(\Omega )$ is the Bose
distribution function and $\psi $ is di-gamma function. The Eliashberg
spectral function is given by
\begin{equation*}
\alpha ^{2}F(\mathbf{k,k}^{\prime },\omega )=N(0)\sum_{\nu }\mid g_{\nu }(%
\mathbf{k,k-k}^{\prime })\mid ^{2}
\end{equation*}%
\begin{equation}
\times \delta (\omega -\omega _{\nu }(\mathbf{k-k}^{\prime }))\gamma
_{c}^{2}(\mathbf{k,k-k}^{\prime })  \label{Elia-tJ}
\end{equation}%
where $g_{\nu }(\mathbf{k,p})$ is the $EPI$ coupling constant for the $\nu $%
-the mode, where the renormalization by long-range Coulomb interaction is
included, i.e. $g_{\nu }(\mathbf{k,p})=g_{ep,\nu }^{0}(\mathbf{k,p}%
)/\varepsilon _{e}(\mathbf{p})$. $\langle ...\rangle _{\mathbf{k}}$ denotes
Fermi-surface average with respect to the momentum $\mathbf{k}$ and $N(0)$
is the density of states renormalized by strong correlations. The effect of
strong correlations (in the adiabatic limit) is stipulated in the \textit{%
charge vertex function} $\gamma _{c}(\mathbf{k,k-k}^{\prime })$ which, as we
shall see below, changes the properties of $V_{ep}(\mathbf{q},\Omega )$
drastically compared to weakly correlated systems. In fact the charge vertex
depends on frequency but in the adiabatic limit ($\omega _{ph}\ll W$) and
for $qv_{F}>\omega _{ph}$ it is practically frequency independent, i.e. $%
\gamma _{c}^{(ad)}(\mathbf{k,q},\omega )\approx \gamma _{c}(\mathbf{k,q}%
,\omega =0)$ where the latter is \textit{real quantity}. For $J=0$ the $1/N$
expansion gives $N(0)=N_{0}(0)/q_{0}$ where $q_{0}=\delta /2$ in the $%
t-t^{\prime }$ model. For $J\neq 0$ the density of states $N(0)$ does not
diverge for $\delta \rightarrow 0$ where $N(0)(\sim 1/J_{0})>N_{0}(0)$. The
\textit{bare density of states} $N_{0}(0)$ is calculated in absence of
strong correlations, for instance by the $LDA-DFT$ scheme.

Depending on the symmetry of the superconducting order parameter $\Delta (%
\mathbf{k},\omega )$ ($s-$, $d-wave$ pairing) various \textit{projected
averages} (over the Fermi surface) of $\alpha ^{2}F(\mathbf{k,k}^{\prime
},\omega )$ enter the Eliashberg equations. Assuming that the
superconducting order parameter transforms according to the representation $%
\Gamma _{i}$ of the point group $C_{4v}$ of the square lattice (in the $%
CuO_{2}$ planes) the appropriate \textit{symmetry-projected spectral function%
} is given by

\begin{equation*}
\alpha ^{2}F_{i}(\mathbf{\tilde{k},\tilde{k}}^{\prime },\omega )=\frac{N(0)}{%
8}\sum_{\nu ,j}\mid g_{\nu }(\mathbf{\tilde{k},\tilde{k}-}T_{j}\mathbf{%
\tilde{k}}^{\prime })\mid ^{2}
\end{equation*}%
\begin{equation}
\times \delta (\omega -\omega _{\nu }(\mathbf{\tilde{k}-}T_{j}\mathbf{\tilde{%
k}}^{\prime }))\gamma _{c}^{2}(\mathbf{\tilde{k},\tilde{k}-}T_{j}\mathbf{%
\tilde{k}}^{\prime })D_{i}(j).  \label{Elia-pr}
\end{equation}%
$\mathbf{\tilde{k}}$ and $\mathbf{\tilde{k}}^{\prime }$ are momenta on the
Fermi line in the irreducible Brillouin zone ($1/8$ of the total Brillouin
zone). $T_{j}$ , $j=1,..8,$ denotes the eight point-group transformations
forming the symmetry group of the square lattice. This group has five
irreducible representations which we distinguish by the label $i=1,2,...5$.
In the following we discuss the representations $i=1$ \ and $i=3$, which
correspond to the $s-$ and $d-wave$ symmetry of the full rotation group,
respectively. $D_{i}(j)$ is the representation matrix of the $j-th$
transformation for the representation $i$. Assuming that the superconducting
order parameter $\Delta (\mathbf{k},\omega )$ does not vary much in the
irreducible Brillouin zone one can average over $\mathbf{\tilde{k}}$ and $%
\mathbf{\tilde{k}}^{\prime }$ in the Brillouin zone. For each symmetry one
obtains the corresponding \textit{pairing spectral function} $\alpha
^{2}F_{i}(\omega )$
\begin{equation}
\alpha ^{2}F_{i}(\omega )=\langle \langle \alpha ^{2}F_{i}(\mathbf{\tilde{k},%
\tilde{k}}^{\prime },\omega )\rangle _{\mathbf{\tilde{k}}}\rangle _{\mathbf{%
\tilde{k}}^{\prime }}  \label{Elia-pr-ave}
\end{equation}%
which governs the transition temperature for the order parameter with the
symmetry $\Gamma _{i}$. For instance $\alpha ^{2}F_{3}(\omega )$ is the
pairing spectral function in the $d$-channel and it gives the coupling for $%
d-wave$ superconductivity (the irreducible representation $\Gamma _{3}$ -
sometimes labelled as $B_{1g}$). Performing similar calculations for the
phonon-limited resistivity one finds that the resistivity is related to the
\textit{transport spectral function} $\alpha ^{2}F_{tr}(\omega )$
\begin{equation}
\alpha _{tr}^{2}F(\omega )=\frac{\langle \langle \alpha ^{2}F(\mathbf{k,k}%
^{\prime },\omega )[\mathbf{v}(\mathbf{k})-\mathbf{v}(\mathbf{k}^{\prime
})]^{2}\rangle \rangle _{\mathbf{kk}^{\prime }}}{2\langle \langle \mathbf{v}%
^{2}(\mathbf{k})\rangle \rangle _{\mathbf{kk}^{\prime }}}.  \label{Elia-tr}
\end{equation}%
The effect of strong correlations on EPI was discussed in
\citep{Kulic2}
within the model where $g_{\nu }(\mathbf{k,p})$ and the phonon frequencies $%
\omega _{\nu }(\mathbf{\tilde{k}-\tilde{k}}^{\prime })$ are weakly momentum
dependent. In order to elucidate the main effect of strong correlations on
EPI and $\alpha ^{2}F_{i}(\omega )$ we consider the latter functions for a
simple model with Einstein phonon, where these functions are proportional to
the (so called) r\textit{elative coupling constant} $\Lambda _{i}$
\begin{equation}
\Lambda _{i}=\frac{1}{8}\frac{N(0)}{N_{0}(0)}\sum_{j=1}^{8}\langle \langle
\mid \gamma _{c}(\mathbf{\tilde{k},\tilde{k}-}T_{j}\mathbf{\tilde{k}}%
^{\prime })\mid ^{2}\rangle \rangle _{\mathbf{\tilde{k}\tilde{k}}^{\prime
}}D_{i}(j)  \label{Lambda-pr}
\end{equation}%
Similarly, the resistivity $\rho (T)(\sim \lambda _{tr}\sim $ $\Lambda
_{tr}) $ is renormalized by the correlation effects where the transport
coupling constant $\Lambda _{tr}$ is given by
\begin{equation}
\Lambda _{tr}=\frac{N(0)}{N_{0}(0)}\frac{\langle \langle \mid \gamma _{c}(%
\mathbf{\tilde{k},\tilde{k}-}T_{j}\mathbf{\tilde{k}}^{\prime })\mid ^{2}[%
\mathbf{v}(\mathbf{k})-\mathbf{v}(\mathbf{k}^{\prime })]^{2}\rangle \rangle
_{\mathbf{kk}^{\prime }}}{2\langle \langle \mathbf{v}^{2}(\mathbf{k})\rangle
\rangle _{\mathbf{kk}^{\prime }}}.  \label{Lambda-tr}
\end{equation}

As we see, all projected spectral functions $\alpha
_{i}^{2}F(\omega )$ depend on the \textit{charge vertex function}
$\gamma _{c}(\mathbf{k,q})$ which describes the screening
(renormalization) of EPI due to strong correlations (suppression
of doubly occupancy) \citep{Kulic1}, \citep{Kulic2}. This
important ingredient (which respects also Ward identities) is
decisive step beyond the MFA renormalization of EPI in strongly
correlated systems which was previously studied in connection with
heavy fermions - see review \citep{FulKelZwick}.

\subsection{Charge vertex and the EPI coupling}

\begin{figure}[!tbp]
\begin{center}
\resizebox{.5 \textwidth}{!} {\includegraphics*
[width=8cm]{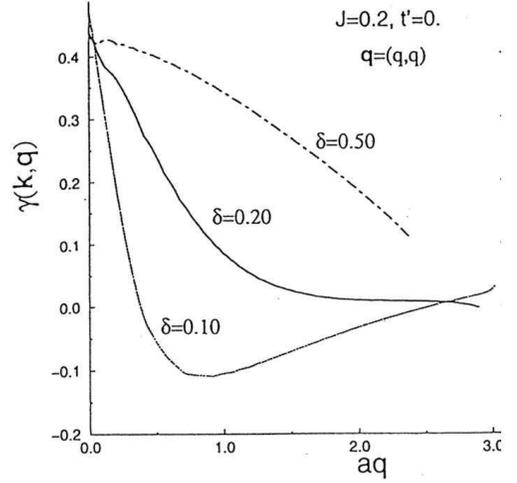}}
\end{center}
\caption{Adiabatic ($\protect\omega =0$) vertex function $\protect\gamma (%
\mathbf{k}_{F},\mathbf{q})$ of the t-J model as a function of the momentum $%
aq$ with $\mathbf{q}=(q,q)$ for three different doping levels $\protect%
\delta $. From \protect\citep{Kulic2}.} \label{Charge-vertex}
\end{figure}

The charge vertex function $\gamma _{c}(\mathbf{k,q})$ (in the adiabatic
approximation) has been calculated in \citep{Kulic1}, \citep{Kulic2}, \citep%
{Kulic3} in the framework of the $1/N$ expansion in the X-method - see also
\citep{KulicReview}, and here we discuss only the main results. Note, that $%
\gamma _{c}(\mathbf{k,q})$ renormalizes all charge fluctuation processes,
such as the $EPI$ interaction, the long range Coulomb interaction,
nonmagnetic impurity scattering etc. In fact $\gamma _{c}(\mathbf{k,q})$
describes specific \textit{screening due to the vanishing of doubly
occupancy in strongly correlated systems}. Note, the latter constraint is at
present impossible to incorporate into the $LDA-DFT$ band structure
calculations, thus making the latter method unreliable in highly correlated
systems. In \citep{Kulic1}, \citep{Kulic2}, \citep{Kulic3} $\gamma _{c}(\mathbf{%
k,q},\omega )$ was calculated as a function of the model parameters $%
t,t\prime ,\delta ,J$ in leading $O(1)$ order of the $t-J$ model
\begin{equation}
\gamma _{c}(\mathbf{k},q)=1-\sum_{\alpha =1}^{6}\sum_{\beta =1}^{6}F_{\alpha
}(\mathbf{k})[\hat{1}+\hat{\chi}(q)]_{\alpha \beta }^{-1}\chi _{\beta 2}(q),
\label{gamma-c}
\end{equation}%
where $\chi _{\alpha \beta }(q)=\sum_{p}G_{\alpha }(p,q)F_{\beta }(\mathbf{p}%
)$, $F_{\alpha }(\mathbf{k})=[t(\mathbf{k}),1,2J_{0}\cos k_{x},2J_{0}\sin
k_{x},2J_{0}\cos k_{y},2J_{0}\sin k_{y}]$, and $G_{\alpha }(p,q)=[1,t(%
\mathbf{p}+\mathbf{q}),\cos p_{x},\sin p_{x},\cos p_{y},\sin p_{y}]\Pi (p,q)$%
. Here, $\Pi (k,q)=-g(k)g(k+q)$ and $q=(\mathbf{q},iq_{n})$, $q_{n}=2\pi nT$%
, $p=(\mathbf{p},ip_{m})$, $p_{m}=\pi T(2m+1)$. The physical meaning of the
vertex function $\gamma _{c}(\mathbf{k},q)$ is following - in the presence
of an external or internal charge perturbation there is a screening due to
the change of the excitation potential $V_{IBS}^{tJ}(\mathbf{x,y})$, i.e. of
the change of the band width, as well as of the local chemical potential.
The central result is that for momenta $\mathbf{k}$\ laying at (and near)
the Fermi surface the vertex function $\gamma _{c}(\mathbf{k},\mathbf{q}%
,\omega =0)$ has very \textit{pronounced\ forward scattering peak}\textbf{\ (%
}at $\mathbf{q}=0$) at very low doping concentration $\delta (\ll 1)$, while
the \textit{backward scattering is substantially suppressed, }as it is seen
in Fig. \ref{Charge-vertex} where $\gamma _{c}(\mathbf{k}_{F},\mathbf{q}%
,\omega =0)$ is shown. The peak at $q=0$ is very narrow at very small doping
since its width $q_{c}$ is proportional to doping $\delta $, i.e. $q_{c}\sim
\delta (\pi /a)$ where $a$ is the lattice constant. It is interesting that $%
\gamma _{c}(\mathbf{k},q)$, as well as the dynamics of charge
fluctuations, depend only weakly on the exchange energy $J$ and
are mainly dominated by the constraint of having no doubly
occupancy of sites, as it is shown in \citep{Kulic1},
\citep{Kulic2}, \citep{Kulic3}.

The existence of the forward scattering peak in $\gamma _{c}(\mathbf{k},%
\mathbf{q})$ at $q=0$ is \textit{confirmed by numerical
calculations in the Hubbard model, }which is very pronounced at
large $U$ \citep{ScalapinoHanke}.
This is important result since it means that $1/N$ expansion in the \textit{%
X-method is reliable method in studying charge fluctuation processes} in
strongly correlated systems. The strong suppression of $\gamma _{c}(\mathbf{k%
},q)$ at large $q(\sim k_{F})$ \ means that the charge fluctuations are
strongly suppressed (correlated) at small distances. Such a behavior of the
vertex function means that a quasi-particle moving in the strongly
correlated medium digs up a \textit{giant correlation hole} with the radius $%
\xi _{ch}(\sim \pi /q_{c})\approx a/\delta $, where $a$ is the lattice
constant. As a consequence of this effect the renormalized \textit{EPI
becomes long ranged} which is contrary to the weakly correlated systems
where it is short ranged.

\begin{figure}[!tbp]
\begin{center}
\resizebox{.5 \textwidth}{!} {\includegraphics*
[width=6cm]{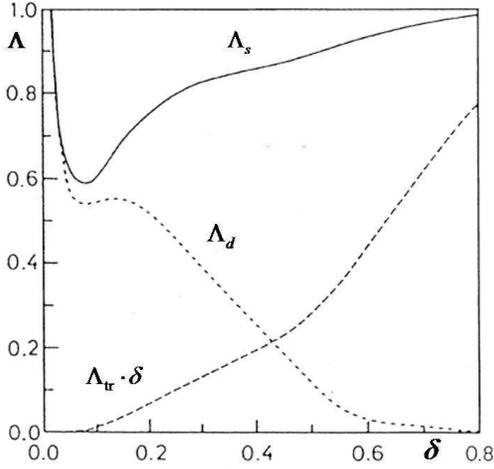}}
\end{center}
\caption{Normalized s-wave $\Lambda _{s}$, d-wave $\Lambda _{d}$
and transport $\Lambda _{tr}\cdot \protect\delta $ coupling
constants as a function of doping $\protect\delta $ for $\
t^{\prime }=0$ and $J=0$. From \protect\citep{Kulic3}.}
\label{Lambdas}
\end{figure}

By knowing $\gamma _{c}(\mathbf{k},q)$ one can calculate the
relative coupling constants $\Lambda _{1}\equiv \Lambda _{s}$,
$\Lambda _{3}\equiv \Lambda _{d}$, $\Lambda _{tr}$ etc. In the
absence of correlations and for an isotropic band one has $\Lambda
_{1}=\Lambda _{tr}=1$, $\Lambda _{i}=0$ for $i>1$. The averages in
$\Lambda _{s}$, $\Lambda _{d}$ and $\Lambda _{tr}$ were performed
numerically in \citep{Kulic2} by using the realistic anisotropic
band dispersion\ in the $t-t\prime -J$ model and the results are
shown in Fig. \ref{Lambdas}. For convenience, the three curves are
multiplied with a common factor so that $\Lambda _{s}$ approaches
$1$ in the empty-band limit $\delta \rightarrow 1$, when strong
correlations are absent. Note, that the superconducting critical
temperature $T_{c}$ in the weak coupling limit and in the $i-th$
channel scales like $T_{c}^{(i)}\sim \exp (-1/(\lambda _{0}\Lambda
_{i}-\mu _{i}^{\ast })$ where $\lambda _{0}$ is some effective
coupling which depends on microscopic details and $\mu _{i}^{\ast
}$ is the effective residual Coulomb repulsion in the i-th
channel. We stress here several interesting results which come out
from the
theory and can be also extracted with the help of Figs. \ref{Charge-vertex}-%
\ref{Lambdas}.

\textbf{(1)} In principle the bare EPI\ coupling constant $g_{\lambda }^{0}(%
\mathbf{k},\mathbf{q})$ depends on the quasi-particle momentum $\mathbf{k}$
and the transfer momentum $\mathbf{q}$. In the $t-J$ model the EPI\ coupling
is dominated by the ionic coupling $\hat{H}_{ep}^{ion}$ (see Eq.(\ref%
{Hep-ion})) and corresponding EPI depends only on the momentum transfer $%
\mathbf{q}$, i.e. $g_{\lambda }^{0}(\mathbf{k},\mathbf{q})\approx $ $%
g_{\lambda }^{0}(\mathbf{q})$ while for the much smaller covalent coupling $%
\hat{H}_{ep}^{cov}$ depends on both $\mathbf{k}$ and $\mathbf{q}$ \citep%
{KulicReview}, \citep{GunnarssonReview2008}. However, the EPI
coupling for most phonon modes are renormalized by the charge
vertex and since the latter
is peaked at small momentum transfer $q=\left\vert \mathbf{k}-\mathbf{k}%
^{\prime }\right\vert $ one expects that the maxima of the corresponding
effective potentials are pushed toward smaller values of $q$. This vertex
renormalization is important result since if in the absence of strong
correlations the bare EPI coupling $\left\vert g^{0}(\mathbf{k},\mathbf{q}%
)\right\vert ^{2}$ for some phonon modes (which enters in the effective $t-J$
model) is detrimental for d-wave pairing it can be less detrimental or even
supports it in the presence of strong correlations. To illustrate this let
us consider the in-plane \textit{oxygen breathing mode} with the frequency $%
\omega _{br}$ which is supposed to be important in HTSC cuprates. The bare
coupling constant (squared) for this mode is approximately given by $%
\left\vert g_{br}^{0}(\mathbf{k},\mathbf{q})\right\vert ^{2}=\left\vert
g_{br}^{0}\right\vert ^{2}[\sin ^{2}(q_{x}a/2)+\sin ^{2}(q_{y}a/2)]$ which
reaches maximum for large $\mathbf{q}=(\pi /a,\pi /a)$. By extracting the
component in the d-channel one has
\begin{equation}
\left\vert g_{br}^{0}(\mathbf{k}-\mathbf{k}^{\prime })\right\vert
^{2}=\left\vert g_{br}^{0}\right\vert ^{2}[1-(1/4)\psi _{d}(\mathbf{k})\psi
_{d}(\mathbf{k}^{\prime })+...]  \label{g-br}
\end{equation}
with
\begin{equation}
\psi _{d}(\mathbf{k})=\cos k_{x}a-\cos k_{y}a.  \label{psi-d}
\end{equation}
This gives the repulsive coupling constant $\lambda _{d}^{0}$ in the
d-channel, i.e. $\lambda _{d}^{0}=(2/\omega _{br})\left\langle \psi _{d}(%
\mathbf{k})\left\vert g_{br}^{0}(\mathbf{k}-\mathbf{k}^{\prime })\right\vert
^{2}\psi _{d}(\mathbf{k}^{\prime })\right\rangle <0$. However, in the
presence of strong correlations one expects that the effective coupling
constant is given approximately by $\left\vert g_{br}^{eff}(\mathbf{k},%
\mathbf{k}-\mathbf{k}^{\prime })\right\vert \approx \left\vert g_{br}^{0}(%
\mathbf{k}-\mathbf{k}^{\prime })\right\vert ^{2}\gamma _{c}^{2}(\mathbf{k}%
_{F},\mathbf{k}-\mathbf{k}^{\prime })$ which is at small doping $\delta $
suppressed substantially at large $q$ since $\gamma _{c}^{2}$ falls off
drastically at $q\sim q_{c}\sim \delta (\pi /a)$. The latter property makes
the effective coupling constant (in the d-channel) $\lambda _{d}^{eff}$for
these modes less negative or even positive (depending on the ratio $\xi
_{ch}/a\sim 1/\delta $), i.e. one has $\lambda _{d}^{eff}>\lambda _{d}^{0}$.
We stress again that this analysis is only qualitative (and
semi-quantitative) since it is based on the $t-J$ model while the better
quantitative results are expected in the strongly correlated ($U_{d}\gg
t,\Delta _{pd}$) three-band Emery model - see Appendix D in \citep%
{KulicReview}. Unfortunately, these calculations are not done until now.

\textbf{(2)} In weakly correlated systems (or for instance in the empty-band
limit $\delta \rightarrow 1$) the relative $d-wave$ coupling constant $%
\Lambda _{d}$ is much smaller than the $s-wave$ coupling constant $\Lambda
_{s}$, i.e. $\Lambda _{d}\ll \Lambda _{s}$ as it is seen in Fig. \ref%
{Lambdas}. Furthermore, $\Lambda _{s}$ decreases with decreasing doping.

\textbf{(3)} It is indicative that independently on the value of $t^{\prime
}\neq 0$ or $t^{\prime }=0$ the coupling constant $\Lambda _{s}$ and $%
\Lambda _{d}$ \textit{meet each other} (note $\Lambda _{s}>$ $\Lambda _{d}$
for all $\delta $) at some small doping $\delta \approx 0.1-0.2$ where $%
\Lambda _{s}\approx $ $\Lambda _{d}$. We would like to stress that such an
unique situation (with $\Lambda _{s}\approx $ $\Lambda _{d}$) was
practically never realized in low temperature and weakly correlated
superconductors and in that respect the strong momentum dependent EPI in
HTSC cuprates is an exclusive phenomenon.

\textbf{(4)} By taking into account residual Coulomb repulsion of
quasi-particles then the $s-wave$ superconductivity (which is governed by $%
\Lambda _{s}$) is suppressed, while the $d-wave$ superconductivity (which is
governed by $\Lambda _{d}$) stays almost unaffected, since $\mu _{s}^{\ast
}\gg \mu _{d}^{\ast }$. In that case the $d-wave$ superconductivity which is
mainly governed by $EPI$ becomes more stable than the $s-wave$ one at
sufficiently low doping $\delta $. This transition between $s$- \ and $d$%
-wave superconductivity is triggered by electronic correlations
because in the model calculations \citep{Kulic1}, \citep{Kulic2}
the bare $EPI$ coupling is assumed to be momentum independent,
i.e. the bare coupling constant contains the $s-wave$ symmetry
only.

\textbf{(5)} The calculations of the charge vertex $\gamma _{c}$ are
performed in the \textit{adiabatic limit}, i.e. for $\omega <\mathbf{q\cdot v%
}_{F}(\mathbf{q})$ the frequency $\omega $ in $\gamma _{c}$ can be
neglected. In the \textit{non-adiabatic regime,} i.e. for $\omega >\mathbf{%
q\cdot v}_{F}(\mathbf{q})$, the function $\gamma _{c}^{2}(\mathbf{k}_{F},%
\mathbf{q,}\omega \mathbf{)}$ may be substantially larger compared to the
adiabatic case because $\gamma _{c}(\mathbf{k}_{F},\mathbf{q,}\omega \mathbf{%
)}$ tends to the bare value $1$ for $q=0$. This means that EPI for
different phonons (with different energies $\omega $) is
differently affected by strong correlations. For a given $\omega $
the EPI coupling to those phonons with momenta $q<$ $q_{\omega
}=\omega /v_{F}$ will be (relatively) enhanced since $\gamma
_{c}(\mathbf{k}_{F},\mathbf{q,}\omega \mathbf{)\approx }1$, while
the coupling to those with $q>q_{\omega }=\omega /v_{F}$ will be
substantially reduced due to the suppression of the backward
scattering by strong correlations \citep{KulicAIP}. These results
are a consequence of the
Ward identities and generally hold in the Landau-Fermi liquid theory \citep%
{Castellani}.

\textbf{(6)} The transport EPI coupling constant $\Lambda _{tr}$ is
significantly reduced in the presence of strong correlations especially for
low doping where $\Lambda _{tr}<\Lambda /3$. This mathematical result is
physically plausible since the resistivity is dominated by the backward
scattering processes (large $q\sim k_{F}$) which are suppressed by strong
correlations - the suppression of $\gamma _{c}(\mathbf{k}_{F},\mathbf{q,}%
\omega \mathbf{)}$ at large $q$.

The theory based on the forward scattering peak in EPI is a good
candidate to explain the linear temperature behavior of the
resistivity down to very small temperatures $T(\sim \Theta
_{D}/30)\approx 10$ $K$ in some cuprates with low $T_{c}(\approx
10$ $K)$ \citep{KulicReview}, \citep{VarelogResist}
\citep{KulicDolgovResistAIP}. A physically rather plausible model
(based on the forward scattering peak in EPI) is elaborated in
\citep{VarelogResist}. It takes into account: (\textit{i}) the
quasi-particle \textit{scattering on acoustic (a) optic (o)
phonons}; (\textit{ii}) the \textit{extended van Hove singularity}
in the quasi-particle density of states $N(\xi )$ which in some
cuprates is very near the Fermi surface; (iii) umklapp and
"undulation" (due
to the flat regions at the Fermi surface) processes with $\mathbf{v}_{%
\mathbf{k}^{\prime }}\cong -$ $\mathbf{v}_{\mathbf{k}}$ - this condition can
partly increase the EPI coupling. The transport Eliashberg function $\alpha
_{tr}^{2}F(\omega )$ is calculated similary to Eq.(\ref{Elia-tr}) by using
the definition of $\alpha ^{2}F(\mathbf{k,k}^{\prime },\omega )$ in Eq.(\ref%
{Elia-pr}) with the \textit{renormalized coupling constant} $g_{\nu }^{(r)}(%
\mathbf{k-k}^{\prime })=g_{\nu }(\mathbf{k-k}^{\prime })\gamma _{c}(\mathbf{%
k-k}^{\prime })$ of the $\nu =a,o$ mode, respectively. The phenomenological
form for the forward scattering peak in $\gamma _{c}(\mathbf{k-k}^{\prime })$
with the cut-off \ $q_{c}\ll k_{F}$ (and which mimics the exact results from
\citep{Kulic1}, \citep{Kulic2}, \citep{Kulic3}) can be found in \citep%
{VarelogResist}. Since the scattering of the quasi-particles on
phonons (with the sound velocity $v_{s}$) is limited to small-$q$
transfer processes (with $q<q_{c}$) then the maximal energy of the
acoustic branch is not the Debye energy $\Theta _{D}(\approx
k_{F}v_{s})$ but the \textit{effective Debye energy} $\Theta
_{A}(\approx q_{c}v_{s})\ll \Theta _{D}$. In the case of $Bi-2201$
in \citep{VarelogResist} it is taken (from the numerical results
in \citep{Kulic1}, \citep{Kulic2}, \citep{Kulic3}) that
$q_{c}\approx k_{F}/10$
which gives $\Theta _{A}\approx (30-50)$ $K$. As the result the calculated $%
\alpha _{tr}^{2}F(\omega )$ gives that $\rho _{ab}(T)\sim T$ down
to very low $T(\sim 0.2$ $\Theta _{A})\approx 10$ $K$. The slope
($d\rho _{ab}/dT$) is governed by the effective EPI\ coupling
constant for acoustic phonons. In systems with the extended
van-Hove singularity (in $N(\xi )$) near the Fermi surface, what
is the case in Bi-2201, the effective coupling constant for
acoustic phonons can be sufficiently large to give experimental
values for the slope $(d\rho _{ab}/dT)\sim (0.5-1)$ $\mu \Omega
cm/K$ - for details see \citep{VarelogResist}. This physical
picture is applicable also to cuprates near and at the optimal
doping but since in these systems $T_{c}$ is large the linearity
of $\rho _{ab}(T)$ down to very low $T$ is "screened" by the
appearance of superconductivity.

(\textbf{7}) The forward scattering peak in the charge vertex $\gamma _{c}(%
\mathbf{k}_{F},\mathbf{q)}$ with the width $q_{c}\sim \delta (\pi /a)$ is
very narrow in underdoped cuprates which may have further interesting
consequences. For instance, HTSC cuprates are characterized not only by
strong correlations but also by relatively small Fermi energy $E_{F}$, which
is in \textit{underdoped systems} not much larger than the characteristic
(maximal) phonon frequency $\omega _{ph}^{\max }$, i.e. $E_{F}\simeq 0.1-0.3$
$eV$, $\omega _{ph}^{\max }\simeq 80$ $meV$. Due to the appreciable
magnitude of $\omega _{D}/E_{F}$ in HTSC oxides it is necessary to correct
the Migdal-Eliashberg theory by the non-Migdal vertex corrections due to the
EPI. It is well-known that these vertex corrections lower $T_{c}$ in systems
with isotropic EPI. However, the non-Migdal vertex corrections in systems
with the forward scattering peak in the EPI coupling with the cut-off $%
q_{c}<<k_{F}$ may increase T$_{c}$ which can be appreciable. The
corresponding calculations \citep{Grimaldi} give two interesting results: (%
\textit{i}) there is a drastic increase of T$_{c}$ by lowering $%
Q_{c}=q_{c}/2k_{F}$, for instance $T_{c}(Q_{c}=0.1)\approx 4T_{c}(Q_{c}=1)$;
(\textit{ii}) Even small values of $\lambda _{ep}<1$ can give large T$_{c}$.
The latter results open a new possibility in reaching high T$_{c}$ in
systems with appreciable ratio $\omega _{D}/E_{F}$ and with the forward
scattering peak. The difference between the Migdal-Eliashberg and the
non-Migdal theory can be explained qualitatively in the framework of an
approximative McMillan formula for T$_{c}$ (for not too large $\lambda $)
which reads $T_{c}\approx \langle \omega \rangle e^{-1/[\tilde{\lambda}-\mu
^{\ast }]}$. The \textit{Migdal-Eliashberg theory} predicts $\tilde{\lambda}%
\approx \lambda /(1+\lambda )$ while the \textit{non-Migdal theory} \citep%
{Grimaldi} gives $\tilde{\lambda}\approx \lambda (1+\lambda )$. For instance
$T_{c}\sim 100$ $K$ in HTSC oxides can be explained by the Migdal-Eliashberg
theory for $\lambda \sim 2$, while in the non-Migdal theory much smaller
coupling constant is needed, i.e. $\lambda \sim 0.5$.

(\textbf{8}) The existence of the forward scattering peak in EPI can in a
plausible way explain the ARPES puzzle that the anti-nodal kink is shifted
by the maximal superconducting gap $\Delta _{\max }$ while the nodal kink is
unshifted. The reason is (as explained in Subsection III.C), that the EPI
spectral function $\alpha ^{2}F(\mathbf{k},\mathbf{k}^{\prime },\Omega
)\approx \alpha ^{2}F(\varphi -\varphi ^{\prime },\Omega )$ is strongly
peaked due to strong correlations at $\varphi -\varphi ^{\prime }=0$ \citep%
{KulicDolgovShift}.

(\textbf{9}) Finally, the scattering potential on non-magnetic
impurities is renormalized by strong correlations giving also the
forward scattering peak in scattering potential (amplitude)
\citep{KuOudo}. The latter effect gives large d-wave channel in
the renormalized impurity potential what is the reason that the
d-wave pairing in HTSC cuprates is robust in the presence of
non-magnetic impurities (and defects) \citep{KulicReview},
\citep{KuOudo}.

\subsection{EPI and strong correlations - other methods}

The calculations of the static (adiabatic) charge-vertex $\gamma _{c}(%
\mathbf{k}_{F},\mathbf{q)}$ in the X-method are done for the case
$U=\infty $ \citep{Kulic1}, \citep{Kulic2}, \citep{Kulic3} where
it is found that it is peaked at $q=0$ - forward scattering peak
(FSP). This result is confirmed by
the numerical Monte-Carlo calculations for the finite-$U$ Hubbard model \citep%
{ScalapinoHanke}, where it is found that FSP exists for all $U$,
but it is especially pronounced in the limit $t\ll U$. These
results are additionally confirmed in the calculations
\citep{Cerruti4SB} within the \textit{four
slave-boson method} of Kotliar-R\"{u}ckenstein where $\gamma _{c}(\mathbf{k}%
_{F},\mathbf{q)}$ is again peaked at $q=0$ and the peak is also pronounced
at $t\ll U$.

There are several calculations of the charge vertex in the \textit{one
slave-boson} method \citep{Castellani}, \citep{KotliarLiu}, \citep{KimTesanovic}%
, \citep{Nagaosa} which is invented to study the limit $U\rightarrow \infty $%
. It is interesting to compare the results for the charge vertex
\textit{in the X-method} \citep{Kulic1}, \citep{Kulic2},
\citep{Kulic3} and \textit{in the
one slave-boson theory} \citep{KotliarLiu} which are calculated in $%
O((1/N)^{0})$ order. For instance, for $J=0$ one has
\begin{equation}
\gamma _{c}^{(X)}(\mathbf{k},\mathbf{q)=}\frac{1+b(\mathbf{q})-a(\mathbf{q}%
)t(\mathbf{k})}{\left[ 1+b(\mathbf{q})\right] ^{2}-a(\mathbf{q})c(\mathbf{q})%
}  \label{X-gamma}
\end{equation}%
\begin{equation}
\gamma _{c}^{(SB)}(\mathbf{k},\mathbf{q)=}\frac{1+b(\mathbf{q})-a(\mathbf{q})%
\left[ t(\mathbf{k})+t(\mathbf{k+q})\right] /2}{\left[ 1+b(\mathbf{q})\right]
^{2}-a(\mathbf{q})c(\mathbf{q})}.  \label{SB-gamma}
\end{equation}%
The explicit expressions for the "bare" susceptibilities $a(\mathbf{q)}$, $b(%
\mathbf{q})$ and $c(\mathbf{q})$ can be found in \citep{Kulic1}, \citep{Kulic2}%
. It is obvious from Eqs.(\ref{X-gamma}-\ref{SB-gamma}) that $\gamma
_{c}^{(X)}(\mathbf{k},\mathbf{q}=0\mathbf{)=}\gamma _{c}^{(SB)}(\mathbf{k},%
\mathbf{q}=0\mathbf{)}$ but the calculations give that $\max
\{\gamma _{c}^{(X)}(\mathbf{k},\mathbf{q)}\}$ is for $q=0$, while
$\max \{\gamma _{c}^{(SB)}(\mathbf{k},\mathbf{q)}\}$ is for $q\neq
0$ \citep{KuAk2}. So, the SB\ vertex is peaked at finite $q$ which
is \textit{in contradiction} with the numerical Monte Carlo
results for the Hubbard model \citep{ScalapinoHanke} and with the
four slave-bosons theory \citep{Cerruti4SB}. The reason for the
discrepancy of the one slave-boson (SB) in studying EPI with the
numerical results and the X-method is not clear and might be due
to the symmetry breaking of the local gauge invariance in leading
order of the SB theory.

\section{Summary of Part II}

The experimental results in HTSC cuprates which are exposed in Part I imply
that EPI coupling constant is large and in order to be conform with d-wave
pairing this interaction must be very nonlocal (long range), i.e. weakly
screened and peaked at small transfer momenta. In absence of quantitative
calculations in the framework of the ab initio microscopic many-body theory
the effects of strong correlations on EPI are studied within the minimal $%
t-J $ model where this \textit{pronounced non-locality} is due to two main
reasons: ($1$) \textit{strong electronic correlations} and ($2$) \textit{%
mixed metallic-ionic layered structure} in these materials. In the case ($1$%
) the pronounced non-locality of EPI, which is found in the $t-J$ model
system, is due to the suppression of doubly occupancy at the $Cu$ lattice
sites in the $CuO_{2}$ planes, which drastically weakens the screening
effect in these systems. The pronounced non-locality and suppression of the
screening is mathematically expressed by the charge vertex function $\gamma
_{c}(\mathbf{k}_{F},\mathbf{q},\omega )$ which multiplies the bare coupling
constants. This function is peaked at $q=0$ and strongly suppressed at large
$q$, especially for low (oxygen) hole doping $\delta \ll 1$ near the
Mott-Hubbard transition. Such a structure of $\gamma _{c}$ gives that the $d$%
-wave and $s$-wave coupling constant are of the same order of magnitude
around and below some optimal doping $\delta _{op}\approx 0.1$, i.e. $%
\lambda _{d}\approx \lambda _{s}$. This is very peculiar situation never met
before. Since the residual effective (low-energy) Coulomb interaction is
much smaller in the $d$-channel than in the $s$-channel, i.e. $\mu
_{s}^{\ast }\gg \mu _{d}^{\ast }$, then the critical temperature for $d$%
-wave pairing is much larger than for the $s$-wave one, i.e.
$T_{c}^{(d)}\gg T_{c}^{(s)}$. Since all charge fluctuation
processes are modified by strong correlations then the
quasi-particle scattering on non-magnetic impurities is
drastically changed, that the pair-breaking effect on $d$-wave
pairing is drastically reduced. This non-local effect, which is
not discussed here - see more \citep{KulicReview} and References
therein, is one of the main reasons for the robustness of $d$-wave
pairing in HTSC oxides in the presence of non-magnetic impurities
and numerous local defects. The
development of the forward scattering peak in $\gamma _{c}(\mathbf{k}_{F},%
\mathbf{q})$ and suppression at large $q(\gg q_{c}=\delta (\pi /a))$ gives
rise to the suppression of the transport coupling constant $\lambda _{tr}$
making it much smaller than the self-energy coupling constant $\lambda $,
i.e. one has $\lambda _{tr}\approx \lambda /3$ near the optimal doping $%
\delta =0.1-0.2$. Thus the behavior of the vertex function and the dominance
of EPI\ in quasi-particle scattering resolves the experimental puzzle that
the transport and and self-energy coupling constant take very different
values, $\lambda _{tr,ep}\ll \lambda _{ep}$. This is not the case with the
SFI mechanism which is dominant at large $q$ thus giving $\lambda
_{tr,sf}\approx \lambda _{sf}$.

We stress that the strength of the EPI coupling constants $\lambda _{ep}$, $%
\lambda _{ep,d}$ is at present impossible to calculate since it is difficult
to incorporate strong correlations and numerous structural effects in a
tractable microscopic theory.

\section{Discussions and conclusions}

Numerous experimental results related to tunnelling, optics, ARPES,
inelastic neutron and x-ray scattering measurements in HTSC cuprates \textit{%
at and near the optimal doping} give evidence for strong electron-phonon
interaction (EPI) with the coupling constant $1<\lambda _{ep}<3.5$. The
tunnelling measurements furnish evidence for strong EPI which give that the
peaks in the bosonic spectral function $\alpha ^{2}F(\omega )$ coincide well
with the peaks in the phonon density of states $F_{ph}(\omega )$. The
tunnelling spectra show\ that \textit{almost all phonons contribute} to T$%
_{c}$ and that no particular phonon mode can be singled out in the spectral
function $\alpha ^{2}F(\omega )$ as being the only one which dominates in
pairing mechanism. In light of these results the small oxygen isotope effect
in optimally doped systems can be partly due this effect, thus not
disqualifying the important role of EPI in pairing mechanism. The
compatibility of the strong EPI with $d$-wave pairing implies an important
constraint on the EPI pairing potential - it must be nonlocal, i.e. peaked
at small transfer momenta. The latter is due to: (\textit{a}) strong
electronic correlations and (\textit{b}) the mixed metallic-ionic structure
of these materials. If EPI is the main player in pairing in HTSC cuprates
then this non-locality implies that at and below some optimal doping ($%
\delta _{op}\sim 0.1$) the magnitude of the EPI coupling constants in $d$%
-wave and $s$-wave channel must be of the same order, i.e. $\lambda
_{ep,d}\approx \lambda _{ep,s}$. This result in conjunction with the fact
that the residual effective Coulomb coupling in $d$-wave channel is much
smaller than in the $s$-wave one, i.e. $\mu _{s}^{\ast }\gg \mu _{d}^{\ast }$
gives that the critical temperature for $d$-wave pairing is much larger than
for s-wave pairing.

The tunnelling, ARPES, optic and magnetic neutron scattering measurements
give sufficient evidence that the spin fluctuation interaction (SFI) plays a
secondary role in pairing in HTSC cuprates. Especially important evidence
for the smallness of SFI (in pairing) comes from the magnetic neutron
scattering measurements which show that by varying doping slightly around
the optimal one there is a huge reconstruction of the SFI spectral function $%
Im\chi (\mathbf{Q},\omega )$ (imaginary part of the spin susceptibility)
while there is very small change in the critical temperature $T_{c}$. These
experimental results imply important constraints on the \textit{pairing
scenario} \textit{for systems at and near optimal doping}: (1) the strength
of the d-wave pairing potential is provided by EPI (i.e. one has $\lambda
_{ep,d}\approx \lambda _{ep,s}$) while the role of SFI, together with the
residual Coulomb interaction, is to trigger $d$-wave pairing; (2) the
Migdal-Eliashberg theory, but with the pronounced momentum dependent of EPI,
is a rather good starting theory.

The ab initio microscopic theory of pairing in HTSC cuprates fails at
present to calculate T$_{c}$ and to predict the magnitude of the d-wave
order parameter. From that point of view it is hardly to expect a
significant improvement of this situation at least in the near future.
However, the studies of some minimal (toy) models, such as the single band $%
t-J$ model, allow us to understand part of the physics in cuprates on a
qualitative and in some cases even on a semi-quantitative level. In that
respect the encouraging results come from the theoretical studies of EPI in
the $t-J$ model by using controllable mathematical methods. This theory
predicts dressing of quasi-particles by strong correlations which dig up a
large scale correlation hole of the size $\xi _{ch}\sim a/\delta $ for $%
\delta \ll 1$. These quasi-particles respond to lattice vibrations in such a
way to produce an effective electron interaction (due to EPI) which is long
ranged, i.e. the effective pairing potential $V_{eff}(\mathbf{q},\omega )$
is peaked at small transfer momenta $q$ - forward scattering peak. This
theory (of the toy model) is conform with the experimental scenario by
predicting the following results: ($i$) the EPI coupling constants in $d$%
-wave and $s$-wave channels are of the same order, i.e. $\lambda
_{ep,d}\approx \lambda _{ep,s}$, at some optimal doping $\delta _{op}\sim
0.1 $; ($ii$) the transport coupling is much smaller than the pairing one,
i.e. $\lambda _{tr}\approx \lambda /3$; ($iii)$ due to strong correlations
there is forward scattering peak in the potential for scattering on
non-magnetic impurities, thus making d-wave pairing robust in materials with
a lot of defects and impurities. Applied to HTSC superconductors at and near
the optimal doping this theory is a realization of the Migdal-Eliashberg
theory but with strongly momentum dependent EPI coupling, which is conform
with the proposed experimental pairing scenario. This scenario which is also
realized in the $t-J$ toy model may be useful in making a (phenomenological)
theory of pairing in cuprates. However, all present theories are confronted
with the unsolved and \textit{challenging task} - the \textit{calculation of}
$T_{c}$. From that point of view we do not have at present a proper
microscopic theory of pairing in HTSC cuprates.

\begin{acknowledgments}
We devote this paper to our great teacher
and friend Vitalii Lazarevich Ginzburg who passed away recently.
His permanent interest in our work and support in many respects
over many years are unforgettable. M. L. K. is thankful to Karl
Bennemann for inspiring discussions on many subjects related to
physics of HTSC cuprates. We thank Godfrey Akpojotor for careful
reading of the manuscript. M. L. K. is thankful to the
Max-Born-Institut f\"{u}r Nichtlineare Optik und
Kurzzeitspektroskopie, Berlin for the hospitality and financial
support during his stay where part of this work has been done.
\end{acknowledgments}

\section{Appendix - Spectral functions}

\subsection{Spectral functions $\protect\alpha ^{2}F(\mathbf{k},\mathbf{k}%
^{\prime },\protect\omega )$ and $\protect\alpha ^{2}F(\protect\omega )$}

The quasi-particle bosonic \ (Eliashberg) spectral function $\alpha ^{2}F(%
\mathbf{k},\mathbf{k}^{\prime },\omega )$ and its Fermi surface average $%
\alpha ^{2}F(\omega )=\left\langle \left\langle \alpha ^{2}F(\mathbf{k},%
\mathbf{k}^{\prime },\omega )\right\rangle \right\rangle _{\mathbf{k},%
\mathbf{k}^{\prime }}$ enter the quasi-particle self-energy $\Sigma (\mathbf{%
k},\omega )$, while the transport spectral function $\alpha
^{2}F_{tr}(\omega )$ enters the transport self-energy $\Sigma _{tr}(\mathbf{k%
},\omega )$ and dynamical conductivity $\sigma (\omega )$. Since the
Migdal-Eliashberg theory for EPI is well defined we define the spectral
functions for this case and the generalization to other electron-boson
interaction is straightforward. In the superconducting state the Matsubara
Green's function $\hat{G}(\mathbf{k},\omega _{n})$ and $\hat{\Sigma}(\mathbf{%
k},\omega _{n})$ are $2\times 2$ matrices with the diagonal elements $%
G_{11}\equiv G(\mathbf{k},\omega _{n}),\Sigma _{11}\equiv \Sigma (\mathbf{k}%
,\omega _{n})$ and off-diagonal elements $G_{12}\equiv F(\mathbf{k},\omega
_{n}),\Sigma _{12}\equiv \Phi (\mathbf{k},\omega _{n})$ which describe
superconducting pairing. By defining $i\omega _{n}\left[ 1-Z(\mathbf{k}%
,\omega _{n})\right] =\left[ \Sigma (\mathbf{k},\omega _{n})-\Sigma (\mathbf{%
k},-\omega _{n})\right] /2$ and $\chi (\mathbf{k},\omega _{n})=\left[ \Sigma
(\mathbf{k},\omega _{n})+\Sigma (\mathbf{k},-\omega _{n})\right] /2$, the
Eliashberg functions for EPI in the presence of the Coulomb interaction (in
the singlet pairing channel) read \citep{AllenMitrovic}, \citep%
{MaksimovEliashberg}, \citep{MarsiglioCarbotteBook}
\begin{equation}
Z(\mathbf{k},\omega _{n})=1+\frac{T}{N}\sum_{\mathbf{p},m}\frac{\lambda _{%
\mathbf{kp}}^{Z}(\omega _{n}-\omega _{m})\omega _{m}}{N(\mu )\omega _{n}}%
\frac{Z(\mathbf{p},\omega _{m})}{D(\mathbf{p},\omega _{m})},  \label{Z-Eli}
\end{equation}%
\begin{equation}
\chi (\mathbf{k},\omega _{n})=-\frac{T}{N}\sum_{\mathbf{p},m}\frac{\lambda _{%
\mathbf{kp}}^{Z}(\omega _{n}-\omega _{m})}{N(\mu )}\frac{\epsilon (\mathbf{p}%
)-\mu +\chi (\mathbf{p},\omega _{m})}{D(\mathbf{p},\omega _{m})},
\label{chi-Eli}
\end{equation}%
\begin{equation}
\Phi (\mathbf{k},\omega _{n})=\frac{T}{N}\sum_{\mathbf{p},m}\left[ \frac{%
\lambda _{\mathbf{kp}}^{\Delta }(\omega _{n}-\omega _{m})}{N(\mu )}-V_{%
\mathbf{kp}}\right] \frac{\Phi (\mathbf{p},\omega _{m})}{D(\mathbf{p},\omega
_{m})},  \label{Fi-Eli}
\end{equation}%
where $N(\mu )$ is the density of states at the Fermi surface, $\omega
_{n}=\pi T(2n+1)$, $\Phi (\mathbf{k},\omega _{n})\equiv Z(\mathbf{k},\omega
_{n})\Delta (\mathbf{k},\omega _{n})$ and $D=\omega _{m}^{2}Z^{2}+\left(
\epsilon -\mu +\chi \right) ^{2}+\Phi ^{2}$. (For studies of optical
properties - see below, it is useful to introduce the renormalized frequency
$i\tilde{\omega}_{n}(i\omega _{n})(\equiv i\omega _{n}Z(\omega _{n}))=\omega
_{n}-\Sigma (\omega _{n})$ (or its analytical continuation $\tilde{\omega}%
(\omega )=Z(\omega )\omega =\omega -\Sigma (\omega )$). These equations are
supplemented with the electron number equation $n(\mu )$ ($\mu $ is the
chemical potential)
\begin{equation*}
n(\mu )=\frac{2T}{N}\sum_{\mathbf{p},m}G(\mathbf{p},\omega _{m})e^{i\omega
_{m}0^{+}}
\end{equation*}%
\begin{equation}
=1-\frac{2T}{N}\sum_{\mathbf{p},m}\frac{\epsilon (\mathbf{p})-\mu +\chi (%
\mathbf{p},\omega _{m})}{D(\mathbf{p},\omega _{m})}.  \label{n-mju}
\end{equation}%
Note that in the case of EPI one has $\lambda _{\mathbf{kp}}^{\Delta }(\nu
_{n})=\lambda _{\mathbf{kp}}^{Z}(\nu _{n})(\equiv \lambda _{\mathbf{kp}}(\nu
_{n}))$ (with $\nu _{n}=\pi Tn$) where $\lambda _{\mathbf{kp}}(\nu _{n})$ is
defined by
\begin{equation}
\lambda _{\mathbf{kp}}(\nu _{n})=2\int_{0}^{\infty }\frac{\nu \alpha _{%
\mathbf{kp}}^{2}F(\nu )d\nu }{\nu ^{2}+\nu _{n}^{2}}  \label{lambda-kp}
\end{equation}

\begin{equation}
\alpha _{\mathbf{kp}}^{2}F(\nu )=N(\mu )\sum_{\kappa }\left\vert g_{\kappa ,%
\mathbf{kp}}^{ren}\right\vert ^{2}B_{\kappa }(\mathbf{k}-\mathbf{p,}\nu )
\label{alpha-kp}
\end{equation}%
where $B_{\kappa }(\mathbf{k}-\mathbf{p;}\nu )$ is the phonon spectral
function of the $\kappa $-th phonon mode related to the phonon propagator
\begin{equation}
D_{\kappa }(\mathbf{q,}i\nu _{n})=-\int_{0}^{\infty }\frac{\nu }{\nu
^{2}+\nu _{n}^{2}}B_{\kappa }(\mathbf{q,}\nu )d\nu .  \label{D-phonon}
\end{equation}%
However, in systems with is measured the generalized phonon density of
states $GPDS(\omega )(\equiv G(\omega ))$ (see Part I Subsection D) defined
by $G(\omega )=\sum_{i}(\sigma _{i}/M_{i})F_{i}(\omega )/\sum_{i}(\sigma
_{i}/M_{i})$, where $\sigma _{i}$ and $M_{i}$ are the cross section and the
mass of the i-th nucleus and $F_{i}(\omega =(1/N)\sum_{q}\left\vert
\varepsilon _{q}^{i}\right\vert ^{2}\delta (\omega -\omega _{q})$ is the
amplitude-weighted density of states. The renormalized coupling constant $%
g_{\kappa ,\mathbf{kp}}^{ren}(\approx g_{\kappa ,\mathbf{kp}}^{0}\varepsilon
_{e}^{-1}\gamma )$ comprises the screening effect due to long-range Coulomb
interaction ($\sim \varepsilon _{e}^{-1}$ - the inverse electronic
dielectric function) and short-range strong correlations ($\sim \gamma $ -
the vertex function) - see more in \textit{Part II.} Usually in the case of
low-temperature superconductors (LTS) with s-wave pairing the anisotropy is
rather small (or in the presence of impurities it is averaged out) which
allows an averaging of the Eliashberg equations \citep{AllenMitrovic}, \citep%
{MaksimovEliashberg}, \citep{MarsiglioCarbotteBook}
\begin{equation}
Z(\omega _{n})=1+\frac{\pi T}{\omega _{n}}\sum_{m}\frac{\lambda (\omega
_{n}-\omega _{m})\omega _{m}}{\sqrt{\omega _{m}^{2}+\Delta ^{2}(\omega _{m})}%
},  \label{Z-iso}
\end{equation}%
\begin{equation*}
Z(\omega _{n})\Delta (\omega _{n})=\pi T\sum_{m}[\lambda (\omega _{n}-\omega
_{m})
\end{equation*}%
\begin{equation}
-\mu (\omega _{c})\theta (\omega _{c}-\left\vert \omega _{m}\right\vert )]%
\frac{\Delta (\omega _{m})}{\sqrt{\omega _{m}^{2}+\Delta ^{2}(\omega _{m})}},
\label{Gap-iso}
\end{equation}%
\begin{equation}
\lambda (\omega _{n}-\omega _{m})=\int_{0}^{\infty }d\nu \frac{2\nu \alpha
^{2}F(\nu )}{\nu ^{2}+(\omega _{n}-\omega _{m})^{2}}.  \label{lambda}
\end{equation}%
Here $\alpha ^{2}F(\omega )=\left\langle \left\langle \alpha ^{2}F(\mathbf{k}%
,\mathbf{k}^{\prime },\omega )\right\rangle \right\rangle _{\mathbf{k},%
\mathbf{k}^{\prime }}$where $\left\langle \left\langle ...\right\rangle
\right\rangle _{\mathbf{k},\mathbf{k}^{\prime }}$ is the average over the
Fermi surface. The above equations can be written on the real axis by the
analytical continuation $i\omega _{m}\rightarrow \omega +i\delta $ where the
gap function is complex i.e. $\Delta (\omega )=\Delta _{R}(\omega )+i\Delta
_{I}(\omega )$. The solution for $\Delta (\omega )$ allows the calculation
of the current-voltage characteristic $I(V)$ and \textit{tunnelling
conductance} $G_{NS}(V)=dI_{NS}/dV$ in the superconducting state of the $NIS$
tunnelling junction where $I_{NS}(V)$ is given by

\begin{equation*}
I_{NS}(V)=2e\sum_{\mathbf{k},\mathbf{p}}\mid T_{\mathbf{k},\mathbf{p}}\mid
^{2}\int_{-\infty }^{\infty }\frac{d\omega }{2\pi }
\end{equation*}%
\begin{equation}
A_{N}(\mathbf{k},\omega )A_{S}(\mathbf{p},\omega +eV)[f(\omega )-f(\omega
+eV)].  \label{I(V)}
\end{equation}%
Here, $A_{N,S}(\mathbf{k},\omega )=-2ImG_{N,S}(\mathbf{k},\omega )$ are the
spectral functions of the normal metal and superconductor, respectively and $%
f(\omega )$ is the Fermi distribution function. Since the angular and energy
dependence of the tunnelling matrix elements $\mid T_{\mathbf{k},\mathbf{p}%
}\mid ^{2}$ is practically unimportant for $s-wave$ superconductors, then in
that case the relative conductance $\sigma _{NS}(V)\equiv
G_{NS}(V)/G_{NN}(V) $ is proportional to the tunnelling density of states $%
N_{T}(\omega )=\int A_{S}(\mathbf{k},\omega )d^{3}k/(2\pi )^{3}$, i.e. $%
\sigma _{NS}(\omega )\approx N_{T}(\omega )$ where
\begin{equation}
N_{T}(\omega )=Re\left\{ \frac{\omega +i\tilde{\gamma}(\omega )}{\sqrt{%
(\omega +i\tilde{\gamma}(\omega ))^{2}-\tilde{Z}^{2}(\omega )\Delta (\omega
)^{2}}}\right\} .  \label{Sigma-NS}
\end{equation}%
Here, $\tilde{Z}(\omega )=Z(\omega )/ReZ(\omega )$, $\tilde{\gamma}(\omega
)=\gamma (\omega )/ReZ(\omega )$, $Z(\omega )=ReZ(\omega )+i\gamma (\omega
)/\omega $ and the \textit{quasi-particle scattering rate }in the
superconducting state $\gamma _{s}(\omega ,T)=-2Im\Sigma (\omega ,T)$ is
given by
\begin{equation*}
\gamma _{s}(\omega ,T)=2\pi \int\limits_{0}^{\infty }d\nu \alpha ^{2}F(\nu
)N_{s}(\nu +\omega )\{2n_{B}(\nu )
\end{equation*}%
\begin{equation}
+n_{F}(\nu +\omega )+n_{F}(\nu -\omega )\}+\gamma ^{imp},  \label{Gamma-T}
\end{equation}%
where $N_{s}(\omega )=Re\{\omega /(\omega ^{2}-\Delta ^{2}(\omega ))^{1/2}$
is the quasi-particle density of states in the superconducting state, $%
n_{B,F}(\nu )$ are Bose and Fermi distribution function respectively. Since
the structure of phononic spectrum is contained in $\alpha ^{2}F(\omega )$,
it is reflected on $\Delta (\omega )$ for $\omega >\Delta _{0}$ (the real
gap obtained from $\Delta _{0}=Re\Delta (\omega =\Delta _{0})$) which gives
the structure in $G_{S}(V)$ at $V=\Delta _{0}+\omega _{ph}$. On the contrary
one can extract the spectral function $\alpha ^{2}F(\omega )$ from $%
G_{NS}(V) $ by the inversion procedure proposed by McMillan and Rowell \citep%
{KulicReview}, \citep{McMillanRowell}. It turns out that in
low-temperature superconductors the peaks of $-d^{2}I/dV^{2}$ at
$eV_{i}=\Delta +\omega
_{ph,i}$ correspond to the peak positions of $\alpha ^{2}F(\omega )$ and $%
F(\omega )$. However, we would like to point out that in HTSC cuprates the
gap function is unconventional and very anisotropic, i.e. $\Delta (\mathbf{k}%
,i\omega _{n})\sim \cos k_{x}a-\cos k_{y}a$. Since in this case the
extraction of $\alpha ^{2}F(\mathbf{k},\mathbf{k}^{\prime },\omega )$ is
difficult and at present rather unrealistic task, then an \textquotedblright
average\textquotedblright\ $\alpha ^{2}F(\omega )$ is extracted from the
experimental curve $G_{S}(V)$. There is belief that it gives relevant
information on the real spectral function such as the energy width of the
bosonic spectrum ($0<\omega <\omega _{\max }$) and positions and
distributions of peaks due to bosons. It turns out that even such an
approximate procedure gives valuable information in HTSC cuprates - see
discussion in \textit{Section III D}.

Note that in the case when both EPI and spin-fluctuation interaction (SFI)
are present one should make difference between $\lambda _{\mathbf{kp}%
}^{Z}(i\nu _{n})$ and $\lambda _{\mathbf{kp}}^{\Delta }(i\nu _{n})$ defined
by

\begin{equation}
\lambda _{\mathbf{kp}}^{Z}(i\nu _{n})=\lambda _{sf,\mathbf{kp}}(i\nu
_{n})+\lambda _{ep,\mathbf{kp}}(i\nu _{n}),  \label{lambdaZ}
\end{equation}%
\begin{equation}
\lambda _{\mathbf{kp}}^{\Delta }=\lambda _{ep,\mathbf{kp}}(i\nu
_{n})-\lambda _{sf,\mathbf{kp}}(i\nu _{n}).  \label{lambdaD}
\end{equation}%
In absence of EPI $\lambda _{\mathbf{kp}}^{Z}(i\nu _{n})$ and $\lambda _{%
\mathbf{kp}}^{\Delta }(i\nu _{n})$ differ by sign i.e. $\lambda _{\mathbf{kp}%
}^{Z}(i\nu _{n})=-\lambda _{\mathbf{kp}}^{\Delta }(i\nu _{n})>0$ since SFI
is repulsive interaction in the singlet pairing-channel.

\subsubsection{Inversion of tunnelling data}

Phonon features in the conductance $\sigma _{NS}(V)$ at $eV=\Delta
_{0}+\omega _{ph}$ makes the tunnelling spectroscopy a powerful method in
obtaining the Eliashberg spectral function $\alpha ^{2}F(\omega )$. Two
methods were used in the past for extracting $\alpha ^{2}F(\omega )$.

The \textit{first} \textit{method} is based on solving the
\textit{inverse problem} of the nonlinear Eliashberg equations.
Namely, by measuring $\sigma _{NS}(V)$, one obtains the tunnelling
density of states $N_{T}(\omega )(\sim \sigma _{NS}(\omega ))$ and
by the inversion procedure one gets $\alpha ^{2}F(\omega )$
\citep{McMillanRowell}. In reality the method is based on the
iteration procedure - the McMillan-Rowell ($MR$) inversion, where
in the first step an initial $\alpha ^{2}F_{ini}(\omega )$, $\mu
_{ini}^{\ast }$ and $\Delta _{ini}(\omega )$ are inserted into
Eliashberg equations (for
instance $\Delta _{ini}(\omega )=\Delta _{0}$ for $\omega <\omega _{0}$ and $%
\Delta _{ini}(\omega )=0$ for $\omega >\omega _{0}$) and then $\sigma
_{ini}(V)$ is calculated. In the second step the functional derivative $%
\delta \sigma (\omega )/\delta \alpha ^{2}F(\omega )$ ($\omega \equiv eV$)
is found in the presence of a small change of $\alpha ^{2}F_{ini}(\omega )$
and then the iterated solution $\alpha ^{2}F_{\mathbf{(1)}}(\omega )=\alpha
^{2}F_{ini}(\omega )+$ $\delta \alpha ^{2}F(\omega )$ is obtained, where the
correction $\delta \alpha ^{2}F(\omega )$ is given by
\begin{equation}
\delta \alpha ^{2}F(\omega )=\int d\nu [ \frac{\delta \sigma _{ini}(V)}{%
\delta \alpha ^{2}F(\nu )}]^{-1}[\sigma _{exp}(\nu )-\sigma _{ini}(\nu )].
\label{Inv-dat}
\end{equation}
The procedure is iterated until $\alpha ^{2}F_{(n)}(\omega )$ and $\mu
_{(n)}^{\ast }$ converge to $\alpha ^{2}F(\omega )$ and $\mu ^{\ast }$which
reproduce the experimentally obtained conductance $\sigma _{NS}^{\exp }(V)$.
In such a way the obtained $\alpha ^{2}F(\omega )$ for $Pb$ resembles the
phonon density of states $F_{Pb}(\omega )$, that is obtained from neutron
scattering measurements. Note that the method depends explicitly on $\mu
^{\ast }$ but on the contrary it requires only data on $\sigma _{NS}(V)$ up
to the voltage $V_{\max }=\omega _{ph}^{\max }+\Delta _{0}$ where $\omega
_{ph}^{\max }$ is the maximum phonon energy ($\alpha ^{2}F(\omega )=0$ for $%
\omega >\omega _{ph}^{\max }$) and $\Delta _{0}$ is the zero-temperature
superconducting gap. One pragmatical feature for the interpretation of
tunnelling spectra (and for obtaining the spectral pairing function $\alpha
^{2}F(\omega )$) in $LTS$ and $HTSC$ cuprates is that the negative peaks of $%
d^{2}I/dV^{2}$ are at the peak positions of $\alpha ^{2}F(\omega )$ and $%
F(\omega )$. This feature will be discussed later on in relation with
experimental situation in cuprates.

The \textit{second method }has been invented in \citep{GDSMethod}
and it is based on the combination of the Eliashberg equations and
dispersion relations for the Greens functions - we call it GDS
method. First, the tunnelling density of states is extracted from
the tunnelling conductance in a more rigorous way
\citep{Ivanshenko}
\begin{equation*}
N_{T}(V)=\frac{\sigma _{NS}(V)}{\sigma _{NN}(V)}-\frac{1}{\sigma ^{\ast }(V)}%
\int_{0}^{V}du
\end{equation*}%
\begin{equation}
\times \frac{d\sigma ^{\ast }(u)}{du}\left[ N_{T}(V-u)-N_{T}(V)\right]
\label{NT(V)}
\end{equation}%
where $\sigma ^{\ast }(V)=\exp \{-\beta V\}\sigma _{NN}(V)$ and the constant
$\beta $ are obtained from $\sigma _{NN}(V)$ at large biases - see \citep%
{GDSMethod}. $N_{T}(V)$ under the integral can be replaced by the BCS
density of states. Since the second method is used in extracting $\alpha
^{2}F(\omega )$ in a number of LTSC as well as in HTSC cuprates - see below,
we describe it briefly for the case of isotropic EPI at T=0 K. In that case
the Eliashberg equations are \citep{AllenMitrovic}, \citep{MaksimovEliashberg}%
, \citep{MarsiglioCarbotteBook}, \citep{GDSMethod}
\begin{equation*}
Z(\omega )\Delta (\omega )=\int_{\Delta _{0}}^{\infty }d\omega ^{\prime }Re
\left[ \frac{\Delta (\omega ^{\prime })}{\left[ \omega ^{\prime 2}-\Delta
^{2}(\omega ^{\prime })\right] ^{1/2}}\right]
\end{equation*}%
\begin{equation}
\times \lbrack K_{+}(\omega ,\omega ^{\prime })-\mu ^{\ast }\theta (\omega
_{c}-\omega )]  \label{Gap-iso2}
\end{equation}%
\begin{equation}
1-Z(\omega )=\frac{1}{\omega }\int_{\Delta _{0}}^{\infty }d\omega ^{\prime
}Re\left[ \frac{\omega ^{\prime }}{\left[ \omega ^{\prime 2}-\Delta
^{2}(\omega ^{\prime })\right] ^{1/2}}\right] K_{-}(\omega ,\omega ^{\prime
})  \label{Z-iso2}
\end{equation}%
where
\begin{equation*}
K_{\pm }(\omega ,\omega ^{\prime })=\int_{\Delta _{0}}^{\omega _{ph}^{\max
}}d\nu \alpha ^{2}F(\nu )(\frac{1}{\omega ^{\prime }+\omega +\nu +i0^{+}}
\end{equation*}%
\begin{equation}
\pm \frac{1}{\omega ^{\prime }-\omega +\nu -i0^{+}}).  \label{Kern}
\end{equation}%
Here $\mu ^{\ast }$ is the Coulomb pseudo-potential, the cutoff
$\omega _{c}$ is approximately $(5-10)$ $\omega _{ph}^{\max }$,
$\Delta _{0}=\Delta (\Delta _{0})$ is the energy gap. Now by using
the dispersion relation for the matrix Greens functions
$\hat{G}(\mathbf{k},\omega _{n})$ one obtains \citep{GDSMethod}
\begin{equation}
ImS(\omega )=\frac{2\omega }{\pi }\int_{\Delta _{0}}^{\infty }d\omega
^{\prime }\frac{N_{T}(\omega ^{\prime })-N_{BCS}(\omega ^{\prime })}{\omega
^{2}-\omega ^{\prime 2}}  \label{ImS}
\end{equation}%
where $S(\omega )=\omega /\left[ \omega ^{2}-\Delta ^{2}(\omega )\right]
^{1/2}$. From Eqs. (\ref{Gap-iso2}-\ref{Z-iso2}) one obtains
\begin{equation*}
\int_{0}^{\omega -\Delta _{0}}d\nu \alpha ^{2}F(\omega -\nu )Re\left\{
\Delta (\nu )\left[ \nu ^{2}-\Delta ^{2}(\nu )\right] ^{1/2}\right\}
\end{equation*}%
\begin{equation*}
=\frac{Re\Delta (\omega )}{\omega }\int_{0}^{\omega -\Delta _{0}}d\nu \alpha
^{2}F(\nu )N_{T}(\omega -\nu )+\frac{Im\Delta (\omega )}{\pi }
\end{equation*}%
\begin{equation}
+\frac{Im\Delta (\omega )}{\pi }\int_{0}^{\infty }d\omega ^{\prime
}N_{T}(\omega ^{\prime })\int_{0}^{\omega _{ph}^{\max }}d\nu \frac{2\alpha
^{2}F(\nu )}{(\omega ^{\prime }+\nu )^{2}-\omega ^{2}}.  \label{IntEq}
\end{equation}

Based on Eqs. (\ref{NT(V)}-\ref{IntEq}) one obtains the scheme for
extracting $\alpha ^{2}F(\omega )$%
\begin{equation}
\sigma _{NS}(V),\sigma _{NN}(V)\rightarrow N_{T}(V)  \label{sigma-ns}
\end{equation}

\begin{equation}
\rightarrow ImS(\omega )\rightarrow \Delta (\omega )\rightarrow \alpha
^{2}F(\omega ).  \label{ImS-F}
\end{equation}%
The advantage in this method is that the explicit knowledge of
$\mu ^{\ast }$ is not required \citep{GDSMethod}. However, the
integral equation for $\alpha ^{2}F(\omega )$ is linear Fredholm
equation of the first kind which is ill-defined - see the
discussion in Section II.B.2.

\subsubsection{Phonon effects in $N_{T}(\protect\omega )$}

We briefly discuss the physical origin for the phonon effects in $%
N_{T}(\omega )$ by considering a model with only one peak, at $\omega _{0}$,
in the phonon density of states $F(\omega )$ by assuming for simplicity $\mu
^{\ast }=0$ and neglecting the weak structure in $N_{T}(\omega )$ at $%
n\omega _{0}+\Delta _{0}$, which is due to the nonlinear structure
of the Eliashberg equations \citep{SSW}.
\begin{figure}[!tbp]
\begin{center}
\resizebox{.45 \textwidth}{!} {\includegraphics*[
width=6cm]{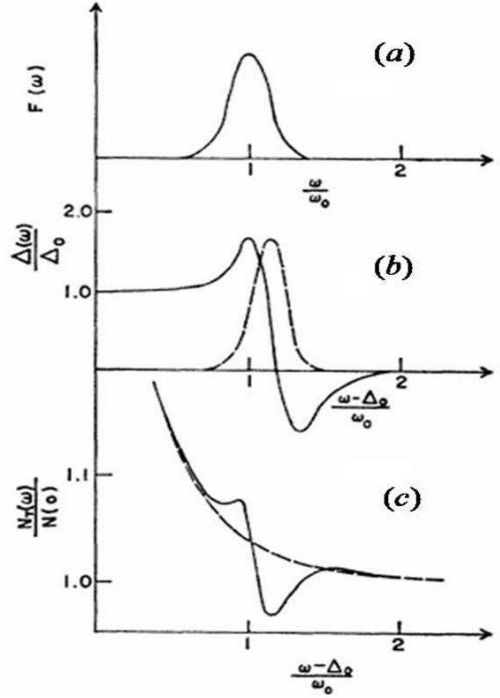}}
\end{center}
\caption{(a) Model phonon density of states $F(\protect\omega )$ with the
peak at $\protect\omega _{0}$. (b) The real (solid) $\Delta _{R}$ and
imaginary (dashed) part $\Delta _{I}$ of the gap $\Delta (\protect\omega )$.
(c) The normalized tunnelling density of states $N_{T}(\protect\omega )/N(0)$
(solid) compared with the BCS density of states (dashed). From \protect\citep%
{SSW}.}
\label{Tunnel-density}
\end{figure}
In Fig.\ref{Tunnel-density} it is seen that the real gap $\Delta _{R}(\omega
)$ reaches a maximum at $\omega _{0}+\Delta _{0}$ then decreases, becomes
negative and zero, while $\Delta _{I}(\omega )$ is peaked slightly beyond $%
\omega _{0}+\Delta _{0}$ that is the consequence of the effective
electron-electron interaction via phonons.

It follows that for $\omega <\omega _{0}+\Delta _{0}$ most phonons have
higher energies than the energy $\omega $ of electronic charge fluctuations
and there is over-screening of this charge by ions giving rise to
attraction. For $\omega \approx \omega _{0}+\Delta _{0}$ charge fluctuations
are in resonance with ion vibrations giving rise to the peak in $\Delta
_{R}(\omega )$. For $\omega _{0}+\Delta _{0}<\omega $ the ions move out of
phase with respect to charge fluctuations giving rise to repulsion and
negative $\Delta _{R}(\omega )$. This is shown in Fig. \ref{Tunnel-density}%
(b). The structure in $\Delta (\omega )$ is reflected on $N_{T}(\omega )$ as
shown in Fig. \ref{Tunnel-density}(c) which can be reconstructed from the
approximate formula for $N_{T}(\omega )$ expanded in powers of $\Delta
/\omega $%
\begin{equation}
\frac{N_{T}(\omega )}{N(0)}\approx 1+\frac{1}{2}\left[ \left( \frac{\Delta
_{R}(\omega )}{\omega }\right) ^{2}-\left( \frac{\Delta _{I}(\omega )}{%
\omega }\right) ^{2}\right] .  \label{NT}
\end{equation}%
As $\Delta _{R}(\omega )$ increases above $\Delta _{0}$ this gives $%
N_{T}(\omega )>N_{BCS}(\omega )$, while for $\omega \gtrsim \omega
_{0}+\Delta _{0}$ the real value $\Delta _{R}(\omega )$ decreases while $%
\Delta _{I}(\omega )$ rises and $N_{T}(\omega )$ decreases giving rise for $%
N_{T}(\omega )<N_{BCS}(\omega )$.

\subsection{Transport spectral function $\protect\alpha _{tr}^{2}F(\protect%
\omega )$}

The spectral function $\alpha _{tr}^{2}F(\omega )$ enters the dynamical
conductivity $\sigma _{ij}(\omega )$ ($i,j=a,b,c$ axis in $HTS$ systems)
which generally speaking is a tensor quantity given by the following formula
\begin{equation*}
\sigma _{ij}(\omega )=-\frac{e^{2}}{\omega }\int \frac{d^{4}q}{(2\pi )^{4}}%
\gamma _{i}(q,k+q)
\end{equation*}%
\begin{equation}
\times G(k+q)\Gamma _{j}(q,k+q)G(q),  \label{Sigma-gen}
\end{equation}%
where $q=(\mathbf{q},\nu )$ and $k=(\mathbf{k}=0,\omega )$ and the bare
current vertex $\gamma _{i}(q,k+q;\mathbf{k}=0)$ is related to the Fermi
velocity $v_{F,i}$, i.e. $\gamma _{i}(q,k+q;\mathbf{k}=0)=v_{F,i}.$ The
vertex function $\Gamma _{j}(q,k+q)$ takes into account the renormalization
due to all scattering processes responsible for finite conductivity \citep%
{SchrieffLTS}. In the following we study only the in-plane conductivity at $%
\mathbf{k}=0$. The latter case is realized due to the long
penetration depth in HTSC cuprates and the skin depth in the
normal state are very large. In the $EPI$ theory, $\Gamma
_{j}(q,k+q)\equiv \Gamma _{j}(\mathbf{q},i\omega _{n},i\omega
_{n}+i\omega _{m})$ is a solution of an approximative integral
equation written in the symbolic form \citep{KMS} $\Gamma
_{j}=v_{j}+V_{eff}GG\Gamma _{j}$. The effective potential
$V_{eff}$ (due to EPI) is given by $V_{eff}=\sum_{\kappa }\mid
g_{\kappa }^{ren}\mid ^{2}D_{\kappa }$, where $D_{\kappa }$ is the
phonon Green's function. In
such a case the Kubo theory predicts $\sigma _{ii}^{intra}(\omega )$ ($%
i=x,y,z$)

\begin{equation*}
\sigma _{ii}(\omega )=\frac{\omega _{p,ii}^{2}}{4i\pi \omega }%
\{\int_{-\omega }^{0}d\nu th(\frac{\omega +\nu }{2T})S^{-1}(\omega ,\nu )
\end{equation*}%
\begin{equation}
+\int_{0}^{\infty }d\nu \lbrack th(\frac{\omega +\nu }{2T})-th(\frac{\nu }{2T%
})]S^{-1}(\omega ,\nu )\},  \label{Sigma-tr}
\end{equation}%
where $S(\omega ,\nu )=\omega +\Sigma _{tr}^{\ast }(\omega +\nu )-\Sigma
_{tr}(\nu )+i\gamma _{tr}^{imp}$, and $\gamma _{tr}^{imp}$ is the impurity
contribution$.$ In the following we omit the tensor index $ii$ in $\sigma
_{ii}(\omega )$. In the presence of several bosonic scattering processes the
transport self-energy $\Sigma _{tr}(\omega )=Re\Sigma _{tr}(\omega
)+iIm\Sigma _{tr}(\omega )$ is given by
\begin{equation}
\Sigma _{tr}(\omega )=-\sum_{l}\int_{0}^{\infty }d\nu \alpha
_{tr,l}^{2}F_{l}(\nu )[K_{1}(\omega ,\nu )+iK_{2}(\omega ,\nu )],
\label{Self-tr}
\end{equation}%
\begin{equation}
K_{1}(\omega ,\nu )=Re[\Psi (\frac{1}{2}+i\frac{\omega +\nu }{2\pi T})-\Psi (%
\frac{1}{2}+i\frac{\omega -\nu }{2\pi T})],  \label{K1}
\end{equation}%
\begin{equation}
K_{2}(\omega ,\nu =\frac{\pi }{2}[2cth(\frac{\nu }{2T})-th(\frac{\omega +\nu
}{2T})+th(\frac{\omega -\nu }{2T})]  \label{K2}
\end{equation}%
Here $\alpha _{tr,l}^{2}F_{l}(\nu )$ is the \textit{transport spectral
function }which measures the strength of the $l$-th (bosonic) scattering
process and $\Psi $ is the di-gamma function. The index $l$ enumerates EPI,
charge and spin-fluctuation scattering processes. Like in the case of EPI,
the transport bosonic spectral function $\alpha _{tr,l}^{2}F(\Omega )$
defined in Eq. (\ref{Elia-tr}) is given explicitly by
\begin{equation*}
\alpha _{tr,l}^{2}F(\omega )=\frac{1}{N^{2}(\mu )}\int \frac{dS_{\mathbf{k}}%
}{v_{F,\mathbf{k}}}\int \frac{dS_{\mathbf{k}^{\prime }}}{v_{F,\mathbf{k}%
^{\prime }}}
\end{equation*}%
\begin{equation}
\times \left[ 1-\frac{v_{F,\mathbf{k}}^{i}v_{F,\mathbf{k}}^{i}}{(v_{F,%
\mathbf{k}}^{i})^{2}}\right] \alpha _{\mathbf{kk}^{\prime },l}^{2}F(\omega ).
\label{Tr-spec-fun}
\end{equation}%
We stress that in the phenomenological SFI theory \citep{Pines} one assumes $%
\alpha _{\mathbf{kk}^{\prime }}^{2}F(\omega )\approx N(\mu )g_{sf}^{2}$Im$%
\chi (\mathbf{k}-\mathbf{p},\omega )$, which, as we have repeated several
times, can be justified only for small $g_{sf}$, i.e. $g_{sf}\ll W_{b}$ (the
band width).

In case of weak coupling ($\lambda <1$), $\sigma (\omega )$ can be written
in the generalized (extended) Drude form as discussed in Section III.B.

\end{document}